\documentclass[a4paper,twocolumn,fleqn,nofootinbib]{revtex4-1}

\usepackage{amsmath}             
\usepackage{txfonts}               
 \usepackage{bm}                  
\usepackage{graphicx}

\usepackage[normalem]{ulem} 
\usepackage{units} 
\usepackage{float} 

\usepackage{color}
\definecolor{dgreen}{rgb}{0.0,0.5,0.0}
\newcommand{\cb}{\color{black} }

\newcommand{\cred}{\color{black} }
\def\nablab{{\bm\nabla}}
\def\figures{.}

\AtBeginDocument{\mathcode`v=\varv} 

\makeatletter
\def\p@subsection{}
\makeatother
\makeatletter
\AtBeginDocument{
\g@addto@macro{\appendix}{}
}
\makeatother

\begin{document}

\title{On the Effect of Beating during Nonlinear Frequency Chirping}


\thanks{This article is based on an invited talk at the 37th JSPF Annual Meeting (2020, Matsuyama, online meeting).}

\author{Andreas~BIERWAGE$^{1}$}
\email[\it Author's electronic address: ]{bierwage.andreas@qst.go.jp}
\author{Roscoe~B.~WHITE$^{2}$}
\author{Vin\'{i}cius~N.~DUARTE$^{2}$}

\affiliation{
  $^{1}$National Institutes for Quantum and Radiological Science and Technology (QST), \\ Rokkasho Fusion Institute, Aomori 039-3212, and Naka Fusion Institute, Ibaraki 311-0193, Japan \\
  $^{2}$Princeton Plasma Physics Laboratory, Princeton University, Princeton, NJ 08543, USA}


\begin{abstract}
Spectroscopic \looseness=-1 analyses of energetic particle (EP) driven bursts of MHD fluctuations in magnetically confined plasmas often exhibit chirps that occur simultaneously in groups of two or more. While the superposition of oscillations at multiple frequencies necessarily causes beating in the signal acquired by a {\it localized} external probe, self-consistent hybrid simulations of chirping EP modes in a JT-60U tokamak plasma have demonstrated the possibility of {\it global beating}, where the mode's electromagnetic field vanishes globally between beats and reappears with opposite phase {\cb [Bierwage {\it et al., Nucl.\ Fusion} {\bf 57} (2017) 016036]}. This implies that there can be a {\it single coherent field mode} that oscillates at multiple frequencies simultaneously when it is resonantly driven by {\it multiple density waves} in EP phase space. Conversely, this means that the EP density waves are mutually coupled and interfere with each other via the jointly driven field, a mechanism ignored in some theories of chirping. In this thesis-style treatise, we study the role of field pulsations in general and beating in particular using the Hamiltonian guiding center orbit-following code {\tt ORBIT} with a reduced wave-particle interaction model in realistic geometry. {\cred {\bf Beating is found to drive the evolution of EP phase space structures.} A key mechanism is the pulsation of effective phase space islands combined with the alternation of their effective O- and X-points due to phase jumps between each beat. Observations:} (1) Beating causes density wave fronts to advance radially in a pulsed manner {\cred and the resulting chirps become staircase-like}. (2) The pulsations facilitate convective transfer of material between neighboring layers of phase space density waves. On the one hand, this may inhibit the early detachment of solitary phase space vortices. On the other hand, it facilitates the accumulation of hole and clump fragments into larger structures. (3) Long-range chirping is observed when massive holes or clumps detach and drift away from the turbulent belt around the seed resonance. It is remarkable that the detached vortices remain robust and, on average, maintain their concentric nested layers while being visibly perturbed by the field's continued beating.
\end{abstract}



\maketitle  

\tableofcontents


\section{Introduction}
\label{sec:intro}

In magnetically confined fusion (MCF) research, bursting and chirping Alfv\'{e}nic fluctuations have attracted much interest since they may affect the confinement of fusion-born alpha particles, which are expected to provide most of the plasma heating in a power plant. Much work has been done on this subject during the last several decades, as can be appreciated from recent reviews and tutorials, such as Refs.~\cite{Chen16, Todo19, Heidbrink20}. However, there is more to be learned about these important and fascinating phenomena, and about the underlying nonlinear self-organization processes that occur when the plasma is driven away from equilibrium under the influence of strong sources, sinks and dissipation. A thorough understanding can help with the construction of reduced models that yield quantitative predictions. {\cb It is possible to adopt analytic methods that have been successfully applied in other research fields, such as particle acceleration and quantum field science, with adaptations made to account for the specific conditions of MCF plasmas, in particular their strong magnetization and field geometry \cite{Zonca15b, Chen16} (more details will follow shortly).}

\begin{figure*}[tb]
  \centering
  \includegraphics[width=13.5cm,clip]{\figures/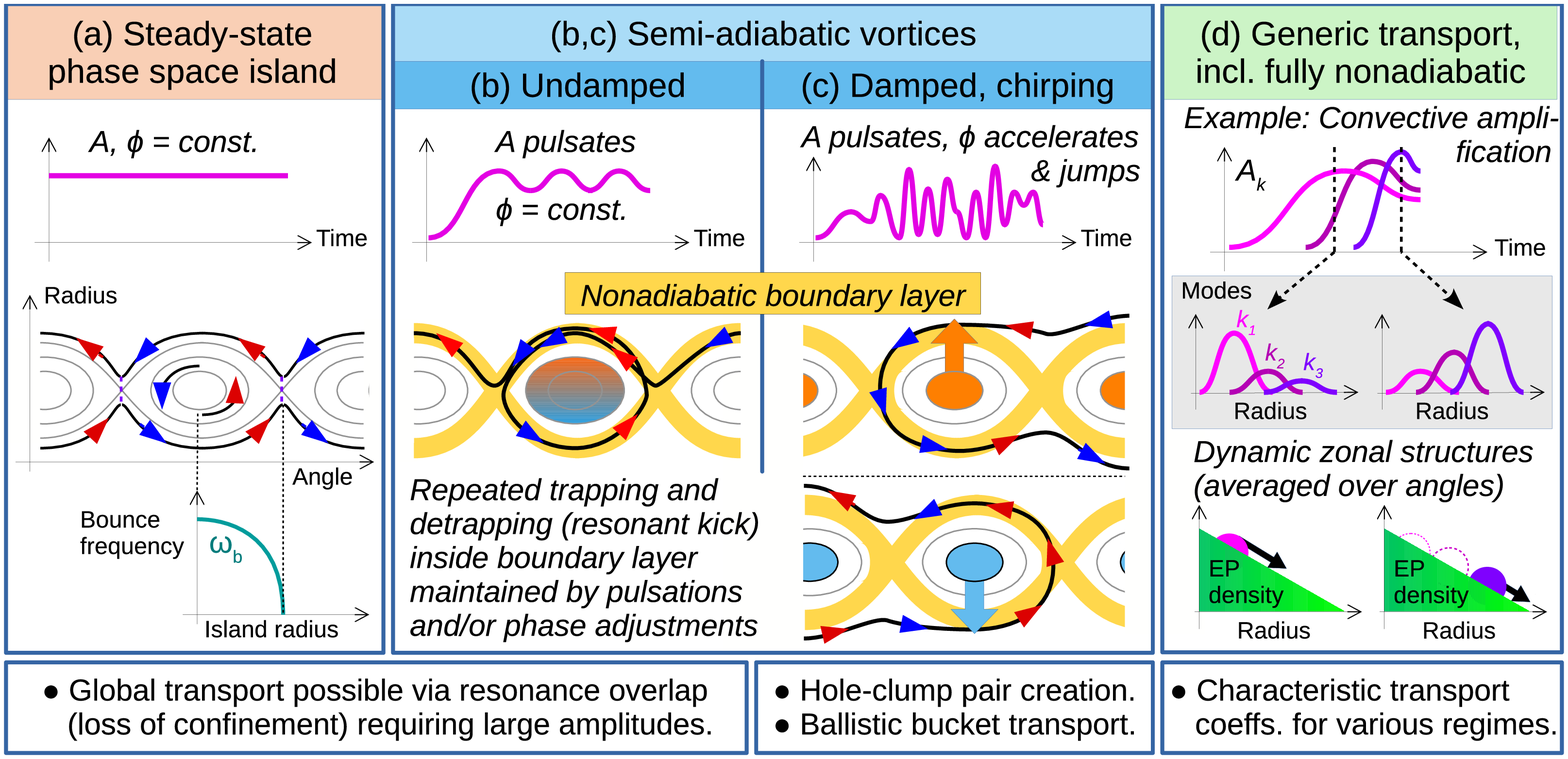}
  \caption{Schematic illustration of processes in resonantly interacting systems; here, energetic particles (EP) with shear Alfv\'{e}n waves. The upper diagrams show time traces of the amplitude $A(t)$ of a symmetry-breaking field perturbation, whose time-dependence has the form $A(t)\sin(-\omega_0 t - \phi(t))$, with initial frequency $\omega_0$ and dynamic phase $\phi(t)$. In columns (a)--(c), only a single mode is considered, giving rise to resonant phase space islands whose internal contours are drawn as gray ellipses. The background EP density decreases with increasing radius. Red and blue arrows indicate uphill and downhill convection, respectively. Here, ``radius'' is a proxy for a generalized action variable in suitable action-angle coordinates. Column (a) illustrates a steady state; i.e., a mode with fixed frequency $A$ and phase $\phi$. The closed separatrix of the perfectly adiabatic steady state (a) becomes a nonadiabatic boundary layer (orange band) in the semi-adiabatic cases (b) and (c) due to a pulsating amplitude $A(t)$ and/or an accelerating phase $\phi(t)$ (i.e., chirping). Faster phase adjustments and larger fluctuation amplitudes yield a wider boundary layer around the adiabatic core, until the entire system becomes nonadiabatic. Column (d) shows an example of convective amplification, where three radially adjacent modes labeled $k=1,2,3$ successively rise to large amplitudes and ``relay'' a resonant particle bunch radially outward \protect\cite{ZoncaTCM99}.}
  \label{fig:intro_x-point}
\end{figure*}

A key issue in many systems is whether or not there exists a separation of time scales between the dynamics of the field (growth, saturation, pulsations, chirping) and the time scales of the dynamics of collective structures in the particle distribution, because this determines whether adiabatic invariants can utilized. In typical cases of practical interest, time scale separation may be vague and exist only in a portion of the system. The problem can be illustrated using the cartoon in Fig.~\ref{fig:intro_x-point}(a), which shows an island in action-angle phase space, with ``radius'' representing the action variable. {\cred The island and surrounding contours appear when taking Poincar\'{e} maps of the perturbed orbits of particles (or guiding centers) in the frame of reference moving with the field wave. Poincar\'{e} maps of particles}\footnote{\cred Hereafter, when we speak loosely of ``particle motion in/around a phase space island'', we mean the motion of the particle's Poincar\'{e} map.}
that are trapped inside the effective potential well of the field wave circulate around the island center with a characteristic bounce frequency $\omega_{\rm b}$, which peaks in the center of the island and drops to zero at the separatrix between the trapped and untrapped domains \cite{Meng18}. No matter how slowly the amplitude $A(t)$ and phase $\phi(t)$ of the field vary, there will always be a boundary layer, where the value of $\omega_{\rm b}$ is smaller than the rate at which $A(t)$ or $\phi(t)$ evolve.

The nonadiabatic boundary layer provides a path for particles outside the remaining island to pass from one side of the effective resonance to the other. In the case of frequency chirping at constant amplitude, a well-defined separatrix still exists in suitable coordinates, but its domain is reduced so that it no longer isolates the regions above and below the resonance \cite{Hsu94}. As illustrated in Fig.~\ref{fig:intro_x-point}(c), this allows the remaining adiabatic island core to propagate through the surrounding phase space fluid.

The situation becomes more complicated when the field amplitude fluctuates. For instance, consider the process of saturation of an undamped resonant instability as shown in Fig.~\ref{fig:intro_x-point}(b). Deriving an analytical formula that predicts the saturation level of such an instability turned out to be a nontrivial task. It was solved by Dewar \cite{Dewar73}, who obtained a fairly accurate estimate by examining two idealized extremes as lower and upper limits. The problem that Dewar encountered was that the system is actually fully nonadiabatic during the saturation process.

The same is true for the spontaneous onset of nonlinear frequency chirping as observed in simulations of instabilities that contain a {\it coherent} energy loss channel. The energy loss may occur in the field (damping) or in the particles (drag), its role being to permit phase slippage between field waves and density waves, as we will discuss in more detail later. An analytic formula for the nonadiabatic onset of chirping {\cred in tokamak geometry} has been derived by Zonca \& Chen \cite{ZoncaTCM99} based on the concept of a convective instability as illustrated in Fig.~\ref{fig:intro_x-point}(d). The chirping rate is predicted to follow the curve of a hyperbolic tangent; i.e., exponential acceleration followed by exponential deceleration. We will confirm that this prediction compares well with simulations during the stage of spontaneous formation and separation of a {\it nonadiabatic} hole-clump wave pair.

Rapid short-lived chirps that last only a millisecond or less are often observed experimentally in well-resolved spectrograms {\cred (e.g., Fig.~6 of Ref.~\cite{Heidbrink21}). These chirps may} remain fully nonadiabatic until their end, but they may also spawn vortices with an adiabatic core. Sufficiently massive vortices may gain a large degree of control over the field and become self-sustained for several milliseconds. Their propagation can be observed as long-range chirps and they may cause ballistic long-range transport. We will analyze some concrete examples where semi-adiabatic vortices form after a few milliseconds of total nonadiabaticity.

Berk, Breizman and Petviashvili \cite{Berk97} have derived a widely used formula that predicts the rate of vortex-mediated chirps in regimes near marginal stability; i.e., when strong damping is overcome by a slightly stronger drive. The predicted overall range and rate of chirping often lies in the right ``ball park'', even in cases with fairly rapid chirps and even relatively far from marginality. However, the actual form of long-range chirps in a realistic setting often deviates from the predicted square-root time dependence. This is to be expected because the assumptions made are, at best, satisfied only at intermediate times, after the nonadiabatic layer has narrowed and before background nonuniformities and collisions become important.

A rigorous treatment of the nonadiabatic processes remains challenging and usually requires numerical simulations, but analytical theory can help to gain insight and minimize the computational effort by reducing the Maxwell-Vlasov system. A comprehensive theoretical framework has been formulated by Chen \& Zonca \cite{Chen16} on the basis of concepts from quantum field theory, invoking a Schr\"{o}dinger-like equation for the intensity evolution (bursts or solitons) and a Dyson-like equation for (driven or spontaneous) emission of (zonal or vortical) phase space structures that break the symmetry of the reference state \cite{Zonca15b}. For instance, this theory captures nonadiabatic avalanches in a process called convective amplification, as illustrated in Fig.~\ref{fig:intro_x-point}(d). The Schr\"{o}dinger-Dyson system of equations is less complex than the original gyrokinetic Maxwell-Vlasov system, but handy formulas for the quantitative prediction of chirping dynamics are not yet easy to come by, except in certain limits, such as short-wavelength modes interacting with magnetically deeply trapped particles \cite{ZoncaTCM99}, or in certain space plasmas \cite{ZoncaChorus}.

It is particularly difficult to reduce analytically the dynamics of {\it long-wavelength modes} in a MCF plasma, because the plasma profiles and the global geometry of the field can have a significant influence on the dynamics. In this case, it can be difficult to justify the assumption of the existence of two disparate spatial scales that underlies some of the reductions used by Chen \& Zonca. However, the long-wavelength regime is of practical interest, because such modes are common in present-day MCF experiments and the associated transport tends to be global.

The present work was motivated by a study of long-wavelength chirping modes with toroidal mode numbers $n=1,2,3$ driven by energetic particles (EP) in JT-60U tokamak plasmas \cite{Bierwage17a}, where experimental and numerical evidence for beating in the field signal due to interference between multiple chirps was reported. Using the {\tt ORBIT} code with a reduced model for the fluctuating field \cite{White20}, we investigate here the physical implications of beating. We examine the wave-particle interactions in realistic tokamak geometry, but use a somewhat simpler scenario than the original JT-60U case that motivated this work, where the primary resonances were located near the magnetic axis, touching the domain of stagnation orbits. Here, we avoid such boundary effects by working with a resonance located at mid-radius in the domain of circulating particles. Our plasma has the dimensions of the conceptual fusion reactor FIRE, but a weaker magnetic field with a flux density of $B_0 = 0.49\,{\rm T}$ at the axis located at $R_0 = 2.15\,{\rm m}$.

The effect of field pulsations in general and beating in particular is relevant for several aspects of nonlinear frequency chirping, {\cred including} the nonadiabatic onset of chirping, hole-clump pair formation, and the applicability of reduced transport models, such as the so-called bucket transport \cite{Hsu94} and waterbag \cite{Berk67, Khain07, Hezaveh21}. In Section~\ref{sec:review}, we introduce essential physical concepts, review methods used to characterize chirping systems, and discuss open questions that may be addressed by elucidating the effects of field pulsations and beating. In Section~\ref{sec:model}, we describe the physical model and numerical methods used in this study. The processes of phase space structure formation, nonlinear saturation\footnote{``Saturation'' is used here as a generic term, referring to a stagnation in the growth of the field amplitude, such as a (quasi-)steady state due to a permanent exhaustion of resonant drive, or temporary peaking due to a transient dynamic balance between resonant drive and damping coming from different regions of phase space that we integrate over in Eq.~(\protect\ref{eq:mdl}).}
and the onset of beating are revisited in Section~\ref{sec:saturation}. The effects of beating during the nonadiabatic onset of chirping, the formation and propagation of massive vortex structures and the particle transport they cause are analyzed in Section~\ref{sec:beating}.

This study makes extensive use of comparisons between cases near and far from marginal stability. Comparisons are also made with simulations where no sustained chirping occurs due to the absence of field damping. The contrast between these examples will help us to highlight the trends. By presenting known and less known effects in a unified systematic treatise, we hope to contribute to a better understanding of nonlinear frequency chirping.

Sections~\ref{sec:review}, \ref{sec:saturation} and \ref{sec:beating} can be regarded as three shorter articles that have been compiled here into a unified thesis-style treatise. We conclude with a summary, discussion and outlook in Section~\ref{sec:summary}. The comprehensive Appendices contain a mathematical theory of two-wave beating, additional simulation results, and a thorough characterization of the model and methods used, including tests for numerical convergence and sensitivity. At the end of the paper, there are links to animated movies for several simulations.

\section{Physical concepts \& open questions}
\label{sec:review}

In this section, we establish the conceptual framework that we will use in our discussion and interpretation of the simulation results. The rationales behind the chosen physical pictures are explained and useful quantities characterizing chirping systems are defined. Finally, we discuss why we expect that studying the effects of beating may improve our understanding of chirping systems. Before we {\cred discuss} the theory, let us show a concrete example of nonlinear chirps as seen in our simulations.

\subsection{Phenomenology of a chirping system}
\label{sec:review_gen}

\begin{figure}[tb]
  \centering
  \includegraphics[width=8cm,clip]{\figures/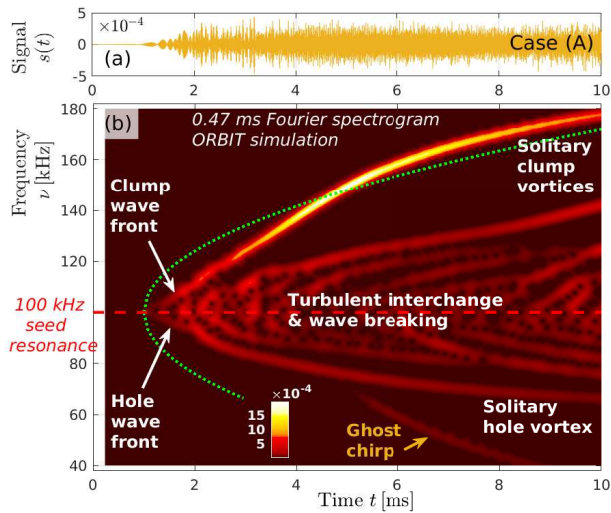}
  \caption{Evolution of ideal MHD fluctuations in one of our numerical simulations of nonlinear frequency chirping using the {\tt ORBIT} code \protect\cite{White84,ChenY99,White20} that we analyze in detail in this paper. Panel (a) shows the raw signal $s(t)$ of Eq.~(\protect\ref{eq:intro_xi}) in code units, and panel (b) shows the Fourier spectrogram obtained with a sliding time window of size $0.47\,{\rm ms}$. This case is marginally unstable and develops semi-adiabatic vortices. The dotted parabola is the theoretically predicted chirp front assuming marginality, adiabaticity, and uniform background \protect\cite{Berk97}. The ``ghost chirp'' seems to be an artifact and is discussed in Appendix~\protect\ref{apdx:misc_ghost}.}
  \label{fig:intro_fire-A_chirp}
\end{figure}

Figure~\ref{fig:intro_fire-A_chirp} shows results from one of the simulations that we will analyze in detail in this paper. {\cred Appendix~\ref{apdx:dg_movie} contains a link to the movie file.} The simulations use a reduced model of EP-driven ideal magnetohydrodynamic (MHD) fluctuations in the frequency band of shear Alfv\'{e}n waves in a tokamak plasma. {\cred Our EPs are fast deuterons with kinetic energies around $80\,{\rm keV}$. Figure~\ref{fig:intro_fire-A_chirp}(a) shows} the raw signal
\begin{gather}
s(t) = A(t)\sin(-\omega_0 t - \phi(t)),
\label{eq:intro_xi}
\end{gather}

\noindent which represents ideal MHD displacements $\xi$. In the present case, the peak values of this signal fluctuate at levels below $10^{-3}$, which corresponds to radial displacements of resonantly trapped particles by $\delta r_{\rm res}/R_0 \lesssim 6\times 10^{-3}$ and ideal MHD bulk displacements $\delta r_{\rm mhd}/R_0 \lesssim 7\times 10^{-4}$ (cf.\ Appendix~\ref{apdx:model_calibration}). Panel (b) shows the corresponding Fourier spectrogram obtained with a sliding time window of size $\Delta t_{\rm win} = 0.47\,{\rm ms}$. The chirps in Fig.~\ref{fig:intro_fire-A_chirp}(b) proceed at rates on the order of $\ddot\phi = \delta\dot\nu \lesssim 10\,{\rm kHz}/1\,{\rm ms}$ ($\dot\phi \equiv {\rm d}\phi/{\rm d}t$) around the seed resonance with initial frequency $\nu_0 = \omega_0/(2\pi) = 100\,{\rm kHz}$. In the present example, the chirping dynamics enhance the displacement of resonant and near-resonant particles by an order of magnitude to $\delta r_{\rm chirp}/R_0 \approx 4\times 10^{-2}$, which illustrates why this phenomenon is of practical interest.

The example in Fig.~\ref{fig:intro_fire-A_chirp} was simulated in the perfectly collisionless limit and without external EP sources, so the chirps tend to continue and dominate the signal for a long time, whereas the original mode is not revived. In reality, rapid nonlinear chirps may be visible for less than a millisecond, and even long-range nonlinear chirps usually last no longer than $10\,{\rm ms}$, so this phenomenon can be taken to be insensitive to changes in the background plasma conditions (magnetic geometry, kinetic profiles, rotation) on global scales, which typically evolves on the time scale of $100\,{\rm ms}$ or more under the influence of heating, fueling, collisional diffusion and turbulent transport.

\begin{figure}[tb]
  \centering
  \includegraphics[width=8cm,clip]{\figures/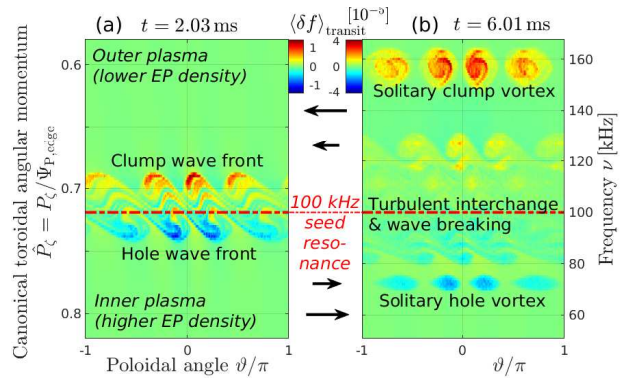}
  \caption{Two snapshots of EP phase space density waves around a seed resonance with $p=4$ poloidal elliptic points found in the case of Fig.~\protect\ref{fig:intro_fire-A_chirp}. The structures are visualized using $\delta f$-weighted kinetic Poincar\'{e} plots, with shades of blue representing decreased density ($\delta f < 0$, ``hole'') and red increased density ($\delta f > 0$, ``clump''). The Poincar\'{e} maps are accumulated during a short interval of one toroidal transit, so, strictly speaking, we plot a time-integrated phase space density perturbation $\left<\delta f\right>_{\rm transit}$. The canonical toroidal angular momentum $P_\zeta$ serves as a radial coordinate, here normalized by the poloidal magnetic flux at the plasma edge $\Psi_{\rm P,edge}$. The initial EP density (not shown) decreases from the bottom (inner plasma) to the top (outer plasma). In the rotating frame of reference chosen here, fluctuations at the seed resonance ($\hat{P}_{\zeta,{\rm res}} = 0.719$) appear stationary, and the poloidal phase velocities increase away from the resonance as indicated by the horizontal arrows. Using the calibration in Eq.~(\protect\ref{eq:pz_nu}), the vertical axis of panel (b) has been converted to frequency $\nu$ for easier comparability with the spectrogram in Fig.~\protect\ref{fig:intro_fire-A_chirp}(b).}
  \label{fig:intro_fire-A_df}
\end{figure}

The $\delta f$-weighted kinetic Poincar\'{e} plots in Fig.~\ref{fig:intro_fire-A_df} are a way to visualize the structure and motion of the incompressible Vlasov fluid that makes up the EP phase space. Density waves in this incompressible fluid are realized by the convective interchange of denser and lighter fluid elements that exist due to an initially imposed density gradient. Here, the gradient is along the canonical toroidal angular momentum $P_\zeta$, which represents the radial direction along the vertical axis in Fig.~\ref{fig:intro_fire-A_df}(a). EP phase space structures causing upward chirps travel upward in Fig.~\ref{fig:intro_fire-A_df} and {\it vice versa}. The vertical axis of Fig.~\ref{fig:intro_fire-A_df}(b) has been converted to frequency units for easier comparability with the spectrogram in Fig.~\ref{fig:intro_fire-A_chirp}(b).

The Poincar\'{e} section in Fig.~\ref{fig:intro_fire-A_df} moves with the phase velocity of the seed wave along the toroidal angle $\zeta$, so $\delta f$ structures at the seed resonance appear stationary. Shades of blue indicate a reduction in EP phase space density ($\delta f < 0$, ``hole'') where lighter fluid has moved into regions previously occupied by denser fluid, and shades of red indicate increased density ($\delta f > 0$, ``clump''). In the present scenario, four cross-sections of the same phase space structure appear at different poloidal angles $\vartheta$ since the underlying resonance has $p=4$ elliptic points along $\vartheta$. Note that we use the terms ``hole'' and ``clump'' in a more general manner than some readers may be used to. Namely, we use them not only to refer to solitary vortices but also for short-lived convective perturbations in EP density.

Some notable dynamic structures seen in the system studied here have been labeled in Figs.~\ref{fig:intro_fire-A_chirp} and \ref{fig:intro_fire-A_df}. Chirping starts with the radial propagation of convective EP density ``wave fronts''. In the wake of these wave fronts, a turbulent belt forms, which consists of more or less stratified layers of differentially rotating hole and clump waves. The interactions between these layers are reminiscent of sheared ``convective interchange'' and ``wave breaking'': wave tops rise, overtake wave bottoms, then collapse back down. Sometimes, we observe the complete detachment of a massive hole or clump from the turbulent belt. Such structures can be seen in the upper and lower parts of Fig.~\ref{fig:intro_fire-A_df}(b) and are referred to here as ``solitary vortices''. The massive solitary clump vortex, whose interior looks like a spiral galaxy, is responsible for the long-range upward chirp in Fig.~\ref{fig:intro_fire-A_chirp}(b) that spurs away from the turbulent domain up to $180\,{\rm kHz}$.

The propagating phase space structures and associated chirps may be affected by but are not necessarily constrained to the linear MHD response spectra of the background plasma, which (in the limit $t \rightarrow \infty$) consist of singular continua or discrete global eigenmodes. In fact, the MHD spectra are entirely ignored in the reduced model that we employ here --- boldly assuming that the plasma responds equally well at any frequency, being subject only to an {\it ad hoc} damping rate $\gamma_{\rm d}$ that is spatially uniform, independent of frequency and constant in time. Our reduced model captures the essential physical mechanisms for chirping with full complexity in the particle dynamics and reduced complexity in the field dynamics, which enhances physical transparency and computational efficiency.

\subsection{$\delta f$ structures = ensemble of pump waves}
\label{sec:review_pump}

In the previous section, we have, in a casual way, related the coherent structures appearing in Poincar\'{e} plots of $\delta f$ in Fig.~\ref{fig:intro_fire-A_df} to the chirps seen in the spectrogram in Fig.~\ref{fig:intro_fire-A_chirp}(b). Let us review the underlying theoretical foundation.

When applying a symmetry-breaking electromagnetic perturbation that displaces our energetic charged particles in the direction of a preexisting gradient in their initial guiding center distribution $F_0 = F(t=0)$, one obtains a perturbed distribution $F(t) = F_0 + \delta f(t)$, whose fluctuating part $\delta f(t)$ consists of wave-like structures. Short-wavelength portions of $\delta f$ that  cease to contribute to the dynamics of interest act like a zonal (angle-independent) structure, which we denote by $\delta\bar f$ and interpret as a modification of the axisymmetric reference state $F_0$. During the ensuing mutual interactions between the phase space density waves and the electromagnetic field, remaining or newly formed gradients in $\bar{F}(t) = F_0 + \delta\bar{f}(t)$ may facilitate the excitation of additional waves or continue to feed existing ones. Although conventionally referred to as ``wave-particle interactions'' --- which is, of course, correct from the energetic point of view --- we interpret these dynamics in terms of interactions between two types of waves:
\begin{itemize}
\item electromagnetic {\it field waves}, here represented by ${\bm E}_\perp$, and
\item collective {\it density waves} represented by $\delta f(t)$.
\end{itemize}

\noindent Speaking of density waves instead of particles is not just a semantic matter. It is motivated by Maxwell's equations, which state that the field responds to collective structures in $\delta f$, whose group velocities generally differ from particle velocities. Let us elucidate this point within the scope of the reduced field model that we will be using in this work.

We express the combined time-dependence of the ideal MHD field waves as $E_\perp(t) \propto A(t)\cos(-\omega_0 t - \phi(t))$, with an amplitude factor $A(t)$ and nonlinear phase shift $\phi(t)$ relative to a reference wave with frequency $\omega_0$.\footnote{$A(t)\cos(\Theta(t))$ corresponds to the real part of the analytic signal $\tilde{s}(t) = s(t) + i\hat{s}(t) = A(t)e^{i\Theta(t)}$, where $\hat{s}(t)$ is the Hilbert transform of $s(t)$.} Our reduced equations governing the field response have the form
\begin{subequations}
\begin{align}\vspace{-0.0cm}
\frac{{\rm d}A}{{\rm d}t} &= -\Omega_{\rm c}\frac{\nu_{\rm A0}^2}{\omega_0^2}\int{\cred {\rm d}^5 Z}_{\rm gc} \delta f \frac{{\bm v}_{\rm gc}\cdot{\bm E}_\perp}{A} - \gamma_{\rm d}A,
\label{eq:dadt}
\\
A\frac{{\rm d}\phi}{{\rm d}t} &= -\Omega_{\rm c}\frac{\nu_{\rm A0}^2}{\omega_0^3}\int{\cred {\rm d}^5 Z}_{\rm gc} \delta f \frac{{\bm v}_{\rm gc}\cdot\partial_t{\bm E}_\perp}{A};
\label{eq:dpdt}
\end{align}
\label{eq:mdl}\vspace{-0.2cm}
\end{subequations}

\noindent (see Section 6.9 of Ref.~\cite{WhiteTokBook3}) where $\nu_{\rm A0} = v_{\rm A0}/R_0$ is the on-axis Alfv\'{e}n frequency, $\Omega_{\rm c}$ the cyclotron frequency, ${\cred {\rm d}^5 Z}_{\rm gc}$ is a volume element in guiding center (GC) phase space, ${\bm v}_{\rm gc}$ is the GC velocity, and $\partial_t \equiv \partial/\partial t$ in Eq.~(\ref{eq:dpdt}) acts only on the ``fast'' oscillations $\omega_0$. All damping mechanisms unrelated to the particle distribution $F_0$ are captured by $\gamma_{\rm d}$, which we take to be constant. The Jacobian has been absorbed in the distribution function $F$, so $\int{\cred {\rm d}^5 Z}_{\rm gc} F = N$ is the number of particles (whose conservation implies $\int{\cred {\rm d}^5 Z}_{\rm gc} \delta f = 0$). The set of GC coordinates ${\bm Z}_{\rm gc} = (P_\zeta,K,\mu,\vartheta,\zeta)$ consists of the canonical toroidal angular momentum $P_\zeta$, kinetic energy $K$, the magnetic moment $\mu$, and poloidal and toroidal angles $\vartheta$ and $\zeta$. For particles with mass $M$, electric charge ${\cred Q}e$ and velocity $v$, in a toroidally axisymmetric magnetic field ${\bm B} = \nablab\zeta\times\nablab\Psi_{\rm P} + B_\zeta\nablab\zeta$, these GC coordinates have the form
\begin{equation}
P_\zeta = -\Psi_{\rm p} + B_\zeta \frac{v_\parallel}{\cred \Omega_{\rm c}}, \quad
\mu = \frac{M v_\perp^2}{2 B}, \quad
K = \frac{M v^2}{2},
\end{equation}

\noindent where $2\pi\Psi_{\rm P}$ is the poloidal magnetic flux, $v_\parallel$ is the GC velocity parallel to the magnetic field, $v_\perp^2 \equiv v^2 - v_\parallel^2$, {\cred and $\Omega_{\rm c} = QeB/M$ with $e$ the electric charge of a positron.}

Note that the phase space density gradients $\partial_{{\bm Z}_{\rm gc}} F$ do not appear in the field equation (\ref{eq:mdl}) explicitly; only the density perturbation $\delta f({\bm Z}_{\rm gc})$ appears. The Vlasov equation
\begin{align}
\partial_t\delta f =& -\dot{\bm Z}_{\rm gc}\cdot\partial_{{\bm Z}_{\rm gc}}F \nonumber
\\
=& {\cred\, -\underbrace{(\dot\vartheta\partial_\vartheta\delta f + \dot\zeta\partial_\zeta\delta f)}\limits_{\text{free streaming}} - \underbrace{(\dot{P}_\zeta\partial_{P_\zeta}F + \dot{K}\partial_K F)}\limits_{\text{acceleration}} }
\label{eq:vlasov}
\end{align}

\noindent states that the {\it instantaneous rate of change} of $\delta f$ is determined by {\cred phase space flows $\dot{\bm Z}_{\rm gc} \equiv {\rm d}{\bm Z}_{\rm gc}/{\rm d}t$ across phase space density gradients $\partial_{{\bm Z}_{\rm gc}} F$.} Meanwhile, the {\it instantaneous structure} of $\delta f$ reflects the {\it time history} of these {\cred dynamics} (until erased by external sources, sinks and collisions). Thus, like the fields in Maxwell's equations, our field governed by Eq.~(\ref{eq:mdl}) sees only the result of that history in the form of {\cred an evolving density perturbation $\delta f(t)$}.

For instance, let us assume that the zonal gradients have been flattened, so that $\bar{F} \approx {\rm const}$., and what remains is a single density wave with frequency $\omega_{\rm pump}$. The absence of zonal gradients means that there is no net guiding center current $\delta f{\bm v}_{\rm gc}$ along the transverse electric field ${\bm E}_\perp$, so the total energy transfer $\int{\cred {\rm d}^5 Z}_{\rm gc}\delta f{\bm v}_{\rm gc}\cdot{\bm E}_\perp$ vanishes, because there is no ``DC'' component and the ``AC'' components $\delta f \propto \sin(-\omega_{\rm pump}t)$ and $E_\perp \propto \cos(-\omega_0t - \phi)$ are out of phase. Ignoring damping, Eq.~(\ref{eq:dadt}) becomes $\dot{A} = 0$. Meanwhile, the time derivative on the right-hand side of the phase equation (\ref{eq:dpdt}) aligns the phases of density and field fluctuations. Using the fact that the integral over $\partial_t(\delta f{\bm v}_{\rm gc}\cdot{\bm E}_\perp)$ vanishes, we can let $\delta f{\bm v}_{\rm gc}\cdot\partial_t{\bm E}_\perp \rightarrow \frac{1}{2}(\delta f {\bm v}_{\rm gc}\cdot\partial_t{\bm E}_\perp - {\bm v}_{\rm gc}\cdot{\bm E}_\perp \partial_t\delta f)$, so that Eq.~(\ref{eq:dpdt}) becomes $\dot\phi = \omega_{\rm pump} - \omega_0$. This means that the field wave responds with oscillations $E_\perp \propto \cos(-\omega_{\rm pump}t)$ of the same frequency as the density wave, which acts as a {\it pump wave} {\cred as it propagates subject to the free streaming terms in Eq.~(\ref{eq:vlasov})}. In the presence of zonal gradients, the pump waves propagate up or down those gradients and grow or decay in magnitude.

Therefore, Eq.~(\ref{eq:mdl}) states that the field responds to the pumping action exerted by an ensemble of density waves, so it behaves as a driven oscillator; albeit an active one, that is capable of shaping the $\delta f$ landscape.\footnote{Parallels with the concept of parametric instabilities are evident.}
One may say that the integral on the right-hand side of Eq.~(\ref{eq:dadt}) measures the {\it correlation} between the structure of the density waves' current density ${\cred Q}e\delta f{\bm v}_{\rm gc}$ and the field wave ${\bm E}_\perp$, so the response is stronger the better the phases match. Conversely, the coherence of density waves is maintained by resonant interactions with the field, where ``resonance'' means that phase-matching is maintained over many field oscillation cycles. Thus, the effect of resonance emerges on intermediate time scales situated between rapid field oscillations and transport, allowing coherent $\delta f$ structures to form and to remain robust.

The total field signal $E_\perp(t)$ cannot be phase-locked to multiple pump waves simultaneously, and as the relative phases shift, the sign of the phase correlation integral in Eq.~(\ref{eq:dadt}) can change.\footnote{\label{fn:grad} For instance, even when the density waves $\delta f$ are still growing, the field may saturate or even vanish if the phase relation is such that the contributions of pump waves with positive and negative $\delta f$ cancel in the correlation integral of Eq.~(\protect\ref{eq:dadt}). This illustrates the difference between the linear regime --- where phases are spontaneously aligned such that $\partial_t\ln A = \partial_t\ln\delta f$ regardless of amplitude and position --- and the nonlinear regime, where the phase relations can become arbitrarily complicated and one has to consider the {\cred flows $\dot{\bm Z}_{\rm gc}$ of GC phase space fluid across phase space density gradients $\partial_{{\bm Z}_{\rm gc}}F$.} The local flow contours, in turn, are determined by the field's instantaneous phase and amplitude.}
The beating phenomenon, whose effect we study in this work, can be interpreted as the temporal interference pattern of two or more pump waves in phase space density, which propagate with different phase velocities around the plasma torus. The math for the special case of two-wave beating is presented in Appendix~\ref{apdx:beat}.

{\cred Variations in the pump wave distribution are observable in spectrograms of the field signal $s(t)$. The spectral} patterns one sees depend on the time window used for the spectral analysis. Short time windows can reveal details about rapid phase adjustments performed by the field, {\cred but may not reflect the existence of multiple pump waves, only their combined instantaneous action. Long time windows show only the overall trend, but allow to discern a larger number of spectral harmonics that can then be attributed to multiple pump waves as in Figs.~\ref{fig:intro_fire-A_chirp} and \ref{fig:intro_fire-A_df}. We will analyze dynamics in our simulations on short and long time scales.

When the EP density waves propagate radially in a tokamak plasma, their transit frequencies change, giving rise to the chirps in the spectrogram of the field waves they drive. We will consider scenarios with a destabilizing ``radial'' gradient $\partial_{P_\zeta} \bar{F} > 0$ and will observe ``clump waves'' ($\delta f > 0$) that tend to propagate downhill and ``hole waves'' ($\delta f < 0$) that tend to propagate uphill. The radial propagation of these phase space structures is facilitated by field damping $\gamma_{\rm d}$ and the reason for this is considered next.}

\begin{figure}[tb]
  \centering
  \includegraphics[width=8cm,clip]{\figures/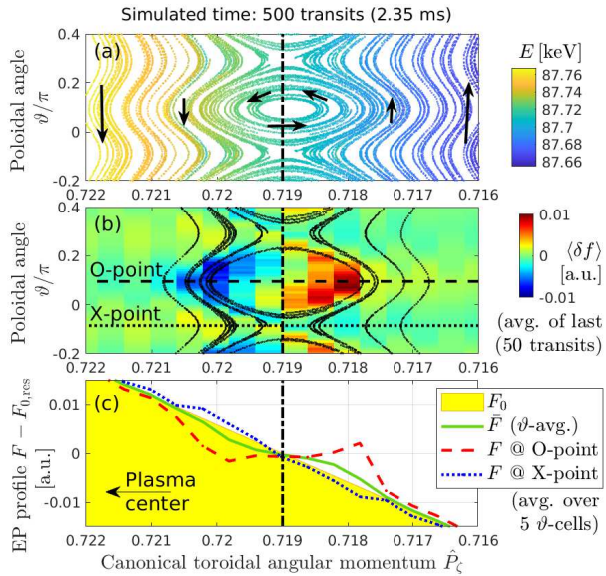}
  \caption{Perturbation of the EP trajectories and EP density in the presence of an imposed oscillating field with fixed amplitude and phase velocity. The kinetic Poincar\'{e} plot in panel (a) shows a phase space island around the resonance at $\hat{P}_{\zeta,{\rm res}} = 0.719$. The colors in (a) represent the instantaneous total (kinetic + potential) energy $E$ of a particle. Panel (b) shows the relaxed phase space density perturbation $\left<\delta f\right>$, averaged over the last 50 transits. Panel (c) shows the radial profile of the axisymmetric EP reference distribution $F_0$ (shaded triangle) and three views of the perturbed EP distribution $F = F_0 + \left<\delta f\right>$: poloidally averaged, O-point profile and X-point profile.}
  \label{fig:intro_fire-A-freeze500_poink-df}
\end{figure}

\begin{figure*}[tb]
  \centering
  \includegraphics[width=14.3cm,clip]{\figures/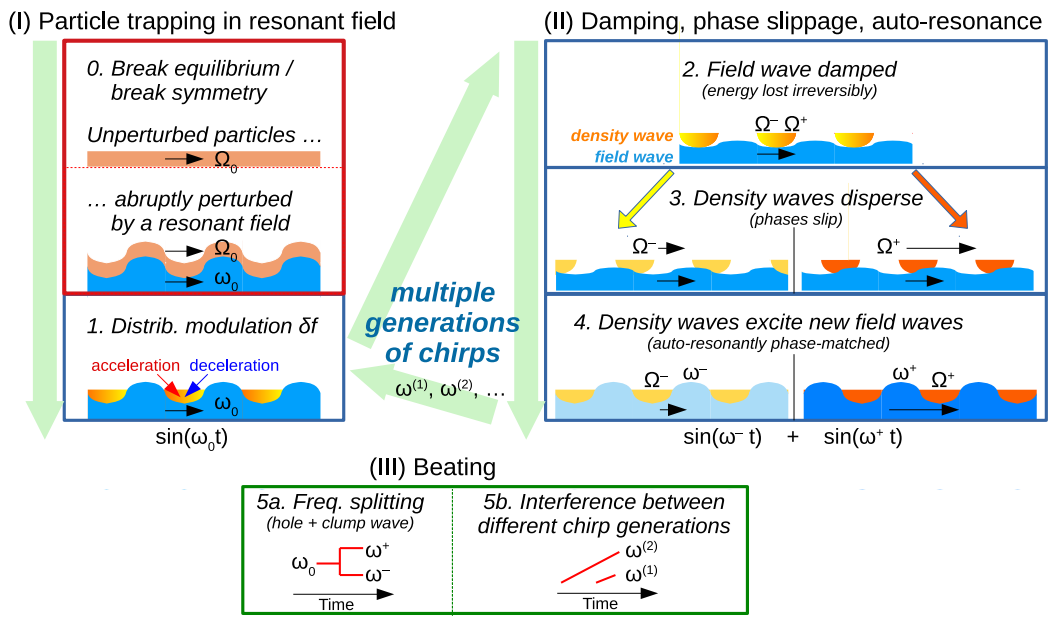}
  \caption{Schematic illustration of the nonlinear frequency chirping process in a {\cred monoenergetic beam} that is abruptly perturbed out of equilibrium. Note that, {\cred in reality, stages 1--4 are not as distinct as they are shown here. For instance,} phase slippage (item 3) occurs before the wave amplitude saturates and decays; in fact, we will show that, in a marginally unstable system, this phase slippage and the resulting beats (item 5a) contribute actively to an early temporary saturation {\cred and decay, and subsequent} revival of the field wave. In the case of EPs interacting with shear Alfv\'{e}n waves, acceleration (deceleration) of particles and resonant damping (drive) of the field is connected with {\cred radially} inward (outward) transport.}
  \label{fig:intro_onset}
\end{figure*}

\subsection{Phase slippage between pump and field}
\label{sec:review_mech}

Interactions between pump waves in EP phase space density and electromagnetic field waves do not necessarily lead to spontaneous chirping. There needs to be a mechanism that causes phase slippage between these density waves and field waves, which we review in this section.

When the field wave is unaffected by the presence of the EP density perturbation $\delta f$ (e.g., when it is externally enforced), one obtains the situation shown in Fig.~\ref{fig:intro_fire-A-freeze500_poink-df}. Around the resonance (vertical dash-dotted line), an island structure forms in {\cred phase space}, which can be visualized using a kinetic Poincar\'{e} plot as in panel (a), using action-angle coordinates {\cred like} $(P_\zeta,\vartheta)$ in the frame of reference moving with the phase velocity of the wave field. The circulation of particles that are trapped inside the island causes a perturbation in phase space density $\delta f$ {\cred with a coarse-scale structure} as shown in (b), with a hole ($\delta f < 0$, blue) on one side and a clump ($\delta f > 0$, red) on the other.

The hole and clump in this configuration are stationary collective modulations of the particle density, not tied to the motion of individual particles. Asymptotically, the system develops a non-axisymmetric EP distribution that is consistent with the perturbed field.\footnote{{\cred It must be noted that} Fig.~\ref{fig:intro_fire-A-freeze500_poink-df} does {\it not} show an equilibrium state. We have fixed the amplitude of the field perturbation at an arbitrary value $A_0 = 10^{-6}$ that is far below the level of about $3\times 10^{-5}$ required for the formation of a (quasi-)steady state that is consistent with the initial EP density gradient (cf.~Fig.~\protect\ref{fig:result_fire-A-A0_saturation}). The non-equilibrium character of this relaxed EP distribution can be inferred from the lack of mirror symmetry of $\delta f$ in Fig.~\ref{fig:intro_fire-A-freeze500_poink-df}(b), which implies that there is still potential for net energy transfer (which we have inhibited here by freezing the amplitude). Thus, the flat O-point profile {\cred in (c)} is misleading! There are still active gradients nearby. The X-point profile is also misleading; the actual direction of the X-point gradient is diagonal in $(\hat{P}_\zeta,\vartheta)$ and the energy stored there is not easily tapped (cf.~Section~\protect\ref{sec:saturation_x}). The result of launching a simulation from this relaxed state is shown in Fig.~\protect\ref{fig:sensitivity_fire-A_freeze500} of Appendix~\protect\ref{apdx:model_freeze500}. Since a primordial hole-clump pair has already been set up, this setup effectively skips the exponential growth phase and first beat.}
If the amplitude of the {\cred symmetry-breaking} field perturbation is reduced, some particles are released from the island and carry with them the previously established density modulation. One will then observe a counter-propagating hole-clump wave pair. When the dynamics of the field's amplitude $A(t)$ and phase $\phi(t)$ are coupled to the dynamics of the density waves $\delta f$ as in Eq.~(\ref{eq:mdl}), the propagating hole and clump structures will act as pump waves that drive field fluctuations, which respond auto-resonantly to the pump frequencies.

{\cred From this we infer} that that instability is not essential for triggering a frequency shift. Although a gradient in the reference state $F_0$ is required to produce density waves in an incompressible fluid, that {\cred initial gradient by itself} does not need to be destabilizing. The energy required to launch off-resonant pump waves can also be provided by an abruptly imposed {\cred perturbation that breaks the symmetry of} the particle distribution. What is essential for nonlinear chirping is a mechanism that facilitates phase slippage by preventing complete trapping of displaced particles; {\cred for instance, a mechanism that damps} the field wave.

The process leading to the onset of chirping is illustrated schematically by the cartoon in Fig.~\ref{fig:intro_onset}. As shown on the left-hand side, the process begins with symmetry breaking and {\cred (partial)} particle trapping in the potential well of the field wave:
\begin{enumerate}
\item[0.] Consider a monoenergetic bunch of particles circulating around the plasma torus at a characteristic transit frequency $\Omega_0$. The particle bunch is abruptly perturbed by a resonant field wave that causes an abrupt finite displacement and oscillates at the same frequency $\omega_0 = \Omega_0$. This perturbation breaks the symmetry of the system and takes it out of equilibrium.

\item[1.] The collective response of the guiding centers produces a modulation $\delta f$ in their distribution function $F = F_0 + \delta f$ in position and velocity space. Depending on the local phase of the field wave, some particles are accelerated, others decelerated.
\end{enumerate}

\noindent When the field and the particles constitute an energy-conserving closed system, the particles oscillate indefinitely inside the field's potential well. A high-resolution spectrogram may show transient frequency splitting, but no sustained chirping occurs.

The situation changes when this field-particle system is not isolated and exchanges energy with other components of the plasma or its surroundings. {\cred On the right-hand side of Fig.~\ref{fig:intro_onset}, we illustrate how} sustained frequency chirping can occur when
\begin{enumerate}
\item[2.] the field wave experiences some form of energy loss that is not related to the resonance at hand and exceeds a certain threshold. Possible damping mechanisms include phase mixing in the wave-carrying medium (a.k.a.\ Landau damping), nonlinear wave-wave coupling, or viscous dissipation. When this happens, the particles are detrapped, and ... 

\item[3.] ... the density waves disperse. For simplicity, this cartoon ignores the continuity of the velocity distribution and shows only the density wave fronts constituted by maximally decelerated and maximally accelerated particles, whose characteristic transit frequencies satisfy $\Omega^- < \Omega_0 < \Omega^+$.\footnote{Like a two-stream approximation \protect\cite{Roberts67} of the continuous system \protect\cite{Armstrong67}.} The consequence of detrapping is phase slippage between the density waves and the (weakened) field wave.

\item[4.] The released density waves excite auto-resonantly phase-matched field waves. In the {\cred present symmetric case,} the result is a pair of new field waves, one down-shifted ($\omega^-$) and one up-shifted ($\omega^+$) in frequency.
\end{enumerate}

\noindent As indicated by the triangle of green arrows in the center of Fig.~\ref{fig:intro_onset}, this cycle of trapping, detrapping, phase slippage and auto-resonance can repeat, spawning multiple generations of chirps that successively appear as time passes by.

In reality, steps 1--4 illustrated simplistically in Fig.~\ref{fig:intro_onset} are not so distinct and may overlap in time. For instance, the field is usually damped continuously, so the phase slippage described in step 3 may already begin {\it before} the field amplitude saturates and decays. The essential point is that, because of damping, the field wave has irreversibly lost a part of the energy that it had received when disturbing the gradients in the particle's phase space density, so it does not possess enough energy to reverse all the displacements that have occurred in the particle distribution. In other words, the (weakened) field is able to trap only a fraction of the particles that have been displaced, and that fraction decreases with increasing damping rate. The particles that the field fails to pull back across the resonance will carry along density modulations that act as pump waves at shifted frequencies $\omega^+$ and $\omega^-$.

The same principles apply in {\cred more complex systems}. The consequences of irreversible field energy loss is particularly manifested in (but not limited to) marginally unstable scenarios subject to strong drive and nearly equally strong damping, like our examples in Figs.~\ref{fig:intro_fire-A_chirp} and \ref{fig:intro_fire-A_df}. In such cases, the onset of chirping occurs early and proceeds {\cred gradually}. In contrast, in cases with strong drive but weak damping, chirping tends to be retarded and begin abruptly.

\subsection{Field amplitude pulsations}
\label{sec:review_instab}

In order to characterize {\cred the evolution of the field amplitude}, we define the instantaneous growth rate $\gamma(t)$ as
\begin{equation}
A(t) = A_0 e^{\gamma t}, \quad \gamma = {\rm d}\ln A/{\rm d}t = \gamma_{\rm k} - \gamma_{\rm d};
\label{eq:gamma}
\end{equation}

\noindent where $\gamma_{\rm d}$ represents some form of damping (here taken to be constant), and $\gamma_{\rm k}(t)$ represents the instantaneous kinetic drive; namely, the integral on the right-hand side of Eq.~(\ref{eq:dadt}) divided by $A(t)$. At the beginning of a simulation with a sufficiently small initial perturbation $A_0$ and an axisymmetric EP distribution $F_0$,\footnote{As shown in Apendix~\ref{apdx:model_freeze500}, exponential growth and the first beat are effectively skipped if one starts with a relaxed non-axisymmeteric EP distribution that has adapted to the initial perturbation of the field as in Fig.~\protect\ref{fig:intro_fire-A-freeze500_poink-df}.} the kinetic drive has an approximately constant value that corresponds to the growth rate of the linearized system: $\gamma_{\rm k} \approx \gamma_{\rm L} = {\rm const}$. The linear growth rate $\gamma_{\rm L}(\partial_{P_\zeta} F_0,\partial_K F_0)$ is determined by the gradients of the reference state $F_0(P_\zeta,K,\mu)$ at the seed resonance. When $\gamma_{\rm L} > \gamma_{\rm d}$, the field amplitude $A(t)$ begins to grow (nearly) exponentially with a rate $\gamma \approx \gamma_{\rm L} - \gamma_{\rm d}$.

\begin{figure}[tb]
  \centering
  \includegraphics[width=8cm,clip]{\figures/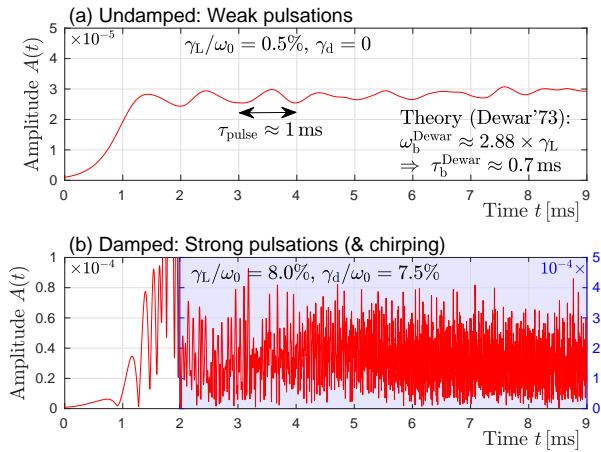}
  \caption{(a) Weak pulsations and (b) strong beating in the amplitude $A(t)$ associated with (a) an undamped and (b) a strongly damped electromagnetic field wave destabilized by resonant EPs. Both cases have the same small initial growth rate $\gamma/\omega_0 = (\gamma_{\rm L} - \gamma_{\rm d})/\omega_0 \approx 0.5\%$ $(\gamma/\nu_0 \approx 0.03)$ with $\nu_0 = \omega_0/(2\pi) = 100\,{\rm kHz}$. The damped case in (b) exhibits strong chirping as was shown in Fig.~\protect\ref{fig:intro_fire-A_chirp}. Panel (b) is split in two parts with different scales (factor 5) in order to show the small amplitudes during the first $2\,{\rm ms}$.}
  \label{fig:intro_fire-A-A0_amp}
\end{figure}

Figure~\ref{fig:intro_fire-A-A0_amp} shows two examples of the amplitude evolution in {\tt ORBIT} simulations in the scenario that we study in this paper. Panel (a) shows an undamped case ($\gamma_{\rm d} = 0$) with relatively weak drive $\gamma_{\rm L}/\omega_0 = 0.5\%$. The rapid oscillation of the field signal $s(t)$ with frequency $\omega_0 = 2\pi\times 100\,{\rm kHz}$ are not visible, since we plot only its temporal envelope $A(t)$. One can see that the amplitude grows to about $A \approx 2.7\times 10^{-5}$. The period $\tau_{\rm pulse} \lesssim 1\,{\rm ms}$ of the subsequent pulsations seen in Fig.~\ref{fig:intro_fire-A-A0_amp}(a) is more or less consistent with Dewar's theoretical prediction \cite{Dewar73}
\begin{equation}
\omega_{\rm b}^{\rm Dewar} = \frac{256}{9\pi^2}\gamma_{\rm L} \approx 2.88\times\gamma_{\rm L},
\label{eq:wb_dewar}
\end{equation}

\noindent which gives a bounce period $\tau_{\rm b}^{\rm Dewar} = 2\pi/\omega_{\rm b}^{\rm Dewar} \approx 0.7\,{\rm ms}$ for our parameters.

Pulsations like those appearing in Fig.~\ref{fig:intro_fire-A-A0_amp}(a) are sometimes said to be caused by the continued over- and undershoots of the gradients around the resonance that, in turn, are caused by the bouncing motion of near-resonant particles that are trapped inside the effective potential well of the field wave. Although this interpretation may appeal to intuition, {\cred it is important to keep in mind that pulsations can be caused not only by the acceleration terms $(\dot{P}_\zeta\partial_{P_\zeta} + \dot{K}\partial_K)F$ of the Vlasov equation (\ref{eq:vlasov}), but also by the free streaming terms $(\dot\vartheta\partial_\vartheta + \dot\zeta\partial_\zeta)\delta f$. The details of these interactions are} complicated, which is why we use numerical simulations. As suggested in Section~\ref{sec:review_pump}, it may be more instructive to think of these amplitude pulsations as being a consequence of the motion of density waves, because it is $\delta f$ (not {\cred $\dot{\bm Z}_{\rm gc}\cdot\partial_{{\bm Z}_{\rm gc}} F$}) that appears in the field equation (\ref{eq:mdl}). The field pulsations cause a repeated trapping and detrapping of density waves in the boundary layer of the phase space island as illustrated schematically in Fig.~\ref{fig:intro_x-point}(b). The resulting variations in the relative phase between density and field waves, in turn, feed back on the evolution of the field via the phase correlation integral between EP current density and electric field in Eq.~(\ref{eq:mdl}). We suspect that these boundary layer dynamics, in combination with radial asymmetries in our simulation setup, are the primary cause for the sustained and somewhat irregular oscillations seen in Fig.~\ref{fig:intro_fire-A-A0_amp}(a). The fact that the bounce frequencies vary across the island radius \cite{Meng18} may also play a role. Limited numerical accuracy can cause spurious growth of the field amplitude (see Appendix~\ref{apdx:model_convergence}). We will analyze these data in more detail in Appendix~\ref{apdx:undamped}, examining the effect of the pulsations and demonstrating that, although there is no sustained chirping, the field's oscillation frequency can vary measurably even in the absence of field damping.

Figure~\ref{fig:intro_fire-A-A0_amp}(b) shows that a very different evolution with strong pulsations of the field amplitude $A(t)$ is seen in the presence of strong damping, which is here nearly equal to the drive. This is the same case as in Figs.~\ref{fig:intro_fire-A_chirp} and \ref{fig:intro_fire-A_df}, with $\gamma_{\rm d}/\omega_0 = 7.5\%$. The damping is marginally overcome by a steep gradient $\partial F_0/\partial P_\zeta > 0$ that gives a linear kinetic drive $\gamma_{\rm L}/\omega_0 \approx 8\%$. Thus, the overall growth rate during the exponential phase is small, $\gamma/\omega_0 \approx 0.5\%$, and equal to that in the undamped example in Fig.~\ref{fig:intro_fire-A-A0_amp}(a). We observe in Fig.~\ref{fig:intro_fire-A-A0_amp}(b) an early saturation of the field amplitude during the first millisecond, at a still tiny amplitude of only $A \approx 6\times 10^{-6}$, not far from the initial value $A_0 = 10^{-6}$. Subsequently, the amplitude grows by nearly two orders of magnitude to about $3\times 10^{-3}$, while performing pulsations with a magnitude of 100\% on time scales of $0.1\,{\rm ms}$ or less, not much longer than the $10\,\mu{\rm s}$ seed wave period.

These amplitude pulsations have been shown in many previous works dealing with nonlinear frequency chirping --- theoretical, numerical and experimental. Clarifying the role of these pulsations in different stages of a chirping simulation is the main subject of the present study.

\subsection{Characterization of chirping systems}

Aside from geometric effects and other nonuniformities of the plasma, the behavior of our reduced system is largely determined by its initial proximity to the state of marginal stability, where drive and damping approximately balance,
\begin{equation}
|\gamma_{\rm L}-\gamma_{\rm d}|/\gamma_{\rm L} \ll 1 \quad \text{(marginal (in)stability)},
\end{equation}

\noindent and by the amount of free energy stored in the initial gradient, which can be measured by the ratio of $\gamma_{\rm L}$ to the initial wave frequency $\nu_0$ at the seed resonance,\footnote{In Eq.~(\protect\ref{eq:weak_strong}), we write $\gamma_{\rm L}/\nu_0$ instead of $\gamma_{\rm L}/\omega_0$ in order to be able to make a clearer distinction between strong ($\lesssim 1$) and weak ($\ll 1$) drive. Apart from this, we usually work with $\gamma_{\rm L}/\omega_0$.}
\begin{align}
\gamma_{\rm L}/\nu_0 \ll 1 \quad \text{(weak drive)}, \\
\gamma_{\rm L}/\nu_0 \lesssim 1 \quad \text{(strong drive)}.
\label{eq:weak_strong}
\end{align}

\noindent The cases discussed in this paper are situated in the strongly driven regime, $\gamma_{\rm L}/\nu_0 \approx 0.5$ ($\gamma_{\rm L}/\omega_0 \approx 8\%$). We consider one case close to marginal stability, $\gamma_{\rm L}/\gamma_{\rm d} \approx 1.07$, and one case with relatively rapid growth, $\gamma_{\rm L}/\gamma_{\rm d} \approx 1.88$.

An important measure that characterizes the phase space dynamics underlying the chirps is the degree of adiabaticity of the dynamics, which is determined by the ratio of {\cred the mean} and maximal displacement of a simulation particle (representing an EP Vlasov fluid element) during a field pulsation period $\tau_{\rm pulse}$,\footnote{As indicated in Eq.~(\ref{eq:adiabatic0}), the variations of a particle's canonical toroidal momentum $P_\zeta$ and energy $E = K + {\cred Q}e\Phi$ are closely related in the system studied here. We will use this relation later (cf., Eq.~(\protect\ref{eq:enr_rot})).}
\begin{equation}
\frac{\big\langle \delta P_\zeta\big\rangle_{\rm pulse}}{{\rm max}|\delta P_\zeta|_{\rm pulse}} \sim \frac{\left<\delta E\right>_{\rm pulse}}{{\rm max}|\delta E|_{\rm pulse}} < 1 \;\;\, \substack{\text{\normalsize (adiabatic} \\ \text{\normalsize regime);}}
\label{eq:adiabatic0}
\end{equation}

\noindent where $\delta P_\zeta = P_\zeta - \langle P_\zeta\rangle_{\rm pulse}$ and $\left<...\right>_{\rm pulse} \equiv \tau_{\rm pulse}^{-1}\int{\rm d}\tau_{\rm pulse}(...)$ denotes the pulse average. $E = K + {\cred Q}e\Phi$ is the total particle energy, consisting of kinetic energy $K$ and electrostatic potential energy ${\cred Q}e\Phi$. In simple terms, Eq.~(\ref{eq:adiabatic0}) means that, in the adiabatic regime, the coherent phase space structures (if any) rotate faster than they advance, so the chirps are associated with the drift of vortex-like structures in phase space, which we have labeled ``solitary hole/clump vortices'' in Fig.~\ref{fig:intro_fire-A_df}(b). In this case, the pulse period $\tau_{\rm pulse}$ may be identified with the bounce period $\tau_{\rm b}$, and the pulse average becomes the bounce average. In contrast, chirps in the nonadiabatic regime are associated with the propagation of convective plume-like structures in phase space density as in Fig.~\ref{fig:intro_fire-A_df}(a). These structures {\cred advance radially faster than they rotate, taking the} form of convective instability \cite{Zonca15b}. In this case, a bounce time $\tau_{\rm b}$ cannot be easily defined, and this is the reason why we have chosen the amplitude pulsation time $\tau_{\rm pulse}$ to measure (non)adiabaticity in Eq.~(\ref{eq:adiabatic0}).

Nonlinear frequency chirping always begins nonadiabatically, with the propagation of convective wave fronts in whose wake dynamics reminiscent of convective interchange and wave breaking can be observed in {\cred plots of} $\delta f(P_\zeta,\vartheta)$, with corresponding chirping patterns appearing in the spectrograms. In Figs.~\ref{fig:intro_fire-A_chirp} and \ref{fig:intro_fire-A_df}, this turbulent belt can be clearly seen in the form of complicated patterns around the seed resonance. The region around the seed resonance tends to remain nonadiabatic, but solitary vortices with an effectively adiabatic core can be emitted from its boundary. Thus, the same system may simultaneously contain adiabatic and nonadiabatic portions. If the evolution of the wave field is dominated by one of the adiabatic vortical structures in phase space, Eq.~(\ref{eq:adiabatic0}) can be written as\footnote{The quantity $\alpha = \delta\dot\omega/\omega_{\rm b}^2$ has been used as an expansion parameter in theoretical treatments. Here, we included a factor $2\pi$ since we feel that this quantifies the upper limit of adiabaticity in a more precise and intuitive way: $2\pi\alpha = \tau_{\rm b}/\tau_{\rm chirp} < 1$ instead of $\alpha \ll 1$. An indication for this can be seen in Fig.~13 of Ref.~\protect\cite{WangGe18} and we will offer further justification in Section~\protect\ref{sec:result_transport} when inspecting the motion of tracer particles in a vortex.}
\begin{equation}
\frac{2\pi\delta\dot\omega}{\omega_{\rm b}^2} = \tau_{\rm b}^2 \delta\dot\nu = \frac{\tau_{\rm b}}{\tau_{\rm chirp}} < 1 \quad \substack{\text{\normalsize (single smooth} \\ \text{\normalsize adiabatic chirp);}}
\label{eq:adiabatic1}
\end{equation}

\noindent where $\tau_{\rm chirp}^{-1} \equiv \delta\nu(\tau_{\rm b}) \approx \tau_{\rm b}\delta\dot\nu$ is the frequency shift that the wave experiences during one bounce time.

In summary, we will refer to a subdomain of phase space as being ``adiabatic'' when the dynamics in that region exhibit a distinct separation of time scales, where a particle's mean nonlinear displacement $\left<\delta P_\zeta\right>_{\rm b}$ during one bounce cycle $\tau_{\rm b}$ is smaller than the maximal displacement ${\rm max}|\delta P_\zeta|_{\rm b}$ during that period. The physical relevance of such a time scale separation is that it facilitates the formation of robust vortex structures in phase space, that can be viewed as a generalization of Bernstein-Greene-Kruskal (BGK) modes formed by electron Langmuir waves \cite{Bernstein57}. We will show that the large vortex that is visible in the upper part of Fig.~\ref{fig:intro_fire-A_df} satisfies the adiabaticity condition only marginally ($\tau_{\rm b} \lesssim \tau_{\rm chirp}/2$), but it nevertheless resembles a BGK mode since its interior consists of concentric nested layers as illustrated in Fig.~\ref{fig:intro_x-point}(c), with the remarkable ability to withstand the strong and incoherent beats of the field.

\subsection{Discussion of the adiabatic limit}
\label{sec:review_adiabat}

The scale separation in the adiabatic limit is useful for analytical treatments and has allowed Berk, Breizman {\it et al}.\ (BB) to develop a model that yields a quantitative prediction of the chirping rate if the values of $\gamma_{\rm L}$ and $\gamma_{\rm d}$ are known \cite{Berk97, Berk99}. With some approximations, their integral relation between $\delta\omega$ and $\omega_{\rm b}$ reduces to the analytical form
\begin{subequations}
\begin{align}
\delta\omega^{\rm BB} = \omega_{\rm b}^{\rm BB} \left(\frac{2\gamma_{\rm d}}{3} t\right)^{1/2} \approx 0.44 \times \gamma_{\rm L} (\gamma_{\rm d} t)^{1/2},
\label{eq:dw_bb}
\\
{\rm with} \quad \omega_{\rm b}^{\rm BB} = \frac{16}{3\pi^2} \gamma_{\rm L} \approx 0.54\times \gamma_{\rm L}.
\label{eq:wb_bb}
\end{align}
\label{eq:bb} \vspace{-0.2cm}
\end{subequations}

\noindent Note that since $A \propto \omega_{\rm b}^2$, Eq.~(\ref{eq:wb_bb}) predicts the field amplitude associated with a chirping BGK mode. Comparison with Eq.~(\ref{eq:wb_dewar}) shows that $\omega_{\rm b}^{\rm BB}/\omega_{\rm b}^{\rm Dewar} \approx 0.18$, so the field amplitude of the BGK wave is predicted to be $30$ times smaller than the saturation amplitude of the undamped wave driven by gradients that give the same value of $\gamma_{\rm L}$. Conversely, this means that a BGK wave is predicted to require a $1/0.18 = 5.6$ times larger value of $\gamma_{\rm L}$ in order to yield the same bounce frequency (and amplitude). We suspect that this is connected with the assumption of marginal stability that underlies $\omega_{\rm b}^{\rm BB}$ but not $\omega_{\rm b}^{\rm Dewar}$. Consequently, we expect to find reasonable agreement with $\omega_{\rm b}$ in simulations where $(\gamma_{\rm L}-\gamma_{\rm d})/\gamma_{\rm L}$ is comparable to $\omega_{\rm b}^{\rm BB}/\omega_{\rm b}^{\rm Dewar} \sim \mathcal{O}(0.1...0.2)$. In our simulations, this marginality parameter will have the values $0.06$ and $0.47$ and we will find bounce periods that lie slightly above and below the prediction $\tau_{\rm b}^{\rm BB} \approx 0.24\,{\rm ms}$.

The dotted green curve in Fig.~\ref{fig:intro_fire-A_chirp}(b) indicates the theoretical prediction $\delta\nu^{\rm BB} \sim 24\,{\rm kHz}\times\sqrt{t[{\rm ms}]}$ from Eq.~(\ref{eq:dw_bb}) for our parameters. One can see that the curve lies in the right ``ball park'', especially for the upward chirp.\footnote{One reason for the smaller extent of the downward chirp is that it propagates away from the peak of the field mode in our setup.}
Similar agreement in the overall extent of the chirps was seen in many other studies, even when the parameters clearly stretch the range of validity of the theoretical model, as in our marginally unstable but strongly driven example.

This remarkable robustness of the model may be attributed to the fact that Eq.~(\ref{eq:bb}) is the result of balancing the rate at which the fluctuating field receives energy from the destabilizing gradients ($\gamma_{\rm L}$) and loses energy via the {\it ad hoc} damping $\gamma_{\rm d}$. The theory simply postulates that the BGK wave is somehow able feed on the gradients at the same rate as the field loses energy, implicitly relying on the existence of a nonadiabatic layer as illustrated in Fig.~\ref{fig:intro_x-point}(c), without modeling the detailed processes inside that layer. Therefore, the overall agreement means only that both theory and simulation satisfy the energy balance constraint. It does not necessarily mean that the theoretical picture on which the theory is built captures the processes that actually occur in the simulation (or in a real plasma).\footnote{As an analogy, consider the phenomenon of so-called magnetic reconnection, which can be realized by different physical mechanisms. All mechanisms break the frozen-in-flux condition of ideal MHD, but the energy conversion rates and side effects (such as wave excitation) differ.}

A generalized model that accounts for nonlinear distortions and shrinking of the field's effective potential well has been derived by Breizman \cite{Breizman10} and further extended to expanding potential wells by Hezaveh {\it et al}.\ \cite{Hezaveh21}. A key point that is captured in these and related studies of long-range chirping \cite{Nyquist12, Nyquist13, Hezaveh17, Hezaveh20}, and which is closely related to the topic of the present paper, is that the generalized BGK-like modes have an active boundary layer that can dynamically expand or shrink as the field amplitude grows or decays.\footnote{Our interpretation of the {\cred results} presented in Ref.~\protect\cite{Breizman10} is that the leading-order effect is the variation of the potential well's depth. The distortions of its shape seem to be corrections of higher order, comparable to various other effects (e.g., geometric nonuniformity) that were ignored.}

\subsection{Some open questions: Formation of solitary vortices and the role of beating}
\label{sec:review_q}

Having reviewed the concepts that will be needed for the discussion of our results, we proceed now with an outline of some open questions that we wish to tackle.

One question that, to our knowledge, has not been satisfactorily answered yet is how solitary vortices are formed and emitted from the turbulent belt around the seed resonance. We believe that some of the results reported {\cred here} can contribute to a better understanding of the nonadiabatic onset of chirping and how it can lead to the detachment of solitary vortices. We will present evidence indicating that these processes are facilitated by pulsations of the field amplitude $A(t)$ like those in Fig.~\ref{fig:intro_fire-A-A0_amp}(b).

Such pulsations and accompanying phase jumps have been reported in a previous analysis of moderate bursts of chirping shear Alfv\'{e}n waves seen in JT-60U tokamak experiments \cite{Bierwage17a}. The patterns were seen both in the experimental data and in self-consistent hybrid simulations with realistic EP sources and collisions, and they were explained (loosely speaking) as the beating between multiple chirps that occur simultaneously at different frequencies. Some examples were shown in Fig.~13 of Ref.~\cite{Bierwage17a}.

Indeed, the chain of events sketched in Fig.~\ref{fig:intro_onset} and discussed in Section~\ref{sec:review_mech} implies that one chirp rarely comes alone. As is illustrated schematically in the bottom part of Fig.~\ref{fig:intro_onset}, we may distinguish two ways for beating to occur:
\begin{enumerate}
\item[5a.] interference of wave pairs $\omega^\pm$ produced by frequency splitting (described in Section~\ref{sec:review_mech});
\item[5b.] interference between different generations of chirps $\omega^{(1)}, \omega^{(2)}, ...$ (like those seen in Fig.~\ref{fig:intro_fire-A_chirp}).
\end{enumerate}

\begin{figure}[tb]
  \centering
  \includegraphics[width=8cm,clip]{\figures/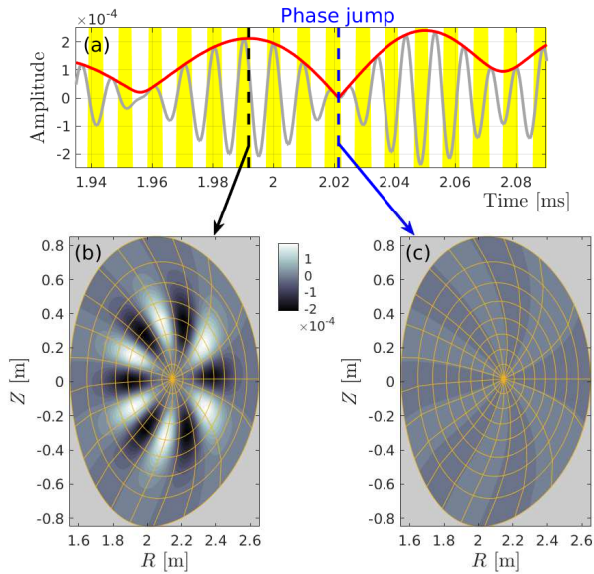}
  \caption{Example of global beating in a short time window around $t \approx 2\,{\rm s}$ of Fig.~\protect\ref{fig:intro_fire-A-A0_amp}(b). Panel (a) shows the raw signal $s(t)$ (gray) and the envelope $A(t)$ (red). The equidistant vertical yellow stripes help to discern changes in the signal's phase. Panels (b) and (c) show, respectively, the global mode structure of the ideal MHD displacement in code units at times of maximal constructive and destructive interference.  While global beating is inevitable in the semi-perturbative model used in the {\tt ORBIT} code with which this signal was computed, it is remarkable that such behavior was also observed in self-consistent hybrid simulations (cf.~Fig.~12 of Ref.~\protect\cite{Bierwage17a}).}
  \label{fig:intro_fire-A_beat}
\end{figure}

\noindent The physical picture we will use may be described as follows. The convective interchange of EP Vlasov fluid produces multiple more or less coherent structures in the perturbed EP phase space density $\delta f$ that are more or less stratified radially along the vertical ($P_\zeta$) axis in Fig.~\ref{fig:intro_fire-A_df}. As indicated by the horizontal arrows in Fig.~\ref{fig:intro_fire-A_df}, these structures propagate at different phase velocities around the plasma torus (horizontal axis and out of plane), depending on the local value of $P_\zeta$. These EP density waves drive the same field mode at different frequencies simultaneously. The field mode responds by beating as shown in Fig.~\ref{fig:intro_fire-A_beat}, as if it consisted of multiple field waves, each locked to a different pump wave. Indeed, the total signal $s(t) = A(t)\sin(-\omega_0 t - \phi(t))$ in Eq.~(\ref{eq:intro_xi}) represents an arbitrary number of harmonic waves, whose superposition causes the combined amplitude $A(t)$ to pulsate and the combined phase $\phi(t)$ to jump by $\pm\pi$ between pulses as in Fig.~\ref{fig:intro_fire-A_beat}. In Appendix~\ref{apdx:beat}, we show that this physical picture is mathematically consistent with the field equations we solve.

When we represent the time-dependence of the combined signal in the form $s(t) = A(t)\sin(-\omega_0 t - \phi(t))$, we demand that its harmonic components must all have the same spatial structure. One may argue that representing EP-driven shear Alfv\'{e}n waves in this form is an oversimplification, because a self-consistent (nonperturbative) field mode may simply decompose into several spatially separated nonlinear modelets, whose amplitudes and phases can evolve more or less independently, each driven by its own {\cred resonant phase space structure}. This is certainly a possible scenario, but it is not a necessity, as one can infer from a remarkable observation made in the above-mentioned self-consistent hybrid simulation of a JT-60U plasma \cite{Bierwage17a}: Fig.~12 of that paper shows that the energetic particle mode (EPM) \cite{Chen94} with toroidal mode number $n=1$ that was responsible for the chirps in that system sometimes vanishes completely between beats and reappears with perfectly opposite phase. In principle, the field in that hybrid simulation could have decomposed into multiple independent EPMs, but this did not occur, at least not at a readily visible level (i.e., not at leading order). The same behavior is reproduced here in idealized form in Fig.~\ref{fig:intro_fire-A_beat}.

We interpret the possibility of {\it global} beating as a consequence of a certain degree of rigidity that is inherent to MHD modes with long wavelengths and which is further enhanced by large magnetic drifts. For instance, in the above-mentioned JT-60U scenario, EP orbits spanned 20...30\% of the minor radius (cf.~Fig.~3 of Ref.~\cite{Bierwage14}). This rigidity preserves the coherence of the wave field across a significant portion of the plasma radius and one might say that this is the reason why we call it a {\it mode}. All chirps studied in the present paper are caused by radially propagating phase space structures that stay within the range of an unperturbed drift orbit, which justifies {\it a posteriori} the assumption of a rigid mode whose evolution is captured by the time-dependence of two scalars, $A(t)$ and $\phi(t)$.

One physical consequence that is highly relevant for the present study is that the resonant phase space density waves are mutually coupled and interfere with each other via the jointly driven MHD field mode. Such an interaction has been explicitly ignored in the model of chirping BGK modes formulated by Berk {\it et al}.\ (Eq.~(23) of Ref.~\cite{Berk99}). We believe that this is a point where interesting unexplored physics may still be lurking and we are aware of one more group of researchers who are tackling this topic by considering the effect of beating on chirps driven by a pair of electron resonances \cite{Hezaveh21}. It is difficult to foresee all the implication of this complex feedback; apart from the perhaps obvious fact that all particles are temporarily detrapped at the times of destructive interference, and that the EP density waves find themselves out-of-phase when the field wave regrows after performing a phase jump as in Fig.~\ref{fig:intro_fire-A_beat}. Thus, instead of independent BGK waves, we have a complex multi-body system consisting of several more or less strongly coupled components.

{\cb Such multi-component interactions can become more regular in cases where {\it chirping is suppressed}, e.g., by processes that {\it scatter} the resonant particles. In such cases, the spectrum of a resonantly driven mode can still broaden, but it does so in a well-structured manner by splitting into a sequence of discrete lines with equal frequency spacing $\Delta\nu$ as reported by Fasoli {\it et al}.\ \cite{Fasoli97}. According to the theory presented by Fasoli {\it et al}.\ \cite{Fasoli98} based on Berk {\it et al}.\ \cite{Berk97}, such phenomena can be explained by a period doubling bifurcation of a near-threshold kinetic instability. The modulation of the field amplitude (beating) through the interference between the waves associated with such a bunch of spectral lines is an integral part of the process, but in that collision-dominated (non-chirping) regime the phase space dynamics as well as the beats of the field become regular, settling into a cyclic pattern. The feature that makes our chirping scenarios (with weak or no collisions) more complicated is that irregular convective motion can persist, especially near the seed resonance. Fascinatingly, however, coherent structures in the form of long-lived phase space vortices do form in this turbulent environment. One purpose of our study is to throw light on the role that beating may play in the formation of such robust vortices.}

One may argue that beating has a zeroth-order effect only on the motion of individual particles, while its effect on the coherent phase space structures is of higher order, because the interaction is nonresonant.\footnote{The beats may resonate with other regions in phase space, which is an interesting story on its own, and not pursued here.}
However, chirping itself is often a higher-order effect, especially in the limit of slow (adiabatic) chirps. Therefore, we think that it is worthwhile to consider the effects of beating on the same footing. In this paper, we take a close look at how beating affects the phenomenon of nonlinear frequency chirping and the underlying motion of Vlasov fluid in EP phase space, which we represent here using discrete particles.

\section{Model and methods}
\label{sec:model}

\subsection{Global beating as a rationale for a semi-perturbative model}
\label{sec:model_justify}

{\cred MHD-PIC hybrid simulations often compare well with experiments, at least for EP-driven Alfv\'{e}n modes with long wavelengths ($n < 10$) \cite{Todo15,Todo16a,Bierwage17a,Bierwage18}. However, such simulations} require high-performance supercomputers, primarily because of the expensive field solver that has to evolve the electromagnetic fields, plasma density and pressure on a dense three-dimensional (3D) mesh consisting of tens of millions of grid points. The {\cred globality of the} beating observed in Ref.~\cite{Bierwage17a} offers a justification for a semi-perturbative approach, where the spatial structure of the Alfv\'{e}nic field modes is prescribed, reducing the MHD solver's large number of degrees of freedom to the evolution of merely two scalars; namely, the amplitude $A(t)$ and phase $\phi(t)$ appearing in Eq.~(\ref{eq:intro_xi}). Numerical studies of nonlinear frequency chirping in realistic tokamak geometry can then be performed on a laptop PC.

In the presence of multiple modes $k=1,...N_{\rm mode}$, with toroidal mode numbers $n_k$ and initial seed frequencies $\omega_{0k}$, the field model in Eq.~(\ref{eq:intro_xi}) can be written as
\begin{subequations}\vspace{-0.05cm}
\begin{align}
\tilde{\bm \xi}(\psi_{\rm P},\vartheta,\zeta,t) &= \sum\limits_{l=1}^{N_{\rm mode}} A_k(t) \sum\limits_m \hat{\bm\xi}_{k,m}(\psi_{\rm P}) e^{i\Theta_{k,m}(t)},
\label{eq:mode_xi}
\\
\Theta_{k,m}(t) &= n_k\zeta - m\vartheta - \omega_{0k}t - \phi_k(t);
\label{eq:mode_phase}
\end{align}
\label{eq:mode}\vspace{-0.35cm}
\end{subequations}

\noindent where $\tilde{\bm\xi}$ is the complex-valued ideal MHD displacement vector and $\hat{\bm\xi}_{k,m}(\psi_{\rm P})$ is the radial profile of each poloidal Fourier harmonic $m$, here written as a function of normalized poloidal flux $0 \leq \psi_{\rm P} \equiv \Psi_{\rm P}/\Psi_{\rm P,edge} \leq 1$ that increases monotonically from the center (0) to the edge (1) of the plasma. The imaginary unit is denoted by $i$, and $\Theta_{k,m}$ is the complex phase at a given point in time and space, here expressed in toroidal coordinates $(\psi_{\rm P},\vartheta,\zeta)$ with poloidal and toroidal angles $\vartheta$ and $\zeta$. Our convention is that the physical electric and magnetic field perturbations ${\bm E}$ and $\delta{\bm B}$ are the real (cosine) component of the complex signal, so that the potentials $\Phi$ and $\delta{\bm A}$, {\cb and, thus,} the ideal MHD displacement $\xi = \Im\{\tilde{\xi}\}$, are the imaginary (sine) components.

Reduced field models based on Eq.~(\ref{eq:mode}) have been implemented in codes such as {\tt ORBIT} \cite{White84,ChenY99,White19}, {\tt HAGIS} \cite{Pinches98, Pinches04} and {\tt MEGA} \cite{Todo03, Nishimura13}, which have then been used to study nonlinear EP-Alfv\'{e}n wave interactions and chirping in realistic geometries. The reduced number of degrees of freedom in the field dynamics enhances physical transparency, and the low computational cost combined with a set of free parameters (Table~\ref{tab:parm_model}), permits detailed physics studies through extensive parameter scans that are not feasible with self-consistent simulations. Meanwhile, the full geometric complexity of the EP orbits is retained, allowing us to explore practically relevant conditions with respect to spatio-temporal scales and geometric nonuniformity.

While global beating gave us an excuse to ignore perturbations of the mode structure in Eq.~(\ref{eq:mode}), the implicit assumption made in the field equation (\ref{eq:mdl}) of the existence of {\it a global ``mode'' that can oscillate well in a wide range of frequencies} is a rather strong and disputable simplification of the Alfv\'{e}nic plasma response. The dynamics in the reduced system (\ref{eq:mdl}) do show some preference for resonant frequencies at the radius of the mode's peak: We observe that an off-peak seed resonance tends to be produce stronger and more rapid chirps in the direction of the mode's peak (hence the asymmetry in Fig.~\ref{fig:intro_fire-A_chirp}(b)). However, {\cred we are missing entirely} the effect of continuous spectra of shear Alfv\'{e}n waves, and their band gaps that may contain discrete eigenmodes. The radial structure $\omega_{\rm A}(\psi_{\rm P})$ of these spectra is known to affect chirping by providing a preferred trajectory for the chirps (e.g., see Refs.~\cite{Bierwage16a, Bierwage16b, WangTao20}) and by contributing to the field damping \cite{Zonca92}. {\cred For the time being, we choose to ignore these effects without further justification other than saying that this simplification} appears to be acceptable for our purposes. Other factors that can play a role but are ignored here include fluid nonlinearities, collisions and sources of EPs. Due to these limitations, the scope of our study is constrained to the qualitative features of chirping in the semi-perturbative limit, with no intention to make quantitative predictions for experiments. Further considerations related to nonperturbative effects can be found in Appendix~\ref{apdx:model_pert}.

\begin{table*}[tb]
\centering
\begin{tabular}{c|cc|cc|ccc|c|l}
\hline\hline
Case & \multicolumn{2}{c}{Harmonic} & \multicolumn{2}{|c|}{Mode profile} & \multicolumn{3}{|c|}{Drive} & Damping & Remarks \\
& $\nu_0\,{\rm [kHz]}$ & $m/n$ & $r_0/a$ & $r_{\rm w}/a$ & $\frac{G_0}{F_0}\partial_{P_\zeta}F_0$ & $\frac{G_0}{F_0}\partial_K F_0$ & $\gamma_{\rm L}/\omega_0$ & $\gamma_{\rm d}/\omega_0$ & \\
\hline
(A) & $100$ & $6/5$ & $0.65$ & $0.15$ & $2.80164$ & $-0.084898$ & $\Rightarrow 0.080$ & $0.075$ & Marginally unstable \\
(B) & $100$ & $6/5$ & $0.65$ & $0.15$ & $2.80164$ & $-0.084898$ & $\Rightarrow 0.075$ & $0.040$ & Strongly unstable \\
\hline
(A0) & $100$ & $6/5$ & $0.65$ & $0.15$ & $0.46227$ & $-0.014008$ & $\Rightarrow 0.005$ & $0$ & (A) without damping \\
(B0) & $100$ & $6/5$ & $0.65$ & $0.15$ & $1.79305$ & $-0.054335$ & $\Rightarrow 0.035$ & $0$ & (B) without damping \\
\hline
(C) & $100$ & $6/5$ & $0.7$ & $0.15$ & $5.53325$ & $-0.16767$ & $\Rightarrow 0.081$ & $0.075$ & (A) with peak shifted outward \\
\hline\hline
\end{tabular}
\caption{Model parameters used in our {\tt ORBIT} simulations. Cases (A) and (B) will be examined in detail. The undamped cases (A0) and (B0) are used for some comparisons in Section~\protect\ref{sec:saturation}, and Appendix~\ref{apdx:undamped} contains further details. Case (C) is used in Appendix~\ref{apdx:model_opt} to examine the effect of changing the mode peak location relative to the resonance, which can affect the effective seed frequency as well as the direction and rate of chirping. The values of the gradients $\frac{G_0}{F_0}\partial_{P_\zeta}F_0$ (dominant) and $\frac{G_0}{F_0}\partial_K F_0$ (negligible) are those of the input variables {\tt df0dpz} and {\tt df0de} in {\tt ORBIT} when $v_{\rm A0} = {\tt falf} = 0.07$ and (for historical reasons) $F_0/G_0 = 1 + {\tt df0dpz}$. The values of $\gamma_{\rm L} = \gamma + \gamma_{\rm d}$ were measured in the simulations during the phase of exponential growth using Eq.~(\protect\ref{eq:gamma}).}
\label{tab:parm_model}
\end{table*}

\subsection{Equations and parameters}
\label{sec:model_eq}

\begin{figure}[tp]
  \centering
  \includegraphics[width=8cm,clip]{\figures/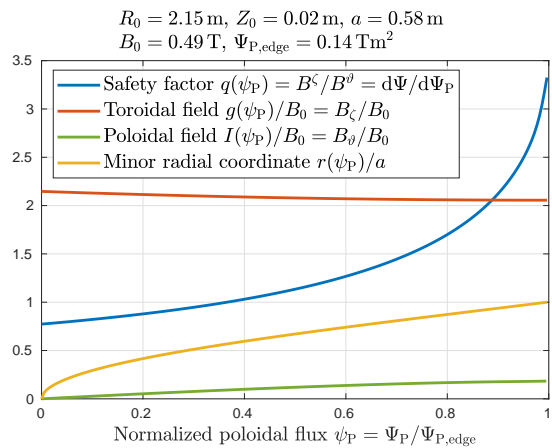}
  \caption{Equilibrium parameters and profiles for the magnetic field in Eq.~(\protect\ref{eq:bfield_eq}). This is the same case as in Ref.~\protect\cite{White20}.}
  \label{fig:model_profiles_fire}
\end{figure}

In the present work, we use the {\tt ORBIT} code \cite{White84} with the reduced $\delta f$ field-particle interaction model derived by Pinches {\it et al}.\ \cite{Pinches98}, extended by Chen {\it et al}.\ \cite{ChenY99}, and reformulated by White {\it et al}.\ \cite{WhiteTokBook3,White20}. First simulations of chirping using {\tt ORBIT} were reported in \cite{White19,White20}. We use the same MHD equilibrium as in {\cred those recent papers, which resembles a plasma of the conceptual reactor FIRE with reduced} field strength $B_0 = 0.49\,{\rm T}$ at the magnetic axis located at $R_0 = 2.15\,{\rm m}$. The profiles characterizing the magnetic geometry are shown in Fig.~\ref{fig:model_profiles_fire}. The following Eqs.~(\ref{eq:bfield})--(\ref{eq:pcan}) are normalized by the deuteron cyclotron frequency $\Omega_{{\rm c}0} = {\cred Q} e B_0 / M$ ({\cred inverse} time unit) and the magnetic flux density $B_0$ (field strength unit) at the magnetic axis.

The time-independent axisymmetric magnetic field ${\bm B}$ of the torus and time-dependent perturbations that resemble ideal incompressible electromagnetic flute modes are represented in Boozer coordinates as
\begin{subequations}
\begin{align}
{\bm B} &= g\nablab\zeta + I\nablab\vartheta + B_{\Psi_{\rm P}}\nablab\Psi_{\rm P} \;\, \text{(equilibrium)},
\label{eq:bfield_eq}
\\
\delta {\bm B} &= \nablab\times\alpha{\bm B} \quad\quad\quad\quad\quad \;\;\;\, \text{(perturbation)},
\label{eq:bfield_pert}
\\
\hat{\alpha}_{k,m} &= \frac{(nq - m)}{(gq + I)} \frac{\hat{\Phi}_{k,m}}{\cred \omega_{0k}} \quad\quad\quad \text{\normalsize (ideal MHD)}, \nonumber
\\
&= \frac{(m - nq)}{(mg + nI)} \frac{\hat{\xi}_{k,m}^\Psi}{q} \quad\quad\;\; \substack{\text{\normalsize (radial} \\ \text{\normalsize displacement),}}
\label{eq:bfield_ideal}
\end{align} \vspace{-0.2cm}
\label{eq:bfield}
\end{subequations}

\noindent where $\alpha = \delta{\bm A}\cdot{\bm B}/B^2 = \delta A_\parallel/B$ {\cred represents} the parallel component of the perturbed vector potential, $\Phi$ is the electrostatic potential, $q(\psi_{\rm P}) = {\rm d}\Psi/{\rm d}\Psi_{\rm P}$ is the field line helicity (tokamak safety factor), and $\Psi$ the toroidal flux.\footnote{Strictly speaking, our $\Psi$ and $\Psi_{\rm P}$ are the fluxes divided by $2\pi$.}
The coordinate surfaces $\psi_{\rm P} = {\rm const}$.\ and $\vartheta = {\rm const}$.\ of the equilibrium field ${\bm B}$ can be seen in Fig.~\ref{fig:intro_fire-A_beat}(b,c) (orange lines). The fluctuating potentials $\alpha$ and $\Phi$ are decomposed like $\xi$ in Eq.~(\ref{eq:mode_xi}). The radial profile of the contravariant {\cred radial} component $\hat\xi_{k,m}^\Psi = \hat{\bm\xi}_{k,m}\cdot\nablab\Psi/B_0$ of the ideal MHD displacement vector defined in Eq.~(\ref{eq:mode}) is given as input (specified below).

The stepping equations for the amplitude and phase appearing in the Fourier representation of Eq.~(\ref{eq:mode}) are\footnote{Our Eqs.~(\protect\ref{eq:evol})--(\protect\ref{eq:d}) correspond to those in Ref.~\protect\cite{White20}, except that here the lengths are not normalized by $R_0$, so the velocity $v_{\rm A0}$ (normalized by $\Omega_{\rm c0}$) appears here instead of the frequency $\nu_{\rm A0} = v_{\rm A0}/R_0$. Moreover, we let $D_k = 1$ since only a single Fourier harmonic $m/n=6/5$ is used.}
\begin{subequations}
\begin{align}
\frac{{\rm d}A_k}{{\rm d}t} &= \frac{-v_{\rm A0}^2}{{\cred \omega_{0k}} D_k} \sum\limits_{j=1}^{N_{\rm p}}\sum\limits_m w_{k,j} \Re\{S_{k,m,j}\} - \gamma_{\rm d}A_k,
\label{eq:evol_amp}
\\
\frac{{\rm d}\phi_k}{{\rm d}t} &= \frac{-v_{\rm A0}^2}{{\cred \omega_{0k}} A_k D_k} \sum\limits_{j=1}^{N_{\rm p}}\sum\limits_m w_{k,j} \Im\{S_{k,m,j}\},
\label{eq:evol_phi}
\end{align} \vspace{-0.2cm}
\label{eq:evol}
\end{subequations}

\noindent with the on-axis Alfv\'{e}n velocity $v_{\rm A0}$, particle index $j$, and
\begin{gather}
S_{k,m,j} = \left[\left(\rho_\parallel B^2 \hat{\alpha}_{k,m}(\psi_{\rm P}) - \hat{\Phi}_{k,m}(\psi_{\rm P})\right)e^{i\Theta_{k,m}}\right]_j,
\label{eq:s}
\\
D_k = 4\pi^2 \sum_m \int{\rm d}\psi_{\rm P}\, [\hat{\xi}_{k,m}^\Psi(\psi_{\rm P})]^2,
\label{eq:d}
\end{gather}

\noindent where $\rho_\parallel = v_\parallel/B = (g\dot{\zeta} + I\dot{\vartheta})/B^2$ with the parallel guiding center (GC) velocity $v_\parallel = {\cred {\bm v}_{\rm gc}}\cdot{\bm B}/B$. The GC phase space is sampled by $j = 1...N_{\rm p}$ particles, whose distribution $G(\psi_{\rm p},\vartheta,\zeta,K,\mu,t)$ satisfies ${\rm d}G/{\rm d}t = 0$ and whose weights $w_j = \delta f_j/G$ are evolved as
\begin{equation}
\frac{{\rm d}w}{{\rm d}t} = \frac{1}{G}\frac{{\rm d}F_0}{{\rm d}t} = -\left(\frac{F_0}{G_0} - w\right)\frac{{\rm d}\ln F_0}{{\rm d}t},
\label{eq:w}
\end{equation}

\noindent beginning with $\delta f(t=0) = 0$. Here, $\delta f$ represents the perturbation around the initial reference distribution $F_0$ of physical {\cred guiding centers}, and $G_0 = G(t=0)$ is the initial distribution of simulation particles (= phase space markers) that will be specified below. The particle trajectories are determined by {\cred solving equations for $\rho_\parallel(t)$, $\Psi_{\rm P}(t)$, $\vartheta(t)$ and $\zeta(t)$, whose explicit form can be found in Ref.~\cite{White84} or Section~3.9.2 of Ref.~\cite{WhiteTokBook3}. These equations of motion can be compactly summarized} in Hamiltonian form,
\begin{gather}
\dot{\vartheta} = \partial_{P_\vartheta} H, \;\; {\cred \dot{\tilde{P}}}_\vartheta = -\partial_\vartheta H, \nonumber
\\
\dot{\zeta} = \partial_{P_\zeta} H, \;\; {\cred \dot{\tilde{P}}}_\zeta = -\partial_\zeta H,
\label{eq:gc}
\end{gather}

\noindent with the Hamiltonian
\begin{equation}
H = {\cred \rho_\parallel^2} B^2/2 + \mu B + \Phi,
\label{eq:h}
\end{equation}

\noindent and the canonical angular momenta {\cred
\begin{equation}
{\cred \tilde{P}}_\zeta = {\cred (\rho_\parallel + \alpha)}g - \Psi_{\rm P}, \quad
{\cred \tilde{P}}_\vartheta = {\cred (\rho_\parallel + \alpha)}I + \Psi.
\label{eq:pcan_alpha}
\end{equation}

\noindent For marker particle loading and phase space diagnostics, we use the canonical toroidal angular momentum of the unperturbed system,}
\begin{equation}
P_\zeta = g\rho_\parallel - \Psi_{\rm P} {\cred\, = \tilde{P}_\zeta - \alpha g}.
\label{eq:pcan}
\end{equation}

\begin{figure*}[tp]
  \centering
  \includegraphics[width=15.9cm,clip]{\figures/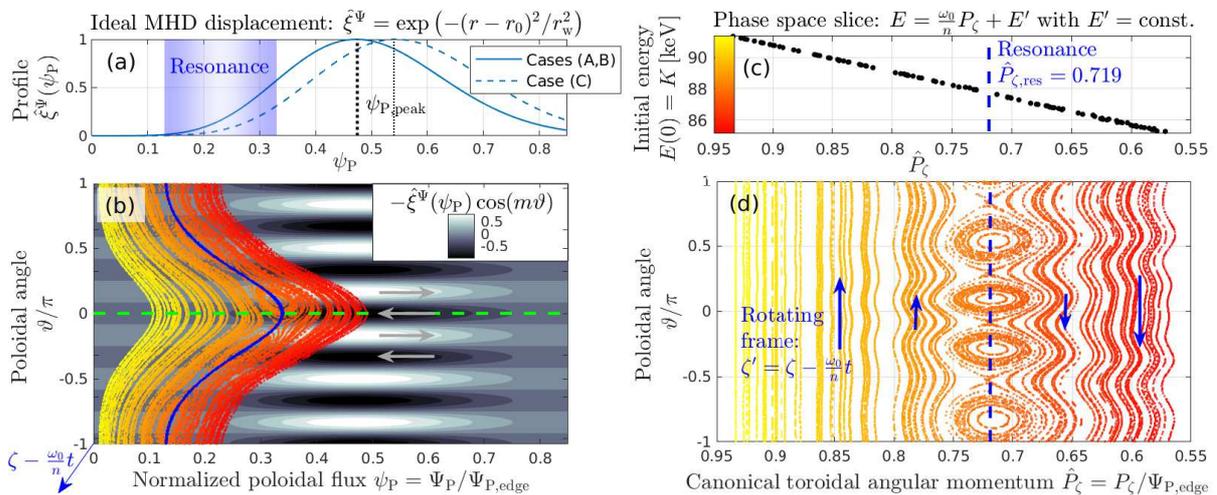} 
  \caption{Structure of the field mode and seed resonance. The panels on the left show (a) the radial profile $\hat{\xi}^{\Psi}(\psi_{\rm P})$ and (b) gray-scale contours of the mode structure in the $(\psi_{\rm P},\vartheta)$-plane in the toroidally rotating frame of reference where $\zeta' = \zeta - \frac{\omega_0}{n} t = 2\pi l$ with integer $l$. The dots in (c) represent a uniformly randomized distribution of 100 particles loaded along the line $E' = E - \frac{\omega_0}{n}P_\zeta = {\rm const.}\, (\approx 75.6\,{\rm keV})$ in the range $0.1 \leq \psi_{\rm P} \leq 0.5$ on the outer midplane $\vartheta = 0$ (horizontal dashed line in (b)). Kinetic Poincar\'{e} plots of the particle trajectories in the perturbed system with fixed amplitude $A_0 = 10^{-3}$ and phase $\phi = 0$ are shown in (b) and (d), with colors representing energy $E = K + {\cred Q}e \Phi$. The seed resonance for this mode with $\omega_0 = 2\pi\times 100\,{\rm kHz}$ and $m/n = 6/5$ is located at $\hat{P}_{\zeta,{\rm res}} = 0.719$. It has $p = 4$ elliptic points in $(\hat{P}_\zeta,\vartheta)$ and $\ell n = 5$ elliptic point in $(\hat{P}_\zeta,\zeta)$ (not shown). In the simulations, particles are loaded uniformly randomized in $\psi_{\rm P}^2$, whereas here they are uniformly randomized in $\psi_{\rm P}$ and followed for 500 transit times $\tau_{\zeta 0}$.}
  \label{fig:model_fire-ABC_poink}
\end{figure*}

Equations (\ref{eq:gc}) and (\ref{eq:h}) for the particle motion can be regarded as exact within the realm of drift-kinetic theory.\footnote{The polarization drift and the ponderomotive force \protect\cite{Cary09} are ignored. All other aspects of GC motion are retained to the extent permitted by a Hamiltonian formulation in Boozer coordinates, which requires neglecting the field component $B_{\Psi_{\rm P}}$ (denoted by ``$\delta$'' in Eq.~(2.26) of Ref.~\protect\cite{WhiteTokBook3}) that arises from the coordinates' nonorthogonality. The model is appropriate for fast ions interacting with shear Alfv\'{e}n modes.}
The field-particle coupling is overestimated by the neglect of the gyroaveraging effect, but otherwise the assumptions made in the derivation hold in the parameter regime considered here. Meanwhile, Eq.~(\ref{eq:evol}) for the field should be regarded as a {\it toy model} because the derivation involves the neglect of terms containing products of more than one time derivative, such as $\partial_t^2 A_k$, $\partial_t^2\phi_k$, $\partial_t A_k\partial_t\phi_k$. These terms are not strictly negligible by mere ordering arguments in a system exhibiting rapid amplitude pulsations and phase jumps that are typical for cases with nonlinear chirping (Fig.~\ref{fig:intro_fire-A_beat}). In addition, the factorization (\ref{eq:mode}) is based on a neglect of the background plasma response (continuous spectra, etc.) as discussed in Section~\ref{sec:model_justify} above. Here, we accept these caveats and simply postulate that the field mode evolves in the manner prescribed by Eq.~(\ref{eq:evol}).

The model parameters are listed in Table~\ref{tab:parm_model}. The fluctuating field in our simulations is initiated with a seed frequency of $\nu_0 = \omega_0/(2\pi) = 100\,{\rm kHz}$ and contains only a single toroidal harmonic $n = 5$ and a single poloidal harmonic $m = 6$, so we omit the mode label $k$. Its radial profile has a Gaussian shape,
\begin{equation}
\hat{\xi}^\Psi(r) = \exp\left((r - r_0)^2/r_{\rm w}^2\right),
\end{equation}

\noindent as shown in Fig.~\ref{fig:model_fire-ABC_poink}(a). Contour plots of the mode structure in the toroidally rotating poloidal plane at $\zeta' = \zeta - \frac{\omega_0}{n} t = 2\pi l$ with integer $l$ are shown in Fig.~\ref{fig:intro_fire-A_beat}(b) in cylinder coordinates $(R,Z,\zeta')$ and Fig.~\ref{fig:model_fire-ABC_poink}(b) in polar coordinates $(\psi_{\rm P},\vartheta,\zeta')$. Due to the large magnetic drifts performed by the circulating energetic deuterons in our simulations, which have kinetic energies $K \geq K_0 = 85\,{\rm keV}$ and magnetic moment $\mu B_0 = 2\,{\rm keV}$, this mode has a rich harmonic content from the point of view of the particles \cite{Bierwage14}, so that efficient transit resonances \cite{White21b}
\begin{equation}
\ell \omega = \ell n\omega_\zeta - p\omega_{\rm pol}
\label{eq:res}
\end{equation}

\noindent with various values of $p$ are possible. As one can infer from the kinetic Poincar\'{e} plot in Fig.~\ref{fig:model_fire-ABC_poink}, our seed resonance has $p=4$ poloidal elliptic points. In the $(\hat{P}_\zeta,\zeta)$ plane (not shown), there are $\ell n = 5$ elliptic points, so $\ell = 1$. The quantities $\omega_{\rm pol} = 2\pi/\tau_{\rm pol}$ and $\omega_\zeta = \Delta\zeta(\tau_{\rm pol})/\tau_{\rm pol} = 2\pi/\tau_\zeta$ in Eq.~(\ref{eq:res}) are the angular frequencies for an unperturbed transit in the poloidal and toroidal directions, respectively. The toroidal transit period on axis,
\begin{equation}
\tau_{\zeta 0} \equiv \frac{2\pi}{\sqrt{2 K_0/M}} \underbrace{\frac{q_0 g_0 + I_0}{q_0 B_0}}\limits_{\approx R_0 = 2.15\,{\rm m}} \approx 4.7\,\mu{\rm s},
\end{equation}

\noindent will sometimes be used here as a unit of time: $\hat{t} \equiv t/\tau_{\zeta 0}$.

\begin{table}[tb]
\centering
\begin{tabular}{ccc}
\hline\hline
Num.\ of marker particles & $N_{\rm p}$ & $10^6$ \\
Loading range @ $\vartheta=0$ & $[\psi_{\rm P,min},\psi_{\rm P,max}]$ & $[0.1,0.5]$ \\
Kinetic energy @ $\psi_{\rm P,max}$ & $K_0\,[{\rm keV}]$ & $85$ \\
Magnetic moment & $\mu B_0\,[{\rm keV}]$ & $2$ \\
Loading duration &  $T_{\rm load}/\tau_{\zeta 0}$ & $10$ \\
Pushing time step & $\Delta t_{\rm step}/\tau_{\zeta 0}$ & $\tfrac{1}{400}$, $\tfrac{1}{1000}$ \\
\hline\hline
\end{tabular}
\caption{Default parameters for particle loading and pushing.}
\label{tab:load}
\end{table}

\subsection{Initialization and stepping}
\label{sec:model_qs}

Table~\ref{tab:load} {\cred contains} the default parameters used for loading and advancing the simulation particles, which is done using a standard 4th-order Runge-Kutta scheme. With time steps of size $\Delta t_{\rm step}/\tau_{\zeta 0} = 1/400$ (damped cases) or $1/1000$ (undamped cases), we obtain an accuracy that is sufficient for our purposes (see Appendix~\ref{apdx:model_convergence}).

The default number of particles in our simulations is $N_{\rm p} = 10^6$. An approximate quiet start is obtained by loading only a portion of the simulation particles at each time step $\Delta t_{\rm step}$ during an interval of 10 transit times $\tau_{\zeta 0}$ (see Appendix~\ref{apdx:model_qs}). The particles are injected at the outer midplane at $\vartheta = 0$ (horizontal dashed line in Fig.~\ref{fig:model_fire-ABC_poink}(b)), where we distribute them along the toroidal angle and normalized poloidal flux as $\zeta_j = 2\pi x_j$ and $\psi_{{\rm P}j} = \psi_{\rm P,min} + (\psi_{\rm P,max}-\psi_{\rm P,min})x_j^{1/2}$ with uniform random numbers $0 \leq x_j \leq 1$. With uniform particle weighting ($F_0/G_0 = {\rm const.}$), this choice yields an approximately uniform gradient $\partial_{P_\zeta}F_0 \approx -\partial_{\Psi_{\rm P}}F_0 = {\rm const}$.\footnote{In the radial direction, the initial marker distribution $G_0$, which measures the number of simulation particles ${\rm d}N$ per cell ${\rm d}\psi_{\rm P}$ is chosen to satisfy $G_0 \propto {\rm d}N/{\rm d}\psi_{\rm P} \propto \psi_{\rm P}$, which is realized by loading particles with ${\rm d}N/{\rm d}x = {\rm const}$.\ and a uniformly randomized variable $x \propto \psi_{\rm p}^2$. For uniform particle weighting $F_0/G_0 = {\rm const}$., this yields $\partial_{\psi_{\rm P}}F_0 = \partial_{\psi_{\rm P}}^2 N \propto \gamma_{\rm L} = {\rm const}$. With $\psi_{\rm P} = (r/a)^2$ this corresponds to an EP density profile of the form $n_{\rm EP}(r) = \frac{{\rm d}N}{r{\rm d}r} \propto \frac{{\rm d}N}{{\rm d}\psi_{\rm P}} \propto 1 - \psi_{\rm P} = 1 - (r/a)^2$.}

\begin{figure}[tp]
  \centering
  \includegraphics[width=8cm,clip]{\figures/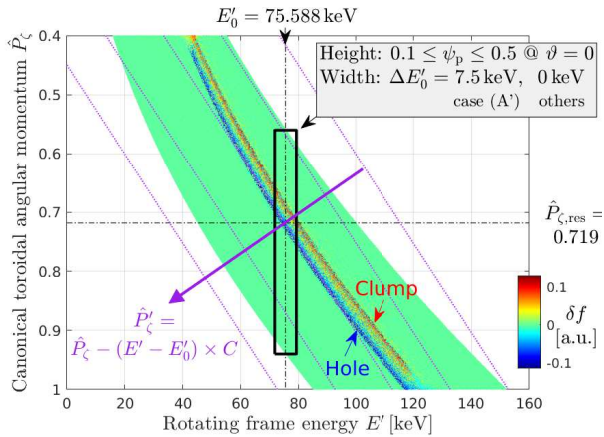}
  \caption{Phase space structures (e.g., holes and clumps) have the form of bent cylinders in $(\hat{P}_\zeta,E',\vartheta)$-space. This plot shows a projection of the marker distribution into the $(\hat{P}_\zeta,E')$-plane, with colors representing $\delta f$. The dynamics are primarily in the vertical direction ($E' = {\rm const}$.), so we simulate only a thin slice of width $\Delta E'_0$ as indicated by the black rectangle. In most cases, only the center line was simulated; i.e., $\Delta E'_0 \approx 0$. The only exception is case (A') in Figs.~\ref{fig:discuss_fire-A-pew_overview} and \ref{fig:discuss_fire-A-pew_C-evol} of Appendix~\ref{apdx:model_pew}, where $\Delta E'_0 = 7.5\,{\rm keV}$. In that region, the phase space structures are approximately perpendicular to the axis of the modified radial coordinate $\hat{P}'_\zeta = \hat{P}_\zeta - (E' - E'_0)\times C$ with $C \approx n/\omega_0$, which is indicated by the long diagonal arrow.}
  \label{fig:model_slice}
\end{figure}

The magnetic moment is fixed ($\mu B_0 = 2\,{\rm keV}$) and the kinetic energies are chosen to satisfy \cite{Hsu92}
\begin{equation}
E' = E - \omega_0 P_\zeta/n = {\rm const}.
\label{eq:enr_rot}
\end{equation}

\noindent as shown in Fig.~\ref{fig:model_fire-ABC_poink}(c). The simulations reported here {\cb were} obtained with $E(t=0) = K_0 = 85\,{\rm keV}$ at $(\psi_{\rm P},\vartheta) = (\psi_{\rm P,max},0)$, so that $E'_0 \approx 75.6\,{\rm keV}$. During the course of the simulations, deviations around this value will be on the order of 3\%, so we can assume $E' \approx E_0' = {\rm const}$. Effectively, we are simulating a thin slice of the phase space structures, which have a cylindrical shape in the $(\hat{P}_\zeta,E',\vartheta)$-space. Figure~\ref{fig:model_slice} shows a 2-D projection in the $(\hat{P}_\zeta,E')$-plane. The black box indicates a slice of width $\Delta E'_0 = 7.5\,{\rm keV}$ as used in a simulation performed for validation purposes (see Appendix~\ref{apdx:model_pew}). All simulations discussed in the main part of this paper were performed with $\Delta E'_0 = 0$.

After loading the markers during the interval $-10 \leq \hat{t} \leq -\Delta\hat{t}$, with $\hat{t} \equiv t/\tau_{\zeta 0}$, the kinetic instability is kick-started at $t=0$ with a small perturbation $A_0$. We use $A_0 = 10^{-6}$ in the marginally unstable case (A). In the strongly unstable case (B), we sometimes use a smaller $A_0 = 10^{-8}$ with the purpose of extending the duration of exponential growth enough to capture the onset of chirping in the Fourier spectrogram.

A comprehensive characterization of our simulation model, including convergence and sensitivity tests, is given in Appendix~\ref{apdx:model}. Note that if one starts the instability simulation with a relaxed non-axisymmetric EP distribution that has adapted to the initial perturbation of the field, as in Fig.~\ref{fig:intro_fire-A-freeze500_poink-df}, the system immediately enters the stage of nonlinear beating, effectively skipping the exponential growth and the first beat. This is demonstrated in Apendix~\ref{apdx:model_freeze500}.

\begin{figure*}[tp]
  \centering
  \includegraphics[width=16cm,clip]{\figures/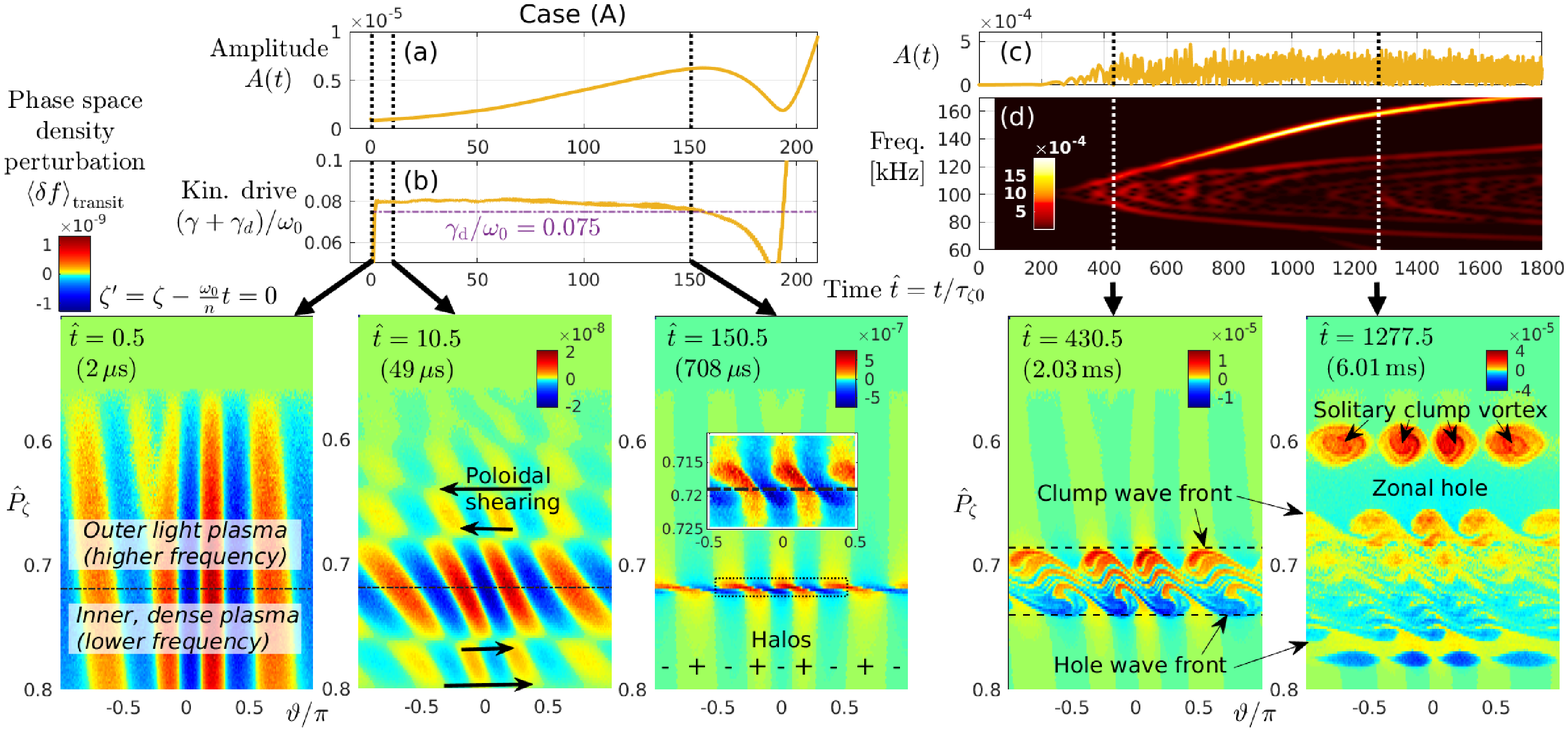}
  \caption{Overview of the dynamic stages in the marginally unstable case (A) (same as in Fig.~\protect\ref{fig:intro_fire-A_chirp}). Panels (a) and (b) show time traces of the field amplitude $A(t)$ and {\cred the instantaneous} kinetic drive (growth rate $\gamma(t)$ plus damping $\gamma_{\rm d}$) during the first 210 transits. Panel (c) shows $A(t)$ for a larger portion of the simulation ($1800\,\tau_{\zeta 0} \approx 8.5\,{\rm ms}$) and panel (d) shows the corresponding Fourier spectrogram obtained with a $0.47\,{\rm ms}$ time window. The vertical dotted lines in (a)--(d) indicate the times at which the five snapshots of the $\delta f$-weighted kinetic Poincar\'{e} plots were taken that are shown in the lower part of the figure. The last two snapshots are the same as those shown in Fig.~\protect\ref{fig:intro_fire-A_df}, except that the contrast at small amplitudes is enhanced here by using a nonlinear color scale (square root). As in all contour plots of $\delta f$ that appear in this work, we have inverted the direction of the $\hat{P}_\zeta$ axis for easier comparison with the spectrograms, so that structures with higher frequencies (smaller $\hat{P}_\zeta$, larger radii) appear at the top (cf.~Fig.~\protect\ref{fig:stages_fire_freq-pzeta}).}
  \label{fig:stages_fire-A}
\end{figure*}

\begin{figure*}[tp]
  \centering
  \includegraphics[width=16cm,clip]{\figures/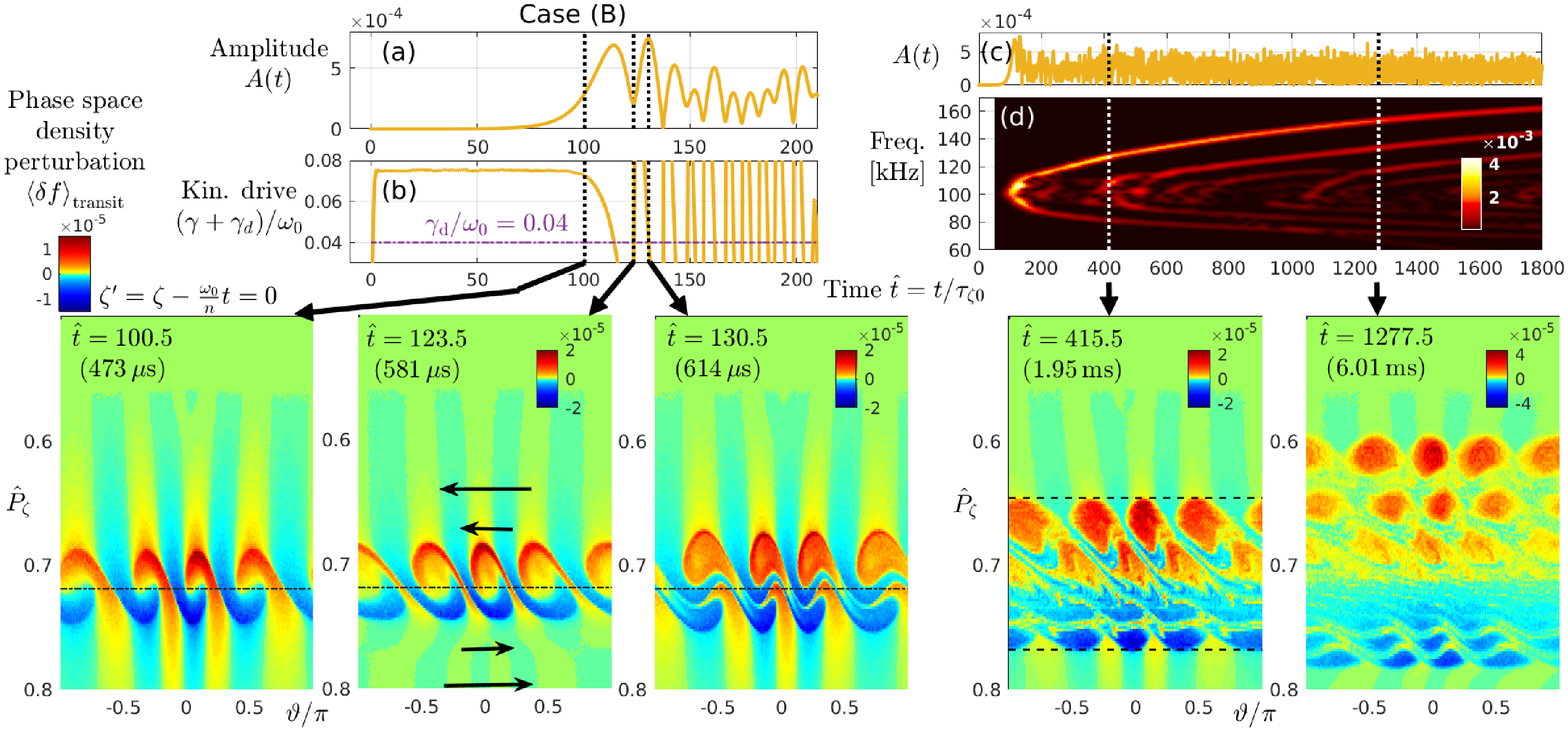}
  \caption{Overview of the dynamic stages in the strongly unstable case (B) with reduced initial perturbation amplitude, $A_0 = 10^{-8}$. Results are arranged as in Fig.~\protect\ref{fig:stages_fire-A}. The first two snapshots of Fig.~\protect\ref{fig:stages_fire-A}, where the resonance gradually emerges, are very similar in the present case, so they are not shown again. Instead, we show additional snapshots of $\delta f$ during the first two beats ($100 \lesssim \hat{t} \lesssim 140$).}
  \label{fig:stages_fire-B}
\end{figure*}

\section{Structure formation and saturation}
\label{sec:saturation}

As a preparation for studying the effects of beating, we examine in this section the processes of structure formation, nonlinear saturation and the onset of beating in cases (A) and (B), whose parameters are given in Table~\ref{tab:parm_model}. Case (A) is marginally unstable, with a linear drive $\gamma_{\rm L}/\omega_0 \approx 8\%$ only slightly larger than the damping $\gamma_{\rm d}/\omega_0 = 7.5\%$. Case (B) is strongly unstable, with $\gamma_{\rm L}/\omega_0 \approx 7.5\%$ nearly twice as large as $\gamma_{\rm d}/\omega_0 = 4\%$. Both cases exhibit strong chirping and an overview of the observations made in different stages of these simulations is given in Figs.~\ref{fig:stages_fire-A} and \ref{fig:stages_fire-B}.

Panels (a) and (c) show the time trace of the amplitude $A(t)$ obtained by solving Eq.~(\ref{eq:evol_amp}). The kinetic drive plotted in panel (b) was measured by computing the exponential running average $\gamma(t) = 0.01\times \frac{\Delta A(t)}{A(t)\Delta t} + 0.99\times\gamma(t-\Delta t)$ of the increment $\Delta A$ on the right-hand side of Eq.~(\protect\ref{eq:evol_amp}) and adding the damping rate $\gamma_{\rm d}$. The {\cred above} values of the linear drive $\gamma_{\rm L}$ correspond to $\gamma_{\rm k} = \gamma + \gamma_{\rm d}$ during the early phase of (nearly) exponential growth. Panel (d) shows the spectrogram obtained by Fourier analyzing the signal $s(t) = A(t)\sin(-\omega t-\phi(t))$ with a sliding time window of size $\Delta t_{\rm win} = 100\,\tau_{\zeta 0} \approx 0.47\,{\rm ms}$.

The lower parts of Figs.~\ref{fig:stages_fire-A} and \ref{fig:stages_fire-B} show five snapshots of the perturbed EP phase space density as $\delta f$-weighted kinetic Poincar\'{e} plots. Our Poincar\'{e} section is the torodially rotating plane $\zeta' = \zeta - \frac{\omega_0}{n}t = 2\pi l$ with integer $l$, where particle weights $w = \delta f/G$ are accumulated in a histogram-like fashion on a mesh consisting of $200$ cells in the poloidal angle $-\pi\leq\vartheta\leq\pi$, and $200...1000$ cells in the {\cred radial} range $0.5 \leq \hat{P}_\zeta \leq 0.9$. The accumulation was performed for a short time interval of one toroidal transit, $\tau_{\zeta 0} \approx 4.7\,\mu{\rm s}$, {\cred during which} each particle crosses the $\zeta' = 2\pi l$ plane approximately $p = 4$ times. The resulting quantity
\begin{equation}
\left<\delta f\right>_{\rm transit}(\hat{P}_\zeta,\vartheta,t) = \int\limits_{t-\tau_{\zeta 0}/2}^{t+\tau_{\zeta 0}/2}{\rm d}t' \delta f(t'|\zeta'=2\pi l)
\end{equation}

\noindent will be referred to simply as ``$\delta f$''.

In the following subsections, we examine the dynamics occurring during the early part of the simulations, until the instabilities saturate; namely, the first three snapshots in Figs.~\ref{fig:stages_fire-A} and \ref{fig:stages_fire-B}. This includes the birth of hole-clump pairs and the onset of beating.

\subsection{Linear response, phase mixing and halos}
\label{sec:saturation_halo}

In the linearized system, the magnitudes of field perturbations, density perturbations and particle displacements bear no meaning. There exists only a trend that is expressed by the complex frequency $\tilde\omega = \omega + i\gamma$ whose value is directly determined by the gradients of the chosen reference state. Strictly speaking, resonant instabilities (also known as ``nonlinear Landau damping'') are never linear because they always involve finite displacements. However, nearly exponential growth does occur initially because the imposed finite-amplitude perturbation spontaneously induces flows with a phase relation such that $\partial_t\ln A \approx \partial_t\ln\delta f$ (independent of amplitude and position). This phase relation changes when nonlinearities begin to dominate over the linear response. This transition can be clearly seen at the beginning of our simulations in the form of a marked change in the structure of $\delta f$. In this subsection, we will take a closer look at this process, because some of the insights gained will be useful later.

\begin{figure}[tb]
  \centering
  \includegraphics[width=8cm,clip]{\figures/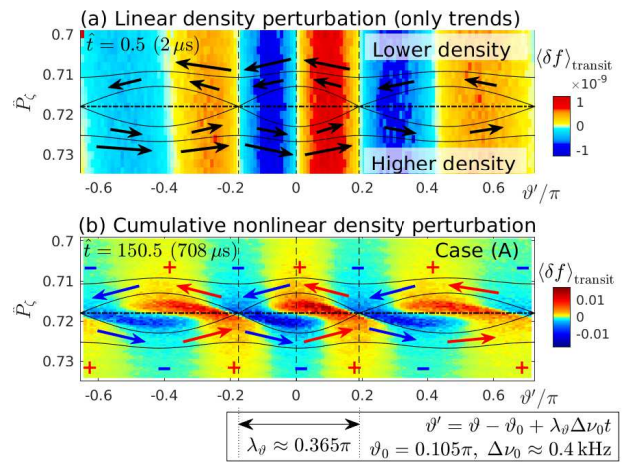}
  \caption{Illustration of the meaning of patterns in kinetic Poincar\'{e} contour plots of the EP phase space density perturbation $\delta f$. The data are taken from (a) the first and (b) the third snapshot of Fig.~\protect\ref{fig:stages_fire-A} for case (A). Here, only three of the resonance's four periods are visible. Black lines indicate schematically the effective positions of phase space islands and neighboring flow contours that one would obtain after waiting infinitely long with the field amplitude $A$ and phase $\phi$ fixed. Bold arrows indicate the flow direction. The island width is sort of realistic in (b) but strongly exaggerated in (a). In fact, the island does not really exist in this fully nonadiabatic stage of the simulation. Here and in the following, we sketch ``effective islands'' only to indicate the phase of the field and the {\cred instantaneous} direction of the associated flow lines. The phases in (a) and (b) have been aligned by using the modified poloidal angle $\vartheta' = \vartheta - \vartheta_0 + \lambda_\vartheta\Delta\nu_0 t$, where $\lambda_\vartheta$ is the wavelength of the central island (2-headed arrow), $\Delta\nu_0$ is the prompt frequency shift (cf.~Appendix~\protect\ref{apdx:model_opt}), and $\vartheta_0 = 0.105\pi$ aligns the O-point with $\vartheta' = 0$. Halos in (b) are indicated by alternating ``+'' and ``--'' signs.}
  \label{fig:model_fire-A_df-interpret}
\end{figure}

The structure of $\delta f$ in the first snapshot ($\hat{t} = 0.5$) in Fig.~\ref{fig:stages_fire-A} shows the {\it linear} short-time response of the EP distribution caused by the ${\bm E}\times{\bm B}$ drift associated with the abruptly applied oscillating field perturbation. The attribute ``linear'' refers to the fact that the structure of $\delta f$ reflects only the instantaneous trend of the motion of EP Vlasov fluid. No appreciable particle displacement has occurred yet, since only a very short time (few $\mu{\rm s}$) has passed and the field amplitude is still very small ($\approx 10^{-6}$). A zoom-up is shown in Fig.~\ref{fig:model_fire-A_df-interpret}(a), where black contour lines indicate schematically the structure of the resonance and the bold black arrows indicate the trend of fluid motion. Along the red stripes ($\delta f > 0$), there is a net outward motion of denser fluid into regions previously occupied by lighter fluid, causing a local increase of the EP phase space density. In the blue stripes ($\delta f < 0$), the density drops due to a net inward flow of lighter fluid.

Returning to the full-size snapshot in Fig.~\ref{fig:stages_fire-A}(left): The periodicity of $\delta f(\vartheta)$ indicates that a $p = 4$ resonance dominates in the region $\hat{P}_\zeta \gtrsim 0.62$ and a $p=5$ resonance dominates in the region $\hat{P}_\zeta \lesssim 0.62$. The $p=4$ seed resonance will appear at $\hat{P}_{\zeta,{\rm res}} = 0.718$ as indicated by a horizontal dashed-dotted line. However, in the first snapshot at $\hat{t} = 0.5$ in Fig.~\ref{fig:stages_fire-A} this resonance is not visible in the structure of $\delta f$ yet. This may be interpreted as a consequence of the fact that the short time interval that has passed corresponds to a large uncertainty $\Delta \nu \sim 1/\tau_{\zeta 0} \approx 200\,{\rm kHz}$ in frequency. Another (closely related) interpretation is that phase mixing (discussed below) has not had enough time to become effective in the frequency band we are looking at; it requires dozens of transits to become noticeable.

Of course, the most efficient coherent accumulation of phase space density perturbations does occur near the resonance, where the particles move more or less in phase with the field mode. For instance, after 10 transit times, the resonance width becomes approximately $\Delta\nu \sim 1/(10\tau_{\zeta 0}) \approx 20\,{\rm kHz}$, which corresponds to $0.7 \lesssim \hat{P}_\zeta \lesssim 0.74$ (cf.~Eq.~(\ref{eq:pz_nu})) and is precisely the region where $\delta f$ is large in the second snapshot at $\hat{t} = 10.5$ in Fig.~\ref{fig:stages_fire-A}.

Outside that (near-)resonant layer, the magnitude of $\delta f$ is smaller by an order of magnitude and represents the nonresonant ideal MHD displacement in Eq.~(\ref{eq:dr_mhd}). The corrugated structure of $\delta f$ in the off-resonant domain as seen in the second snapshot at $\hat{t} = 10.5$ in Fig.~\ref{fig:stages_fire-A} may be interpreted as a superposition of patterns produced by two effects that reflect the emergence of nonlinearities:
\begin{itemize}
\item  One effect is {\it phase mixing} of the remnants from the initial perturbation in $\delta f$ that was produced by the abruptly imposed field at $t=0$. Phase mixing is realized by the radially sheared poloidal propagation (horizontal arrows in Fig.~\ref{fig:stages_fire-A}) of the phase space structures due to the convective nonlinearity $\dot{\vartheta}\partial_\vartheta\delta f$.
\item  The second effect is that $\delta f$ in our Poincar\'{e} plots increasingly shows the {\it cumulative response}; namely, the time integral of the weight equation (\ref{eq:w}).
\end{itemize}

\noindent Phase mixing, also known as ``Landau damping'', causes a progressive tilting and thinning of the off-resonant red and blue stripes in our snapshots of $\delta f$. However, this is visible only for a short time because the second effect --- the cumulative $\delta f$ response --- eventually dominates.

Consequently, the meaning of the patterns we see in $\delta f$-weighted Poincar\'{e} plots changes when the system leaves the linear and enters the nonlinear regime. Let us elucidate this change of meaning using Fig.~\ref{fig:model_fire-A_df-interpret}, which shows zoom-ups of (a) the first snapshot ($\hat{t} = 0.5$) and (b) the third snapshot ($\hat{t} = 150.5$) of Fig.~\ref{fig:stages_fire-A}, highlighting the differences between the linear and nonlinear $\delta f$ structures.

When compared to the linear short-time response in Fig.~\ref{fig:model_fire-A_df-interpret}(a), one can see that the vertical stripes outside the resonant layer in Fig.~\ref{fig:model_fire-A_df-interpret}(b) are shifted to the left by $-\lambda_\vartheta/4$ above the resonance ($\hat{P}_\zeta < 0.718$) and to the right by $+\lambda_\vartheta/2$ below the resonance ($\hat{P}_\zeta > 0.718$), so that the sign of $\delta f$ inside those stripes flips across the resonance. Here, $\lambda_\vartheta$ is the poloidal wavelength of the $p=4$ resonance, whose value is about $\lambda_\vartheta \approx 0.365\pi$ in the center of Fig.~\ref{fig:model_fire-A_df-interpret}(b) as indicated by the two-headed arrow. (For our GC drift orbits, the value of $\lambda_\vartheta$ in Boozer coordinates increases from the outer midplane $\vartheta = 0$ towards $\vartheta = \pm \pi$.)

While the structure of the $\delta f$ pattern has changed in the way we have just described, the structure and phase of the resonance, which is schematically illustrated by black contour lines in Fig.~\ref{fig:model_fire-A_df-interpret}, remains the same in panels (a) and (b) in the frame where the (promptly shifted) resonance is stationary. Thus, the $\delta f$-weighted Poincar\'{e} plots in the nonlinear regime (b) do not represent the instantaneous trend as in (a), but the {\it accumulated} modulations in the EP phase space density relative to the background. In other words, the nonlinear $\delta f$ patterns in (b) tell us where dense fluid (red) and light fluid (blue) has been accumulated until that time; i.e., the history of transport. The red arrows in Fig.~\ref{fig:model_fire-A_df-interpret}(b) indicate where dense fluid has recently moved radially outward (resulting in a local density increase), and the blue arrows indicate where light fluid has moved radially inward (resulting in a local density reduction).

Another consequence of this cumulative nature of the nonlinear $\delta f$ patterns is that they are fairly robust. A particularly impressive example of this robustness is the fact that the off-resonant stripes indicated by alternating ``+'' and ``--'' signs in the third snapshot ($\hat{t} = 150.5$) in Fig.~\ref{fig:stages_fire-A} and in Fig.~\ref{fig:model_fire-A_df-interpret}(b) are almost perfectly vertical. These robust nonlinear density modulations, which seem to emanate radially from the nearest hole and clump wave fronts, will in the following be referred to {\cb as} {\it halos}.

The values of $\delta f$ inside the halos is small, so we use a nonlinear color scale (square root) when we want to visualize them clearly. Their amplitude tends to decrease with increasing distance from the resonant $\delta f$ structures, which can be attributed to phase mixing. However, the shearing effect of phase mixing is almost invisible. This implies that the density modulations that constitute these halos are decoupled from the particle motion along $\vartheta$. Thus, the halos are of an entirely collective nature.

Instead, the halos exhibit a large degree of {\it phase-locking} with respect to the field oscillations that, in turn, are controlled by the dominant resonant $\delta f$ structures. This implies that the halos can tell us the effective locations of O- and X-points as in Fig.~\ref{fig:model_fire-A_df-interpret}(b), so they are a welcome substitute for kinetic Poincar\'{e} plots like that in Fig.~\ref{fig:intro_fire-A-freeze500_poink-df}(a), which are not applicable when the field varies rapidly in time. For instance, the halos will help us to analyze the effect of phase jumps associated with a beating (Section~\ref{sec:result_beat}).

The processes that we have described in this subsection are universal for resonant field-particle interactions (regardless of growth) and, therefore, are qualitatively the same in all cases studied here. Due to this reason, we have omitted the phase mixing stage in Fig.~\ref{fig:stages_fire-B} and, instead, included additional snapshots of the saturation stage that will be discussed next. The subsequent dynamics depend on the parameters, so we choose to proceed with a comparison between the saturation processes with and without damping in the marginally unstable and strongly unstable case.

\begin{figure*}[tb]
  \centering
  \includegraphics[width=15.9cm,clip]{\figures/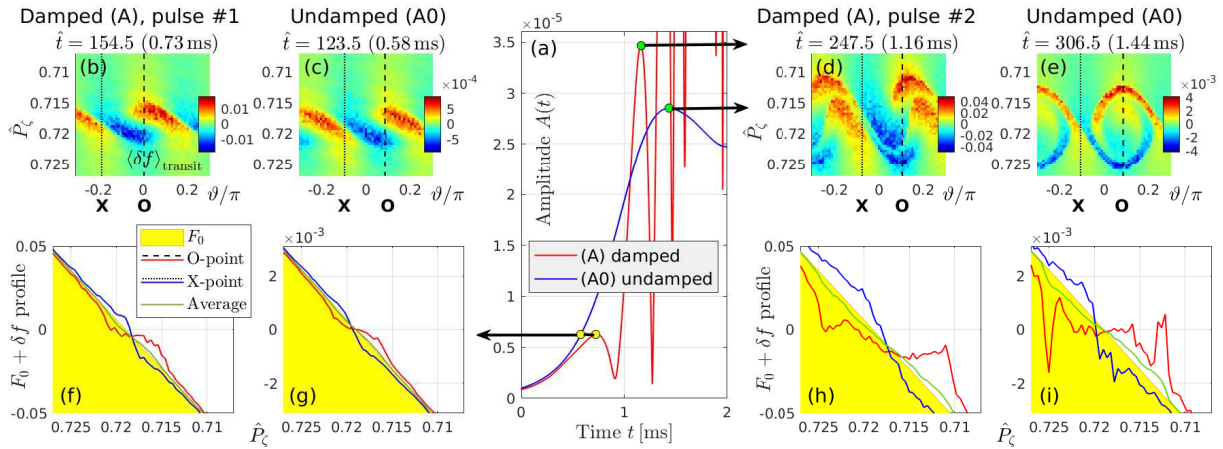}\vspace{-0.08cm}
  \caption{EP phase space density perturbation during the saturation of (A) damped and (A0) undamped instabilities in the marginally unstable cases with $\gamma/\omega_0 = 0.5\%$. Panel (a) in the center shows the time traces of the field amplitude $A(t)$. The situation at low amplitudes $A \approx 6\times 10^{-6}$, where case (A) peaks for the first time, is shown on the left-hand side. The situation at higher amplitudes around $3\times 10^{-5}$, where case (A) peaks for the second and case (A0) for the first time, is shown on the right-hand side. Panels (b)--(e) show contour plots of $\delta f(\hat{P}_\zeta,\vartheta)$, and panels (f)--(i) show the corresponding perturbed EP density profiles $F(\hat{P}_\zeta) = F_0 + \delta f$: one measured at the poloidal angle of the O-point, one at the X-point, and one poloidal average. The effective locations of O- and X-points were inferred from the halos (cf.~Section~\protect\ref{sec:saturation_halo}) and are indicated, respectively, by dashed and dotted vertical lines in panels (b)--(e).}
  \label{fig:result_fire-A-A0_saturation}
\end{figure*}

\begin{figure*}[tb]
  \centering
  \includegraphics[width=15.9cm,clip]{\figures/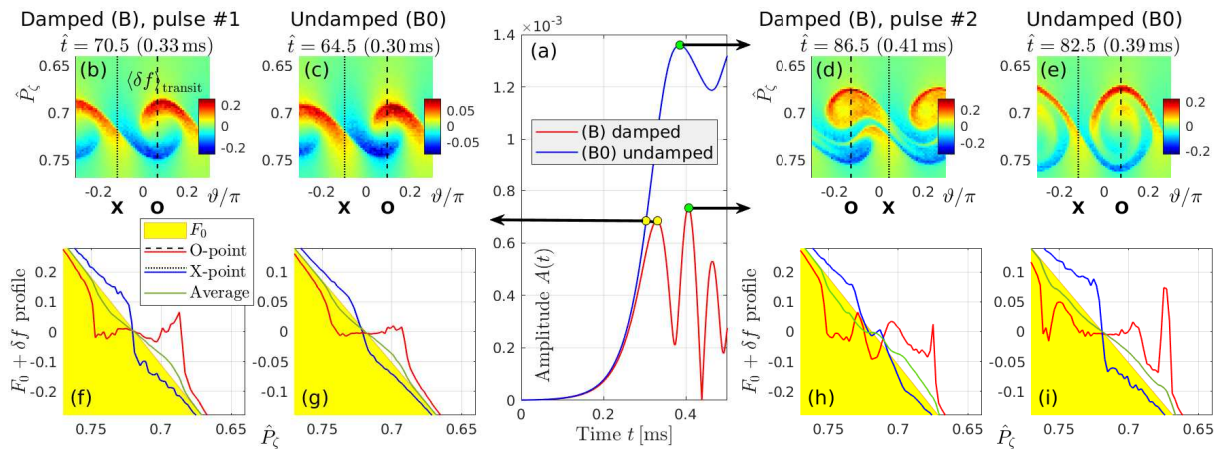}
  \caption{EP phase space density perturbation during the saturation of (B) damped and (B0) undamped instabilities in the strongly unstable cases with $\gamma/\omega_0 = 3.5\%$. The results are arranged as in Fig.~\protect\ref{fig:result_fire-A-A0_saturation}. The initial perturbation amplitude in these simulations was $A_0 = 10^{-6}$, so the first saturation in case (B) occurs earlier here than in Figs.~\protect\ref{fig:stages_fire-B}, \protect\ref{fig:result_spec-fronts}, \protect\ref{fig:result_fire-B_detach} and \protect\ref{fig:result_fire-AB_tracers}, where we used $A_0 = 10^{-8}$.}
  \label{fig:result_fire-B-B0_saturation}
\end{figure*}

\subsection{First saturation with and without damping}
\label{sec:saturation_damp}

Beating and nonadiabatic chirping begin during the first peak of a resonant instability, so it is instructive to examine the saturation process in some detail. We will do so here by comparing damped cases (A) and (B) that do chirp with undamped cases (A0) and (B0) that do not. The latter are obtained by setting $\gamma_{\rm  d} = 0$ and reducing the gradient $F_0' \propto \gamma_{\rm L}$ such that the growth rate $\gamma = \gamma_{\rm L} - \gamma_{\rm d}$ remains the same.

For the marginally unstable cases (A) and (A0), with $\gamma/\omega_0 \approx 0.5\%$, Figure~\ref{fig:result_fire-A-A0_saturation}(a) shows the evolution of the field amplitude $A(t)$ during the first $2\,{\rm ms}$. After growing at the same exponential rate, the instability in case (A) saturates at a low amplitude of $A \approx 6\times 10^{-6}$ at time $\hat{t} = 154.5$ ($0.73\,{\rm ms}$). Panels (b) and (c) show the structure of the phase space density perturbation $\delta f$ in a portion of the $(\hat{P}_\zeta,\vartheta)$ {\cred plane} when $A \approx 6\times 10^{-6}$ in both cases (A) and (A0). The corresponding perturbations of the radial profile of the EP density distribution, $F(\hat{P}_\zeta) = F_0 + \delta f$, measured across the O-point (red) and across the X-point (blue) are shown in panels (f) and (g). A $\vartheta$-averaged density profile (green) is also plotted.

One can see that the O-point profiles in Fig.~\ref{fig:result_fire-A-A0_saturation}(f) for the damped case (A) and in panel (g) for the undamped case (A0) have both significantly reduced gradients at the seed resonance, with the profile of (A) appearing somewhat flatter than in (A0). Note that the vertical axes of panels (f) and (g) have different scales, which differ by a factor $F_0'{\rm (A)}/F_0'{\rm (A0)} \approx 16$. Flattening a steeper gradient requires a larger perturbation $\delta f$, which is indeed an order of magnitude larger in case (A) than in case (A0) on the left-hand side of Fig.~\ref{fig:result_fire-A-A0_saturation}, where the field amplitudes are equal. Recalling that the energy released during the interchange of denser and ligher EP Vlasov fluid that is represented by $\delta f$ has all been transferred to the field wave, the fact that the values of $\delta f$ in cases (A) and (A0) differ by an order of magnitude for equal field amplitudes, reflects the fact that the field in the damped case (A) has lost nearly all the energy that it had previously received from the initial EP density gradient.

In contrast to the undamped case, the field in the damped case does not possess enough energy to reverse the radial displacement of the particles. Equivalently, one may say that only a small amount of trapping occurs before the field's energy is exhausted and its amplitude drops to a small value. This explains the differences seen in the structure of $\delta f$ in panels (d) and (e) on the right-hand side of Fig.~\ref{fig:result_fire-A-A0_saturation}: While the EP phase space fluid revolves around the resonance in the undamped case (A0) in panel (e), the first-generation hole and clump waves in the damped case (A) in panel (d) has departed {\cred radially} from the resonance, closely followed by a second-generation hole-clump pair that has formed during the second pulse of the field.

This multi-layered structure of the near-resonant $\delta f$ patterns in panel (d) makes it difficult to determine the effective locations of O- and X-points, and this is where the halos introduced in Section~\ref{sec:saturation_halo} above come in handy. The halo-based O- and X-points are indicated in panel (d) by dashed and dotted lines, respectively. If one takes into account that the effective resonance has shifted by $\Delta\vartheta = -\lambda_\vartheta\Delta\nu_0\Delta t = -0.365\pi \times 0.4\,{\rm kHz} \times 0.43\,{\rm ms} \approx -0.06\pi$ (cf.~Fig.~\ref{fig:model_fire-A_df-interpret}) during the time between snapshots (b) and (d), it becomes clear that the O- and X-point locations are effectively reversed. This is the consequence of a phase jump $\Delta\phi = \pi$ that was performed by the field during the amplitude minimum between snapshots (b) and (d), around $t \approx 0.9\,{\rm ms}$, which will be examined in detail in Section~\ref{sec:result_beat}.

The dynamics in the strongly unstable cases (B) and (B0) in Fig.~\ref{fig:result_fire-B-B0_saturation}, with $\gamma/\omega_0 \approx 3.5\%$, seem to be qualitatively similar to the marginally unstable cases (A) and (A0), but everything is faster and larger in both magnitude and spatial extent. Here, the saturation amplitudes differ only by about a factor of 2, and the magnitudes of $\delta f$ differ by a factor of about 2...3, consistently with the ratio of the initial gradients, $F_0'{\rm (B)}/F_0'{\rm (B0)} \approx 2$.

After the undamped instabilities in cases (A0) and (B0) saturate, their amplitudes remain near the level of the first peak. At the instant of saturation, Figs.~\ref{fig:result_fire-A-A0_saturation}(i) and \ref{fig:result_fire-B-B0_saturation}(i), show an O-point profile that is flat around the resonance. {\cred In fact, the profile near the O-point can be expected to flatten for the first time after half a bounce cycle. Indeed, the first flattening of the O-point profile near the resonance can be seen long before the amplitude $A(t)$ reaches its first peak. In case (A0), with the resolution of our diagnostics,} this occurs at about 30\% of the peak amplitude (shortly after the profile snapshot in Fig.~\ref{fig:result_fire-A-A0_saturation}(g)), and in case (B0) this occurs at about 18\% (before the snapshot in Fig.~\ref{fig:result_fire-B-B0_saturation}(g)). At first glance, this might look like an overshoot, but that is not necessarily the case because, in order for the net outward flow of EPs (= net energy transfer to the mode) to cease, it does not suffice to flatten the gradient only along a thin line across the O-point. A sufficient amount of profile flattening has to occur in a sufficiently wide portion of the effective island's poloidal length $\lambda_\vartheta$, which is the distance between X-points along the resonance line as indicated in Fig.~\ref{fig:model_fire-A_df-interpret}(b). The structure of $\delta f$ inside the phase space islands in Figs.~\ref{fig:result_fire-A-A0_saturation}(e) and \ref{fig:result_fire-B-B0_saturation}(e) implies that $A(t)$ stops growing only after the deeply trapped particles have completed about 1 to 1\nicefrac{1}{2} revolutions\footnote{The first revolution is part of the resonant trapping process, but its duration is longer than the bounce period $\tau_{\rm b}$ because the first revolution includes the formation process of the field's effective potential well.}
around the nearby O-point (see also Fig.~\ref{fig:discuss_fire-A0-B0_spec-df}). {\cred In other words, the steady-state level around which the field amplitude $A(t)$ subsequently oscillates is reached when a majority of the particles have} more or less returned to or passed their initial positions. This observation is corroborated by the time traces of individual particles as shown in Fig.~\ref{fig:discuss_fire-A0-B0_tracers}(a,b) of Appendix~\ref{apdx:undamped} and is consistent with Dewar's theoretical prediction \cite{Dewar73}. 

Another noteworthy observation is that the saturation in the damped cases (A) and (B) occurs only after the density profile across the O-point has been flattened substantially. This is especially remarkable in the strongly damped marginally unstable case (A) in Fig.~\ref{fig:result_fire-A-A0_saturation}(f): since that case would be linearly stable if the initial density gradient had been just a little less steep, one might na\"{i}vely think that the instability {\cred would} saturate as soon as the gradient at the resonance has been reduced below that threshold, which should require only a small perturbation in $\delta f$. This, however, does not happen and it reflects the fact that, even during the phase of approximately exponential growth, the instantaneous stability of the perturbed system is a more complex problem than the linear stability of the initial reference state, whose gradient is uniform along the resonance. In the perturbed state, the gradient $\partial F/\partial P_\zeta$ with respect to the action variable $\hat{P}_\zeta$ varies along the angle variable $\vartheta$, which is precisely the feature we discussed in the previous paragraph, where we were saying that saturation in the undamped cases required the gradient to flatten in a certain range of $\vartheta$ around the O-point. In the damped cases, that range of $\vartheta$ is shorter, requiring only a small amount of revolution (trapping) around the resonance.

\subsection{Potentially misleading density profiles and the cause of pulsations and phase jumps}
\label{sec:saturation_x}

As mentioned in Section~\ref{sec:review_pump}, we emphasize again that density gradients $\partial_{P_\zeta} F$ can have a stabilizing or destabilizing effect only when there are flow lines in phase space that are directed uphill or downhill, respectively. Those lines of flow are not always easy to determine when the field's amplitude and phase evolve rapidly, and without that information, it can be difficult to interpret the role of the gradients correctly. For instance, the gradients seen in a radial cross-section at an O-point are relatively ineffective for resonant drive or damping, because they are oriented transversely to the lines of flow at that location (even after phase jumps by $\pm\pi$). Instead, it is primarily the degree of profile flattening in the region {\it between} the O- and X-points that determines at what stage a resonant instability ceases to grow (saturates), as mentioned in the previous {\cred section}. A simple direct link between destabilizing gradients and the growth of the field perturbation exists only in the linear regime, where the gradient is still uniform and the relative phases are such that $\partial_t\ln A \approx \partial_t\ln\delta f$ regardless of magnitude and position. The relative phases between the field and the density waves in $\delta f$ at different radii begin to change well before the instabilities saturate, and the evolution of the field is then determined by the field's spatial correlation with $\delta f$; namely, the integral in Eq.~(\ref{eq:mdl}), where gradients do not appear explicitly. The gradients enter the field equations only in a complicated implicit way through the time integration of the equations of motion --- or, equivalently, the Vlasov equation (\ref{eq:vlasov}) --- that yield $\delta f$.

\begin{figure}[tp]
  \centering
  \includegraphics[width=7.95cm,clip]{\figures/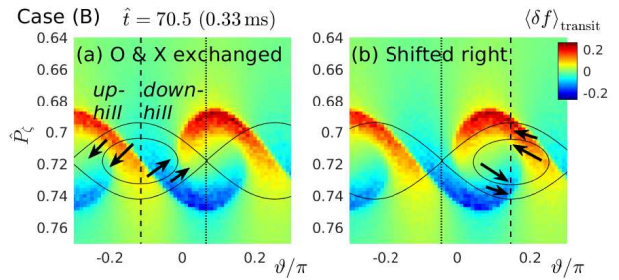}
  \caption{Schematic illustration of how the perturbed EP distribution $\delta f$ near the effective X-point in Fig.~\protect\ref{fig:result_fire-B-B0_saturation}(b) would be affected by the hypothetical phase space flows (arrows) that one would obtain (a) by exchanging O- and X-points, or (b) by slightly shifting the flow contours rightward ($\dot{\vartheta} > 0$, $\dot{\phi} < 0$). The situation in (a) corresponds to a phase jump by $\pm\pi$ and the situation in (b) corresponds to a negative phase shift by about $\Delta\phi \approx -\pi/3$.}
  \label{fig:result_fire-B_x-grad}
\end{figure}

These considerations show that radial profiles like those shown in the lower parts of Figs.~\ref{fig:result_fire-A-A0_saturation} and \ref{fig:result_fire-B-B0_saturation} should be interpreted with great caution. For instance, it might be tempting to hypothesize that the steep gradient at the X-point in Figs.~\ref{fig:result_fire-A-A0_saturation}(f) and \ref{fig:result_fire-B-B0_saturation}(f) is the force that causes the $\pm\pi$ jump in the field's phase $\phi(t)$ during the subsequent amplitude minimum {\cred and allows its amplitude $A(t)$ to grow again to a large value during the next pulse, where the O- and X-point positions are switched \cite{White20}. However, upon closer inspection} this intuitively appealing hypothesis turns out to be self-contradictory: As illustrated in Fig.~\ref{fig:result_fire-B_x-grad}(a), the form of $\delta f$ around the X-point is such that the gradients would actually be {\it stabilizing} when the X-point would be replaced by an O-point. This is because the counter-clockwise flows would have to move dense fluid (red) uphill and light fluid (blue) downhill. Even if one considers a gradual phase shift, a contradiction arises: Energy may be extracted from the X-point gradient if the flow contours would shift to the right ($\dot\vartheta > 0$) as illustrated in Fig.~\ref{fig:result_fire-B_x-grad}(b), which corresponds to a negative phase shift in the field ($\dot\phi < 0$). However, as we will show later, the first phase jump in our simulations is towards the left ($\dot\vartheta < 0$, $\dot\phi > 0$; cf.~Fig.~\ref{fig:result_fire-A_beating}(b)), which can be attributed to an asymmetry in our simulation setup that strengthens the leftward moving pump waves; {\cred namely, the clumps traveling downhill towards the mode's peak.}

In other words, from the energetic point of view, one may say that phase jumps and subsequent pulses occur not because but in spite of the steep X-point gradient. This statement holds not only at the time of the amplitude peak shown in Fig.~\ref{fig:result_fire-B_x-grad}, where we illustrated the situation for a hypothetical phase shift. The X-point gradient continues to play a stabilizing role all the way to the amplitude minimum, where the actual phase jump occurs (cf.~Fig.~\ref{fig:result_fire-A_beat2_island}).

This indicates that something else must be driving the amplitude pulsations and phase jumps, and our suspects are the counter-propagating EP fluid packets of increased ($\delta f > 0$) and decreased ($\delta f < 0$) density that can be seen in panel (b) of Figs.~\ref{fig:result_fire-A-A0_saturation} and \ref{fig:result_fire-B-B0_saturation}. In this work, we choose to interpret the rapid field pulsations and phase jumps as a consequence of the {\it beats} that result from the superposition of multiple {\cred density waves in $\delta f$ that simultaneously drive the field wave.} This is more easily said than proven in general, but at least in the limit of two-wave beating, one can demonstrate that our {\it ansatz} is consistent with Eq.~(\ref{eq:evol}), as shown in Appendix~\ref{apdx:beat}.

\subsection{Primordial hole-clump pairs, frequency splitting and the onset of beating}
\label{sec:saturation_beat}

The packets of dense and light fluid in Figs.~\ref{fig:result_fire-A-A0_saturation}(b) and (c) may be regarded as a {\it primordial hole-clump pair}. The hole wave propagates faster than the seed wave, and the clump wave lags behind, which means that there are now two pump waves in EP phase space density $\delta f$ with elevated and reduced frequencies $\omega^+$ and $\omega^-$ as was illustrated schematically in Fig.~\ref{fig:intro_onset}. The similarity between panels (b) and (c) in Fig.~\ref{fig:result_fire-A-A0_saturation} implies that such a pair of pump waves exists transiently also in the undamped case during the first cycle of hole-clump revolution around the O-point. Their effect is analyzed in Appendix~\ref{apdx:undamped}, where we present evidence for transient frequency splitting. In cases with strong damping, the frequency splitting is sustained and the superposition of the two pump waves produces a beating pump wave signal, to which the electromagnetic field wave responds likewise.

In the limit of symmetric beating between a pair of pump waves with equal amplitudes but opposite signs (cf.~Eq.~(\ref{eq:beat}), their combined signal is $\sin(\omega^+ t) - \sin(\omega^- t) = 2 \cos(\overline\omega t)\sin(\tfrac{1}{2}\Delta\omega t)$ with base frequency $\overline{\omega} = \tfrac{1}{2}(\omega^+ + \omega^-) \approx \omega_0$ and pulse frequency $\Delta\omega = \omega^+ - \omega^-$. Using Eq.~(\ref{eq:pz_nu}) in Appendix~\ref{apdx:model_calibration}, we can estimate the magnitude of frequency splitting and the initial pulse length of the beat,
\begin{equation}
\tau_{\rm pulse} = 1/\Delta\nu = 2\pi/\Delta\omega \approx 2\times 10^{-3}\,{\rm ms}/\Delta\hat{P}_\zeta,
\label{eq:tpulse}
\end{equation}

\noindent from the hole-clump separation in phase space, $\Delta\hat{P}_\zeta$. The results are summarized in Table~\ref{tab:beat1}.

\begin{table}[tbp]
\centering
\begin{tabular}{@{\hspace{0.11cm}}c@{\hspace{0.11cm}}|@{\hspace{0.11cm}}c@{\hspace{0.11cm}}|@{\hspace{0.11cm}}c@{\hspace{0.11cm}}}
\hline\hline Case & (A) & (B) \\
\hline
Hole-clump separation $\Delta\hat{P}_\zeta$ & $\lesssim 0.005$ & $\lesssim 0.07$ \\
Frequency split $\Delta\nu$ (upper bound) & $\lesssim 2.5\,{\rm kHz}$ & $\lesssim 35\,{\rm kHz}$ \\
Pulse length $\tau_{\rm pulse}$ (lower bound) & $\gtrsim 0.4\,{\rm ms}$ & $\gtrsim 0.03\,{\rm ms}$ \\
\hline\hline
\end{tabular}
\caption{Estimates based on Eq.~(\protect\ref{eq:tpulse}) of the frequency shift and beat pulse length corresponding the maximal radial separation $\Delta\hat{P}_\zeta$ of the primordial hole-clump pair during the first pulse in cases (A) and (B).}
\label{tab:beat1}
\end{table}

In the marginal case (A), we have $\Delta\hat{P}_\zeta \lesssim 0.005$ at the time of amplitude peaking in Fig.~\ref{fig:result_fire-A-A0_saturation}(b), which corresponds to a frequency split of $\Delta\nu \lesssim 2.5\,{\rm kHz}$ and a pulse length of $\tau_{\rm pulse} \gtrsim 0.4\,{\rm ms}$. The interval between the first peak and the first amplitude minimum in Fig.~\ref{fig:result_fire-A-A0_saturation}(a) is about $(0.91 - 0.73)\,{\rm ms} = 0.18\,{\rm ms}$, which corresponds to a beat pulse length of $0.36\,{\rm ms}$ close to that estimated from the hole-clump separation. This observation supports our interpretation of the first amplitude minimum as being a consequence of interference between the two pump waves constituted by the primordial hole-clump pair.

In the strongly unstable case (B), we have $\Delta\hat{P}_\zeta \lesssim 0.07$ in Fig.~\ref{fig:result_fire-B-B0_saturation}(b), which corresponds to an upper bound of $\Delta\nu \lesssim 35\,{\rm kHz}$ for the frequency splitting, and a lower bound of $\tau_{\rm pulse} \gtrsim 0.03\,{\rm ms}$ on the pulse length. A glance at the spectrogram in Fig.~\ref{fig:stages_fire-B}(d) shows that the actual frequency splitting at the time where the instability in case (B) saturates is significantly smaller than the upper limit of $35\,{\rm kHz}$. A detailed measurement shows that $\Delta\nu$ at that time is only about $10\,{\rm kHz}$ (cf.~Fig.~\ref{fig:result_spec-fronts}(f)), which corresponds to a pulse length of $\tau_{\rm pulse} \approx 0.1\,{\rm ms}$. This value is consistent with the interval between the first peak and the first amplitude minimum in Fig.~\ref{fig:result_fire-B-B0_saturation}(a), which is about $(0.58 - 0.54)\,{\rm ms} = 0.04\,{\rm ms}$ and corresponds to a beat pulse length of $\approx 0.08\,{\rm ms}$. As in case (A), this supports our interpretation of the first amplitude minimum as being a consequence of pump wave beating. The fact that the actual values of $\Delta\nu$ and $\tau_{\rm pulse}$ in case (B) differ significantly from their respective upper and lower bounds in Table~\ref{tab:beat1} also makes sense. The upper and lower bounds should yield accurate estimates only in the limit of marginal instability, which was indeed true for case (A) discussed in the previous paragraph. With increasing departure from marginality due to reduced damping as in case (B), the initial frequency splitting should become smaller as particle trapping becomes stronger and inhibits the phase slippage of the hole and clump waves with respect to the seed resonance. Eventually, in the absence of damping, sustained frequency splitting does not occur {\cred ($\Delta\nu \rightarrow 0$), whereas $\Delta\hat{P}_\zeta$ (= island width)} can be large.

The small (retarded) frequency splitting at early times means that weakly damped scenarios like our case (B) should undergo a relatively abrupt large frequency splitting at later times, when the large primordial hole-clump pair escapes the grasp of the seed resonance. In contrast, the onset of chirping should be a more gradual process in the marginally unstable case (A) with strong damping, where nearly no trapping occurs and phases slip when the hole-clump pair is still small. These expectations are confirmed numerically and supported by theory in Section~\ref{sec:result_front} below.

\section{Analysis of the role of beating}
\label{sec:beating}

In the above overview Figures~\ref{fig:stages_fire-A} and \ref{fig:stages_fire-B}, the snapshots of the EP phase space density perturbation $\delta f$ taken after the saturation of the instabilities show the following processes:
\begin{itemize}
\item[(i)] the formation of hole-clump wave pairs (until the third snapshot),
\item[(ii)] the radial propagation of hole and clump wave fronts, in whose wake a turbulent belt forms (fourth snapshot), and
\item[(iii)] solitary hole and clump vortices (fifth snapshot).
\end{itemize}

\noindent Here, the word ``front'' will be used to refer to the steep $\delta f$ gradients at the inner and outer edges of the turbulent belt around the seed resonance. These fronts are indicated by horizontal dashed lines in the fourth snapshot in Figs.~\ref{fig:stages_fire-A} and \ref{fig:stages_fire-B}. The distinction between interchanging hole \& clump waves on the one hand and ``turbulence'' on the other is somewhat arbitrary, based mainly on appearance in scale and complexity. In both cases, there is a convective interchange of light and dense packets of EP Vlasov fluid, with plumes of dense fluid advancing outward and {\it vice versa}. These phase space structures are nonadiabatic and may be regarded as a semi-perturbative manifestation of what Wang Tao {\it et al}.\ \cite{WangTao20} call ``convective branch''.\footnote{In Ref.~\protect\cite{WangTao20}, the ``convective branch'' was only responsible for relatively minor transient chirps, while the dominant signal was associated with an MHD eigenmode that was referred to as ``relaxation branch''. In our semi-perturbative model, there are no {\cred discrete} eigenmodes, but the seed wave may be regarded as the counterpart of the ``relaxation branch'', and the early nonadiabatic chirps may be regarded as manifestations of the ``convective branch''. The (semi-)adiabatic phase space structures that are responsible for long-range chirping may then be referred to as ``vortex branch''. Here, all three branches have the same field mode structure.}

All these structures propagate poloidally at characteristic frequencies that depend on their instantaneous radial location measured by $\hat{P}_\zeta$. Their (increasingly uncoordinated) pumping action drives the oscillations of the field. As soon as two or more pump waves are present in the EP phase space density, their interference can cause the field to beat. Among other interesting features, we will show shortly that, at least during the first dozen or so beats, the wave fronts advance radially in pulses that {\cred correlate well} with the beating of the field wave.

In Section~\ref{sec:result_beat}, we characterize the beats and their feedback on the phase space dynamics at the beginning of the simulations, where the first hole-clump wave pair forms (i). The ensuing wave front propagation and nonadiabatic onset of chirping (ii) is analyzed in Section~\ref{sec:result_front}, followed by an analysis of the formation and detachment of solitary vortices (iii) in Section~\ref{sec:result_detach}. While the first three subsections deal with the collective structures in EP phase space density, the last Section~\ref{sec:result_transport} is dedicated to an analysis of the motion of individual {\cred simulation} particles, which yields information about global transport and the internal dynamics of the large vortical structures that manage to survive in the presence of a beating field.

\begin{figure*}[tp]
  \centering
  \includegraphics[width=16.5cm,clip]{\figures/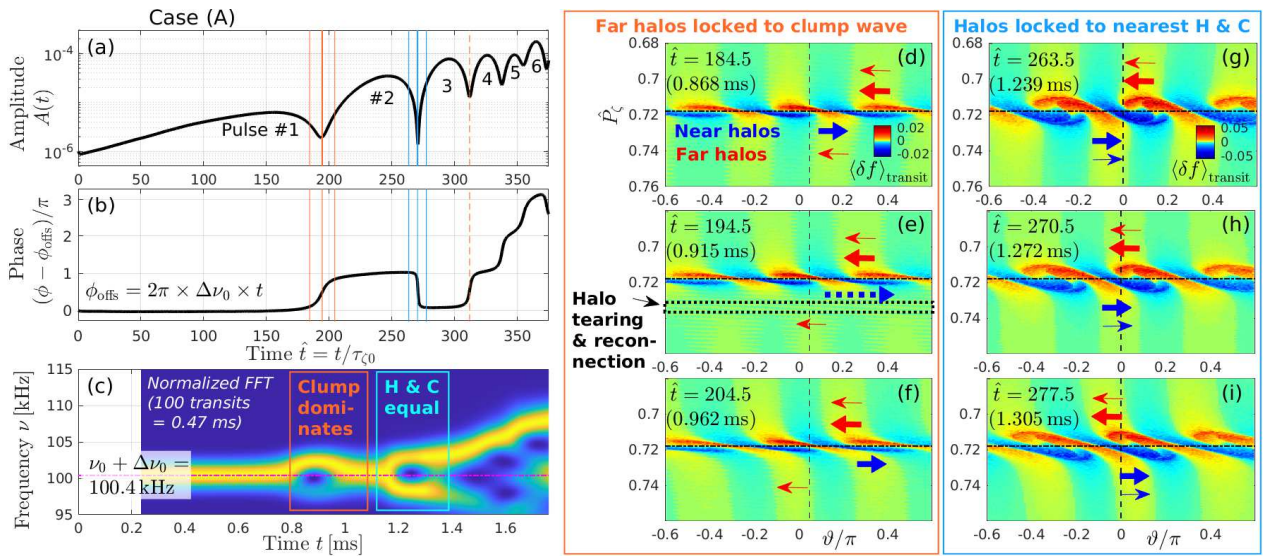}
  \caption{Onset of beating due to the interference of hole \& clump (H\&C) pump waves in the marginally unstable case (A). Panels (a) and (b) show, respectively, the time trace of the field amplitude $A(t)$ (log scale) and phase $\phi(t) - \phi_{\rm offs}(t)$, where the offset $\phi_{\rm offs} = 2\pi\Delta\nu_0 t$ cancels the phase drift associated with the small prompt frequency shift $\Delta\nu_0 \approx 0.4\,{\rm kHz}$ (cf.~Appendix~\protect\ref{apdx:model_opt}). Panel (c) shows the Fourier spectrogram whose amplitude has been normalized at each time step. Panels (d)--(f) and (g)--(i) show snapshots of the phase space density perturbation $\delta f$ around the first two minima of the amplitude in (a), where phase jumps by $+\pi$ and $-\pi$ can be observed in (b). The dash-dotted horizontal lines in (c)--(i) indicate the effective resonance location ($\nu_0 + \Delta\nu_0 = 100.4\,{\rm kHz}$, $\hat{P}_{\zeta,{\rm res}} = 0.718$). A nonlinear color scale (square root) is used to make the halos visible in (d)--(i). It is important to note that the arrows in (d)--(i) indicate halo motion, which are collective structures, not tied to particle motion. Arrows are drawn with different thickness for easier identification. A detailed view of the third amplitude minimum at $\hat{t} = 312$ ($1.467\,{\rm ms}$) between pulses \#3 and \#4 can be gained from Fig.~\protect\ref{fig:result_fire-A_beat3}. The phase jump in the raw signal between pulses \#10 and \#11 (out of range) was shown in Fig.~\protect\ref{fig:intro_fire-A_beat}.}
  \label{fig:result_fire-A_beating}
\end{figure*}

\subsection{Phase jumps and halo dynamics during symmetric and asymmetric beats}
\label{sec:result_beat}

Amplitude pulsations and phase jumps are ubiquitous in experimental measurements of chirping modes and in simulated chirps. The first explicit experimental reports that we are aware of came from JET \cite{Heeter00}, where these dynamics were interpreted as a signature of chaos. Here, we do not go to such lengths and interpret the pulsations and phase jumps simply as a manifestation of beating (as in Ref.~\cite{Bierwage17a} and Appendix~\ref{apdx:beat}). In this section, we perform a detailed analysis of the phase jumps in our simulations, with focus on the associated phase space dynamics and feedback loops. Our results are summarized in Fig.~\ref{fig:result_fire-A_beating}, which shows the dynamics of the field (left) and EP phase space density perturbations (right) during the first few beats. {\cred The beats} consist of amplitude pulsations as shown in panel (a) and phase jumps between the pulses as shown in panel (b). The situation is representative for all our simulations with strong damping, both near marginal stability and strongly unstable. Here, we discuss the marginally unstable case (A).

Before we examine the results, we should point out that, in most simulations, we observe a prompt frequency shift during the first few transits. This phenomenon is attributed to a self-optimization process and is discussed in Appendix~\ref{apdx:model_opt}. The oscillation frequency in case (A) becomes $\nu_0 + \Delta\nu_0 \approx (100 + 0.4)\,{\rm kHz}$ and the resonance location $\hat{P}_{\zeta,{\rm res}}$ shifts from $0.719$ to $\approx 0.718$. {\cred Similarly to the adjustment made to the poloidal angle $\vartheta$ in Fig.~\ref{fig:model_fire-A_df-interpret} above, the phase $\phi(t)$ in Fig.~\ref{fig:result_fire-A_beating}(b)} has also been adjusted by a corresponding offset $\phi_{\rm offs} = 2\pi\Delta\nu_0 t$. The location of the promptly shifted {\it effective} resonance is indicated by dash-dotted lines in panels (c)--(i) of Fig.~\ref{fig:result_fire-A_beating}.

During pulse \#1, the field amplitude in Fig.~\ref{fig:result_fire-A_beating}(a) saturates around $\hat{t} = 150$ ($0.7\,{\rm ms}$), and then it rapidly drops, here by about a factor of 3. Around the instant of the first amplitude minimum, at about $\hat{t} = 194.5$ ($0.9\,{\rm ms}$), we observe in panel (b) a arctan-like phase jump $\Delta\phi = +\pi$ that takes about 20 transits ($0.1\,{\rm ms})$ to complete. Around the same time, the normalized Fourier spectrum in panel (c) shows transient frequency splitting into $\nu^+ \approx 102.4\,{\rm kHz}$ and $\nu^- \approx 98.1\,{\rm kHz}$, which corresponds to a radial separation of $|\Delta \hat{P}_\zeta| \lesssim 5\times 10^{-3}$ around the resonance and is comparable to the extent of the phase space density perturbations seen in the contour plots of $\delta f$ in panels (d)--(f). A qualitatively similar behavior can be seen between pulses \#2 and \#3, during the second amplitude minimum around $\hat{t} = 270.5$ ($1.3\,{\rm ms}$), where the spectral peaks are located at $\nu^+ \approx 103.0$ and $\nu^- \approx 98.2$.

The upper branch $\nu^+$ is pumped by a clump wave in the region $\hat{P}_\zeta < \hat{P}_{\zeta,{\rm res}}$ (upper part of panels (d)--(i)) and the lower branch $\nu^-$ is pumped by a hole wave in the region $\hat{P}_\zeta > \hat{P}_{\zeta,{\rm res}}$ (lower part of panels (d)--(i)). According to the theory in Appendix~\ref{apdx:beat}, the interference between these two pump waves causes beating in the kinetic drive on the right-hand side of the field equation (\ref{eq:evol}), to which the field responds with identical beats. Thus, the amplitude minima highlighted in Fig.~\ref{fig:result_fire-A_beating} correspond to instances of {\it destructive interference} that occurs when pump waves with {\it opposite signs} of $\delta f$ (namely, a hole-clump pair) are {\it aligned}. Consistently with Eq.~(\ref{eq:mode_phase}), which yields the condition $p\vartheta + \phi = {\rm const}$.\ in the Poincar\'{e} section $n\zeta - \omega t = 2\pi l$ with integer $l$, we observe positive phase jumps $\Delta\phi = +\pi$ when the clump wave dominates ($\dot\vartheta < 0$), whereas $\Delta\phi = -\pi$ when the hole wave dominates ($\dot\vartheta > 0$).

There are quantitative differences between the first and second phase jumps in Fig.~\ref{fig:result_fire-A_beating}. During the second event, panel (a) shows a deeper and sharper amplitude minimum, and panel (b) shows a phase jump that is more abrupt and in the opposite direction ($\Delta\phi = -\pi$). The spectrogram in panel (c) shows us the reason for these differences between the first and second amplitude minima: the frequency splitting is asymmetric during the first event, with a dominant upper branch $\nu^+$, whereas relatively symmetric splitting is seen during the second event, with an only slightly (2\%) stronger lower branch $\nu^-$. This means that the interfering pump waves have different strengths during the (asymmetric) first beat, while they are comparable during the (symmetric) second beat.

The beating dynamics of the field wave become increasingly complicated as time advances. For instance, in the presence of more than two pump waves --- pairs of which may dominate and control the field oscillations at different times --- pulses may overlap and slow phase jumps may be cut short ($|\Delta \phi| < \pi$) by faster ones.

So far, we have discussed the beating phenomenon only from the point of view of the field behaving as a driven oscillator. However, in our system there is a feedback loop, by which the pulsations and phase jumps of the field wave acts back on the motion of individual particles as well as on their collective density waves. One manifestation of this feedback are the off-resonant {\it halos} introduced in Section~\ref{sec:saturation_halo}, which appear in the form of vertical stripes of red and blue shade in the snapshots of $\delta f$ on the right-hand side of Fig.~\ref{fig:result_fire-A_beating}. We emphasize again that these halos are nonresonant {\it collective} modulations of EP density, which are not tied to the motion of individual particles. Our observations indicate that the halo patterns are largely phase-locked to the field wave (that, in turn, is controlled by the resonant phase space structures), which makes these halos useful for the present analysis of phase jumps.

We distinguish ``near halos'' (meaning: near-resonant) and ``far halos'' (meaning: far off-resonant). Different halo dynamics are observed during phase jumps associated with symmetric and asymmetric beats:
\begin{itemize}
\item[(i)] {\it Symmetric beat:} Positive (red) and negative (blue) halos abruptly switch places globally during instants of perfect or nearly perfect destructive interference of two pump waves. This is realized in Fig.~\ref{fig:result_fire-A_beating}(g)--(i).

\item[(ii)] {\it Asymmetric beat:} Far halos remain locked to the dominant pump wave, while the subdominant pump wave controls only the nearest halo. Phase slippage is realized here by halos being torn and reconnected with their neighbors as seen in Fig.~\ref{fig:result_fire-A_beating}(d)--(f).
\end{itemize}

\noindent Scenario (i) is simpler, so we start with Fig.~\ref{fig:result_fire-A_beating}(g)--(i). While the hole and clump waves counter-propagate along $\vartheta$ at a more or less steady pace (with only minor jolting), the far halos tend to stay locked in place during the preceding pulse \#2, so that they stay behind the resonant structures that they are attached to. This results in some bending of the near halos. Towards the end of pulse \#2, the far halos begin to accelerate as indicated by the arrows in panel (g): The halos above the resonance are attached to the clump wave and tend to shift to the left as indicated by the pair of red arrows. Similarly, the halos below the resonance are attached to the hole wave and tend to shift to the right as indicated by the pair of blue arrows. Here, an arrow's thickness is used merely to identify it in different snapshots. When the field disappears at the time of snapshot (h), the halos jump ahead of the respective clump or hole wave. Then they decelerate and more or less freeze at the new positions seen in snapshot (f). The motion of the halos corresponds closely to the time trace of the field phase $\phi(t)$ in panel (b), whose sign is determined by that of the dominant phase space structure. Here, the hole wave is only about 2\% stronger, so the pump waves are almost equally strong, causing a very abrupt and radially global phase jump of the halos, which remain attached to the nearest hole or clump wave front.

\begin{figure}[tp]
  \centering
  \includegraphics[width=8cm,clip]{\figures/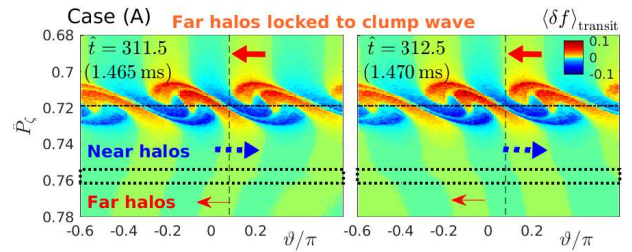}
  \caption{Evolution of $\delta f$ during the third amplitude minimum at $\hat{t} = 312$ ($1.467\,{\rm ms}$) in Fig.~\protect\ref{fig:result_fire-A_beating}. This event is similar to that in Fig.~\protect\ref{fig:result_fire-A_beating}(e), but here the signal-to-noise ratio is larger. As in Fig.~\protect\ref{fig:result_fire-A_beating}, the dotted rectangles indicate the region where halo tearing and reconnection occurs. The arrows indicate halo motion (decoupled from particle motion).}
  \label{fig:result_fire-A_beat3}
\end{figure}

More or less symmetric beats do occur fairly often in our simulations, but more frequently we observe asymmetric beats, where one of the pump waves dominates, so that destructive interference is not complete, and amplitude minima are smoother and shallower. This situation, namely scenario (ii), is realized during the first amplitude minimum in all of our chirping simulations. Panels (d)--(f) of Fig.~\ref{fig:result_fire-A_beating} show the corresponding phase space dynamics. In the present case, the clump wave dominates and we observe that not only the halos above the resonance follow the clump wave, but even the far halos below the resonance do so, as indicated by the red arrows in panel (d). The weaker hole wave controls only the near halo as indicated by the blue arrow in panel (d). During pulse \#1, all structures are more or less locked in phase, so the arrows in panel (d) indicate only the trend of their relative motion. When the field is weakened due to partial destructive interference at the time of snapshot (e), these relative drifts in $\vartheta$ become significant, and a phase jump of magnitude $\pi$ occurs, causing the red and blue halos in panel (f) to exchange places compared to panel (d). However, this time, all far halos have moved to the left with the clump, as one can infer from the positions of individual arrows in the snapshots. This means that around snapshot (e) the halos below the resonance have been torn and reconnected in the region that is roughly indicated by the dotted rectangle. Unfortunately, the signal-to-noise ratio is somewhat low in Fig.~\ref{fig:result_fire-A_beating}(e), so we show another example of an asymmetric beat in Fig.~\ref{fig:result_fire-A_beat3}. Here one can see snapshots taken just before and just after the instant of halo reconnection during the third amplitude minimum in Fig.~\ref{fig:result_fire-A_beating}(a).

In summary, we have shown in this subsection how phase jumps occur during amplitude minima. Hole and clump waves and nearby halos propagate {\cred poloidally} at a relatively uniform pace that depends on their distance from the resonance. In contrast, the motion of the far halos is nonuniform, with a arctan-like time trace that resembles the evolution of the phase $\phi(t)$ in Fig.~\ref{fig:result_fire-A_beating}(b). In the case of symmetric beating, with two pump waves of similar intensity, the phase jump is global along $\hat{P}_\zeta$. In the case of asymmetric beating, in our example dominated by the clump wave above the resonance, the phase jump is realized by tearing and reconnecting near and far halos on the side of the subdominant pump wave, in our example on the hole side.\footnote{Halo reconnection always occurs somewhere, but during symmetric beats it occurs far away from the resonance, outside our field of vision.} The phase jumps have important consequences that we will examine next.

\subsection{{\cred Feedback loop for} pulsed front propagation and nonadiabatic onset of chirping}
\label{sec:result_front}

\begin{figure}[tp]
  \centering
  \includegraphics[width=8cm,clip]{\figures/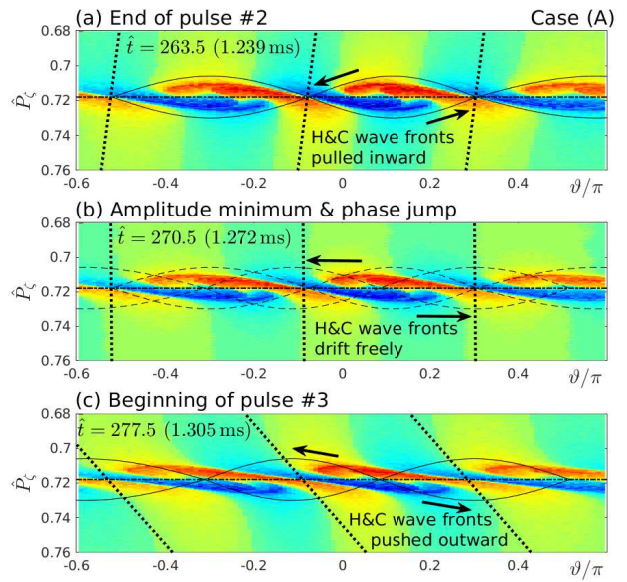}
  \caption{Illustration of the effect of a phase jump, using the snapshots of $\delta f$ in Fig.~\protect\ref{fig:result_fire-A_beating}(g)--(i) around the second amplitude minimum in Case (A). The solid lines in (a) and (c) indicate schematically the effective positions of phase space islands (similarly to Fig.~\protect\ref{fig:model_fire-A_df-interpret}). The island width is somewhat exaggerated. Islands are absent during the amplitude minimum in (b), so the dashed lines indicate their past and future forms. Bold arrows indicate the flow direction. Dotted lines highlight the inclination of the halos in the vicinity of the resonance (far halos above and below the {\cred visible} $\hat{P}_\zeta$ range are almost vertically straight).}
  \label{fig:result_fire-A_beat2_island}
\end{figure}

In Section~\ref{sec:saturation_halo}, we have asserted that the halos that appear in our $\delta f$-weighed kinetic Poincar\'{e} plots can be used to identify the effective instantaneous location of O- and X-points. Here, this procedure is applied to the three snapshots from Fig.~\protect\ref{fig:result_fire-A_beating}(g)--(i) that we analyzed in the previous subsection. Figure~\ref{fig:result_fire-A_beat2_island} shows schematically the effective phase space islands (black sinusoidal curves) plotted on top of the phase space density perturbation $\delta f$ at the times of those snapshots: (a) at the end of pulse \#2, (b) at the instant of the phase jump by $\Delta\phi = -\pi$, and (c) at the beginning of pulse \#3.

The arrows in Fig.~\ref{fig:result_fire-A_beat2_island}(a) show how, at the end of a pulse in the field amplitude $A(t)$, the fronts of hole \& clump (H\&C) waves are approaching the effective X-point, which means that they are pulled ``inward'' (towards the resonance line). However, this process is terminated by the disappearance of the field during the amplitude minimum at the time of snapshot (b), before any significant revolution around the effective O-point can occur. Around that time, the H\&C waves are free to drift poloidally at their characteristic speeds, as indicated by the arrows in Fig.~\ref{fig:result_fire-A_beat2_island}(b). When the field is revived in the next pulse, it does so with opposite phase. This means that the new O-points appear at (or near) the locations of the former X-points and {\it vice versa}, yielding a situation as illustrated schematically in Fig.~\ref{fig:result_fire-A_beat2_island}(c). The H\&C wave fronts that were previously drawn towards the resonance have now begun to travel ``outward'' (away from the resonance).

{\cred This constitutes an important {\it feedback mechanism}}. The new location of the effective island contours (representing flow lines) with respect to the location of the H\&C pair at the time of the phase jump facilitates a further departure of the H\&C structures from the seed resonance. The further outward motion of the clump and inward motion of the hole transfers energy to the field. {\cred The growing field amplitude} makes the effective island larger and the flow contours steeper, allowing the H\&C structures to depart even further. This positive feedback continues until the poloidal drift of the H\&C takes them across the effective O-point, {\cred reversing the trend:} The flow contours at the head of the H\&C structures point towards the resonance, making clumps move inward and holes outward, so that the mode transfers energy to the particles and is, therefore, resonantly damped. As mentioned previously in Section~\ref{sec:saturation_damp}, the additional damping $\gamma_{\rm d}$ makes the process irreversible since the mode's remaining energy is already exhausted after a partial reversal of radial H\&C displacement.

\begin{figure*}[tb]
  \centering
  \includegraphics[width=8cm,clip]{\figures/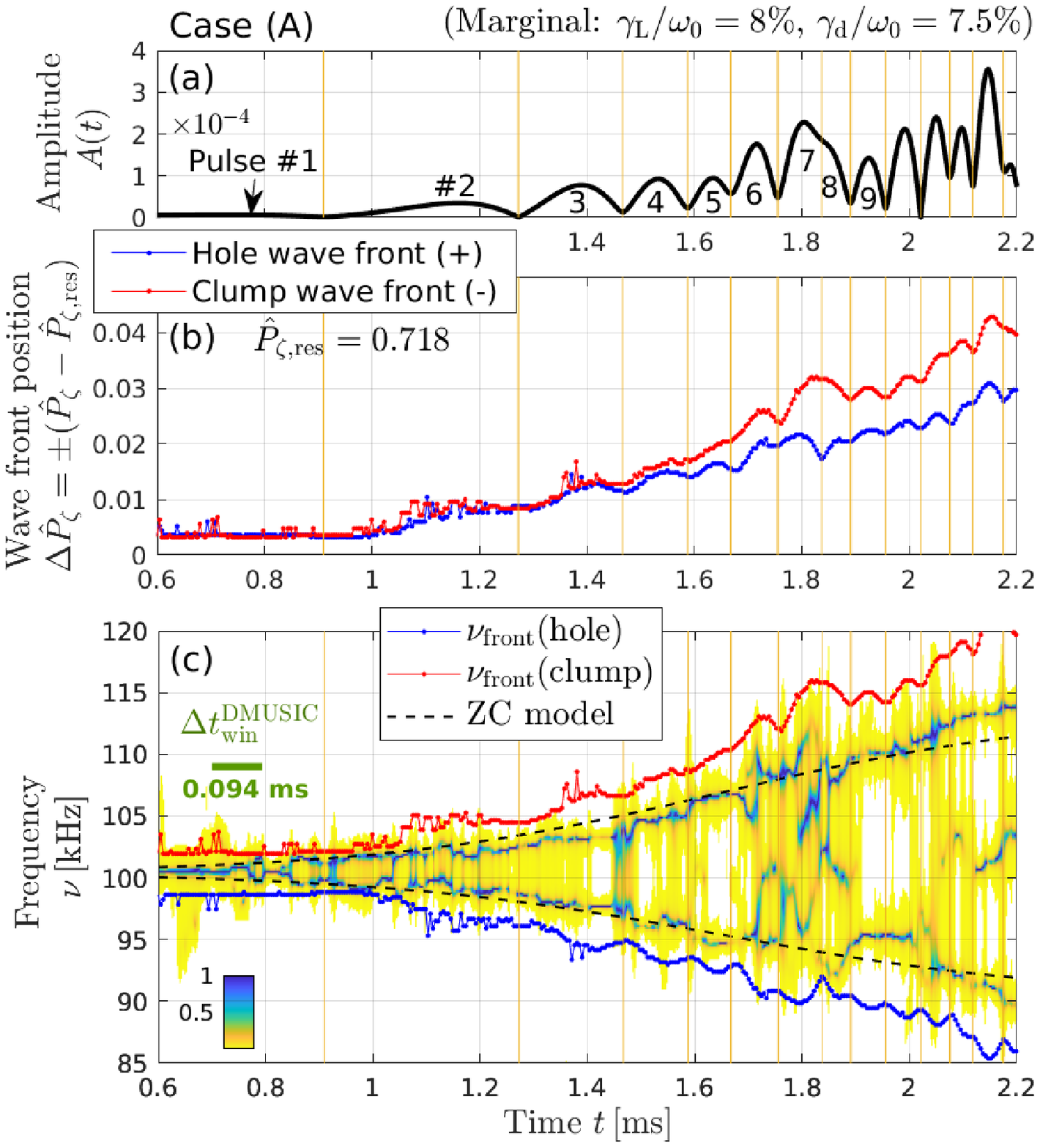}
  \includegraphics[width=8cm,clip]{\figures/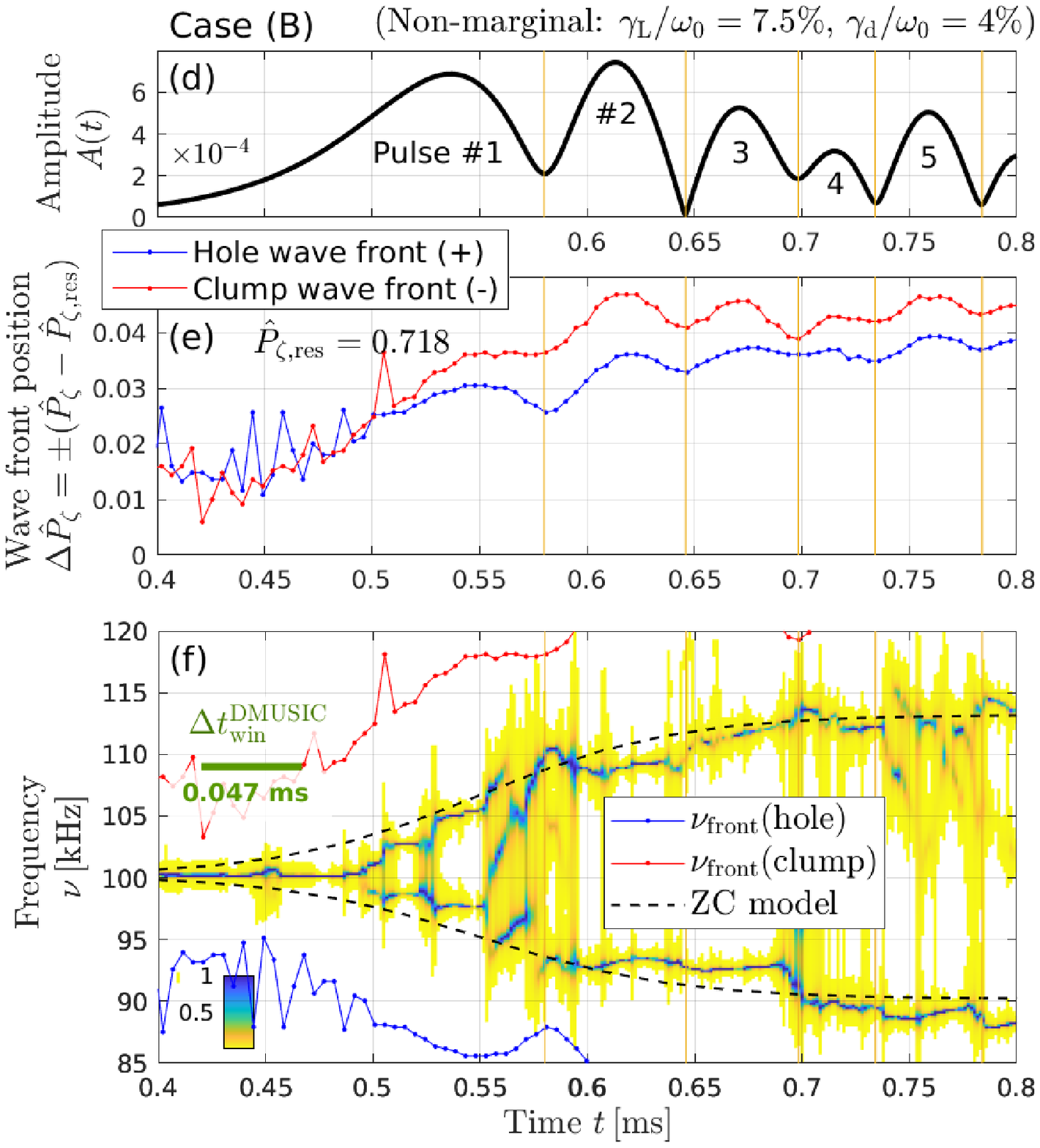}
  \caption{Propagation of EP phase space density wave fronts and nonadiabatic onset of chirping during the first dozen pulses of the marginally unstable case (A) (left) and first half dozen pulses in the strongly unstable case (B) (right). Panels (a) and (d) show the time traces of the field amplitudes $A(t)$. Panels (b) and (e) show the distance $|\Delta\hat{P}_\zeta|$ of the hole wave fronts (blue) and clump wave fronts (red) from the seed resonance $\hat{P}_{\zeta,{\rm res}} = 0.718$. The wave front positions in the $\delta f$-weighed Poincar\'{e} data --- which can be seen indicated by dashed horizontal lines in the fourth snapshot of $\delta f(\hat{P}_\zeta,\vartheta)$ in Figs.~\protect\ref{fig:stages_fire-A} and \protect\ref{fig:stages_fire-B} --- were determined using an automatic threshold-based algorithm. Panels (c) and (f) show high-resolution spectrograms obtained with the DMUSIC algorithm (see Appendix~\protect\ref{apdx:dg_dmusic}), using a time window of size $\Delta t_{\rm win} = 0.094\,{\rm ms}$ in case (A) and $\Delta t_{\rm win} = 0.047\,{\rm ms}$ in the more rapidly evolving case (B). The horizontal green bars indicate $\Delta t_{\rm win}$ for easy comparison with the spectral patterns. The $\Delta\hat{P}_\zeta(t)$ curves from panels (b) and (e) have also been converted to frequency using Eq.~(\protect\ref{eq:pz_nu}) as $|\Delta\nu| \approx |\Delta\hat{P}_\zeta| \times 486.8\,{\rm kHz}$, and then plotted as $\nu_{\rm front} \approx 100.4\,{\rm kHz} \pm |\Delta\nu|$ together with the spectrograms in (c) and (f). Vertical orange lines indicate the times of amplitude minima, where phase jumps $\pm\pi$ occur. The dashed black curves in (c) and (f) represent fits of the ZC model in our parametrization (\protect\ref{eq:zc}) using the values in Table~\protect\ref{tab:zc}.}
  \label{fig:result_spec-fronts}
\end{figure*}

This process is repeated during each beat cycle. The resulting evolution of the system during the first dozen or so beats of the marginally unstable case (A) is shown in the left half of Fig.~\ref{fig:result_spec-fronts}. Panel (a) shows the evolution of the field amplitude $A(t)$, with the first 9 pulses labeled (the small peak of pulse \#1 is hardly visible). Panel (b) shows that the fronts of the hole and clump waves advance radially in a pulsed manner that is perfectly correlated with the pulsations of the field seen in panel (a). One can see that the radial expulsion of the wave fronts is partially reversed at the end of most pulses, but the net displacement clearly increases with time, which is consistent with the discussion in the previous paragraph. Note that there is often a significant difference in the magnitude of the displacement of the hole and clump wave fronts during a given pulse, especially when the amplitude does not drop very far due to asymmetric beats or due to an overlap between two pulses.

Qualitatively similar behavior is seen in the strongly unstable case (B) shown in the right half of Fig.~\ref{fig:result_spec-fronts}. The main differences are the shorter time scale (factor 2...3) and the much larger amplitude of the first pulse (factor 100). In the marginal case (A), the magnitude of the displacement of the wave fronts during pulse \#1 is below the accuracy of our diagnostics, whereas pulse \#1 in the strongly unstable case (B) does more displacement work than all the subsequent pulses that are visible in Fig.~\ref{fig:result_spec-fronts}.

The high-resolution DMUSIC spectrograms in panels (c) and (f) of Fig.~\ref{fig:result_spec-fronts} show the frequencies that are induced by those phase space structures. A striking feature of these spectrograms is that the evolution of the field's oscillation frequencies is far from smooth. The interpretation of these spectrograms requires caution (cf.~Appendices~\ref{apdx:undamped_spec} and \ref{apdx:dg}), but at least some features seem to be correlated with the wave fronts' radial propagation in a pulsed manner. {\cred In many instances, the frequency varies rapidly during an amplitude minimum between successive pulses, so that the early chirping has a {\it staircase-like} appearance.}\footnote{In case (B), {\cred the first few frequency steps in Fig.~\ref{fig:result_spec-fronts}(f) during the interval $0.50\,{\rm ms} \lesssim t \lesssim 0.55\,{\rm ms}$ occur} during pulse \#1, so on a time scale that is shorter by about a factor of 3 compared to beating and particle bouncing. So far, we have not been able to attribute these steps to any physical feature of $\delta f$, so their cause remains unknown. Statistical noise associated with the imperfect quiet start may play a role.}

The spectrograms in Fig.~\ref{fig:result_spec-fronts}(c) and (f) also show copies of the $\Delta\hat{P}_\zeta(t)$ curves from panels (b) and (e), that have been converted to frequency using Eq.~(\ref{eq:pz_nu}) as $|\Delta\nu| \approx |\Delta\hat{P}_\zeta| \times 486.8\,{\rm kHz}$. One can see that the signals in the spectrogram --- and, thus, the effective pump wave frequencies --- lie well within the range $\nu_{\rm front} = 100.4\,{\rm kHz} \pm |\Delta\nu|$ spanned by the wave fronts. The separation between $|\nu_{\rm front}|$ and the spectral peaks is smaller in the marginally unstable case (A) than in the strongly unstable case (B). One reason is simply that the radial width of the primordial holes and clumps in case (B) is much larger than in case (A) during the onset of chirping.

In addition, the poloidal drift speed of the pump waves relative to the seed wave is initially reduced due to stronger resonant trapping in case (B), whose damping rate $\gamma_{\rm d}/\omega_0 = 4\%$ is smaller than the $7.5\%$ in case (A). Eventually, the pump waves in (B) escape from the seed resonance and chirping proceeds similarly to case (A). Thus, the onset of chirping in case (B) is first retarded with respect to mode growth and then occurs in a more abrupt manner than in case (A). This can be seen clearly seen in Fig.~\ref{fig:result_spec-fronts}(c) and (f): while upward chirp takes about ${\rm 1 ms}$ to reach $110\,{\rm kHz}$ in case (A), it takes only about $0.1\,{\rm ms}$ in case (B). This trend should continue {\cred with decreasing $\gamma_{\rm d}$ until, in the absence of damping ($\gamma_{\rm d} = 0$), the spectral peaks remain at or very near the seed frequency (i.e., no sustained chirping), whereas $\Delta\hat{P}_\zeta$ remains large and represents the half-width of the phase space island as in Fig.~\ref{fig:discuss_fire-A0-B0_spec-df}.}\footnote{Here, we have expressed once more in different words what has already been mentioned near the end of Section~\protect\ref{sec:saturation_beat} above.}

The pulsed radial propagation of the wave fronts and the more or less discrete stepping in frequency in Fig.~\ref{fig:result_spec-fronts} during the nonadiabatic phase of the chirp may be regarded as one possible realization of the {\it relay runner} paradigm, a phenomenological model proposed by Zonca \& Chen \cite{ZoncaTCM99,ZoncaFEC02} on the basis of the mode pumping mechanism discussed by White {\it et al}.\ \cite{White83}. In the original model, the relay runners were identified with the poloidal harmonics and EP-driven distortions of a shear Alfv\'{e}n wave packet in toroidal geometry, as illustrated schematically in Fig.~\ref{fig:intro_x-point}(d).

The physics of evolving mode structures are absent in our semi-perturbative model, but this caveat can be tolerated in the present cases, since we have long-wavelength modes and large magnetic drifts. The successive (and phase shifted) pulses in our simulations can play the role of the relay runners who are taking turns to carry the {\it batons} in the form of phase space density waves. Moreover, in the marginally unstable case (A), the first dozen or so pulses have gradually increasing amplitudes, which corresponds to the process of convective amplification \cite{Zonca15b}, except that here the radial convection occurs only in phase space, not in the mode structure.  In addition, recall that we have placed the peak of our field mode at a larger radius than the resonance (cf.~Fig.~\ref{fig:model_fire-ABC_poink}). From the perspective of the clumps, which propagate radially outward towards the mode's peak, the field appears to get stronger with increasing radius, even when $A(t)$ fluctuates around the same level. This process may be referred to as ``preconvective amplification'' as discussed in Appendix~\ref{apdx:model_pert}. Preliminary results of parameter scans indicate that this arrangement significantly affects the preferred direction of chirping. In the present setup, we observe stronger (more intense and more rapid) upward chirping, which means radially outward, towards the peak of the mode.

The Zonca-Chen (ZC) model in the form presented in Ref.~\cite{ZoncaTCM99} predicts that the chirping rate should first accelerate exponentially, then pass through a phase of linear-in-time chirping, and eventually decelerate exponentially; in other words, it should evolve like a hyperbolic tangent:
\begin{equation}
\delta\nu^{\rm ZC}(t) \propto \frac{\exp(\gamma_0 t)}{1 + \exp(\gamma_0 t)} = \frac{1}{2} \left[1 + \tanh\left(\frac{\gamma_0 t}{2}\right)\right],
\label{eq:tanh}
\end{equation}

\noindent where $\gamma_0$ is the linear growth rate, and the time variable $t$ is such that the chirping begins at $t \rightarrow -\infty$, and the transition from accelerating to decelerating chirping occurs at $t = 0$.

\begin{figure}[tb]
  \centering
  \includegraphics[width=8cm,clip]{\figures/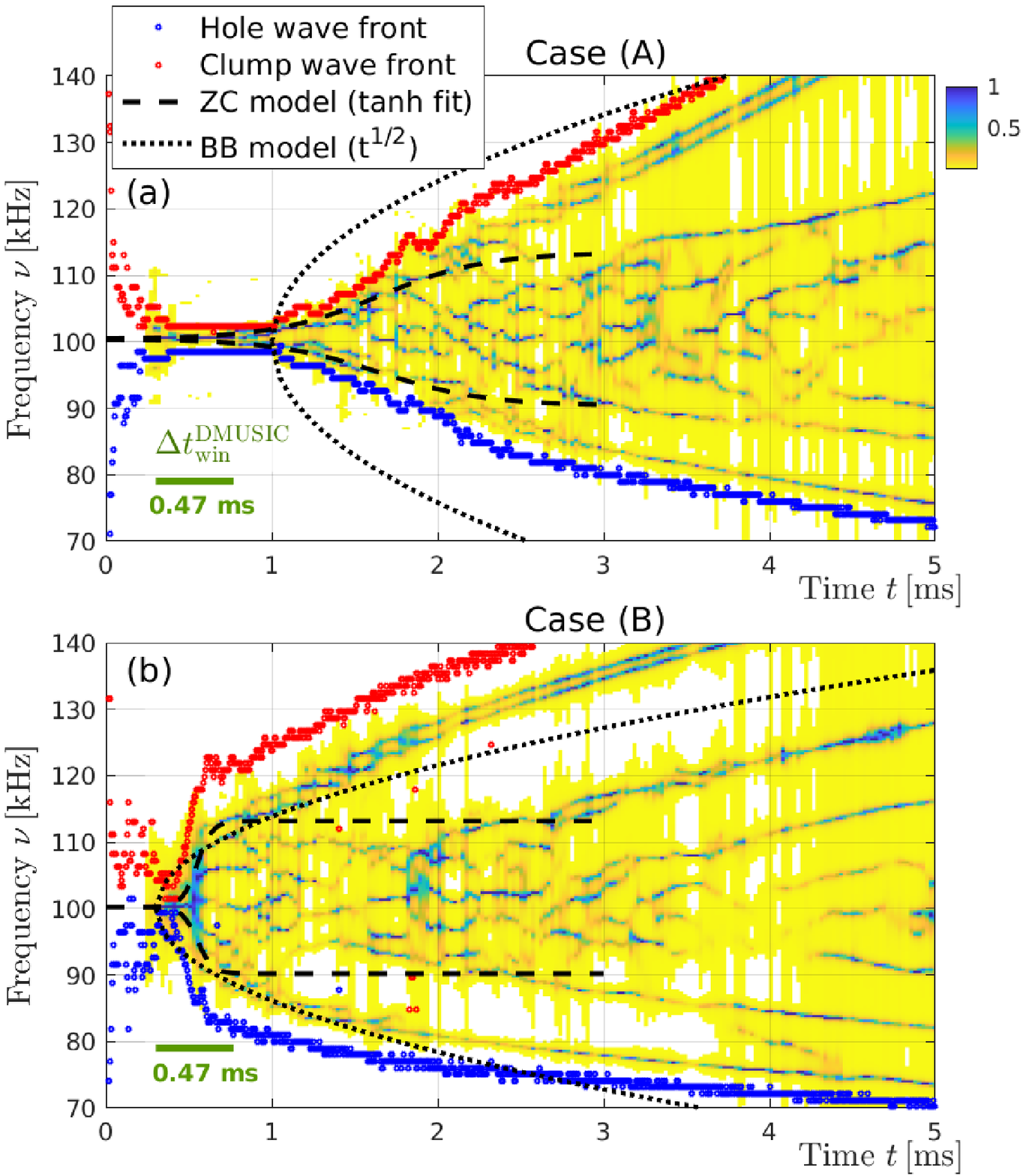}
  \caption{DMUSIC spectra for the first $5\,{\rm ms}$ in cases (A) and (B). The spectral analysis was performed with a longer time window than in Fig.~\protect\ref{fig:result_spec-fronts}; here $\Delta t_{\rm win} = 0.47\,{\rm ms}$. Blue and red lines represent the time traces of the hole and clump wave front frequencies $\nu_{\rm front} = 100.4\,{\rm kHz} \pm |\Delta\nu|$, as in Fig.~\protect\ref{fig:result_spec-fronts}(c) and (f). The dotted black curves represent the prediction of the BB model (\protect\ref{eq:bb}) with the actual values of $\gamma_{\rm L}$ and $\gamma_{\rm d}$ for the respective case, but a somewhat arbitrarily chosen pole location: $1.0\,{\rm ms}$ in case (A) and $0.3\,{\rm ms}$ in case (B). The dashed black curves represent fits of our parametrization of the ZC relay runner model (\protect\ref{eq:zc}) using the values in Table~\protect\ref{tab:zc}.}
  \label{fig:result_fire-AB_spec-fit}
\end{figure}

The ZC relay runner model was formulated for magnetically deeply trapped toroidally precessing particles in the short-wavelength limit with small magnetic drifts, where the particles are passed on from one poloidal harmonic to the next. However, here we deal with a single-harmonic long-wavelength mode and EPs with large magnetic drifts, so we cannot simply adopt the coefficients from Ref.~\cite{ZoncaTCM99}. Instead, we use the parametrization
\begin{equation}
\delta\nu^{\rm ZC} = \nu_{-\infty} + \frac{\Delta\nu_{\pm\infty}}{2}\left[1 + \tanh\left(\frac{\gamma_0}{2}(t - t_0)\right)\right];
\label{eq:zc}
\end{equation}

\noindent where the height $\Delta\nu_{\pm\infty}$ of the hyperbolic tangent is treated as an unknown parameter that is fitted separately to the up- and downward chirping branches in our spectrograms. The location $t_0$ of the inflection point is also fitted, whereas the promptly shifted seed frequency $\nu_{-\infty} = \nu_0 + \Delta\nu_0$ (Appendix~\ref{apdx:model_opt}) and $\gamma_0 = \gamma_{\rm L} - \gamma_{\rm d}$ are measured in the simulations. The parameter values are given in Table~\ref{tab:zc} and the fitted curves are shown as dashed black lines in Fig.~\ref{fig:result_spec-fronts}(c) and (f). One can see that both in the marginally unstable case (A) and in the strongly unstable case (B), the time scale of the chirps is captured well by the $\tanh(\gamma_0 t/2)$ function with the nominal growth rate values. The fitted curves are shown once more in Fig.~\ref{fig:result_fire-AB_spec-fit}, which covers the first $5\,{\rm ms}$ of the simulations and shows DMUSIC spectrograms obtained with a larger time window, $\Delta t_{\rm win} = 0.47\,{\rm ms}$. For comparison, the result of the BB model (\ref{eq:bb}) is also shown.

\begin{table}[tb]
\centering
\begin{tabular}{c|c|c}
\hline\hline Parameter & Case (A) & Case (B) \\
\hline
$\gamma_0 = \gamma_{\rm L} - \gamma_{\rm d}\,{\rm [kHz]}$ & 3.14 & 21.99 \\
$t_0\,{\rm [ms]}$ & $1.65$ & $0.55$ \\
$\nu_{-\infty}\,{\rm [kHz]}$ & $100.4$ & $100.2$ \\
$\Delta\nu_{\pm\infty}\,{\rm [kHz]}$ & \multicolumn{2}{|c}{$+13$ (clump), $-10$ (hole)} \\
\hline\hline
\end{tabular}
\caption{Parameters for our adaptation of the Zonca-Chen (ZC) relay runner model in Eq.~(\protect\ref{eq:zc}).}
\label{tab:zc}
\end{table}

After the hyperbolic tangent levels off, it effectively measures the width of the turbulent belt that has formed around the resonance (although that width continues to evolve). Here, we simply matched the amplitude $\Delta\nu_{\pm\infty}$ of the hyperbolic tangent function to the observed spectrum, but a quantitative model based on readily available parameters such as $\gamma_{\rm L}$, $\gamma_{\rm d}$, drift orbit width, field geometry and mode structure would be desirable. DMUSIC spectrograms with high temporal resolution as in Fig.~\ref{fig:result_spec-fronts} show only the dominant chirps and tend to follow the vortices. In contrast, spectrograms obtained with relatively long time windows as in Fig.~\ref{fig:result_fire-AB_spec-fit} show the turbulent belt as a region characterized by multiple fluctuating spectral peaks that may be attributed to the nonadiabatic interchange dynamics of hole and clump waves of various sizes as seen in the region around $\hat{P}_\zeta \approx 0.7$ of the last two snapshots in Fig.~\ref{fig:stages_fire-A}.

The relay runner model (\ref{eq:tanh}) does not capture the long-range chirps produced by the emission of solitary vortices. In our simulations, chirping continues beyond the turbulent belt as massive solitary vortices detach and propagate away from the seed resonance. The detachment process and the role that beating may play there are examined in the following subsection.

\subsection{Growth, detachment and interference of solitary vortices}
\label{sec:result_detach}

\begin{figure}[tp]
  \centering
  \includegraphics[width=7.9cm,clip]{\figures/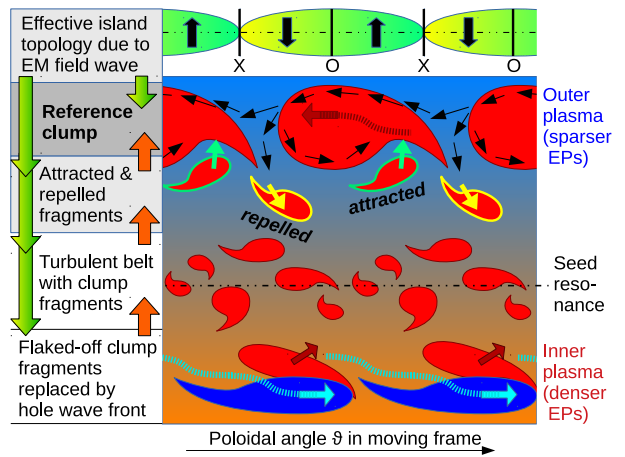}
  \caption{Schematic illustration of the formation of EP clump fragments in the inner plasma (bottom) and their accumulation into a massive clump in the outer plasma (top). The horizontal axis corresponds to the poloidal angle $\vartheta$. As in the phase space plots shown in Figs.~\protect\ref{fig:result_fire-A_detach} and \protect\ref{fig:result_fire-B_detach}, we are in the frame of the electromagnetic (EM) seed wave, so structures above the seed resonance drift to the left and those below drift to the right. The yellow-green-shaded arrows in the text boxes on the left indicate the action of the EM field wave. The red arrows indicate the motion of clump fragments. The black arrows indicate flow patterns in and around the reference clump.}
  \label{fig:result_detach}
\end{figure}

\begin{figure*}[tp]
  \centering
  \includegraphics[width=16cm,clip]{\figures/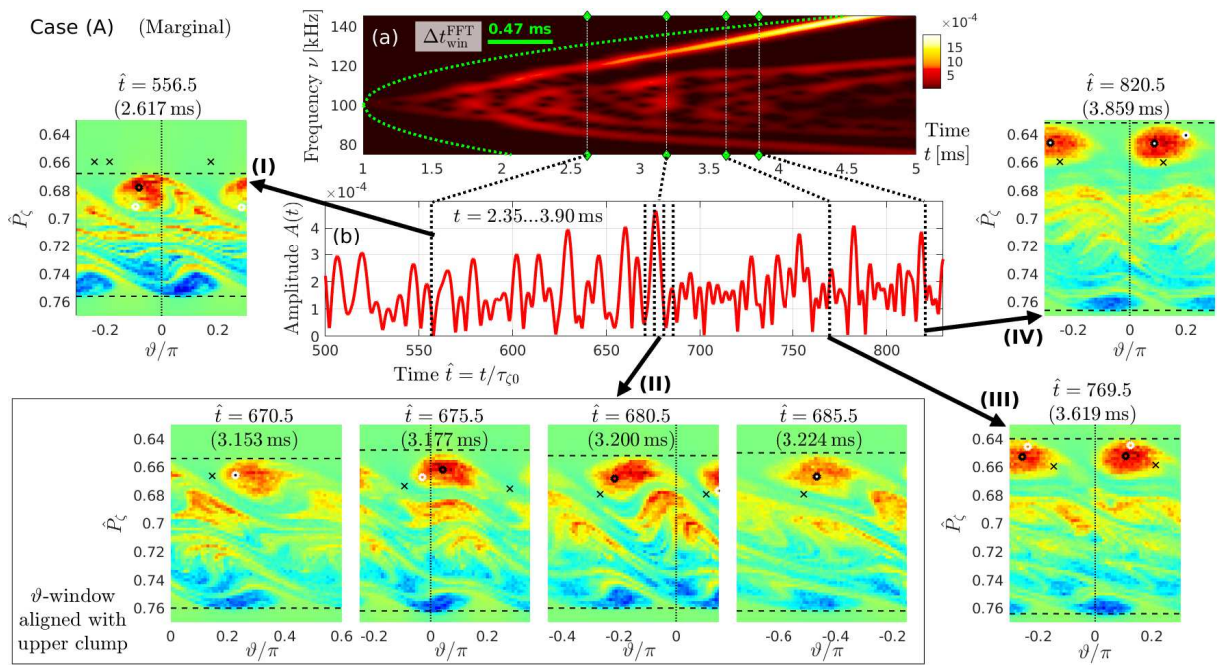}
  \caption{Detachment of solitary vortices in the marginally unstable case (A). Panel (a) shows the Fourier spectrogram obtained with large time window $\Delta t_{\rm win} = 0.47\,{\rm ms}$. The dotted green parabola is the prediction of the BB model (\protect\ref{eq:bb}). Panel (b) shows the evolution of the field amplitude $A(t)$ during the interval $2.35\,{\rm ms} \leq t \leq 3.9\,{\rm ms}$. Four sets of snapshots labeled (I)--(IV) show the structure of the phase space density perturbation $\delta f$ before, during and after the detachment of a massive solitary clump vortex. The $\delta f$ snapshots in set (II) have been aligned with the center of the uppermost clump, so the $\vartheta$-window shown varies, moving leftward from positive to negative values. Horizontal dashed lines indicate the locations of the hole and clump wave fronts, whose evolution was shown in Figs.~\protect\ref{fig:result_spec-fronts} and \protect\ref{fig:result_fire-AB_spec-fit}. The white and black circles and black cross symbols in the $\delta f$ snapshots indicate the locations of particles whose trajectories we analyze in Figs.~\protect\ref{fig:result_fire-AB_tracers} and \protect\ref{fig:result_fire-AB_tracers-spec} below. Each particle appears approximately 4 times during the $\tau_{\zeta 0} = 4.7\,\mu{\rm s}$ interval where the $\delta f$ Poincar\'{e} maps were {\cred accumulated}.}
  \label{fig:result_fire-A_detach}
\end{figure*}

\begin{figure*}[tp]
  \centering
  \includegraphics[width=16cm,clip]{\figures/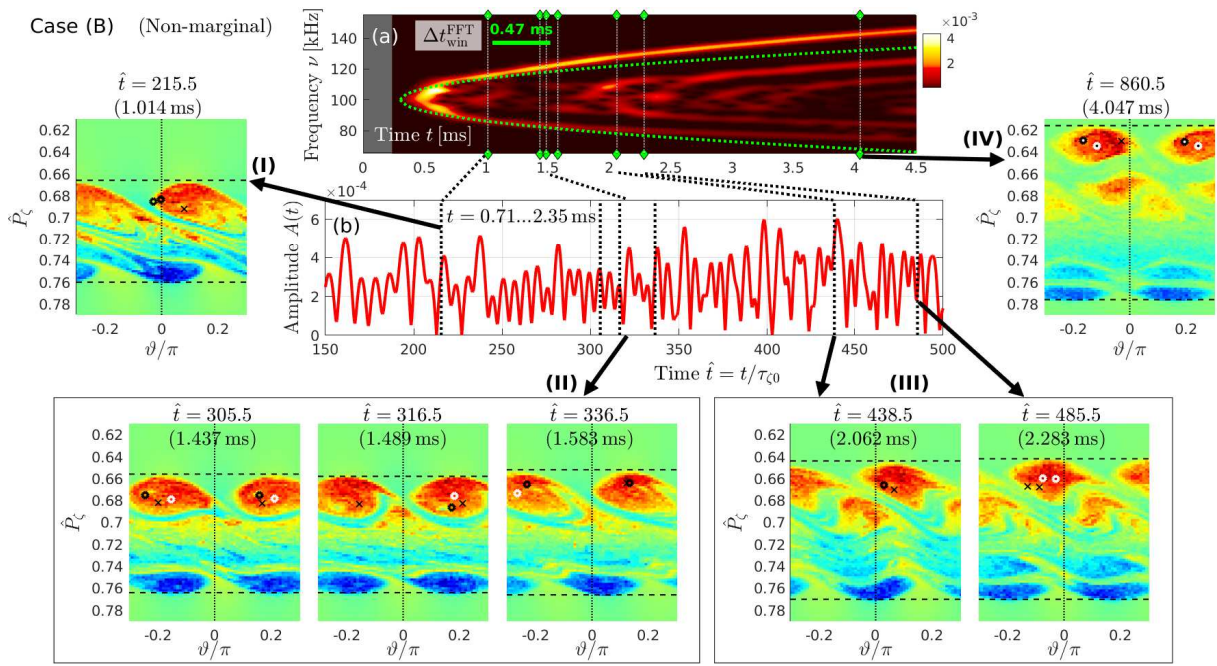}
  \caption{Detachment of solitary vortices in the strongly unstable case (B). The results are arranged as in Fig.~\protect\ref{fig:result_fire-A_detach}. Here, panel (b) shows the time interval $0.71\,{\rm ms} \leq t \leq 2.35\,{\rm ms}$.}
  \label{fig:result_fire-B_detach}
\end{figure*}

When we discussed phase jumps in Section~\ref{sec:result_beat} above, we noted that, at the end of a pulse, the far halos tend to lag behind their H\&C wave fronts, so that the near halos tend to be tilted to the right as indicated by the dotted lines in Fig.~\ref{fig:result_fire-A_beat2_island}(a). At the beginning of the next pulse, the halos are located ahead of their H\&C waves fronts, so that the near halos are tilted to the left as indicated by the dotted lines in Fig.~\ref{fig:result_fire-A_beat2_island}(c). This has an important implication: Since the clump in Fig.~\ref{fig:result_fire-A_beat2_island}(c) moves into a halo region where the EP density has already increased above the reference ($\delta f > 0$, represented by the halo's red shade) and since {\cred the} field amplitude is growing at that time, the clump grows in size as additional dense fluid gets trapped in its neighborhood. The same holds in reverse for the hole. This leads us to the topic of this subsection: the formation of massive holes and clumps and their detachment from the turbulent belt. The detachment process is gradual and difficult to quantify, so this will be a mostly observation-based discussion of the results summarized in Figs.~\ref{fig:result_detach}--\ref{fig:result_fire-B_detach}.

We will be telling the story primarily from the point of view of clumps because, in our setup, they move into regions of higher field amplitude, so they tend to dominate the picture. Apart from this imposed asymmetry, the story may be told in the same way for the holes because here the only difference between holes and clumps is the sign of $\delta f$. It must be kept in mind that packets of Vlasov fluid that are part of a clump at one time may be part of a hole at another time, especially when they are located inside the turbulent belt, where they may be advected up and down between the belt's boundaries and become clump-like or hole-like depending on whether the density at that location was lower or higher initially, at $t=0$. After a vortex has detached from the turbulent belt, its interior may become a ``pure'' clump or hole, but its surface (boundary layer) remains transformative when the field fluctuates (cf.~Fig.~\ref{fig:intro_x-point}(c)).

With the conceptual framework put in place, let us now discuss the vortex formation and detachment process in some detail. Figure~\ref{fig:result_detach} illustrates schematically how we understand the growth of a clump while it is still embedded in the turbulent belt. The turbulent belt in the center of Fig.~\ref{fig:result_detach} can be thought of as consisting of clump and hole fragments of various sizes. Here, only the clump fragments are indicated, which are gradually advected downhill; i.e., upward in Fig.~\ref{fig:result_detach}, towards lower EP density and higher frequency. The clump fragments originate from the lower part of Fig.~\ref{fig:result_detach}, where new clump fragments are flaked off the EP density profile with every new pulse of the field (recall also Fig.~\ref{fig:result_fire-A_beat2_island}). Whenever a clump fragment is flaked off, its place is taken by a hole, which is shown in blue in the lower part of Fig.~\ref{fig:result_detach}. Animations of this process (Appendix~\ref{apdx:dg_movie}) make it appear as if a hole wave front is digging into the EP profile, advancing uphill with each new pulse of the field as indicated by the wavy cyan-colored arrow drawn towards the center of the hole structure.

While being advected downhill, the clump fragments may merge or decompose as they are being continuously interchanged inside the turbulent belt. Large clumps may already form along the way, but in Fig.~\ref{fig:result_detach} we assume that this occurs at the upper rim of the turbulent belt. The wavy dark red arrow inside this massive clump structure indicates how it advances outward with each new pulse. At the same time, neighboring clump fragments may be absorbed into the larger structure.

Whether or not our reference clump repels or absorbs a nearby fragment depends on the {\it instantaneous relative phases} of three players: (i) the reference clump, (ii) the fragment, and (iii) the field wave. The phase of the field wave is indicated at the top of Fig.~\ref{fig:result_detach}, where the black ellipses indicate the separatrix of the effective phase space island. O- and X-point locations are also marked. The bold black arrows depict the direction of the vertical component of phase space flows in different parts of the field wave. The corresponding pattern of phase space flow in the boundary layer of our reference clump are indicated by thin black arrows. The green-outlined clump fragments that happen to be located just behind the center of our reference clump are attracted and can be absorbed, if they are able to reach the same altitude on time. Meanwhile, the yellow-outlined fragments that lie ahead (in the direction of the poloidal drift) are repelled by the downward flows at this time. Fragments that are repelled at one time may be attracted and absorbed later. Overall the reference clump grows due to a net upward flow of denser EP Vlasov fluid.

\begin{figure}[tb]
  \centering
  \includegraphics[width=8cm,clip]{\figures/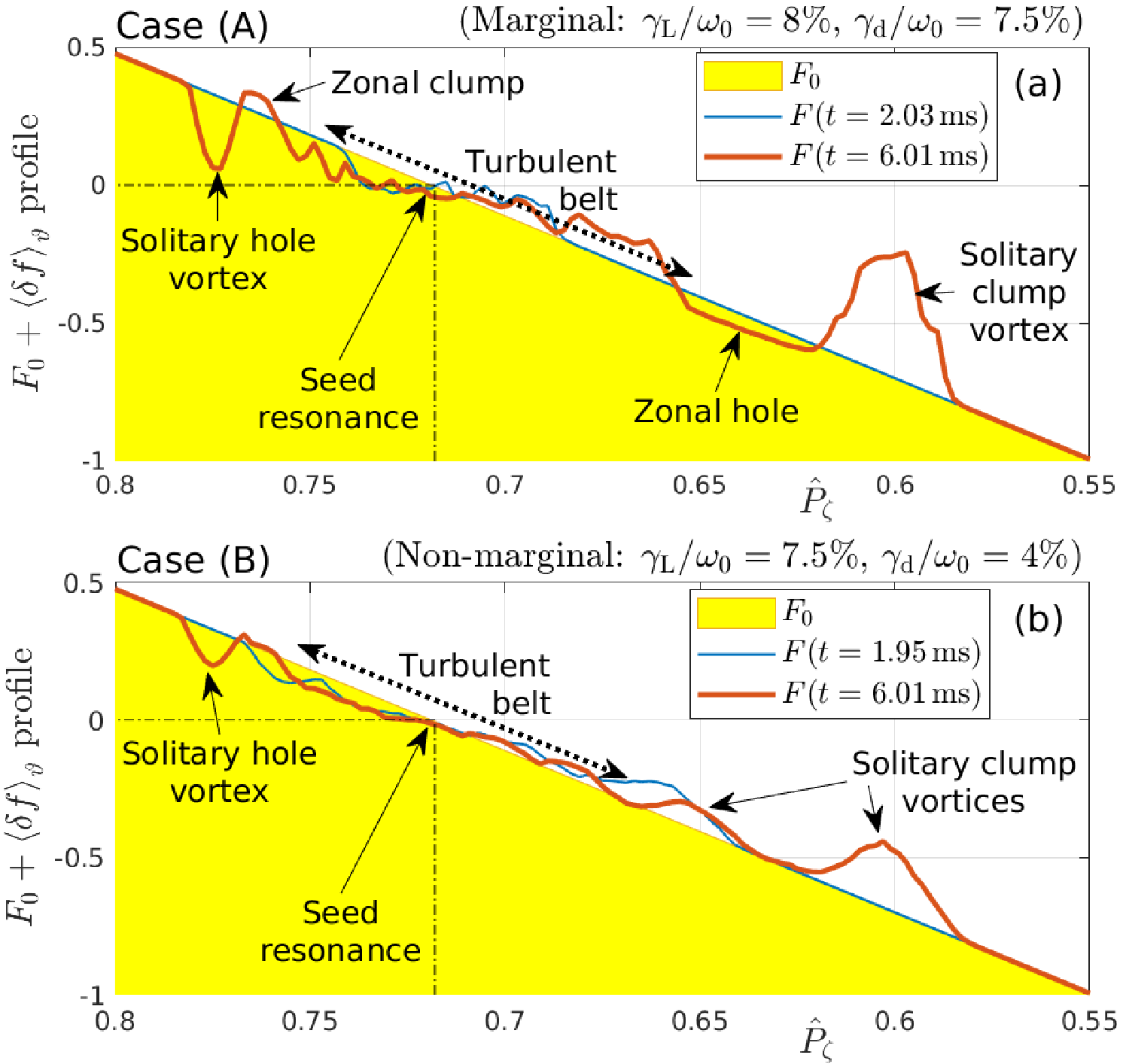}
  \caption{Poloidally averaged EP density profiles for the last two snapshots of Figs.~\protect\ref{fig:stages_fire-A} and \protect\ref{fig:stages_fire-B} for cases (A) and (B). The snapshot at $t = 6.01\,{\rm ms}$ ($\hat{t} = 1277.5$) was taken after solitary hole and clump vortices had detached and became well-separated from the turbulent belt. In case (A), a band of reduced density (zonal hole) lies behind the solitary clump, and a band of increased density (zonal clump) lies behind the solitary hole vortex (see also Fig.~\protect\ref{fig:stages_fire-A}). Note that the labels ``solitary clump/hole vortices'' are not quite accurate: Since we are plotting the poloidally averaged density, the mean value of $F(P_\zeta) = F_0 + \left<\delta f\right>_\vartheta$ at the radius of a hole or clump vortex includes fluid elements at neighboring $\vartheta$-values that originate from different radii. Of course, the local value of $F(P_\zeta,\vartheta)$ of each EP Vlasov fluid element is conserved, as required by the Liouville theorem, but this is not visible in the profiles shown here.}
  \label{fig:result_fire-AB_hc-profiles}
\end{figure}

Snapshots of this accumulation phase as it occurs in the actual simulation can be seen in Fig.~\ref{fig:result_fire-A_detach} for case (A) and Fig.~\ref{fig:result_fire-B_detach} for case (B), where they are labeled as stage (I). The accumulation process still continues for some time during stage (II), which we consider to be the detachment phase. The detachment of a clump or hole vortex from the turbulent belt is a gradual process, during which interactions with neighboring fragments become less intense and strong interactions become less frequent. Note that the $\delta f$ snapshots taken during stage (II) still show a clear correlation between the radial location of the wave fronts (horizontal dashed lines) and the field pulsations in panel (b). Thus, the fronts' radial advancement still occurs in a pulsed manner (recall Fig.~\ref{fig:result_spec-fronts}(b) and (e)), suggesting that beating plays a role in this process.

Although we cannot identify a precise timing, we consider the detachment to be more or less complete in stage (III). In stage (III) of the strongly unstable case (B), the large first-generation clump is subject to strong distortions, which can be attributed to the presence of a second-generation clump vortex that has grown large near the first. These two structures continue to strongly interfere with each other for some time. Finally, the snapshots for stage (IV) show the regime where the outermost clump and innermost hole can be regarded as completely detached vortices as they have separated far from and are hardly distorted by other clumps or holes.

\begin{figure}[tp]
  \centering
  \includegraphics[width=8cm,clip]{\figures/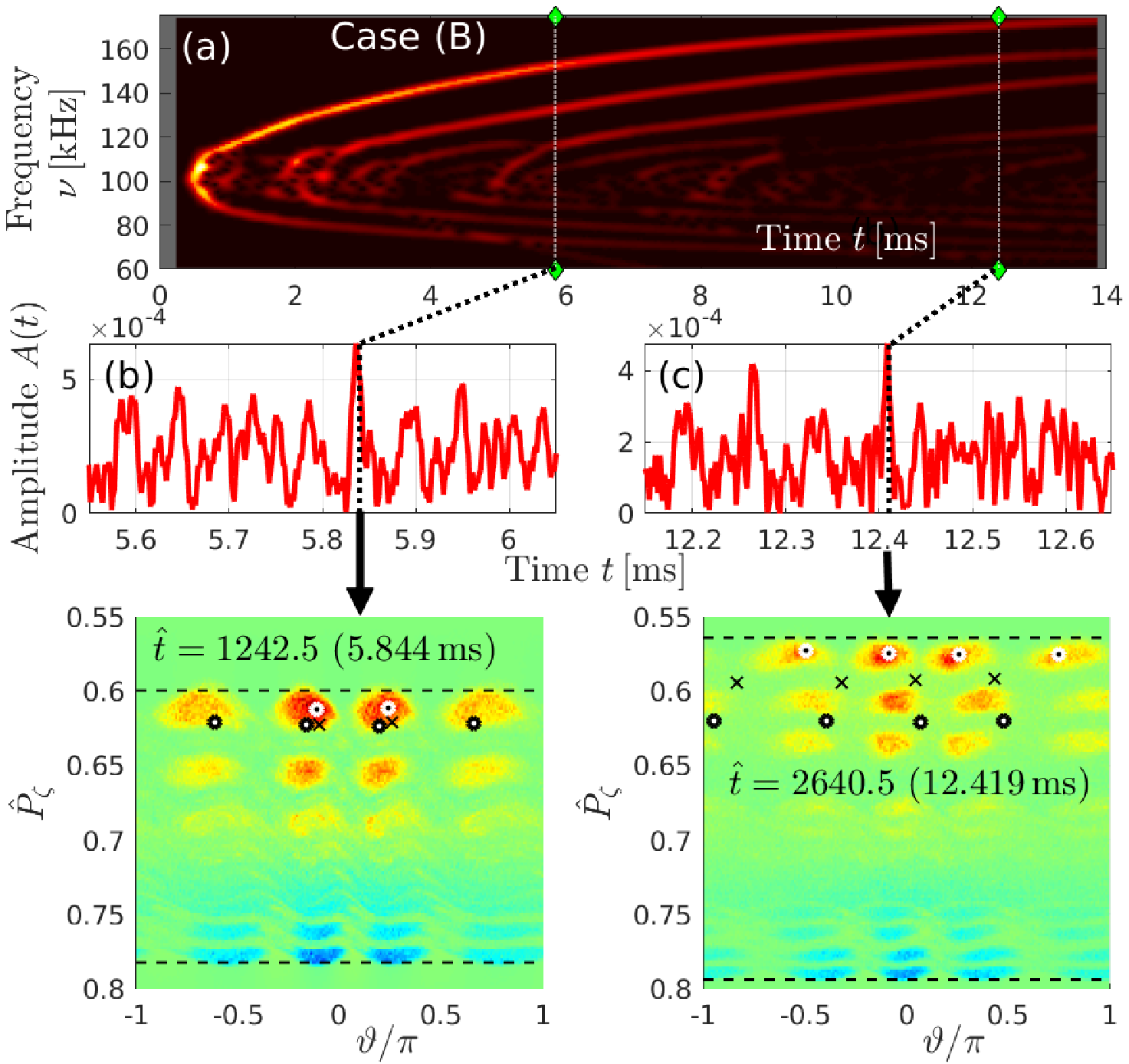}
  \caption{Instances of constructive interference causing large-amplitude spikes through spontaneous alignment of phase space structures in the advanced stages of the strongly unstable case (B), where multiple detached solitary hole and clump vortices are present. Panel (a) shows the Fourier spectrogram ($\Delta t_{\rm win} = 0.47\,{\rm ms}$), with vertical dotted lines indicating the times where relatively large spikes in the amplitude $A(t)$ are observed. The spikes are shown in panels (b) and (c), and coincide with the spontaneous phase alignment of multiple pump waves as shown in the snapshots of $\delta f$ in the lower part of the figure.}
  \label{fig:result_fire-B_interference}
\end{figure}

The last snapshots in Figs.~\ref{fig:stages_fire-A} and \ref{fig:stages_fire-B} (Section~\ref{sec:saturation}) were also taken during the fully detached stage (IV), and the poloidally averaged radial profiles of $\delta f$ for those snapshots are shown in Fig.~\ref{fig:result_fire-AB_hc-profiles}. Here, the detached H\&C vortices have the appearance of radially propagating solitary waves. An interesting observation can be made in the marginally unstable case (A): In the wake of the solitary clump vortex on the right-hand side of Fig.~\ref{fig:result_fire-AB_hc-profiles}(a), there is a depression in the EP density profile that we have labeled {\it zonal hole}. The attribute ``zonal'' refers to the nearly featureless uniformity of that region, which appears in the last snapshot of Fig.~\ref{fig:stages_fire-A} as a cyanish band that contains no obvious macroscopically coherent structures and separates the solitary clump from the turbulent belt. The same is true for the detached solitary hole vortex on the other side, which is separated from the turbulent belt by a {\it zonal clump} that appears as a yellowish band in the lower part of the last snapshot of Fig.~\ref{fig:stages_fire-A}. The presence of a zonal hole implies that the solitary clump vortex has not merely plowed through the surrounding EP Vlasov fluid but also absorbed some of it until the time of this snapshot ($t \approx 6\,{\rm ms}$). Conversely, the solitary hole vortex has deepened. In the next subsection, we will see that the clump begins to disintegrate thereafter.

It must be noted that even after detachment, the interference between the spatially separated phase space structures has noticeable effects. Besides ubiquitous beating, one manifestation of such an interference is the observation of relatively large spikes in the field amplitude that occur when multiple phase space structures happen to have their phases aligned. According to Appendix~\ref{apdx:beat}, we expect large spikes when multiple clumps or multiple holes are aligned, and it is beneficial when the clumps are out-of-phase with respect to the holes, or simply more intense. Two instances of such transient phase alignments can be observed in Fig.~\ref{fig:result_fire-B_interference}. Although transient, such large spikes may have an influence on rate at which the hole and clump vortices advance radially and accumulate (or lose) material. Similar processes may be relevant for triggering Abrupt Large-amplitude Events (ALE) \protect\cite{Bierwage18} {\cred whose precise timing, according to our current understanding, depends sensitively on the trajectories of multiple players,} just like the events of spontaneous phase alignment in Fig.~\ref{fig:result_fire-B_interference}.

It is remarkable that the outermost solitary clump vortex in Fig.~\ref{fig:result_fire-B_interference} retains its structure with $p=4$ elliptic points for a long time even after entering the domain of the $p=5$ resonance; namely, the region $\hat{P}_\zeta \lesssim 0.6$ (cf.~the first snapshot in Fig.~\ref{fig:stages_fire-A}).\footnote{The vortex is even able to travel beyond $\hat{P}_\zeta = 0.55$, where no simulation particles have been loaded. Although not shown here, we find that this initially empty region of phase space is subsequently filled by particles that leak from the massive clump vortex. However, the $\delta f$ simulation becomes invalid in that region because outward propagating Vlasov fluid elements can no longer be replaced with inward propagating ones. {\cred When this boundary effect feeds back on the phase space marker trajectories, we expect that it breaks the Liouville theorem. Our sensitivity tests indicate that this effect remains tolerable for the duration of our simulations.}}

The robustness of the solitary vortices that survive in spite of the persistent beating of the field wave (which is caused by the interference between all the pump waves in the system) indicates that, after detaching from the turbulent belt, their interiors may have attained a certain degree of adiabaticity, with more or less conserved actions as assumed in so-called waterbag models (e.g., see Refs.~\cite{Berk67, Khain07, Hezaveh21}). This can be confirmed by examining the motion of individual simulation particles, which constitutes the final part of this study.

\begin{figure*}[tp]
  \centering
  \includegraphics[width=16cm,clip]{\figures/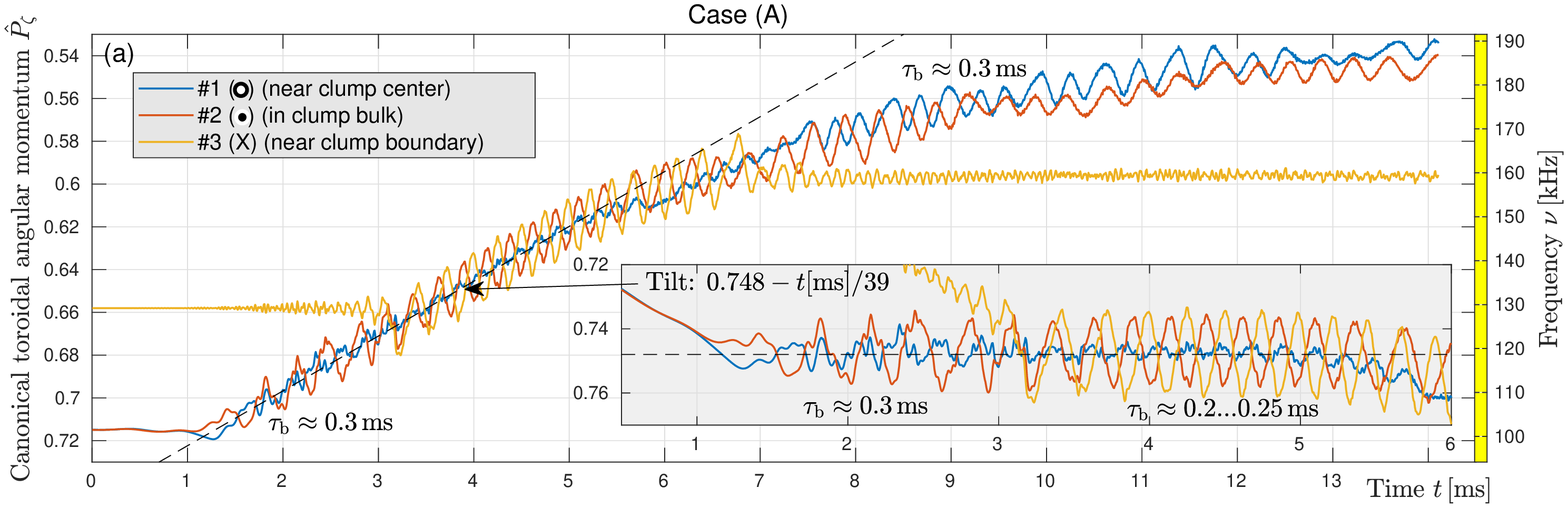}
  \includegraphics[width=16cm,clip]{\figures/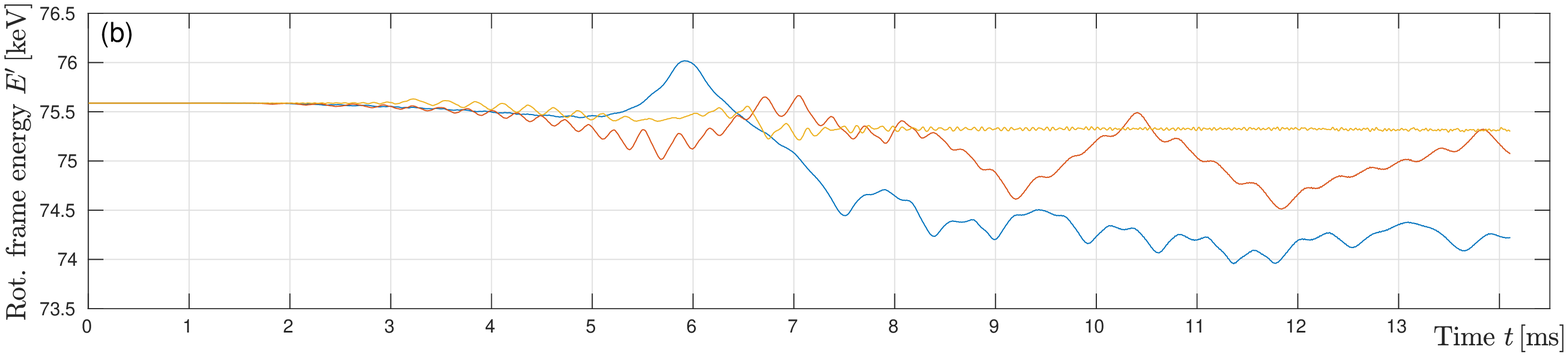}
  \includegraphics[width=16cm,clip]{\figures/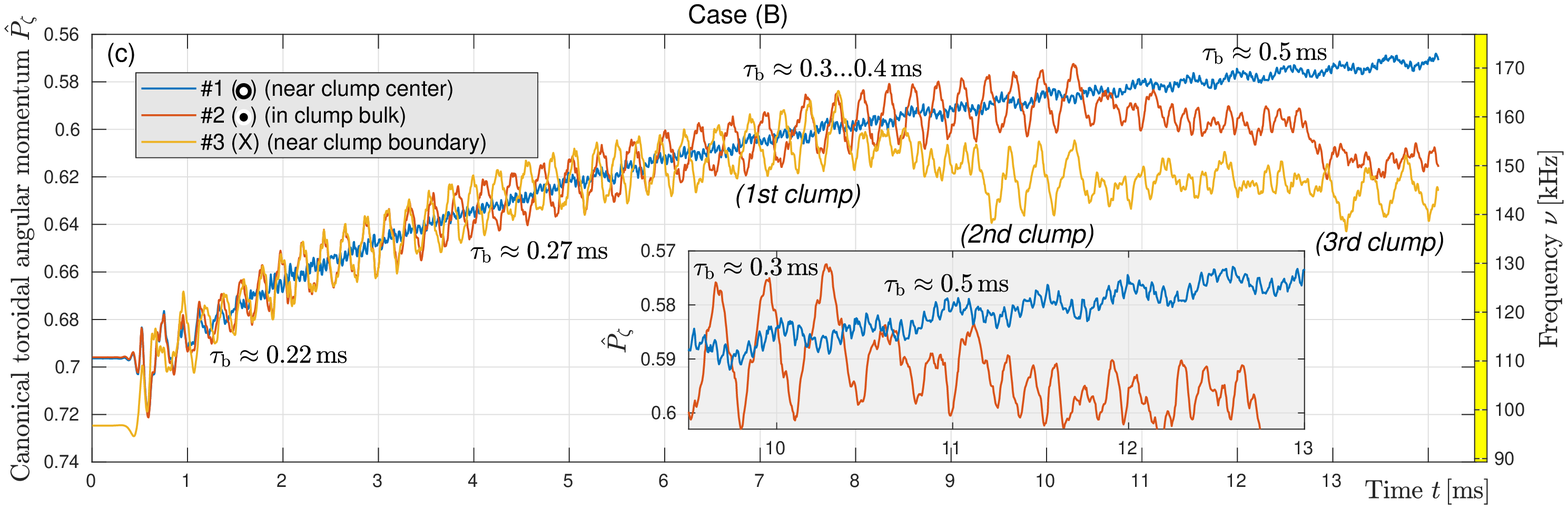}
  \includegraphics[width=16cm,clip]{\figures/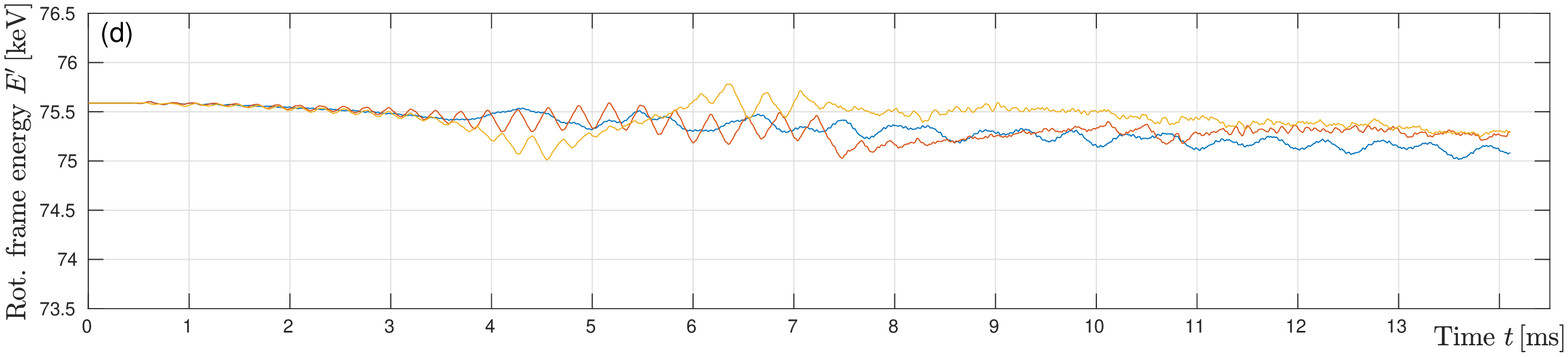}
  \caption{Time traces of the radial position $\hat{P}_\zeta(t)$ and rotating frame energy $E'(t)$ of three tracer particles labeled \#1, \#2 and \#3 in case (A) (top) and case (B) (bottom). One tracer is trapped in the center, one in the bulk, and one {\cred near the boundary} of the first-generation solitary clump vortex. The inset in panel (a) shows a portion of the data tilted such that the dashed line becomes horizontal to highlight the constancy of the clump's radial propagation in that time window. The zoomed-up inset in panel (c) is meant to show more clearly the rapid modulation caused by the field's beating. For the reader's convenience, the approximate bounce periods $\tau_{\rm b}$ during some stages of the evolution are shown as text labels. See also Fig.~\protect\ref{fig:result_fire-AB_tracers-spec} for the full time traces of $\nu_{\rm b}(t)$ for each tracer particle. The vertical axis on the right-hand side of panels (a) and (c) have been converted to frequency $\nu$ using Eq.~(\protect\ref{eq:pz_nu}).}
  \label{fig:result_fire-AB_tracers}
\end{figure*}

\begin{figure*}[tp]
  \centering
  \includegraphics[width=8cm,clip]{\figures/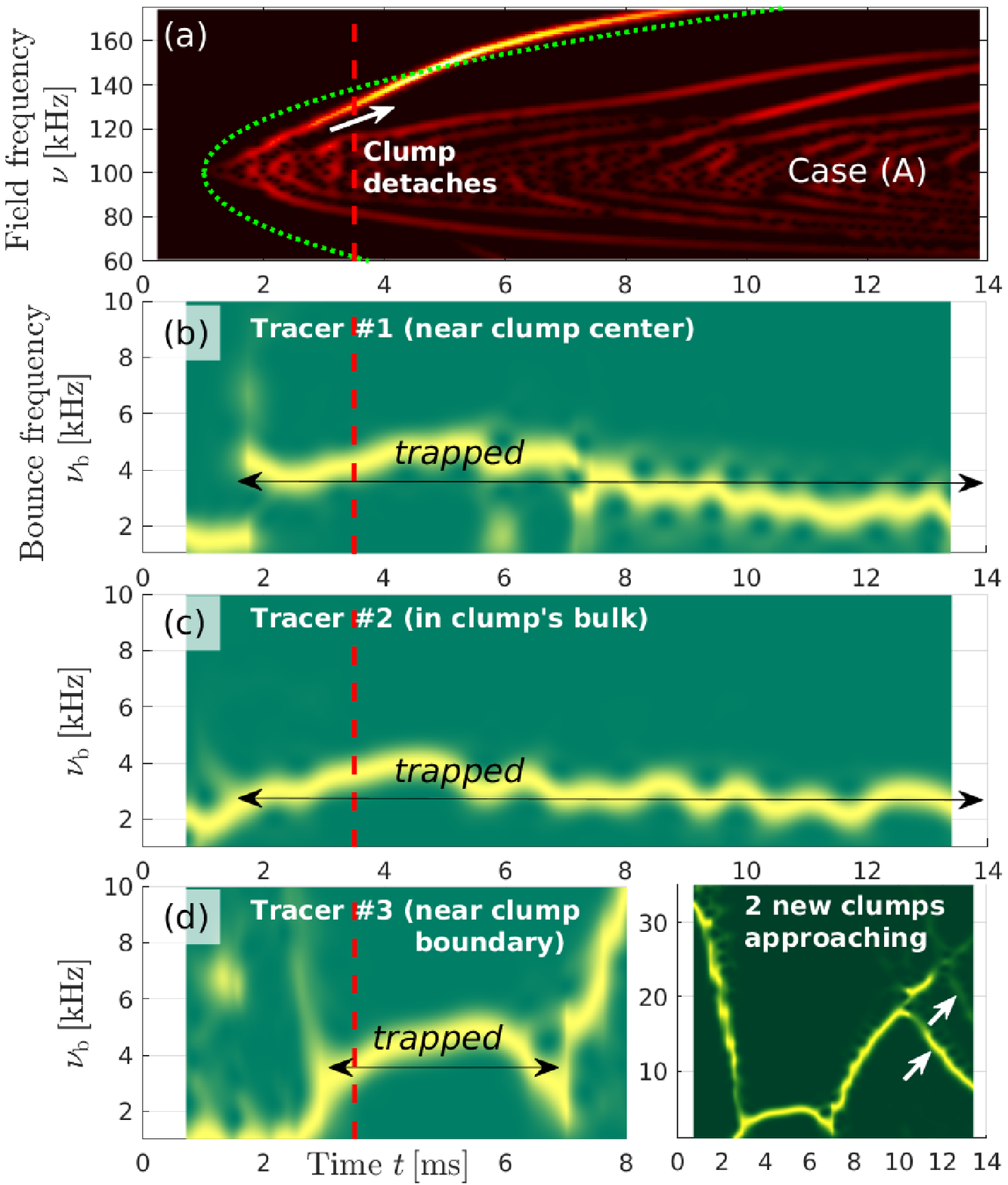}
  \includegraphics[width=8cm,clip]{\figures/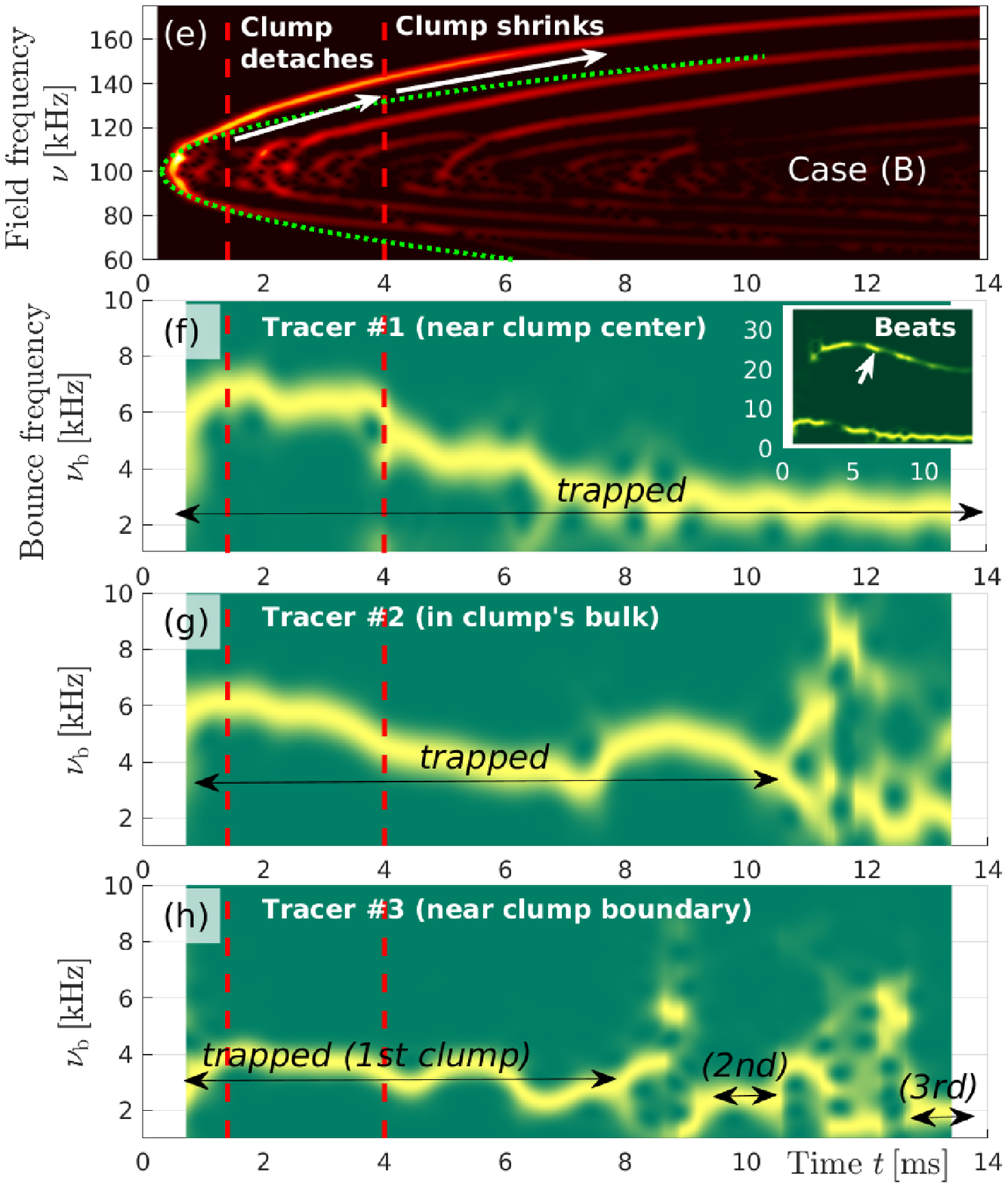} 
  \caption{Evolution of the bounce frequencies $\nu_{\rm b}$ of tracers \#1, \#2 and \#3 whose time traces $\hat{P}_\zeta(t)$ were shown in Fig.~\protect\ref{fig:result_fire-AB_tracers}. For orientation and comparison, panels (a) and (e) show again the chirps observed in case (A) (left) and case (B) (right) using a Fourier time window of size $\Delta t_{\rm win} = 0.47\,{\rm ms}$. Panels (b)--(d) and (f)--(h) show Fourier spectrograms of $\hat{P}_\zeta(t)$ using $\Delta t_{\rm win} = 3\times 0.47\,{\rm ms} = 1.41\,{\rm ms}$.}
  \label{fig:result_fire-AB_tracers-spec}
\end{figure*}

\subsection{Particle trapping and transport}
\label{sec:result_transport}

The $\delta f$ snapshots shown in Figs.~\ref{fig:result_fire-A_detach} and \ref{fig:result_fire-B_detach} contain black and white circles and black crosses, which indicate the locations of three selected simulation particles. During the $\tau_{0\zeta} = 4.7\,\mu{\rm s}$ time interval used to accumulate $\delta f$ Poincar\'{e} maps, each particle appears approximately 4 times at different $\vartheta$. In Fig.~\ref{fig:result_fire-A_detach} for the marginally unstable case (A), one can see that the particles represented by the black and white circles are part of the massive clump structure at the time of snapshot (I), whereas the particle represented by a black cross is absorbed by that clump at some later time.

The time traces $\hat{P}_\zeta(t)$ for these three tracer particles in case (A) are plotted in Fig.~\ref{fig:result_fire-AB_tracers}(a), where one can see that the black cross, here labeled as tracer \#3, is trapped near the clump's boundary at $t \approx 3\,{\rm ms}$, then transported outward with the clump for several milliseconds, before being released and deposited at a larger radius at $t \approx 7\,{\rm ms}$, indicating that the clump has begun to shrink. The time $t \approx 6\,{\rm ms}$ where the clump begins to shrink coincides with the time where the clump enters the domain $\hat{P}_\zeta \lesssim 0.6$, where the presence of the next resonance with $p/n=5/5$ becomes noticeable in the $\delta f$ halos (see the first snapshot in Fig.~\ref{fig:stages_fire-A}). That timing also coincides with a reduction in the clump's radial propagation speed (= chirping rate), which had been remarkably linear at earlier times, as can be seen more clearly in the inset panel in Fig.~\ref{fig:result_fire-AB_tracers}(a).\footnote{The linearity of the first half of the long-range chirp in Fig.~\protect\ref{fig:result_fire-AB_tracers}(a) does not seem to be a robust feature of the simulation. The occurrence, forms and number of long-range chirps is sensitive to details. It can vary significantly when the physical parameters (such as damping and drive) are changed slightly, and it is even affected by numerical parameters. This can be seen by comparing the spectrogram of case (A) in Fig.~\protect\ref{fig:stages_fire-A}(b) with those in Figs.~\protect\ref{fig:discuss_fire-A-pew_overview}, \protect\ref{fig:num_fire-A_qs-spec} and \protect\ref{fig:num_fire-A_dt-spec}(a) in Appendix~\protect\ref{apdx:model}. All these simulations were performed with the same values of $\gamma_{\rm L}$ and $\gamma_{\rm d}$ as in case (A).}

The time traces $\hat{P}_\zeta(t)$ for three tracer particles in case (B) are shown in Fig.~\ref{fig:result_fire-AB_tracers}(c). Here the rate of radial propagation (and chirping) can be seen to decrease continuously. As in case (A), the clump begins to lose particles as soon as it enters the domain $\hat{P}_\zeta < 0.6$. First, the outermost tracer \#3 (orange curve) is detrapped at $t \approx 8\,{\rm ms}$, followed by tracer \#2 (red curve) at $t \approx 10.5\,{\rm ms}$. The innermost tracer \#1 (blue curve) remains trapped until the end of the simulation. Note how tracer \#3 gets temporarily trapped by the second-generation clump around $t \approx 10\,{\rm ms}$ and by the third-generation clump after $t \approx 13\,{\rm ms}$, as these follow closely behind the first clump. As it happens, the fate of tracer \#3 was that the secondary trapping event moved this particle radially back inward.

These observations show the extent of global transport caused by the propagating solitary phase space vortices, which bears similarity to the physical picture of so-called ``bucket transport'', with the addition that --- besides the chirping $\ddot{\phi}$ \cite{Hsu94} --- the pulsations of the field amplitude $A(t)$ assist with the transmission of untrapped particles around the bucket by widening the nonadiabatic boundary layer that would constitute only a singular separatrix in the limit of a field wave with fixed amplitude and phase, and an isolated single resonance (see Fig.~\ref{fig:intro_x-point}).

Another closely related and important piece of information that we can gather from Fig.~\ref{fig:result_fire-AB_tracers} is the following. In both cases (A) and (B), one can see that particles that are located near the vortex center (namely, tracer \#1) remain there for the entire duration of the simulation. Similarly, particles in the bulk of the vortex and particles in the boundary layer remain there, with the peripheral particles being the first to escape when the vortex begins to shrink. This means that these vortices maintain a nested structure.

What we find particularly remarkable about this observation is that the nested structure of the vortices is maintained in spite of the persistent beating of the field wave, and in spite of the fact that the dynamics in case (A) are only marginally adiabatic. Let us elaborate these important points in some more detail.

The effect of the persistent violent fluctuations of the field in the form of amplitude pulsations and phase jumps can be clearly seen in Fig.~\ref{fig:result_fire-AB_tracers} in the form of high-frequency modulations of all time traces $\hat{P}_\zeta(t)$. Typically, we see between 5 and 7 oscillations per bounce period $\tau_{\rm b}$, so the time scale separation is not all that large.\footnote{The inset in panel (h) of Fig.~\protect\ref{fig:result_fire-AB_tracers-spec} shows the spectral peak of the beat signal in case (B) at about $20...25\,{\rm kHz}$, whereas $\nu_{\rm b} \approx 3...7\,{\rm kHz}$.} Moreover, in the case of tracer \#1 (blue curve), which is located near the clump center in both cases (A) and (B), the perturbations caused by the {\cred beating} have amplitudes that are comparable to the radial excursion of the particle due to its revolution around the vortex center. In Fig.~\ref{fig:result_fire-AB_tracers}(a) it is sometimes even difficult to see the bouncing signal of tracer \#1 within the ``beat noise''. Although we are looking only at very few particles, the overall integrity of the vortices suggest that other particles behave likewise. The possibility of maintaining such a nested structure in the presence of strong perturbations is fascinating and strengthens the stance of reduced adiabatic models such as the ``waterbag'' \cite{Berk67, Khain07}; in particular, when cast in a generalized form with a nonadiabatic boundary layer \cite{Breizman10, Hezaveh21}. Thus motivated, let us examine the degree of adiabaticity in our vortices.

Although it is possible to determine the characteristic bounce periods $\tau_{\rm b}$ by-eye when looking closely at the time traces in Fig.~\ref{fig:result_fire-AB_tracers}, the evolution of the bounce frequencies $\nu_{\rm b}$ of our tracers can be conveniently seen at a glance from the Fourier spectra that are summarized in Fig.~\ref{fig:result_fire-AB_tracers-spec}. With $\Delta t_{\rm win} = 1.41\,{\rm ms}$, we have used a fairly large sliding time window for the Fourier transforms, but one can nevertheless see significant variations of the bounce frequencies over time. Various observations that are interesting but not immediately relevant here are discussed in Appendix~\ref{apdx:misc_fbounce}. What we would like to emphasize here is how the bounce frequency compares to the chirping rate: The bounce frequency near the clump center, namely for tracer \#1 in Fig.~\ref{fig:result_fire-AB_tracers-spec}(b) and (f), is approximately $\nu_{\rm b} \approx (4...6)\,{\rm kHz}$ during the first few milliseconds after the clump detaches. The chirping rates in that part of the simulation can be readily inferred from Fig.~\ref{fig:result_fire-AB_tracers}, where the vertical axis on the right-hand side shows the radial position $\hat{P}_\zeta$ converted to frequency $\nu$ using Eq.~(\ref{eq:pz_nu}). In case (A), the slope is $\delta\hat{P}_\zeta/\delta t = \frac{1}{39}{\rm ms}^{-1}$ (as shown explicitly in the inset panel), so we have $\overline{\delta\dot{\nu}} = \frac{486.8\,{\rm kHz}}{39\,{\rm ms}} = 12.5\,{\rm kHz/ms}$. In case (B), we have roughly $\overline{\delta\dot{\nu}} \approx 20\,{\rm kHz}/2.5\,{\rm ms} = 8\,{\rm kHz/ms}$. With this, the adiabaticity parameter $\delta\dot\nu/\nu_{\rm b}^2$ in Eq.~(\ref{eq:adiabatic1}) becomes

\begin{equation}
\overline{\delta\dot\nu}/\nu_{\rm b}^2 = \left\{\begin{array}{lcl}
\frac{12.5\,{\rm kHz/ms}}{(4.5\,{\rm kHz})^2} \approx 0.6 & : & \text{Case (A)}, \\
\frac{8\,{\rm kHz/ms}}{(6\,{\rm kHz})^2} \approx 0.2 & : & \text{Case (B)};
\end{array}\right.
\label{eq:adiabaticity}
\end{equation}

\noindent where the overbar $\overline{(...)}$ means that we are looking only at the overall trend of the chirp, ignoring rapid fluctuations in $\delta\dot{\nu}$, such as phase jumps. Equation~(\ref{eq:adiabaticity}) highlights in a quantitative manner what may already have been obvious to experienced eyes from the time traces in Fig.~\ref{fig:result_fire-AB_tracers}: Our strongly unstable case is more adiabatic than the marginally unstable case (A), with the latter being only marginally adiabatic since its $\overline{\delta\dot\nu}/\nu_{\rm b}^2$ value is not far from unity. In fact, if it had not been for the observed nested structure, one may have been tempted to classify case (A) as nonadiabatic. Instead, it seems that an adiabatic treatment of the detached solitary clump vortices (and presumably also the solitary hole vortices) can be justified in both cases that were analyzed here, near and far from marginal stability, and with fairly strong drive $\gamma_{\rm L}/\omega_0 \approx 8\%$.

\begin{figure}[tb]
  \centering
  \includegraphics[width=8cm,clip]{\figures/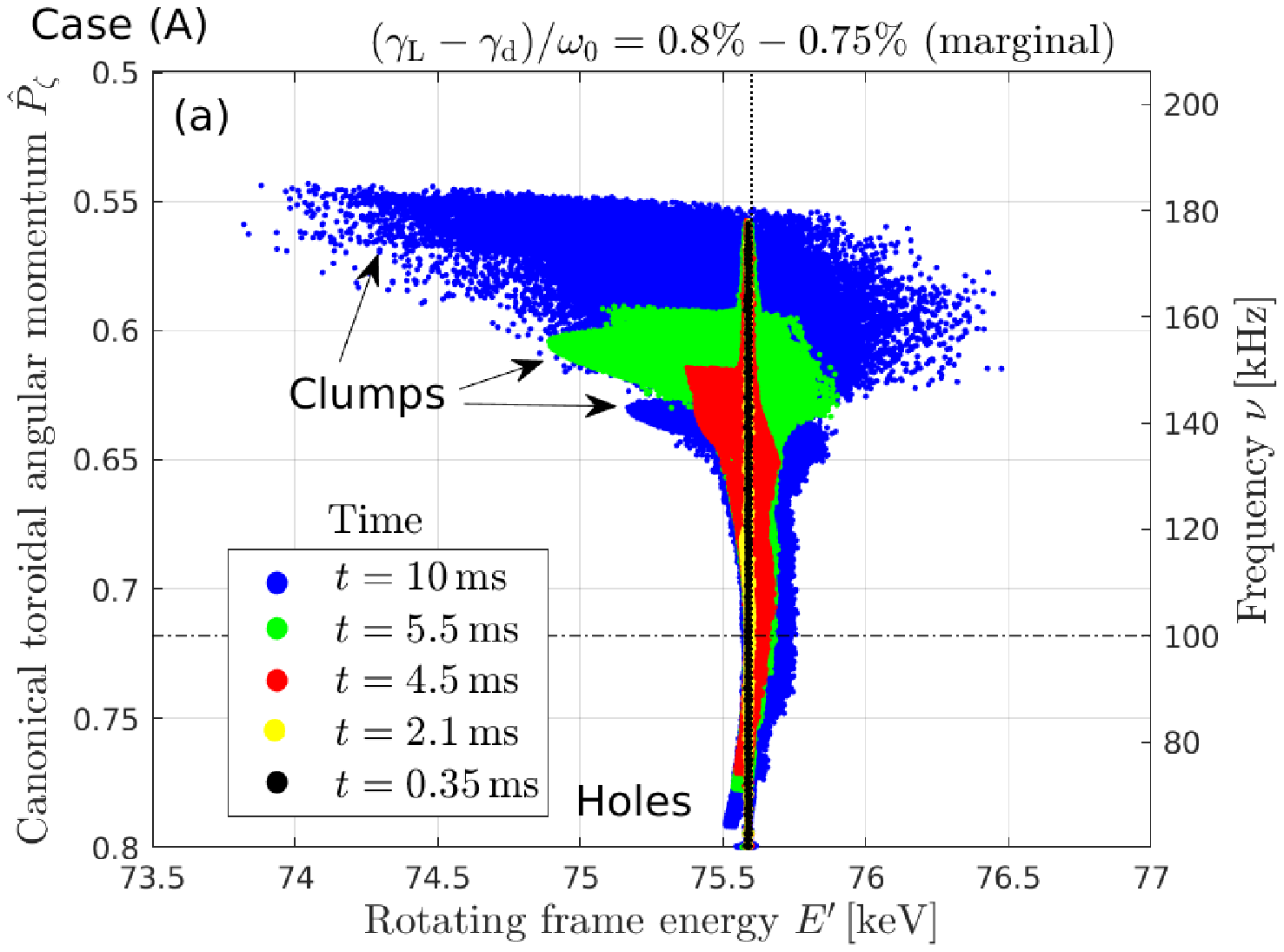}
  \includegraphics[width=8cm,clip]{\figures/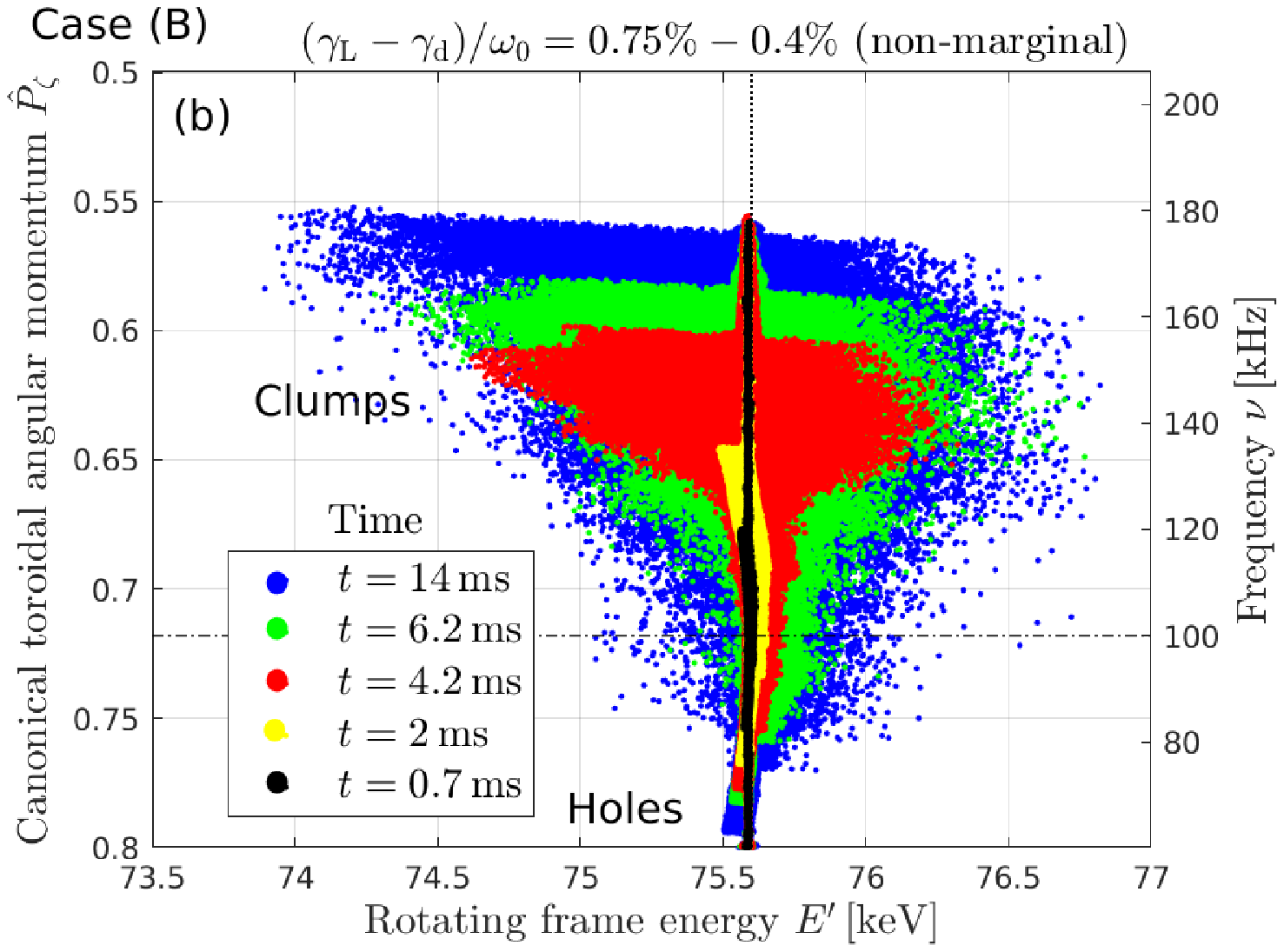}
  \caption{EP transport in the $E'$-$\hat{P}_\zeta$ plane, where $E' = E - \omega_0 \hat{P}_\zeta/n$ is the particle energy in the frame moving with the seed wave. Several snapshots of the particle distribution are shown for case (A) (top) and case (B) (bottom). More details can be found in Appendix~\protect\ref{apdx:misc_distr}, Figs.~\protect\ref{fig:result_fire-A_distr-E-P-th_t2-6ms} and \protect\ref{fig:result_fire-B_distr-E-P-th_t07-14ms}.}
  \label{fig:result_fire-AB_C-evol}
\end{figure}

Finally, we examine to what {\cred extent} the simulation particles depart from their initial position $E' = 75.6\,{\rm keV}$ in the rotating frame energy. Figure~\ref{fig:result_fire-AB_C-evol} shows that the particle distribution becomes slightly crescent-shaped but remains close to the initial $E' = 75.6\,{\rm keV}$ line during the first $\approx 4\,{\rm ms}$ in case (A) and during the first $\approx 2\,{\rm ms}$ in case (B). After that, the distribution broadens and reaches a width of about $\delta E' \approx 2.5\,{\rm keV}$ in both cases. The particle clusters in these broadened energy bands can be clearly associated with the upward propagating clumps. A more detailed view of the distribution's 4-D structure in $(\delta f,\hat{P}_\zeta,E',\vartheta)$ and $(\hat{P}_{\zeta 0},\hat{P}_\zeta,E',\vartheta)$ space, and how it evolves in time is given in Figs.~\protect\ref{fig:result_fire-A_distr-E-P-th_t2-6ms} and \protect\ref{fig:result_fire-B_distr-E-P-th_t07-14ms} of Appendix~\protect\ref{apdx:misc_distr}.

Time traces of $E'(t)$ for a few tracer particles are plotted in panels (b) and (d) of Fig.~\ref{fig:result_fire-AB_tracers}. For case (A), Fig.~\ref{fig:result_fire-AB_tracers}(b) shows that $E'$ is well conserved during the first 4 or 5 bounces. Oscillations, whose frequency seems to match the bounce frequency $\nu_{\rm b}$, become clearly noticeable at about $t \approx 3\,{\rm ms}$, which coincides with the stage where the first massive clump starts to detach from the turbulent belt (cf.~Fig.~\ref{fig:result_fire-A_detach}). Starting around $t \approx 5...6\,{\rm ms}$, where the clump begins to shrink and interact with the $p/n = 5/5$ resonance, we observe nonsinusoidal oscillations with a magnitude on the order of $1\,{\rm keV}$ and long periods of about $3...4\,{\rm ms}$, the cause of which is still unknown. In Fig.~\ref{fig:result_fire-AB_tracers}(b) and (d), there seems to be an overall trend for $E'$ to decrease in time, and the crescent-shaped bending of the full distribution as seen in Fig.~\ref{fig:result_fire-AB_C-evol} is probably related to this trend.

Note that the magnitude of high-frequency beat noise is smaller in the time traces of $E'(t)$ than in the time traces of $\hat{P}_\zeta(t)$, so that the curves in panels (b) and (d) of Fig.~\ref{fig:result_fire-AB_tracers} appear significantly smoother than in panels (a) and (c). This implies that there is a significant cancellation between beat-induced fluctuations in $E$ and $\hat{P}_\zeta$. This observation leads us to conclude that beating may {\it not} be the primary mechanism that leads to the decoupling of $E$ and $\hat{P}_\zeta$ and, thus, the breaking of the $E' = E - \omega P_\zeta/n = {\rm const}$.\ condition. Therefore, we suspect that the solitary vortices themselves may be the primary cause for the perturbations in $E'$; in particular, their radial propagation (chirping and plowing through surrounding material) and direct mutual interactions (e.g., when their nonadiabatic boundary layers interfere or overlap).

The $E'$ values of the three tracers in Figs.~\ref{fig:result_fire-AB_tracers}(b) and (d) vary less in case (B) than in case (A), but this trend is not seen in Fig.~\ref{fig:result_fire-AB_C-evol}, where the full particle distribution is scattered more broadly in case (B) with weaker damping than in case (A) with stronger damping. Presumably this is due to the larger number of more closely spaced vortices in case (B). If conditions were equal, we expect that the variations in $E'$ should decrease with decreasing damping rate, because $\gamma_{\rm d} = 0$ yields effectively constant $E'$ as can be confirmed in Fig.~\ref{fig:result_fire-A0_tracers_en_small} of Appendix~\ref{apdx:undamped} for case (A0).

The $E'$ line broadening seen in Fig.~\ref{fig:result_fire-AB_C-evol} has been reproduced with 2.5 times shorter time steps, which gives better energy conservation, so we may assume that its cause is not a numerical one. In particular, the crescent-shaped distortion that dominates during the first few milliseconds of the simulations seems to be a robust physical feature. We suspect that this (small) modification of $E'$ is a consequence of chirping, as it affects the domains of holes and clumps alike. However, this hypothesis remains to be tested.

At present, a systematic limitation of our simulations prohibits a meaningful study of the observed $E'$ line broadening. Recall from Fig.~\ref{fig:model_slice} that the phase space structures have the form of elongated cylinders in $(P_\zeta,E',\vartheta)$-space, which we have effectively truncated by loading particles only along a line with $E' \approx E'_0 = 75.6\,{\rm keV} = {\rm const}$. This means that our simulations using the $\delta f$ method do not accurately represent dynamics in the $E'$ direction, where no marker particles have been loaded, because the continuity of GC phase space and, thus, the condition ${\rm d}G/{\rm d}t = 0$ underlying Eq.~(\ref{eq:w}) is violated. This does not immediately make the entire simulation unphysical, but it can produce artifacts, and the ``ghost chirp'' in Fig.~\ref{fig:intro_fire-A_chirp}(b) may be one such artifact (see Appendix~\ref{apdx:misc_ghost}). For the scope of the present work, this does not seem to be a problem. The line broadening reaches only $\delta E'/E' \approx 3\%$ in $10\,{\rm ms}$, and we show in Appendix~\ref{apdx:model_pew} that similar dynamics are obtained when marker particles are loaded in a band of width $\Delta E'_0 = 7.5\,{\rm keV}$ that fully encompasses the broadened distribution in Fig.~\ref{fig:result_fire-AB_C-evol}.

\section{Summary, discussion and outlook}
\label{sec:summary}

{\cb When multiple field fluctuations with the same polarization are simultaneously present in the same spatial domain of a plasma, their linear superposition causes beating in the combined signal $s(t) = A(t)\sin(-\omega_0 t - \phi(t))$, here measured relative to a reference wave with frequency $\omega_0$. The beats consist of pulsations in the amplitude $A(t)$ and jumps by $\pm\pi$ in the phase $\phi(t)$. The pulse length of the beats, their magnitude, and the time scale of the phase jumps depend on the separation between the frequencies and the relative amplitudes of the interfering waves.

The amplitude and phase modulations associated with beating may contain useful information about possible nonlinear couplings between the interfering waves. This has been exploited, for instance, in studies of coexisting EP-driven Alfv\'{e}n modes and rotating kink/tearing modes \cite{Hole09, ChenW17, Zhu18}.\footnote{\cb When the frequency difference is large, as for high-frequency Alfv\'{e}n and low-frequency tearing modes, the slowly evolving modes may also be treated as a stationary non-axisymmetric perturbation of the reference state (or equilibrium) around which the high-frequency modes oscillate.}
Since different modes generally have different spatial structures, such cases exhibit not only temporal modulation but also spatial modulation. The spatial interference may add to the complexity of the dynamics.

Beating can occur as a passive side-effect, merely producing spectral sidebands. However, the beats of the field can also play an active role in the dynamics through direct or indirect feedback on the particles and/or field perturbations. One prominent example is spectral line splitting in regimes where chirping is suppressed by strong scattering \cite{Fasoli97, Fasoli98}. Beating is an integral part of that process, where spectral broadening occurs in a well-organized fashion through limit cycle bifurcations (see also the introductory discussion in the last paragraph of Section~\ref{sec:review_q}).

In the present study, we examined the role of beating in cases with strong chirping in a ``collisionless'' plasma; i.e., in the absence of drag or scattering. In contrast to the discrete line splitting seen in the above-mentioned scattering-dominated cases, scenarios with strong chirping exhibit a high degree of temporal incoherence, as a large portion of the phase space tends to be dominated by transient convective dynamics. Coherent vortex structures tend to form only after a significant delay (millisecond scale).

Our beats can be viewed as the linear modulation of one field mode that is driven by multiple sources (density waves) in the phase space of one particle species (here, energetic deuterons).} Although the fundamental process underlying this phenomenon is the exchange of energy between particles and field waves, the observations are perhaps most easily interpreted as a collective effect; namely, as the result of {\it beating} between multiple pump waves in phase space density that jointly drive the field oscillation. This physical picture was introduced in Section~\ref{sec:review_pump} and supported by the analysis in Appendix~\ref{apdx:beat}. The purpose of this study was to clarify how the beating of the field acts back on the pump waves, and what role beating plays in the formation and evolution of solitary vortices in phase space, which are responsible for long-range frequency chirping and associated particle transport.

The system considered was an ideal incompressible electromagnetic flute mode in realistic tokamak geometry and with an initial frequency in the shear Alfv\'{e}n band ($\nu_0 = \omega_0/(2\pi) \sim 100\,{\rm kHz}$) that interacts with circulating energetic deuterons with kinetic energies around $80\,{\rm keV}$. The use of a semi-perturbative model, namely a fixed spatial mode structure, constrains the validity of our numerical analysis to the long-wavelength limit, namely low toroidal mode numbers $n \sim \mathcal{O}(1)$, whose poloidal harmonics tend to have a large radial width and would undergo only relatively little distortion. Such low-$n$ modes are typically driven by energetic particles (EP) with large magnetic drifts. Although the phase space structures responsible for chirping remain within the range of these magnetic drifts, the drifts' large radial {\cb extent} makes the chirp-induced transport global and, therefore, practically relevant.

In the following subsections, we summarize the main insights that we have gained in this work and discuss possible implications and directions for further study. {\cred Concluding remarks are made in the last Section~\ref{sec:summary_exec}.}

\subsection{Halos reveal phase relations and may fuel the growth of holes and clumps}

We demonstrated the existence of robust nonresonant modulations in EP phase space density $\delta f$ that we call {\it halos}. They are visible as radially elongated stripes of alternating sign ($\delta f \gtrless 0$) in $\delta f$-weighted kinetic Poincar\'{e} plots accumulated over a short interval of one toroidal transit time, {\cred here} $\tau_{\zeta 0} = 4.7\,\mu{\rm s}$. These halos are interesting for several reasons, which we summarize here.

In order to correctly interpret the phase space dynamics, it is often essential to compare the phase space structures seen in contour plots of $\delta f$ with the instantaneous flow lines of the EP Vlasov fluid. In principle, the flows are determined by the instantaneous amplitude $A(t)$ and phase $\phi(t)$ of the field, but reconstructing the flow lines from that information is difficult in realistic geometry and with large magnetic drifts. For instance, when trying to estimate locations of X- and O-points along $\vartheta$ at a certain time $t$, we cannot simply shift the initial field pattern forward in time using the relation $\Delta\vartheta = \phi(t)/m$, because the magnetic drifts turn that relation into something more complicated. Conventional (long-time) kinetic Poincar\'{e} plots are not applicable in a rapidly varying field, and inferring the locations of effective O- and X-points through visual inspection of near-resonant $\delta f$ patterns also becomes difficult after the first beat, because those patterns become {\cred complicated}.

Our observations indicate that far off-resonant halos (Fig.~\ref{fig:model_fire-A_df-interpret}(b)) are strongly correlated with the instantaneous phase of the field, so they can be used to track the effective locations of O- and X-points. On that basis, it is possible to determine at least the radial direction of the flows at each poloidal angle. Here, this technique has been used mainly during the early stages of a chirping simulation, where solitary vortices have not formed yet. It {\cred helped} us to estimate flow lines in phase space and {\cred interpret the processes} associated with the onset of beating and chirping. The effect of more or less abrupt phase jumps between successive pulses of the beating field can be observed clearly, and the halos have {\cred revealed details} about that process (Fig.~\ref{fig:result_fire-A_beating}).

Besides being an indicator of the phase relation between field and density waves, we suspect that the halos may also play an active role in the dynamics: The phase relation between halos and density wave fronts is such that, during the beginning of a pulse, a halo of increased density ($\delta f > 0$) lies ahead of the clump wave, and a halo of reduced density ($\delta f < 0$) lies ahead of a hole. We have argued based on Fig.~\ref{fig:result_fire-A_beat2_island} that this may facilitate the inflation of these structures, making them grow larger with each pulse of the beating field.

Although we can explain halos as collective density modulations, the reasons for why they exist in this particular form and why they evolve in the way they do remains to be understood. On the one hand, the halos are attached to the H\&C wave fronts, so we {\cred usually see} two counter-propagating halos. On the other hand, the halos are closely linked to the phase $\phi$ of the field. In that respect, the halos are similar to the equilibrated stationary modulations of phase space density in Fig.~\ref{fig:intro_fire-A-freeze500_poink-df}(b), which are also phase-locked collective structures, not tied to the motion of individual particles. The close connection with the field phase is also manifested in the fact that halos can move ahead of or lag behind the H\&C wave fronts that they are more or less attached to. Overall, the halos seem to be a manifestation of the feedback between collective phase space structures and the field. It would be interesting to take this ``explanation'' from the philosophical to the scientific level.

\subsection{\cred Feedback loop for staircase-like nonadiabatic onset of chirping}
\label{sec:summary_staircase}

One of the most pronounced effects of beating was observed during the nonadiabatic onset of chirping during the first few milliseconds of our simulations (Fig.~\ref{fig:result_spec-fronts}). Based on our interpretation of the field dynamics as being a consequence of a superposition of multiple pump waves in EP phase space, we have identified the following {\it feedback mechanism} (Section~\ref{sec:result_front}, Fig.~\ref{fig:result_fire-A_beat2_island}):
\begin{itemize}
\item  The emergence of two pump waves with frequencies $\omega^+$ and $\omega^-$ causes the field to beat, {\cred with amplitude pulsations of order 100\% and phase jumps by $\pm\pi$}.
\item  {\cred These beats} drive the pump waves further away from the seed resonance {\cred by making the effective phase space islands (= flow contours) pulsate with alternating effective O- and X-point positions.}
\item  {\cred The resulting increase in the frequency difference $\Delta\omega = \omega^+ - \omega^-$ causes more beats per time. Moreover, new pump waves (hole-clump pairs) are generated with each beat.}
\item  {\cred The increased radial transport also pumps more energy into the field perturbation, whose growth increases the size of the islands (= steeper flow contours) and the step size of the radial displacements.}
\end{itemize}

\noindent Initially, this feedback causes an exponential acceleration of the up- and downward chirps $\omega^+(t)$ and $\omega^-(t)$. High-resolution spectrograms obtained with the DMUSIC algorithm indicate that the frequency advances in a somewhat discontinuous fashion (Fig.~\ref{fig:result_spec-fronts}).\footnote{Qualitatively similar behavior known as ``parametric ladder climbing'' has been reported in other systems \protect\cite{Barth14}.} The staircase-like onset of chirping is consistent and often correlated with the pulsed radial expansion of the pump wave fronts. However, the increasing frequency difference $\Delta\omega$ also leads to a shorter pulse length $\tau_{\rm pulse} = 2\pi/\Delta\omega$ {\cred and reduced energy transfer to the field. The shorter pulses and saturating field amplitude yield smaller steps in frequency, the consequence being} a deceleration and saturation of the chirps.

Exponential acceleration followed by exponential deceleration is the nature of a hyperbolic tangent. In a simplified limit of their general theory \cite{Chen16}, Zonca \& Chen have constructed a phenomenological ``relay runner'' model \cite{ZoncaTCM99}, that predicts nonadiabatic chirping of the form
\begin{equation}
\delta\omega(t) \propto \tanh(\gamma_0 (t-t_0)/2);
\end{equation}

\noindent where $t_0$ is the point of inflection and $\gamma_0 = \gamma_{\rm L} - \gamma_{\rm d}$ the initial growth rate (linear drive $\gamma_{\rm L}$ minus field damping $\gamma_{\rm d}$). This formula matches the overall trend and time scale of the early staircase-like nonadiabatic chirps in our simulations, both in the marginally unstable and strongly unstable cases (Figs.~\ref{fig:result_spec-fronts} and \ref{fig:result_fire-AB_spec-fit}). The magnitude of the chirping range had to be fitted with a free parameter, because the original relay runner model was formulated for a different scenario than ours: The model coefficients in Ref.~\cite{ZoncaTCM99} were derived for magnetically deeply trapped EPs and short wavelength modes, where the {\it batons} in the form of resonant pump waves in phase space density are {\it relayed} by neighboring poloidal harmonics of a toroidal Alfv\'{e}n {\cred wave packet} in a process of convective amplification \cite{Zonca15b}, as illustrated in Fig.~\ref{fig:intro_x-point}(d). In the future, it would be beneficial to formulate reduced quantitative models of nonadiabatic chirping for various scenarios.

In our case, there was only a single long-wavelength mode ($m/n = 6/5$) interacting with circulating EPs that perform large magnetic drifts. The relay runners were realized here by successive pulses (beats) of the same mode {\cred with alternating phase $\phi$}. In close proximity to marginal stability ($\gamma_{\rm L} - \gamma_{\rm d} \ll \gamma_{\rm L}$), namely in case (A), the first saturation of the resonant instability occurred at a very small amplitude (Fig.~\ref{fig:result_fire-A-A0_saturation}), and successive pulses gradually rose to higher levels (Figs.~\ref{fig:result_fire-A_beating} and \ref{fig:result_spec-fronts}). This can be regarded as a proxy of the above-mentioned convective amplification process, except that in our semi-perturbative model the convection occurred only in EP phase space, without shifting the peak of the field mode. However, from the perspective of the clumps, the local field did appear to be ``pre-convectively'' amplified, because we placed the seed resonance off-peak, so that the clumps' radial propagation took them into regions with a stronger field. See Appendix~\ref{apdx:model_pert} for further discussion on nonperturbative effects.

\subsection{Zonal and vortical structures}
\label{sec:summary_struct}

In the wake of the pump wave fronts that are observed during the nonadiabatic onset of chirping, a turbulent belt forms, whose dynamics are reminiscent of radially sheared convective interchange and wave-breaking on various scales. Although complicated in their detailed appearance, the phase space structures exhibit a certain degree of radial stratification (e.g., see the last snapshots in Figs.~\ref{fig:stages_fire-A} and \ref{fig:stages_fire-B}). Considering that the overall dynamics are fed by background gradients (generally, in combination with sources and sinks), one may describe the overall transport of material and energy with equations that govern only the zonal component and ignore the angular dependence. That idea has been realized in the theory by Zonca \& Chen \cite{Zonca15b,Chen16}, a specific form of which describes the evolution of so-called ``phase space zonal structures (PSZS)''. A quantitative comparison with the transport occurring in the turbulent belt in our simulations would be interesting. One complication that might cause discrepancies is that the assumption of two disparate spatial scales (background plasma and mode structure) may be violated in our case, since we deal with long-wavelength instabilities comparable to system size.

Generally speaking, a representation in terms of PSZS only as illustrated in Fig.~\ref{fig:intro_x-point}(d), without any angular dependencies, would yield reliable quantitative predictions for transport of particles and energy only if suitable transport coefficients can be calculated. Obviously, these coefficients will be very different in strongly turbulent regimes involving many modes, in cases with intermittent soliton-like avalanches, and in scenarios with propagating vortices subject to a beating field mode as we have simulated. In fact, the effective transport coefficients may even vary within one scenario. For instance in our case, transport coefficients are likely to differ between the near-resonant turbulent belt and the inter-resonant region occupied by detached solitary vortices. The results reported in this work may contain information that can be used in the construction of suitable transport coefficients for reduced models, such as the PSZS paradigm.

From an observational perspective, the formation of massive vortical clump structures involves a gradual aggregation of smaller fragments that flake off on the high-density side of the turbulent belt and are advected downhill towards the clump wave front (see the cartoon in Fig.~\ref{fig:result_detach}), with the reverse being true for holes. Beating plays a key role in this process and a {\it feedback loop} similar to that identified during the nonadiabatic onset of chirping (summarized in Section~\ref{sec:summary_staircase} above) is present here as well:
\begin{itemize}
\item  Radially sheared poloidal motion of pump waves {\cred (holes and clumps)} causes the field to beat.
\item  The beating field causes the hole and clump wave fronts to advance radially in a pulsed manner {\cred and spawns} new clump and hole fragments with each pulse. {\cred The broadening of the spectrum and increasing number of pump waves causes} more (and more complicated) beats.
\item  {\cred By driving radial convection of the fragments, the beats also facilitate} their accumulation into ``massive'' (wide and deep) holes and ``massive'' (wide and dense) clumps.
\end{itemize}

In other words, the beating enhances mixing between radially stratified layers of pump waves (hole and clump fragments), with net transport occurring due to the initial density gradient. One the one hand, the stronger mixing tends to inhibit the early detachment of vortical structures. On the other hand, it can fuel the growth of massive holes and clumps, especially at the boundaries of the turbulent belt. {\cred The larger value of $|\delta f|$ inside a massive structure makes it a more powerful pump wave (cf.~Eq.~(\ref{eq:dadt})). When driven by a few dominant massive structures, the beats of the field may become more regular and more intense.} Ultimately, this may facilitate the detachment of a massive clump or hole, because it is likely that some material will have to stay behind during the detachment. A massive vortex structure is more likely to remain intact and detached after loosing some material than a smaller one, because it has more influence over the auto-resonant field oscillations (due to larger integrated $|\delta f|$ in Eq.~(\ref{eq:dadt})), thus, perpetuating its own stability by prolonging phase locking.

Although a higher degree of on-average phase locking to the field makes a vortex more robust, too much phase locking would lead to enhanced resonant particle trapping and, thus, inhibit chirping. Here, phase slippage is facilitated by strong field damping. In addition, beating (due to the presence of other pump waves) and chirping \cite{Hsu94} both contribute to the maintenance of a relatively wide nonadiabatic boundary layer (Fig.~\ref{fig:intro_x-point}(c)), {\cred which prevents the} vortices from falling into a state of adiabatic stagnation.

We also observed the formation of truly zonal structures (indicated in Figs.~\ref{fig:stages_fire-A} and \ref{fig:result_fire-AB_hc-profiles}). After the first solitary clump vortex in the marginally unstable case (A) had departed from the turbulent belt, it left behind a region of reduced density that we have called ``zonal hole''. Similarly, a ``zonal clump'' was observed behind the first solitary hole vortex on the other side of the {\cred turbulent belt around the} seed resonance. No such structures were visible in the more strongly unstable (less damped) case (B). It may be worthwhile to investigate in a future case study under what conditions such zonal holes and clumps form and what role they play. For instance, one may speculate that such zonal modulations of the EP density profile play a role in the formation and radial propulsion of subsequent generations of hole and clump vortices. Meanwhile, the leading solitary vortices that have produced those zonal perturbations will experience a reduced EP density gradient across their radial width. This may contribute to the reduction of their radial propagation speed and, thus, their chirping rates.

\subsection{Robust nested vortices in the presence of strong beating}
\label{sec:summary_waterbag}

Even before complete detachment, the internal structure of hole and clump waves was found to resemble, on average, concentric nested layers of circulating flows. Especially in the marginally unstable case (A), this observation was somewhat surprising, because the adiabaticity parameter defined in Eq.~(\ref{eq:adiabaticity}) has a value close to unity in this case: $\overline{\delta\dot\nu}/\nu_{\rm b}^2 \approx 0.6$. {\cred It would not have been surprising to see this case dominated by nonadiabatic dynamics,} where long-lived vortices do not form and where phase space structures tend to be of collective nature, in the sense that different particles would constitute these structures at different times. In such a scenario, there would be no ballistic ``bucket'' transport.

{\cred Moreover}, we expected that the amplitude pulsations and phase jumps associated with ubiquitous beating would scramble the interior of any apparently coherent phase space structures and, if not prevent their formation, at least reduce their life time. Therefore, we were fascinated by the fact that prior simulations of such chirping systems (with varying degrees of complexity) routinely show the presence of long-lived phase space vortices. In particular, our curiosity was raised by the fact that the observed beats of the field are characterized by instantaneous growth rates $\gamma = {\rm d}\ln A/{\rm d}t = \dot{A}/A$ and phase shift rates $\dot{\phi}$ that can be comparable to and even larger than the oscillation frequency of the seed wave,
\begin{equation}
{\rm max}\left|\gamma\right| \sim {\rm max}\left|\dot\phi\right| \gtrsim \nu,
\label{eq:adiabat_g_phi}
\end{equation}

\noindent which seems to contradict assumptions like $|\gamma| \sim |\dot\phi| \ll \nu$ that are attached to theories of adiabatic chirping; namely, the separation of time scales needed to construct adiabatic invariants. In the theoretical derivations, this problem had been sidestepped by considering only one solitary vortex structure at a time (see Eq.~(23) in Ref.~\cite{Berk99}). By ignoring explicitly the interference between multiple pump waves, beating is taken out of the picture, leaving only slowly evolving amplitudes and phases.

One of our main results, reported in Section~\ref{sec:result_transport}, is that we (re)confirmed the robustness of the phase space vortices in our simulations and showed that, on average, their interiors {\cred maintains a nested} structure (i.e., good adiabatic constants of motion), in spite of persistent strong beating.

One may argue that this is not surprising since the largest and fastest changes in $A(t)$ and $\phi(t)$ occur at times of destructive interference, when $A(t)$ is small and, therefore, may hardly affect the particle motion. In addition, untrapped phase space structures in a collisionless plasma can be destroyed only by phase mixing, and one might think that the phase mixing times would be much longer than the typical beat periods. However, we did not find order-of-magnitude estimates sufficiently convincing, because for realistic tokamak geometry and with realistic parameters for the particle motion and field oscillations, the separation of time scales is often vague, and it was certainly vague in the cases we have simulated here. For instance, Fig.~\ref{fig:intro_fire-A_beat}(a) shows that the pulse period $\tau_{\rm pulse}$ associated with the beats is only a couple of wave periods $T_0 = 2\pi/\omega_0$ long, and the same counts for the ratio of $\tau_{\rm pulse}$ and the nonlinear bounce times $\tau_{\rm b}$ in Fig.~\ref{fig:result_fire-AB_tracers}: in our cases, we typically have
\begin{equation}
\tau_{\rm pulse}/T_0 \sim \tau_{\rm b}/\tau_{\rm pulse} \approx 5...7.
\end{equation}

\noindent which is less than an order of magnitude and numerical factors of ``order unity'' (like $\pi$) begin to matter. Moreover, for our largest solitary clump vortices (Fig.~\ref{fig:stages_fire-A}), whose width $\Delta\hat{P}_\zeta \approx 0.25$ corresponds to a frequency difference $\Delta\nu = 12\,{\rm kHz} \approx \nu_0/8$ (Eq.~\ref{eq:pz_nu}) between the vortice's inner and outer edge, has a phase mixing time (cf.~Section~\ref{sec:saturation_halo}) that is even comparable to the {\cred beat} pulse period,
\begin{equation}
\tau_{\rm pulse}\Delta\nu \lesssim 1.
\end{equation}

\noindent On top of that, the time traces of the canonical toroidal angular momentum (radial position) $\hat{P}_\zeta(t)$ for individual particles in Fig.~\ref{fig:result_fire-AB_tracers} demonstrated that the ``beat noise'' can be comparable to and even exceed the magnitude of the ``signal'' of the nonlinear bouncing motion, especially for particles that are (on average) trapped deeply inside the field's effective potential well; i.e., near the center of the vortex.

These qualitative and quantitative observations give the impression that the phase space vortices in our simulations live in a fairly hostile environment. Therefore, we think that it is remarkable that, even under these conditions with (at best) vague scale separation, the vortices remain robust and (on average) maintain their nested structure over the course of many milliseconds (a hundred beats and more). This strengthens the stance of ``semi-adiabatic'' models. Our results confirm the applicability of the idea underlying the ``bucket transport'' model, with the addition that, besides the chirping $\ddot{\phi}$ \cite{Hsu94}, the beating of the field assists with the transmission of particles through the nonadiabatic boundary layer. Similarly, our results support extended ``waterbag'' models that include an active boundary layer as in Ref.~\cite{Hezaveh21}, where the effect of multiple interfering long-range chirps is also considered.

If one of the solitary clump vortices would be able to gain full control over the field, the resonance and the associated effective phase space island would be centered around that clump permanently. We expect that, in such a case, the field would stop beating and we should obtain a single smooth long-range chirp. Such non-beating chirps were also observed in JT-60U experiments as shown in Fig.~14 of Ref.~\cite{Bierwage17a}. With $\delta\dot{\nu} \approx 2\,{\rm kHz}/{\rm ms}$, those lone chirps are $4...6$ times slower than those studied in the present work. Efforts are currently underway to reproduce such a single (non-beating) long-range chirp using {\tt ORBIT} with JT-60U equilibria. Those plasmas were driven by on-axis negative-ion-based neutral beams, so it is likely that the seed resonances are located near the axis and nonstandard (stagnation) orbits contribute. Such boundary effects are often important in real-world problems, but have been avoided in the present study.

\subsection{Chirp-induced resonance overlap}
\label{sec:summary_overlap}

The solitary clump vortices in our simulations were seen to travel far enough to reach the domain of a neighboring resonance. This can be interpreted as a manifestation of ``nonlinear resonance overlap'' that does not require a direct overlap of neighboring phase space islands. The chirp-induced (or vortex-mediated) resonance overlap allows transfer of material between neighboring resonances even at fairly low amplitudes of the field perturbation.

The chirp-induced resonance overlap process requires further study. We observed that the chirping rate is reduced and the $p/n = 4/5$ clump vortex begins to disintegrate when it enters the domain of the neighboring resonance. It would be interesting to see whether the structure evaporates completely, or whether a new $p/n = 5/5$ structure forms spontaneously. Such a study may have to deal more carefully with the boundary effects discussed at the end of Section~\ref{sec:result_transport} and in Appendix~\ref{apdx:model_pew}; namely, the $E'$ line broadening and associated problems with the $\delta f$ method.

\subsection{{\cred Concluding remarks} on chirping \& beating}
\label{sec:summary_exec}

Amplitude pulsations in general and beats in particular are part of and can have an impact on various processes in systems governed by resonant interactions and nonlinear frequency chirping. In this treatise, we have begun to elucidate {\cred the influence that beating has on the dynamics (formation and propagation) of collective structures and on the motion of individual particles in numerical simulations of strong chirping with different strengths of field} damping.

In the presence of a symmetry-breaking perturbation in the field (a seed wave) and a gradient in the EP distribution, the EP distribution spontaneously develops collective density perturbations that we refer to as a primordial (nonadiabatic) hole-clump wave pair. Sustained chirping occurs when there is a mechanism that facilitates phase slippage between the hole-clump wave pair and the resonant seed wave. In our simulations, phase slippage resulted from field damping: When the field constantly loses energy, it cannot fully reverse the particle displacements, so that some particles remain untrapped and carry with them {\cb the} above-mentioned collective density modulations.

{\cb Once phase slippage has set in, beating is both a consequence and driver of chirping, where chirping reflects the evolution of EP phase space structures. The feedback mechanism by which beating drives the radial advancement of hole and clump waves consists of
\begin{itemize}
\item  pulsating effective island widths, and
\item  alternating O- and X-points locations,
\end{itemize}

\noindent with field damping making the processes irreversible.

A manifestation of this feedback on a longer time scale (millisecond), comprising many beats and associated frequency steps (``micro-chirps''), is that the nonadiabatic onset of chirping takes the form of a hyperbolic tangent: first accelerating, then decelerating. This behavior has, albeit in a different setting, been predicted theoretically \cite{ZoncaTCM99} and we find qualitative agreement with our simulations.}

Although the beats become increasingly irregular as the number of pump waves increases, each beat constitutes a {\it global} perturbation of the field that moves particles in a globally coherent manner (unlike a stochastic process that scatters individual particles independently). Perhaps it is this large degree of spatial coherence {\cred (here exaggerated by the use of a semi-perturbative model)} and small degree of temporal coherence of the beating field that makes it an effective mechanism for driving the evolution of {\it multiple} coherent phase space structures, from the first primordial (nonadiabatic) hole-clump pair, all the way to the semi-adiabatic solitary vortices of the advanced stages of our simulations. The dynamics are self-sustained as the pumping action of the phase space structures with their different characteristic frequencies cause the field to beat.

It appears that the overall integrity of vortices and their radial propagation is maintained not only in spite of but also owing to ongoing kicks associated with the rapid amplitude pulsations and phase jumps. Together with mechanisms that damp the field fluctuations, chirping and beating prevent the system from falling into a nonlinear equilibrium state of adiabatic stagnation {\cred by maintaining} a nonadiabatic boundary layer around quasi-adiabatic vortex structures. The dynamics in that boundary layer erode nearby density gradients and facilitate the radial propagation of the vortices and, with that, long-range chirping and long-range ballistic transport. Like biological lifeforms --- which tend to die or hibernate in the absence of external stimuli and without repeated throughput of material --- coherent resonant structures become dynamic when receiving stimuli that are partially incoherent and off-resonant. And like social animals, the resonant structures can stimulate each other by jointly causing the field to beat.

The system we have studied here was simplified in many respects --- ignoring dynamic adaptations of the field mode structure, external sources, sinks and collisions --- which can often be important in real MCF plasmas on the long time scales considered (up to $\sim 10\,{\rm ms}$). However, these simplifications have made our simulations physically more transparent and yielded insights that would be difficult to obtain with more realistic but numerically expensive codes. {\cb Even the simplified system poses challenges, some of which have been addressed (e.g., with our approximate quiet start technique), while others remain to be tackled (e.g., dealing with phase space loading boundaries). On the modeling side, there seems to be potential for a more realistic description of the plasma response (MHD spectra), field damping, and dynamic changes in the mode structure. A few ideas were discussed in this paper.

There are also open questions on the physics side, such as the following. A prompt frequency shift was observed and characterized, but a full understanding remains to be attained. The role of beating in maintaining a nonadiabatic boundary layer around solitary vortices was emphasized, but there may be more to be uncovered. For instance, the role of the boundary layer dynamics for the radial propagation of a vortex may deserve further attention. It is conceivable that the rate at which gradients are eroded depends on the rate and magnitude at which the boundary layer around a phase space vortex pulsates due to the beating field. One may then ask whether semi-adiabatic vortices have energetic or evolutionary (dis)advantages compared to nonadiabatic spreading of phase space turbulence, elaborating on related thoughts presented in this work.}

Our results are relevant for the understanding and prediction of particle transport and confinement in MCF plasmas. We have presented evidence showing that the use of quasi-adiabatic models of ballistic ``bucket'' transport \cite{Hsu94} may be justified even in parameter regimes where the formal adiabatic ordering begins to fail. {\cred This seems to suggest that the effect of beating may be accounted for in a simplified manner in the transport coefficients of reduced models.}

\section*{Acknowledgments}

A.B.\ thanks Xin Wang for sharing her {\sc Python} scripts for DMUSIC analysis. Enlightening discussions with Fulvio Zonca are gratefully acknowledged, and we also thank the members of the PPPL EP Group as well as Herbert L.\ Berk and Yasushi Todo for stimulating communications during the early stages of this work (noting that our views expressed here may differ from theirs). The work by R.B.W.\ and V.N.D.\ was supported by the US Department of Energy (DOE) under contract DE-AC02-09CH11466.

The workstation used for many of the simulations reported here was funded by QST President's Strategic Grant (Creative Research). Some of the simulations were carried out using the JFRS-1 supercomputer system at Computational Simulation Centre of International Fusion Energy Research Centre (IFERC-CSC) in Rokkasho Fusion Institute of QST (Aomori, Japan) (Project ID: EPTRANS).

\section*{Appendices}

\appendix

\section{Theory of two-wave beating}
\label{apdx:beat}

Consider the integral on the right-hand side of Eq.~(\ref{eq:dadt}) or, equivalently, the sums in Eq.~(\ref{eq:evol}). {\cred Express $\alpha$ in terms of $\Phi$ using Eq.~(\ref{eq:bfield_ideal}) as $\alpha B^2 = k_\parallel\Phi$ with $k_\parallel \propto nq - m$, and write the electrostatic potential in complex form as} $\tilde{\Phi} = A(t)\hat{\Phi}(\psi_{\rm P})e^{i\Theta(t)}$ with $\Theta(t) \equiv n\zeta - m\vartheta - \omega_0 t - \phi(t)$. Ignoring damping, the equations for the amplitude (\ref{eq:evol_amp}) and phase (\ref{eq:evol_phi}) can be combined as
\begin{equation}
\dot{A} + iA\dot{\phi} = \int{\cred {\rm d}^5 Z}_{\rm gc} C_0 (k_\parallel v_\parallel - \omega_0)\hat{\Phi} e^{i\Theta}\delta f;
\label{eq:dadt_dphidt}
\end{equation}

\noindent where $C_0$ is a real-valued function of $\psi_{\rm p}$. The Jacobian has been absorbed in the distribution $F = F_0 + \delta f$, so $\int{\cred {\rm d}^2 Z}_{\rm gc} F = N$ is the number of particles ($\int{\cred {\rm d}^5 Z}_{\rm gc}\delta f = 0$). Taking the physical density wave signal $\delta f$ to be the imaginary (sine) component of the complex function $\delta\tilde{f} = \frac{2C}{C_0} e^{i\Theta_{\rm orb}}$, we have $\delta f \equiv \Im\{\delta\tilde{f}\} = -\frac{i}{2}(\delta\tilde{f} - \delta\tilde{f}^*)$, {\cred where the asterisk indicates a complex conjugate.} The form of the phase $\Theta_{\rm orb}$ of the density wave in GC phase space will be specified later. Substitution into Eq.~(\ref{eq:dadt_dphidt}) yields
\begin{align}
\dot{A} + iA\dot{\phi} =& -\int{\cred {\rm d}^5 Z}_{\rm gc} (ik_\parallel v_\parallel - i\omega_0 - i\dot{\phi}) C \hat{\Phi}e^{i(\Theta + \Theta_{\rm orb})} \nonumber
\\
& +\int{\cred {\rm d}^5 Z}_{\rm gc} (ik_\parallel v_\parallel - i\omega_0 - i\dot{\phi}) C \hat{\Phi}e^{i(\Theta - \Theta_{\rm orb})} \nonumber
\\
& -i\dot{\phi} \int{\cred {\rm d}^5 Z}_{\rm gc} C \hat{\Phi}e^{i(\Theta + \Theta_{\rm orb})} \nonumber
\\
& +i\dot{\phi} \int{\cred {\rm d}^5 Z}_{\rm gc} C \hat{\Phi}e^{i(\Theta - \Theta_{\rm orb})}.
\end{align}

\noindent The integrals are nonzero only when the complex exponential is a constant; namely, for phase-matched waves. Let us assume that this is the case for the difference between $\Theta$ and $\Theta_{\rm orb}$, so that $\Theta - \Theta_{\rm orb} = \beta_0 + 2\pi l$ with $\beta_0 = {\rm const}$.\ and integer $l$. The factor $(k_\parallel v_\parallel - \omega_0 - \dot{\phi})$ is readily recognized as the time derivative of the phase difference between the field and density waves, so that we can write
\begin{align}
\dot{A} + iA\dot{\phi} =& \frac{\rm d}{{\rm d}t}\int{\cred {\rm d}^5 Z}_{\rm gc} C \hat{\Phi} e^{i(\Theta - \Theta_{\rm orb})} \nonumber
\\
& +i\dot{\phi} \int{\cred {\rm d}^5 Z}_{\rm gc} C \hat{\Phi}e^{i(\Theta - \Theta_{\rm orb})}.
\label{eq:a}
\end{align}

\noindent Evidently, the integral expression corresponds to $A(t)$.

For simplicity, we ignore all the radial dependencies, so that $\hat{\Phi} = 1$. Moreover, the distribution is assumed to contain particles with only one value of the magnetic moment $\mu$ and to be monoenergetic in the rotating frame with energy $E' = E - \frac{\omega}{n} P_\zeta$. We can then integrate the distribution of phase space density waves $\delta f$ over $P_\zeta$ along the line $E' = {\rm const}$. In this way, the time-dependencies of all pump waves are merged into a single signal and we obtain a combined complex pump wave function
\begin{equation}
\delta\tilde{g}(t) e^{in\zeta - ip\vartheta} \equiv \iiint{\rm d}P_\zeta {\rm d}K {\rm d}\mu\, \delta\tilde{f}(P_\zeta,\mu,K,\vartheta,\zeta, {\cred t});
\label{eq:g}
\end{equation}

\noindent where $p$ is the number of elliptic points of the resonance along $\vartheta$.

To be more concrete, suppose that the perturbed component of the distribution function $\delta f$ consists of two sinusoidal pump waves located at different radii $P_\zeta$: a clump wave with frequency $\omega^+$ and a hole wave with frequency $\omega^-$, which can be written as $A_0w^\pm\frac{1\pm\epsilon}{2}\left[1 + \sin(n\zeta - p\vartheta -\omega^\pm t)\right] \geq 0$ with weights $w^\pm = \pm 1$. The parameter $-1 \leq \epsilon \leq 1$ measures the difference in the pump wave amplitudes. Here, the clump will dominate for $\epsilon > 0$ and the hole will dominate for $\epsilon < 0$. The resulting combined complex pump wave function $\delta\tilde{g}$ is then
\begin{align}
\delta\tilde{g}(t) &= \tfrac{A_0}{2(1 + \epsilon^2)}\left[w^+(1 + \epsilon)e^{-i\omega^+ t} + w^-(1 - \epsilon)e^{-i\omega^- t}\right] \nonumber \\
&= \underbrace{\frac{A_0}{1 + \epsilon^2} e^{-i\bar{\omega}t}}\limits_{\rm base\, wave} \underbrace{\left[\epsilon\cos(\tfrac{\Delta\omega}{2} t) - i\sin(\tfrac{\Delta\omega}{2} t)\right]}\limits_{{\rm beat} = b(t) \exp(-i\beta(t))},
\label{eq:beat}
\end{align}

\noindent with mean frequency $\bar\omega = \frac{1}{2}(\omega^+ + \omega^-)$, and frequency difference  $\Delta\omega = \omega^+ - \omega^-$. The amplitude is normalized to $A_0$ and constant offsets are ignored. The complex exponential factor in Eq.~(\ref{eq:beat}) represents an effective base wave oscillating at the mean frequency $\bar\omega$, which is not present in the original signal.\footnote{In signal processing, Eq.~(\protect\ref{eq:beat}) is known as ``double-sideband modulation'' and the effective base wave is called ``suppressed carrier''.}
The beat consists of a modulation amplitude $b(t)$ and phase $\beta(t)$, given by
\begin{subequations}
\begin{align}
b(t) &= \left[\sin^2(\Delta\omega t/2) + \epsilon^2\cos^2(\Delta\omega t/2)\right]^{1/2},
\label{eq:beat_pump_amp}
\\
\beta(t) &= \arctan\left(\epsilon^{-1}\tan(\Delta\omega t/2)\right)
\label{eq:beat_pump_phase}
\end{align}
\label{eq:beat_pump}\vspace{-0.35cm}
\end{subequations}

\begin{figure}[tb]
  \centering
  \includegraphics[width=8cm,clip]{\figures/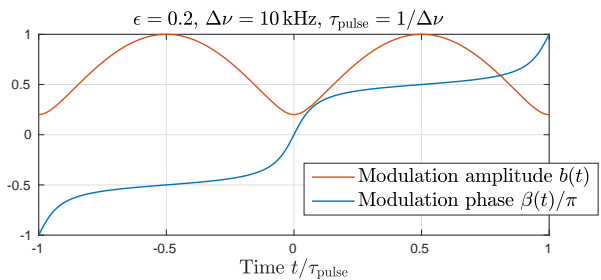}
  \caption{Example for asymmetric beating of two pump waves; namely a hole and a clump separated by $\Delta\nu = 10\,{\rm kHz}$ and with relative amplitudes differing by $2\epsilon = 0.4$. The plot shows the time traces of {\cred $b(t)$} and $\beta(t)$ from Eq.~(\protect\ref{eq:beat_pump}).}
  \label{fig:discuss_2beat}
\end{figure}

\noindent in the case of a superposition of a hole-clump pair. The frequency of the beats is $\Delta\omega = \omega^+ - \omega^-$ (without the factor 1/2), so the pulse length is $\tau_{\rm pulse} = 2\pi/\Delta\omega$. An example with $\epsilon = 0.2$ and $\Delta\omega = 2\pi\times 10\,{\rm kHz}$ is shown in Fig.~\ref{fig:discuss_2beat}.

Times $t_0$ satisfying $\Delta\omega t_0/2 = \pi l$ with integer $l$ correspond to instants of maximal destructive interference, where the modulation factor $b(t)$ has a minimum and the rate of phase shift $|\dot\beta(t)|$ is maximal. Maximal constructive interference occurs at times $t_1$ satisfying $\Delta\omega t_1/2 = \pi/4 + \pi l$. To avoid confusion, one should keep in mind that destructive interference between a hole wave and a clump wave occurs when their phases are aligned at $t = t_0$, because their weights $w^\pm$ have opposite signs. If we superimpose two clump waves, the functions $i\sin(\Delta\omega t/2)$ and $\cos(\Delta\omega t/2)$ would switch places in Eq.~(\ref{eq:beat}), and times $t = t_0$ would be the instants of constructive interference.

Equation~(\ref{eq:beat_pump_phase}) shows that the total phase shift between successive pulses of two-wave beating is always $\pm\pi$ and that the direction of a phase shift is determined by the dominant pump wave at that time; i.e., the sign of $\epsilon$. In our setup, clump and hole waves yield positive and negative phase shifts, respectively. In the symmetric case $\epsilon = 0$, with two pump waves of identical amplitude, the phase shift becomes instantaneous and has no specific direction (degenerate sign); the sign of the modulation factor simply flips abruptly when $b = 0$ and remains constant at other times. In the present {\cred paper}, the term ``phase jump'' refers collectively to both smooth phase shifts and sign flips.

Finally, we determine the response of the combined field amplitude that, according to Eq.~(\ref{eq:a}), obeys
\begin{align}
A(t) = \Re\int{\cred {\rm d}^5 Z}_{\rm gc} C \hat{\Phi} e^{i(\Theta - \Theta_{\rm orb})}.
\end{align}

\noindent {\cred Based on Eq.~(\ref{eq:beat}), the phase of the combined pump waves} with resonance number $p$ can be written
\begin{equation}
{\cred \Theta_{\rm orb} = n\zeta - p\vartheta - \bar{\omega}t - \beta(t)}.
\end{equation}

\noindent Note that the integration along GC drift orbits turns the single poloidal harmonic $e^{-im\vartheta}$ into a broader spectrum $\sum_{m_{\rm orb}}e^{-im_{\rm orb}\vartheta}$, which includes the resonance $m_{\rm orb} = p$ \cite{Bierwage14}. Letting again $\hat{\Phi} = 1$ and substituting $\iiint{\rm d}P_\zeta{\rm d}K{\rm d}\mu\,C e^{-i\Theta_{\rm orb}} = \delta\tilde{g}^* e^{ip\vartheta - in\zeta}$, we obtain
\begin{align}
A &= \Re \sum_{m_{\rm orb}} \iint{\rm d}\vartheta{\rm d}\zeta\, \delta{\cred \tilde{g}^*} e^{i[(p - m_{\rm orb})\vartheta  - \omega_0 t - \phi(t)]}
\label{eq:beat_a_g}
\\
&= \frac{A_0 b(t)}{1 + \epsilon^2} \Re\left\{e^{i(\bar{\omega}-\omega_0)t} e^{i(\beta(t) - \phi(t))}\right\}.
\label{eq:beat_a_exp}
\end{align}

\noindent When the hole and clump wave frequencies are situated symmetrically around the seed frequency, we have $\bar{\omega} = \omega_0$. Assuming phase matching, $\phi(t) = \beta(t) + \beta_0 + 2\pi l$, and ignoring $\beta_0$,\footnote{The value of $\beta_0$ determines the phase lag between the field and the mean density wave: $e^{i(\bar\omega - \omega_0)t + i\beta_0}$. Presumably, one would have to invoke the Vlasov equation in order to determine the proper value of $\beta_0$, but it is irrelevant for the amplitude modulations and phase jumps.} we obtain the result
\begin{subequations}
\begin{align}
A(t) &= \frac{A_0}{1 + \epsilon^2} b(t),
\label{eq:beat_fld_a}
\\
\phi(t) &= \beta(t) + 2\pi l.
\label{eq:beat_fld_phi}
\end{align}
\label{eq:beat_fld}\vspace{-0.35cm}
\end{subequations}

\noindent This shows that the field wave beats in precisely the same way as the combined pump wave function $\delta g$. Since the latter was a superposition of multiple pump waves, we may view the combined field wave also as a linear superposition of multiple harmonic waves with amplitudes $A_k(t)$ and phases $\phi_k(t)$, each driven by a different pump wave:
\begin{equation}
A(t)e^{i\Theta(t)} = \sum_k A_k(t)e^{in\zeta - im\vartheta - i\omega_0 t - i\phi_k(t)}.
\label{eq:multi-pump}
\end{equation}

The response of the field to the concert of pump waves, as in the process of two-wave beating analyzed above, may be viewed as an auto-resonance phenomenon; albeit with one additional twist: Usually, the term ``auto-resonance'' is used to refer to the automatic phase locking of particle bunches or waves in response to (chirped) external drive, as in a synchrotron \cite{Veksler45, McMillan45}. However, in the system we are considering, the prefix ``auto'' can also be interpreted as ``self'' because the combined field-particle system contains feedback loops that make it self-resonant and self-propelled in the presence of a source of free energy in the form of destabilizing gradients. In other words, we are dealing with auto-resonance in an active medium. When this field-particle system is ``opened'', in our case through the introduction of field damping, a directional flow of energy is established, and nonlinear frequency chirping is one manifestation of that.

We would like to point out that there exists an alternative, albeit purely phenomenological, interpretation of the beats: Loosely speaking, one may also interpret each pulse of $A(t)$ as a new instability whose phase is shifted by $\pm\pi$ with respect to its neighbors. This picture exemplifies the connection between the beats in our simulations and the so-called ``relay runner'' model of nonadiabatic chirping proposed by Zonca \& Chen \cite{ZoncaTCM99}, which we discussed in Sections~\ref{sec:result_front} and \ref{sec:summary_staircase}. However, the purely phenomenological nature of this picture in the present context is apparent from the time trace of $A(t)$ in Fig.~\ref{fig:intro_fire-A_beat}, which drops sharply before and recovers equally sharply after a near-zero minimum. This corresponds to a hyperbolic time trace of the growth rate $\gamma(t)$ (cf.~Fig.~\ref{fig:num_fire-A_qs-growth}), and we see no way to reconcile this with any known form of resonant drive or damping, where ${\rm d}A/{\rm d}t$ should decrease (not increase) when the value of $A$ goes to zero, since the {\cred amount of resonantly transferred energy per unit time should decrease with decreasing amplitude $A$ as the interaction becomes weaker.} At present, the only physically viable interpretation for the observed signals that we can offer is the beating picture, where the pulses and phase jumps are the consequence of a superposition of multiple coexisting pump waves whose amplitudes do not vanish even when the overall field amplitude does so temporarily during instants of destructive interference. Supported by the above derivations for the case of two-wave interference, the physical picture of beating is the foundation for this work.

An important implication of the beating picture is that the instantaneous growth rate $\gamma(t)$ of the beating signal does {\it not} represent resonance width. While the growth or damping rate $|\gamma(t)|$ associated with the combined amplitude $A(t)$ can become very large (e.g., see Fig.~\ref{fig:discuss_fire-A-pew_overview}(c)), the amplitudes of individual components $A_k(t)$ in Eq.~(\ref{eq:multi-pump}) usually vary relatively slowly. Thus, instead of one $\gamma$-broadened resonance, we have multiple resonances whose widths correspond to the size of a quasi-adiabatic island core (if any) plus the width of a nonadiabatic boundary layer, as illustrated in Fig.~\ref{fig:intro_x-point}(b,c). The beats (or any pulsations) of the field contribute to this resonance width through their influence on the boundary layer, where particles are detrapped and retrapped repeatedly, as shown for the undamped case in Appendix~\ref{apdx:undamped_boundary} below. Clearly, this resonance width depends on the magnitude of the pulsations.

\begin{figure*}[tb]
  \centering
  \includegraphics[width=16cm,clip]{\figures/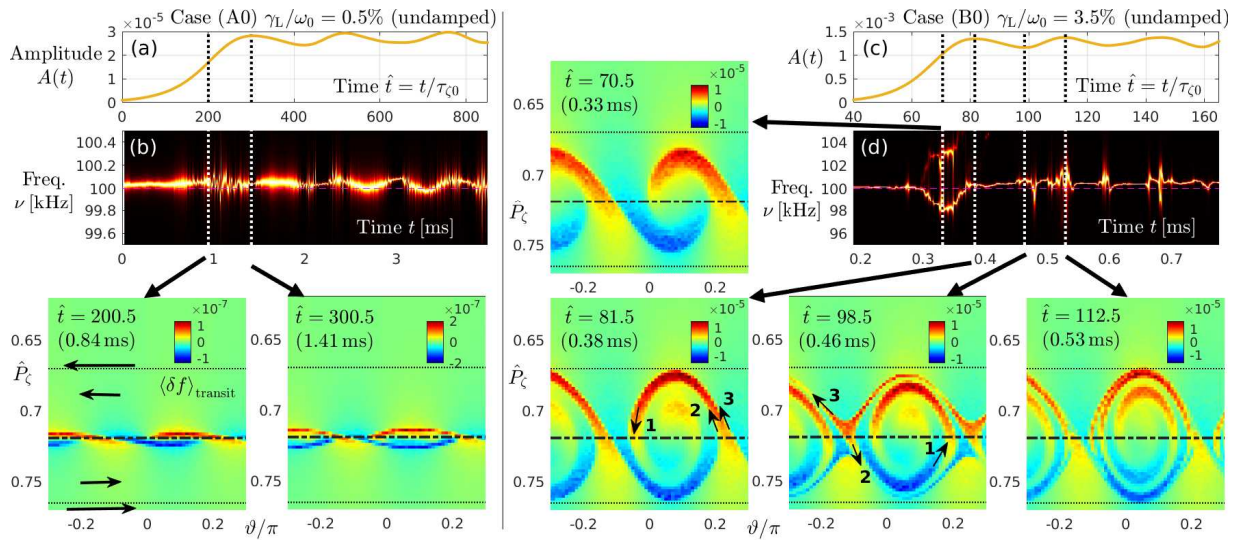}
  \caption{Overview of the field and phase space dynamics during and after saturation of a weakly driven [case (A0), left] and a strongly driven resonant instability [case (B0), right] in the absence of damping. These simulations were performed with shorter time step $\Delta\hat{t} = 10^{-3}$ (default: $1/400$) to reduce spurious growth. Panels (a) and (c) show the respective time traces of the field amplitude $A(t)$, and panels (b) and (d) show the corresponding high-resolution DMUSIC spectrograms ($t_{\rm win} = 0.047\,{\rm ms}$). For each case, several snapshots of the phase space density perturbation $\delta f$ are shown as colored contour plots. For case (B0), the three arrows labeled ``1'', ``2'' and ``3'' roughly indicate the observed flow of EP Vlasov fluid in the boundary layer during about half of a bounce period.}
  \label{fig:discuss_fire-A0-B0_spec-df}
\end{figure*}

\section{Pulsations of undamped resonances}
\label{apdx:undamped}

\begin{figure*}[tb]
  \centering
  \includegraphics[width=8cm,clip]{\figures/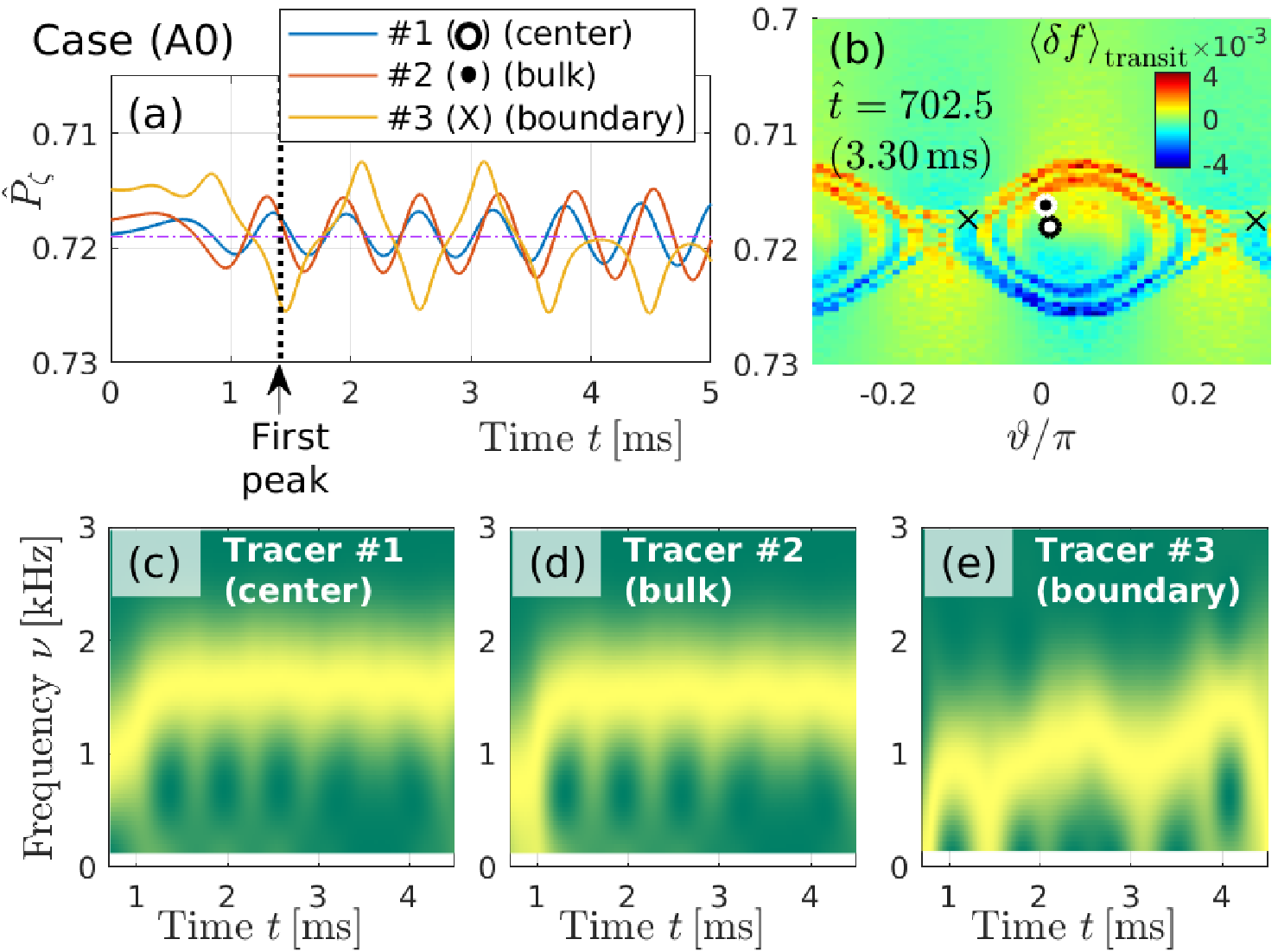}
  \includegraphics[width=8cm,clip]{\figures/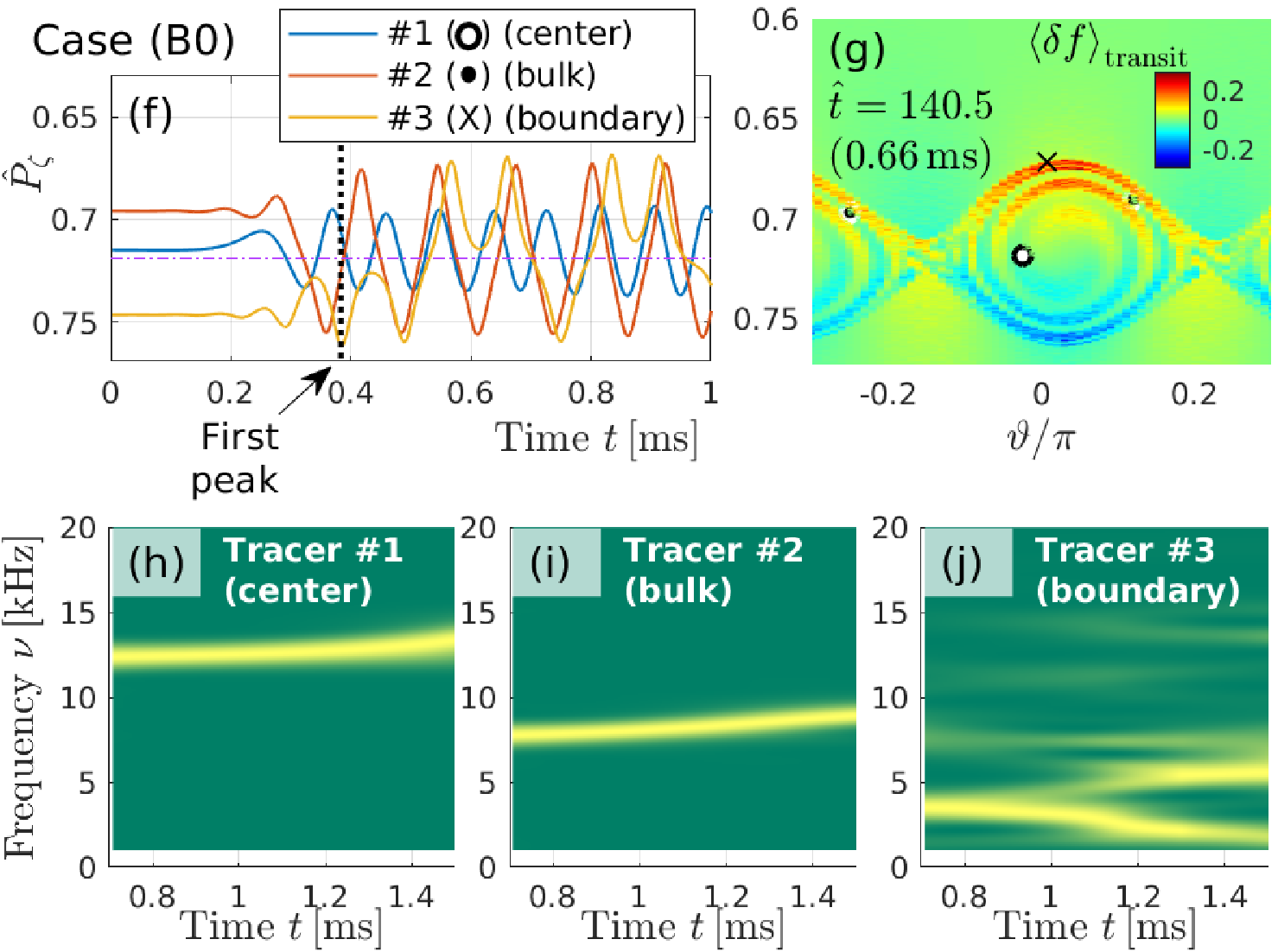}
  \caption{Motion of well-trapped (``center'', ``bulk'') and temporarily trapped (``boundary'') particles in the undamped cases (A0) (left) and (B0) (right). Panels (a) and (f) show the time traces of the radial position $\hat{P}_\zeta(t)$. Panels (b) and (g) show the locations of the particles inside the $\delta f$ phase space structures during an advanced stage of the simulation. These $\delta f$-weighted Poincar\'{e} plots were accumulated during one transit period $\tau_{\zeta 0} = 4.7\,\mu{\rm s}$, during which each tracer particle appears in the Poincar\'{e} section approximately 4 times. Panels (c)--(e) and (h)--(j) show Fourier spectra of the $\hat{P}_\zeta(t)$ signals obtained with sliding time windows of size $\Delta t_{\rm win} = 1.41\,{\rm ms}$, where one can infer the bounce frequencies $\nu_{\rm b}$ of resonantly trapped particles.}
  \label{fig:discuss_fire-A0-B0_tracers}
\end{figure*}

Figure~\ref{fig:discuss_fire-A0-B0_spec-df} shows an overview of the field and phase space dynamics in the absence of damping. The weakly driven case (A0) on the left and the strongly driven case (B0) on the right are qualitatively similar, so snapshots of $\delta f$ are shown mostly for case (B), where the fluctuations are larger, both in magnitude and spatial extent, so they are also {\cred better resolved with our diagnostic mesh}. Figure~\ref{fig:discuss_fire-A0-B0_tracers} shows the motion of a few tracer particles in the domain of these resonant structures.

In Section~\ref{sec:review_instab}, we have already given a brief review of the saturation process of such undamped resonant instabilities and demonstrated that the subsequent pulsations of the mode amplitude correlate well with the bouncing motion of particles trapped in the effective potential well of the field wave. Here, we analyze these dynamics in some more detail and highlight two interesting features that are related to the topic of the present work; namely:
\begin{enumerate}
\item[(i)]  We find evidence for variations in the field's oscillation frequency and even {\it transient frequency splitting}.

\item[(ii)]  The continuing pulsations of the field amplitude $A(t)$ maintain a {\it nonadiabatic boundary layer}, inside which marginally resonant particles are repeatedly detrapped and retrapped.
\end{enumerate}

\subsection{Spectral fluctuations}
\label{apdx:undamped_spec}

Transient frequency splitting can be seen in the high-resolution DMUSIC spectrogram in Fig.~\ref{fig:discuss_fire-A0-B0_spec-df}(d), around $t = 0.3...0.4\,{\rm ms}$, just before the field amplitude reaches its peak value. During that short interval, the initial $100\,{\rm kHz}$ signal seems to split temporarily into $\nu^{\rm -} \approx 98\,{\rm kHz}$ and $\nu^{\rm +} \approx 103\,{\rm kHz}$. The magnitude of the observed frequency shifts by $\Delta\nu^\pm \approx \pm 2...3\,{\rm kHz}$ relative to the seed wave frequency $\nu_0 = 100\,{\rm kHz}$, is more or less consistent with the poloidal group velocity $v_\vartheta^\pm$ of the primordial hole and clump wave fronts that can be seen to pass above and below the resonance's O-point around the time of the first $\delta f$ snapshot for case (B) in Fig.~\ref{fig:discuss_fire-A0-B0_spec-df}, at $\hat{t} = 70.5$ ($0.33\,{\rm ms}$): our measurement gives
\begin{equation}
\frac{v_\vartheta^\pm}{\lambda_\vartheta} = \frac{\Delta\vartheta^\pm}{\lambda_\vartheta\Delta t} \approx \frac{\mp 0.045\pi}{0.365\pi\times 0.05\,{\rm ms}} = \mp 2.5\,{\rm kHz};
\end{equation}

\noindent where $\lambda_\vartheta \approx 0.365\pi$ is the poloidal island length in the region around $\vartheta = 0$ (cf.~Fig.~\ref{fig:model_fire-A_df-interpret}).

Thus, one may speculate that the pumping action of these phase space structures is the cause of the observed transient frequency splitting. The subsequent persisting frequency fluctuations seen in the two spectrograms in panels (b) and (d) of Fig.~\ref{fig:discuss_fire-A0-B0_spec-df} may also be associated with the dynamics of density waves, especially those in the island's nonadiabatic boundary layer.

Although the spectral fluctuations appear to be consistent with the dynamics of phase space structures, the spectral data {\cred have} to be interpreted with care. We were not able to confirm the frequency splitting phenomenon seen in the DMUSIC analysis of case (B0) using conventional Fourier analysis. At present, we cannot rule out the possibility that frequency splitting as seen in Fig.~\ref{fig:discuss_fire-A0-B0_spec-df}(d) has origins other than the motion of the primordial hole and clump waves.

\subsection{Nonadiabatic boundary layer}
\label{apdx:undamped_boundary}

Consider the second and third snapshots of $\delta f$ for case (B0) in Fig.~\ref{fig:discuss_fire-A0-B0_spec-df} at $\hat{t} = 81.5$ and $98.5$, where we have drawn a set of three short arrows that roughly indicate the flow of EP Vlasov fluid during about half of a bounce period. The first arrow, labeled ``1'', passes near an effective X-point at $(\hat{P}_\zeta,\vartheta) \approx (0.719,-0.1\pi)$ while the field amplitude is large. The pair of arrows labeled ``2'' and ``3'' approaches the same X-point during an amplitude minimum. The inner arrow ``2'' circulates back around the effective O-point of the resonant phase space structure, while the outer arrow ``3'' passes above the effective X-point and continues its poloidal drift towards the left. This process repeats during each pulse and can also be seen in the motion of individual particles, which we discuss in the following.

For both cases (A0) and (B0), we have chosen three particles, whose trajectories $\hat{P}_\zeta(t)$ are plotted in panels (a) and (f) of Fig.~\ref{fig:discuss_fire-A0-B0_tracers}. Fourier spectrograms of these time traces are plotted in panels (c)--(e) and (h)--(j), which effectively show the bounce frequencies (if any). In panels (b) and (g), we show snapshots of the phase space structures in $\delta f$ after several bounces. The circles and crosses indicate the positions where the three tracer particles have appeared in that Poincar\'{e} section (which rotates toroidally with the seed wave) during one transit time $\tau_{\zeta 0} \approx 4.7\,\mu{\rm s}$.
\begin{itemize}
\item Tracer \#1 labeled ``center'' circulates not far from the O-point. It has the largest bounce frequency, which is particularly clear in panel (j) for case (B0).

\item Tracer \#2 labeled ``bulk'' is also well-trapped, but circulates closer to the boundary, with a smaller bounce frequency than the ``center'' particle.

\item Tracer \#3 labeled ``boundary'' lives in the boundary layer and undergoes repeated detrapping and retrapping. Depending on whether it approaches an X-point during an amplitude maximum or minimum, it is reflected or passes. The Fourier spectra in this case are unintelligible, but the affective bounce time can be inferred from the raw $\hat{P}_\zeta(t)$ data in (a) and (f).
\end{itemize}

\noindent The behavior of the ``boundary'' particle was illustrated schematically in Fig.~\ref{fig:intro_x-point}(b). The nonadiabatic boundary layer is maintained by the amplitude pulsations seem in panels (a) and (c) of Fig.~\ref{fig:discuss_fire-A0-B0_spec-df}. Judging by the size of the structures seen around the X-points in panels (b) and (g) of Fig.~\ref{fig:discuss_fire-A0-B0_tracers}, the radial width of this layer seems to be about 1/5...1/3 of the width of the effective phase space island.

Dewar's theoretical prediction for the bounce frequency of deeply trapped particles \cite{Dewar73}, $\nu_{\rm b}^{\rm Dewar} = 2.88\times\gamma_{\rm L}/(2\pi)$ in Eq.~(\ref{eq:wb_dewar}), gives $1.44\,{\rm kHz}$ for case (A0) and $10.1\,{\rm kHz}$ for case (B0). Given the approximations made in the theory, we consider these estimates to be in good agreement with the simulation results in Figs.~\ref{fig:discuss_fire-A0-B0_spec-df} and \ref{fig:discuss_fire-A0-B0_tracers}.

\begin{figure}[tb]
  \centering
  \includegraphics[width=8cm,clip]{\figures/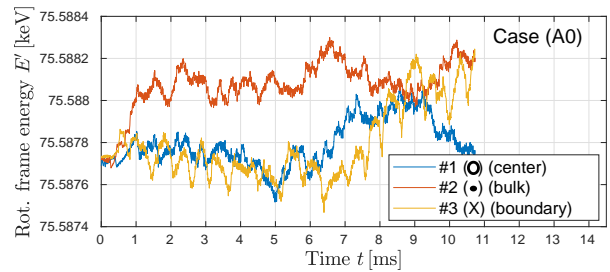}
  \caption{Time trace of the rotating frame energy $E'(t)$ for the same three tracers as in Fig.~\protect\ref{fig:discuss_fire-A0-B0_tracers}(a) for case (A0).}
  \label{fig:result_fire-A0_tracers_en_small}
\end{figure}

Finally, Figure~\ref{fig:result_fire-A0_tracers_en_small} demonstrates the high degree of energy conservation in the undamped simulation: $|\delta E'|/E' \sim 10^{-5}$, which is two orders of magnitude better than in the simulations of damped cases (cf.~Fig.~\ref{fig:result_fire-AB_C-evol}).

\begin{figure*}[tb]
  \centering
  \includegraphics[width=16cm,clip]{\figures/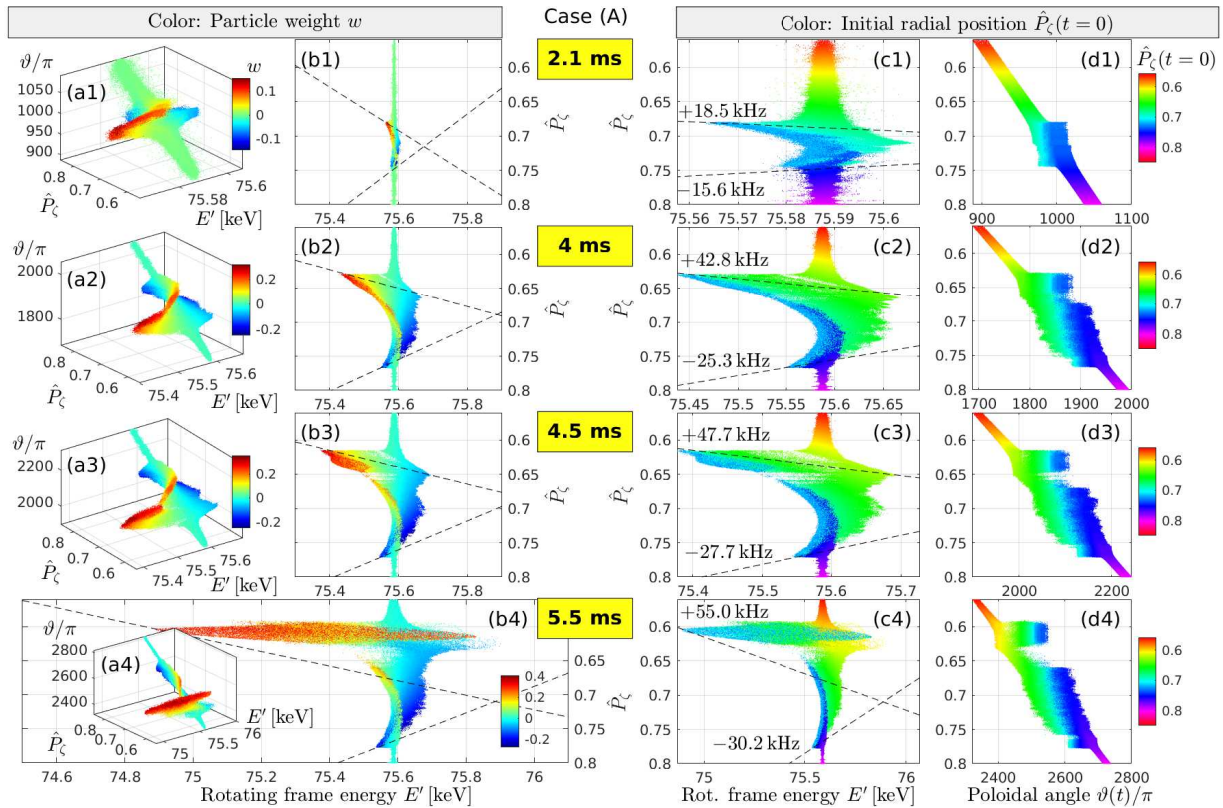}
  \caption{Particle transport in $(\hat{P}_\zeta, E', \vartheta)$-space in the marginally unstable case (A). This figure provides a more detailed view of the particle distribution data in Fig.~\protect\ref{fig:result_fire-AB_C-evol}(a) at four snapshot times: $t = 2.1$, $4.0$, $4.5$ and $5.5\,{\rm ms}$ (top to bottom). The colors in columns (a) and (b) on the left-hand side represent marker weights $w_j = \delta f_j/G$, so red means that a marker represents an increased phase space density ($\delta f > 0$) and blue means reduced density ($\delta f < 0$) at that instant of time. The colors in columns (c) and (d) on the right-hand side represent the initial radial position $\hat{P}_\zeta(t=0)$. The dashed lines in columns (b) and (c) represent the {\cred inclination of $E' - (\omega - \omega_0)\hat{P}_\zeta/n = {\rm const}$.\ lines for frequencies $\omega = 2\pi\nu$ at the edges of the crescent-shaped distortion of the distribution or the tip of the outermost massive  clump structure. The values of $\nu$ associated with these dashed lines are shown in column (c).} Note that column (b) has a fixed scale in both $E'$ and $\hat{P}_\zeta$, so that one can clearly see the broadening of the distribution from snapshot (b1) to (b4). In columns (a), (c) and (d), the limits of the $E'$-axis vary between snapshots, so that one can clearly see the structures.}
  \label{fig:result_fire-A_distr-E-P-th_t2-6ms}
\end{figure*}

\section{Other interesting observations}
\label{apdx:misc}

\subsection{Splitting of spectral lines {\cred (in DMUSIC)}}
\label{apdx:misc_split}

Panels (a) and (b) of Fig.~\ref{fig:result_fire-AB_spec-fit} both show a splitting of some spectral lines, especially those associated with the upward chirping solitary clump vortices. {\cred Another view is given in the second zoom-up of Fig.~\ref{fig:dmusic_fire-A_compare-fft}(b).

So far, we have seen this phenomenon only in DMUSIC spectrograms exploiting the complex frequency plane as described in Appendix~\ref{apdx:dg_dmusic}, not in FFT spectrograms or when DMUSIC is applied only along the real frequency axis. Since the split strands here are separated by only about $2\,{\rm kHz}$, the splitting is seen only with time windows $\Delta t_{\rm win}$ that have width of about half a millisecond or longer. More strands tend to appear with larger time windows. The line splitting seems to become stronger with increasing chirping rate, so it is most easily seen along the spectral lines of massive clumps when they chirp rapidly upward, while traveling radially outward, towards the peak of the mode. But with sufficiently large time windows, line splitting can also observed on downward chirps, especially around times where the chirping rate changes relatively abruptly. As the chirping rate or intensity of the signal varies, the line splitting pattern may change. For instance, it may alternate between odd and even numbers of strands, or exhibit an oscillatory pattern.}

It remains to be clarified what factors (physical, numerical, post-processing) cause these spectral patterns. For instance, one may check whether there is a relation to Fresnel ripples known to affect chirp spectra.\footnote{\cred When we apply a Hann window before the DMUSIC algorithm, the pattern becomes smoother, but the line splitting as such remains.}
{\cred The patterns vary depending on DMUSIC parameters, such as the time window size and the number of damped sinusoids. However, that does not necessarily mean that it is entirely a post-processing artifact. The spectrum in complex frequency space may truly (for physical reasons) be so complicated that its appearance changes drastically depending on how it is being processed and visualized.

If this kind of splitting has physical reasons, a better understanding of such spectral patterns may enable us to extract information about processes such as the accumulation and detachment of massive hole and clumps, or the internal dynamics of structures such as the solitary clump vortex that resembles a spiral galaxy in Fig.~\ref{fig:intro_fire-A_df}(b). At present, however, this is mere speculation.}

\begin{figure*}[tb]
  \centering
  \includegraphics[width=16cm,clip]{\figures/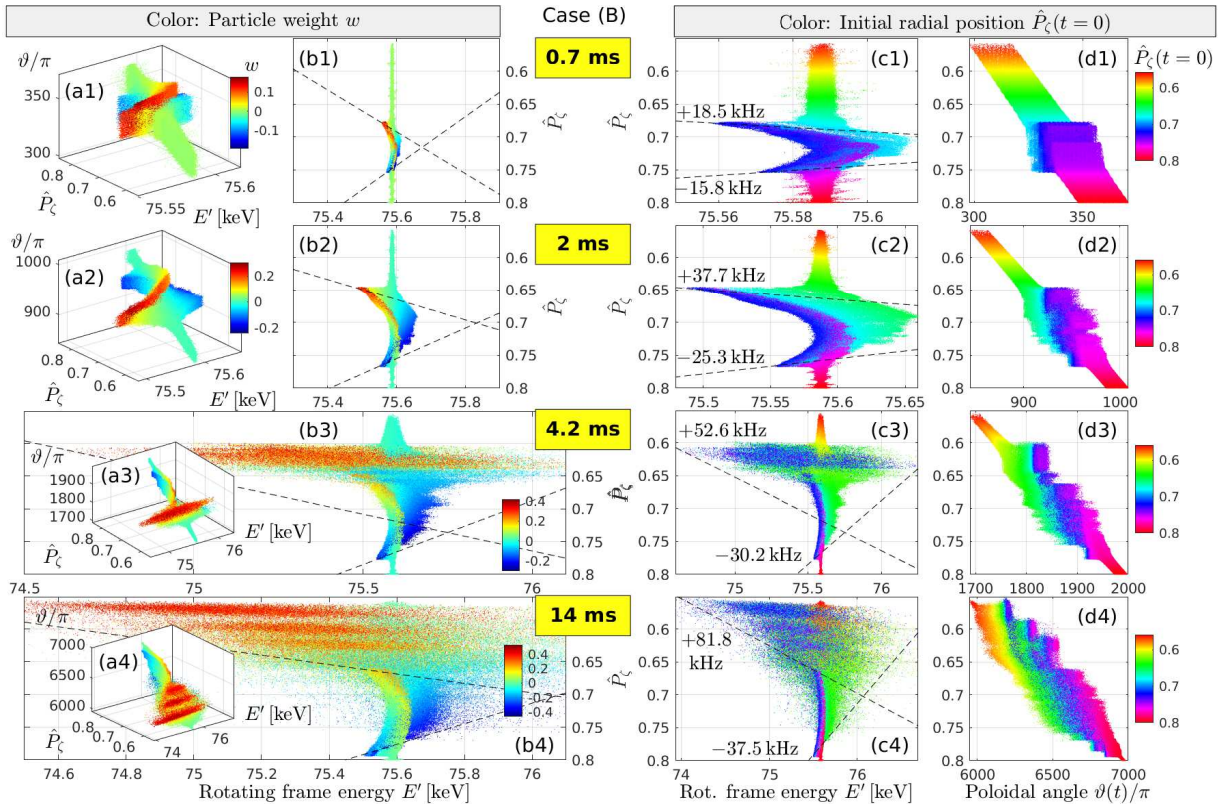}
  \caption{Particle transport in $(\hat{P}_\zeta, E', \vartheta)$-space in the strongly unstable case (B). Arranged in the same way as Fig.~\protect\ref{fig:result_fire-A_distr-E-P-th_t2-6ms}, this figure provides a more detailed view of the particle distribution data in Fig.~\protect\ref{fig:result_fire-AB_C-evol}(b) at four snapshot times: $t = 0.7$, $2.0$, $4.2$ and $14\,{\rm ms}$ (top to bottom).}
  \label{fig:result_fire-B_distr-E-P-th_t07-14ms}
\end{figure*}

\subsection{Anomalies in bounce frequency spectra}
\label{apdx:misc_fbounce}

The spectrograms of the bouncing motion of resonantly trapped particles that were shown in Fig.~\ref{fig:result_fire-AB_tracers-spec} exhibit rich dynamics, and it may be interesting to analyze the underlying reasons. Here, we merely speculate.

In the stationary case, with fixed amplitude $A$ and phase $\phi$, it is known that the bounce frequency decreases monotonically from the O-point to the boundary (separatrix) of a resonant phase space structure. This has recently been reconfirmed using {\tt ORBIT} \cite{Meng18}. However, our simulations show interesting anomalies that violate this rule in the case of chirping solitary vortices. On the left-hand side of Fig.~\ref{fig:result_fire-AB_tracers-spec} for case (A), one can see that, during the period $4\,{\rm ms} \lesssim t \lesssim 6\,{\rm ms}$, tracer \#3 near the boundary bounces approximately as rapidly as tracer \#1 in the core of the clump with $\nu_{\rm b} \approx 4...5\,{\rm kHz}$; i.e., faster than tracer \#2 in the bulk with $\nu_{\rm b} \approx 3...4\,{\rm kHz}$. Similarly, on the right-hand side of Fig.~\ref{fig:result_fire-AB_tracers-spec} for case (B), one can see that, during and around the interval $t = 8...10\,{\rm ms}$, the tracer \#2 in the bulk bounces with a significantly higher frequency $\nu_{\rm b} \approx 5\,{\rm kHz}$ than the centralized tracer \#1, which has $\nu_{\rm b} \approx 3\,{\rm kHz}$.

The reason for this behavior is unclear. We suspect a connection to the processes that make these vortices grow, detach from the turbulent belt and propagate radially (chirp) as discussed in Section~\ref{sec:result_detach}. This speculation is motivated by the fact that the bounce frequencies vary significantly during the course of a simulation. The largest bounce frequencies can be observed in Fig.~\ref{fig:result_fire-AB_tracers-spec} during the detachment process and when the chirping rate is large.

The overall trend is that the bounce frequencies decrease in time. This observation is in itself surprising, because these clumps propagate radially outward towards the peak of the mode, so they are actually experiencing stronger field oscillations of progressively larger amplitude {\cred when} $A(t)$ fluctuates more or less at the same level. This contradicts the expectation that $\omega_{\rm b}^2 \propto A(t)\times\hat{\xi}(\psi_{\rm P})$. However, this prediction is based on linear drive, $\omega_{\rm b} \propto \gamma_{\rm L}$ (cf.~Section~\ref{sec:review_adiabat}), so it should be expected to hold only during the first bounces after the resonant instability saturates. {\cred Further study is necessary for clarification.} One may also investigate the cause of the relatively rapid oscillation of the bounce frequencies $\nu_{\rm b}(t)$ in Fig.~\ref{fig:result_fire-AB_tracers-spec}, which have a time scale comparable to the bounce period itself.

\subsection{EP distribution during long-range chirps}
\label{apdx:misc_distr}

In Section~\ref{sec:result_transport}, we have shown that particles in our chirping simulations depart from the line $E' = 75.6\,{\rm keV}$ in the $(P_\zeta,E)$-plane where they have initially been loaded (Fig.~\ref{fig:result_fire-AB_C-evol}). To motivate further study, we include here a more detailed view of the distribution's 4-D structure in $(\delta f,\hat{P}_\zeta,E',\vartheta)$ and $(\hat{P}_{\zeta 0},\hat{P}_\zeta,E',\vartheta)$ space, four snapshots of which are shown in Figs.~\ref{fig:result_fire-A_distr-E-P-th_t2-6ms} and \ref{fig:result_fire-B_distr-E-P-th_t07-14ms} for cases (A) and (B), respectively. The left half of these figures shows the domains occupied by structures with different values of $\delta f$; in particular, holes ($\delta f > 0$, red) and clumps ($\delta f < 0$, blue). The right half of these figures shows the radial transport and mixing of phase space.

The crescent-shaped distortion that is centered around the seed resonance $\hat{P}_{\zeta,{\rm res}} \approx 0.72$ seems to be physical, since it is also present in simulations with particles loaded in a band of width $\Delta E'_0 = 7.5\,{\rm keV}$ (see Appendix~\ref{apdx:model_pew} below). We suspect that this crescent is due to chirping, but this hypothesis remains to be confirmed. A physics study of this and other features seen in Figs.~\ref{fig:result_fire-A_distr-E-P-th_t2-6ms} and \ref{fig:result_fire-B_distr-E-P-th_t07-14ms} requires more careful considerations of boundary effects associated with the $\delta f$ method, as discussed at the end of Section~\ref{sec:result_transport}.

\subsection{Ghost chirps and resonance overlap}
\label{apdx:misc_ghost}

The spectrogram in Fig.~\ref{fig:intro_fire-A_chirp}(b) contains a downward chirp labeled ``ghost'', which differs from the other down-chirps in that it does not seem to have a corresponding inward propagating hole vortex in EP phase space. Further examples can be seen in Fig.~\ref{fig:num_fire-A_dt-spec}. These ghost chirps seem to be somewhat correlated with the long-range upward chirps. According to our observations, a downward chirping ghost appears whenever an upward chirp exceeds $150\,{\rm kHz}$. This coincides with the time after which we observe enhanced $E'$ line broadening in Figs.~\ref{fig:result_fire-AB_C-evol}, \ref{fig:result_fire-A_distr-E-P-th_t2-6ms} and \ref{fig:result_fire-B_distr-E-P-th_t07-14ms} discussed in the previous Section~\ref{apdx:misc_distr}. Indeed, we do not see any ghost chirps in a simulation with $\Delta E'_0 = 7.5\,{\rm keV}$ that we performed for validation purposes (see Fig.~\ref{fig:discuss_fire-A-pew_overview}(d) below). Therefore, we suspect the ghosts to be a consequence of boundary effects caused by letting $\Delta E'_0 = 0$.

Although the cause of the ghosts seems to be an artifact {\cred of sampling only a part of the GC phase space}, the mechanism of their generation in that particular form may be physical. The first snapshot of Fig.~\ref{fig:stages_fire-A} shows that a neighboring resonance with $p/n=5/5$ elliptic points dominates the structure of the $\delta f$ landscape in the region $\hat{P}_\zeta \lesssim 0.62$, which corresponds to $\nu \gtrsim 150\,{\rm kHz}$. When the upward-chirping solitary clump vortices with $p/n = 4/5$ enter that domain, we expect that these structures are modulated by the $p/n = 5/5$ resonance, and that this modulation produces an image of the original oscillation at a different frequency. Poloidal transit frequency estimates suggest that the resonances are separated by about $\omega_\vartheta/(2\pi) \approx 210\,{\rm kHz}$. This is consistent with the appearance of ghosts around $60\,{\rm kHz}$ when a clump hits the mark of $150\,{\rm kHz}$. Thus, the ghosts may also be an indication of the chirp-mediated nonlinear resonance overlap discussed in Section~\ref{sec:summary_overlap}. In fact, even the enhanced spreading of particles away from the initial line $E'_0 = 75.6\,{\rm keV}$ in Figs.~\ref{fig:result_fire-AB_C-evol}, \ref{fig:result_fire-A_distr-E-P-th_t2-6ms} and \ref{fig:result_fire-B_distr-E-P-th_t07-14ms} may at least partially be a consequence of this kind of resonance overlap.

\section{Characterization and verification of the model}
\label{apdx:model}

\subsection{Background plasma response}
\label{apdx:model_pert}

The semi-perturbative approach used in this work, where we prescribe the radial profile of the field mode, may be regarded as an application of the Rayleigh-Ritz method: We are probing the system's frequency response with an approximate (eigen)mode, thus, reducing the infinite degrees of freedom of the real system to a few variables; here, the field amplitude $A(t)$, the phase $\phi(t)$, and a bunch of marker particles that sample the EP guiding center phase space. 

The process of convective amplification \cite{Zonca15b} in the form illustrated in Fig.~\ref{fig:intro_x-point}(d) could {\cred also} be simulated in our model using Eq.~(\ref{eq:mode}) with the mode index $k$ identifying independent pairs of Fourier harmonics $(m,n)$; i.e., ignoring the geometric (toroidal, elliptic, etc.) coupling between different poloidal harmonics $m$ for a given toroidal harmonic $n$. This procedure may allow to simulate resonant instabilities with relatively short wavelengths (which usually interact best with EPs that have relatively small drift orbits). Nevertheless, as discussed in Section~\ref{sec:summary_staircase}, even our present simulations with a single harmonic $(m,n) = (6,5)$ can realize a form of convective amplification:
\begin{itemize}
\item  Firstly, although the peak of the mode does not shift, the phase space structures are convected radially with each pulse of the beating field. Successive pulses can be amplified, especially in cases close to marginal stability (Fig.~\ref{fig:result_spec-fronts}, left).
\item  Secondly, by placing the seed resonance off-peak (cf.~Fig.~\ref{fig:model_fire-ABC_poink}), some phase space structures (in our case the clumps) travel into regions with stronger fluctuations even if $A(t)$ remains around the same level.
\end{itemize}

\noindent In some sense, the use of an off-peak resonance can be said to exchange the cause and effect of convective amplification: instead of a new peak growing at a new location after the resonant drive has propagated radially (post-convective amplification), here the resonant phase space structures experience a larger field amplitude on one side than on the other and their resonant drive effectively pre-amplifies the field further in the direction of the peak (pre-convective amplification). This may accelerate the growth and radial drift of the phase space structures that are predisposed to propagate towards the peak; in our case the clumps, because our peak is located in a region of lower EP density (at a larger radius) than the resonance. We suspect that the process of pre-convective amplification is not merely a curiosity of our simulation setup, but may bear practical relevance, because off-peak resonances are common and since Alfv\'{e}n mode profiles seem to have a certain degree of rigidity, at least in the case of long wavelength modes.

Motivated by the observation of global beating in hybrid simulations \cite{Bierwage17a} as discussed in Section~\ref{sec:review_q}, we assume that, for our current purposes, it is acceptable to ignore perturbations of the amplitude profile for long-wavelength modes. However, it is not evident to what extent our ignoring of the dynamic adjustments of the {\it phase profile} is justified, since this directly touches the quantity $\phi(t)$ responsible for chirping. This may require further consideration, which we choose to postpone. Discussions concerning the role of the phase profile during interaction between EPs and shear Alfv\'{e}n waves can be found, for instance, in Section III D of Ref.~\cite{Bierwage16a} and in Refs.~\cite{Kramer19,Meng20}.

While perturbations of the field mode structure are ignored, the frequency is entirely determined by the resonant drive in our simulations. We even observe a prompt frequency shift that we attribute to an auto-optimization process, which will be discussed in  Section~\ref{apdx:model_opt} below. Although it is possible to include the MHD plasma response without directly solving MHD equations, the required extensions are not trivial. Wang Ge {\it et al}.\ \cite{WangGe18} have developed and successfully applied a reduced model for Alfv\'{e}nic chirping that includes the effect of continuous spectra. We have made no attempt to adopt this model yet, because it requires sophisticated numerical methods to deal with the small scale structures that inevitably develop in an ideal MHD formulation. {\cred In the end, however,} these troublesome micro-structures seem to have little effect on the chirping dynamics when the EP's magnetic drift is significant. Perhaps, it is worthwhile to develop a reduced kinetic description of the continua, where the ion Larmor radius would impose a lower bound on the spatial scales. This may reduce the numerical obstacles and give an even more realistic representation of the continua in the form of kinetic Alfv\'{e}n waves, whose wave vector contains also a radial component (radiative damping) \cite{Zonca96,Zonca98}. Another possibility one may consider is to use a precomputed damping rate $\gamma_{\rm d}(r,\omega)$ that depends on radius and frequency, for instance using a gyrokinetic solver for the plasma dispersion relation, such as {\tt LIGKA} \cite{Lauber07,Bierwage17b}. {\cred Moreover}, it has been shown that radial electric fields can also have an impact on the structure of Alfv\'{e}n continua in the frequency range of interest \cite{Koenies20} and, thus, may affect chirping dynamics.

\begin{figure}[tb]
  \centering
  \includegraphics[width=8cm,clip]{\figures/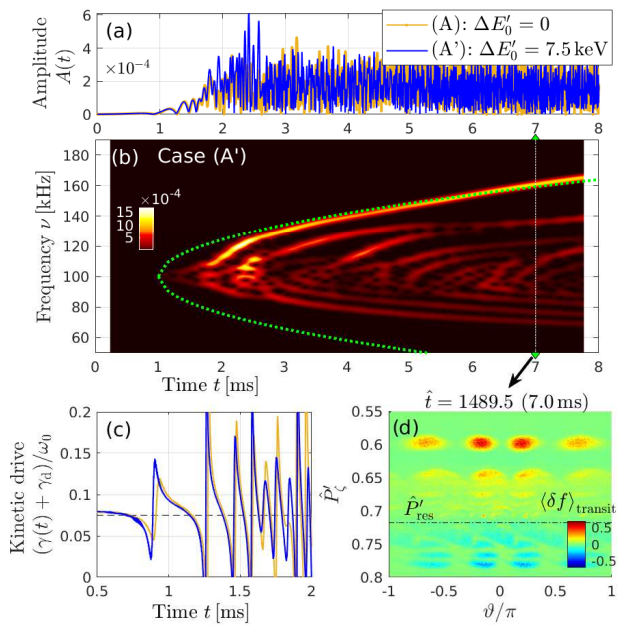}
  \caption{Evolution of case (A'), where particles are loaded in an energy band of width $\Delta E'_0 = 7.5\,{\rm keV}$ around $E'_0 = 75.6\,{\rm keV}$ as shown in Fig.~\protect\ref{fig:discuss_fire-A-pew_C-evol}. All other parameters are the same as in the marginally unstable case (A), where $\Delta E'_0 = 0$. Panels (a) and (c) show, respectively, the time traces of the amplitude $A(t)$ and growth rate $\gamma(t)$ plus damping rate $\gamma_{\rm d}/\omega_0 = 7.5\%$. The data of case (A) is also plotted for comparison. Panel (b) shows a long-time Fourier spectrogram obtained with large sliding time window $\Delta t_{\rm win} = 0.47\,{\rm ms}$, and the dotted green curves represent the prediction of the BB model (\protect\ref{eq:bb}). Panel (d) shows a snapshot of $\delta f(\hat{P}'_\zeta,\vartheta)$ taken at time $t = 7\,{\rm ms}$, using the modified radial coordinate $\hat{P}'_\zeta$ as shown in Fig.~\protect\ref{fig:model_slice}. The full 3-D distribution $\delta f(\hat{P}_\zeta,E',\vartheta)$ is shown in Fig.~\ref{fig:discuss_fire-A-pew_distr}.}
  \label{fig:discuss_fire-A-pew_overview}
\end{figure}

\begin{figure}[tb]
  \centering
  \includegraphics[width=8cm,clip]{\figures/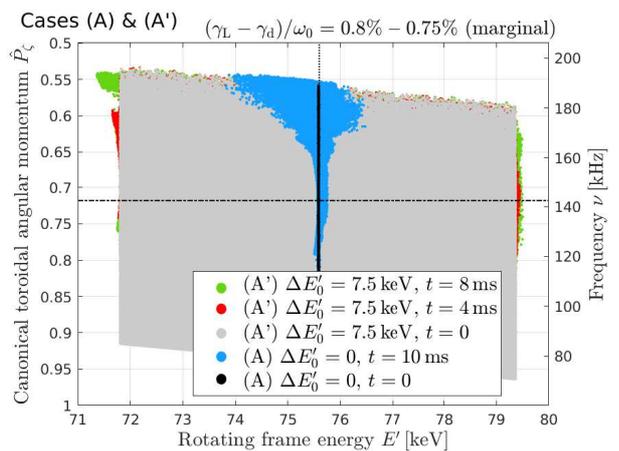}
  \caption{Comparison between the marker particle distributions in our default case (A), where $\Delta E'_0 = 0$, and the modified case (A'), where $\Delta E'_0 = 7.5\,{\rm kHz}$. The number of particles is $N_{\rm p} = 10^6$ in both cases.}
  \label{fig:discuss_fire-A-pew_C-evol}
\end{figure}

\begin{figure}[tb]
  \centering
  \includegraphics[width=8cm,clip]{\figures/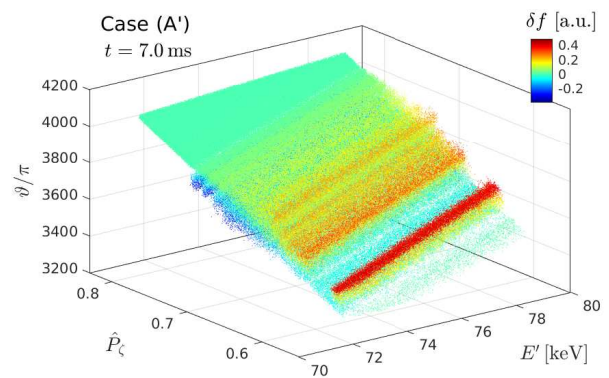}
  \caption{Snapshot of the 3-D particle distribution $\delta f(\hat{P}_\zeta,E',\vartheta)$ at $t = 7\,{\rm ms}$ in case (A'), where $\Delta E'_0 = 7.5\,{\rm kHz}$.}
  \label{fig:discuss_fire-A-pew_distr}
\end{figure}

\subsection{Case (A'): Nonzero loading width $\Delta E'_0$}
\label{apdx:model_pew}

The condition ${\rm d}G/{\rm d}t = 0$ underlying our implementation of the $\delta f$ method, namely Eq.~(\ref{eq:w}), does not strictly hold in our simulations, since particles are found to travel outside the region that was initially filled with phase space markers. This was shown in Figs.~\ref{fig:result_fire-AB_C-evol}, \ref{fig:result_fire-A_distr-E-P-th_t2-6ms} and \ref{fig:result_fire-B_distr-E-P-th_t07-14ms}. Nevertheless, {\cred our choice to load marker particles only along the line $E' = E'_0 = 75.588\,{\rm kHz} = {\rm const}$.\ in the present study is justified because the errors remain small and because we are able to reproduce the essential features of earlier simulations, where markers were loaded in an energy band of nonzero width $\Delta E'_0$ \cite{White19,White20}.

This is demonstrated here using the example in Fig.~\ref{fig:discuss_fire-A-pew_overview}, which} was obtained with the same parameters as our marginally unstable case (A), except that the initial distribution of marker particles covered a band of width $\Delta E'_0 = 7.5\,{\rm keV}$ around the $E' = 75.588\,{\rm kHz}$ line. This modified case is called (A'). Figure~\ref{fig:discuss_fire-A-pew_overview} shows that the early evolution of the field is nearly identical in cases (A) and (A'). Differences in the long-term evolution can be attributed to the high sensitivity of these dynamics with respect to physical and numerical parameters (e.g., see Figs.~\ref{fig:num_fire-A_qs-spec} and \ref{fig:num_fire-A_dt-spec} below).

Figure~\ref{fig:discuss_fire-A-pew_C-evol} shows that the broadened distribution of case (A) at $t=10\,{\rm ms}$ (light blue) lies well within the particle loading range of case (A') (gray). The 3-D view of $\delta f(\hat{P}_\zeta,E',\vartheta)$ in Fig.~\ref{fig:discuss_fire-A-pew_distr} shows the cylinder-like structure of the clumps and (albeit partially hidden) the holes. Taking advantage of this quasi-2-D structure, the simulations analyzed in this paper were performed for only a vanishingly thin slice of this distribution (cf.~Fig.~\ref{fig:model_slice}).

Figure~\ref{fig:discuss_fire-A-pew_C-evol} also shows that the small crescent-shaped distortion of the distribution that we observed in case (A) extends all the way to the edges of the $7.5\,{\rm keV}$ wide $\Delta E_0'$ band in case (A'). This supports our assertion made in Section~\ref{sec:result_transport}, that this crescent-shaped distortion breaking the $E' = E - \omega P_\zeta/n = {\rm const}$.\ condition may be a consequence of chirping; i.e., variations in the frequency $\omega$. Meanwhile, enhanced scattering of particles seen in Figs.~\ref{fig:result_fire-AB_C-evol}, \ref{fig:result_fire-A_distr-E-P-th_t2-6ms} and \ref{fig:result_fire-B_distr-E-P-th_t07-14ms}, especially in the domain of the clumps, is not visible in Figs~\ref{fig:discuss_fire-A-pew_C-evol} and \ref{fig:discuss_fire-A-pew_distr}. Such a diffuse spreading may be occurring inside the $\Delta E'_0 = 7.5\,{\rm keV}$ window, but we have not checked this yet. The absence of such diffuse spreading in case (A') would suggest that it is a (small) boundary artifact of the $\delta f$ method as discussed at the end of Section~\ref{sec:result_transport}.

Finally, returning once more to the overview in Fig.~\ref{fig:discuss_fire-A-pew_overview}, note how the up-chirping clump wave front is close to the $\sqrt{t}$ curve (green dots) of the BB model (\protect\ref{eq:bb}) for several milliseconds. We would like to emphasize that one should not interpret too much into this ``agreement'', which seems to be largely accidental, because we have repeatedly observed that a small change in the parameters can significantly alter the chirping pattern; for instance, giving rise to rapid chirps as in Fig.~\ref{fig:intro_fire-A_chirp} (another example will appear in Ref.~\cite{WhiteFEC21}). Disagreements between the BB model and our simulation results (here the downward chirps, but generally also the upward chirps) may be easily explained: too strong drive, too rapid chirping, nonuniform mode structure and other asymmetries, plasma geometry, large magnetic drifts. The point is that an (apparent) agreement in a single case can be misleading. Sensitivity checks are important and constitute a large portion of this Appendix.

\begin{figure}[tb]
  \centering
  \includegraphics[width=8cm,clip]{\figures/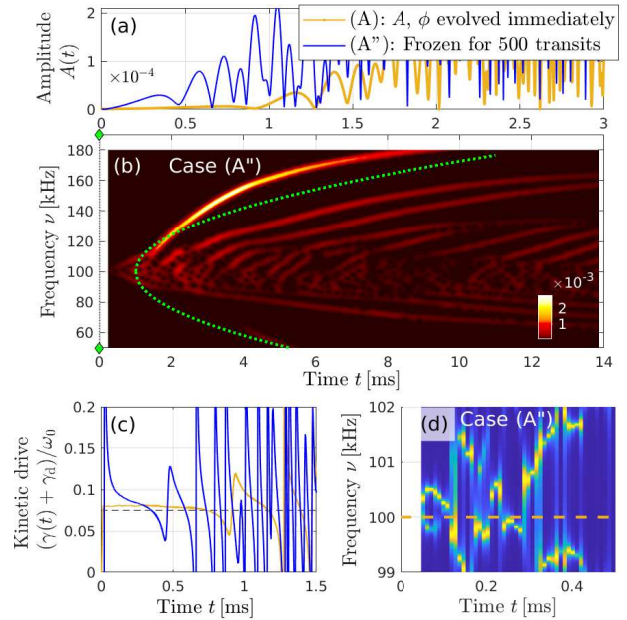}
  \caption{Evolution of case (A"), where the amplitude and phase velocity of field were fixed for 500 transits ($2.35\,{\rm ms}$) before launching the simulations with the same parameters as the marginally unstable case (A). The EP distribution at the beginning of the simulation was shown in Fig.~\protect\ref{fig:intro_fire-A-freeze500_poink-df}. Panels (a) and (c) show, respectively, the time traces of the amplitude $A(t)$ and growth rate $\gamma(t)$ plus damping rate $\gamma_{\rm d}/\omega_0 = 7.5\%$. The data of case (A) are also plotted for comparison. Panel (b) shows a long-time Fourier spectrogram obtained with sliding time window $\Delta t_{\rm win} = 0.47\,{\rm ms}$. For the first $0.5\,{\rm ms}$, panel (d) shows a high-resolution DMUSIC spectrogram obtained with $\Delta t_{\rm win} = 0.047\,{\rm ms}$.}
  \label{fig:sensitivity_fire-A_freeze500}
\end{figure}

\subsection{Case (A"): Relaxed initial EP distribution}
\label{apdx:model_freeze500}

Our standard initialization procedure was to use an unperturbed axisymmetric particle distribution and a field that contains a non-axisymmmetric perturbation in the form of a traveling wave with initial frequency $\nu_0 = 100\,{\rm kHz}$, poloidal mode number $m=6$ and toroidal mode number $n=5$. This means that, when a simulation is started with a destabilizing gradient, two things happen simultaneously:
\begin{itemize}
\item  The perturbation amplitude grows and, with that, the size of the resonantly perturbed phase space domain.
\item  The particle distribution adapts to the {\it growing} non-axisymmetric perturbation.
\end{itemize}

\noindent We were curious to see what happens if we let the particle distribution adapt before we let the instability grow, so we have launched a simulation from the state shown in Fig.~\ref{fig:intro_fire-A-freeze500_poink-df}.\footnote{Here, we have strong symmetry breaking {\cred in the relaxed initial EP distribution. A related problem, albeit with an unmodulated phase space density plateau perturbed only by noise,} has been considered in Ref.~\protect\cite{Lilley14}.}
That state was produced by fixing the field amplitude at a small value of $A = 10^{-6}$ and letting the EP distribution adapt to the perturbed field for 500 transits ($2.35\,{\rm ms}$). The subsequent simulation with evolving amplitude $A(t)$ and phase $\phi(t)$ used the same parameters as the marginally unstable case (A), and we call this modified case (A").

The results are shown in Fig.~\ref{fig:sensitivity_fire-A_freeze500}. In panel (a), one can see that the instability in case (A") is significantly enhanced compared to the default case (A). Exponential growth is not observed, as can be confirmed in panel (c). Instead, the amplitude grows almost in a straight line. The first pulse, which saturated at a tiny amplitude of $A \approx 6\times 10^{-6}$ in case (A), is effectively skipped. {\cred Instead}, case (A") saturates for the first time at the level $A \approx 3\times 10^{-5}$, similar to the peak value $3.5\times 10^{-5}$ of the second pulse in case (A). The high-resolution DMUSIC spectrogram in panel (d) of Fig.~\ref{fig:sensitivity_fire-A_freeze500} shows effectively immediate frequency splitting, as may be expected from the preestablished hole-clump pair seen in Fig.~\ref{fig:intro_fire-A-freeze500_poink-df}. The long-time Fourier spectrogram in panel (d) of Fig.~\ref{fig:sensitivity_fire-A_freeze500} shows long-range chirping dynamics {\cred similar to those observed} in case (A).

In summary, the main consequence of initializing a simulation of nonlinear frequency chirping with a relaxed non-axisymmetric EP distribution is that one effectively skips the exponential growth phase and the first pulse. Since the primordial (nonadiabatic) hole-clump wave pair has already been established in case (A"), the simulation goes straight into beating and chirping. The two initialization methods used in cases (A) and (A") --- axisymmetric vs. relaxed EP distribution --- are perhaps both unrealistic, and the real situation may lie somewhere in between. The choice of initialization method does not seem to play a significant role for most of the dynamics examined in this work, except for the processes of early structure formation and first saturation described in Section~\ref{sec:saturation}.

\subsection{Auto-optimization of resonant drive}
\label{apdx:model_opt}

\begin{figure}[tb]
  \centering
  \includegraphics[width=8cm,clip]{\figures/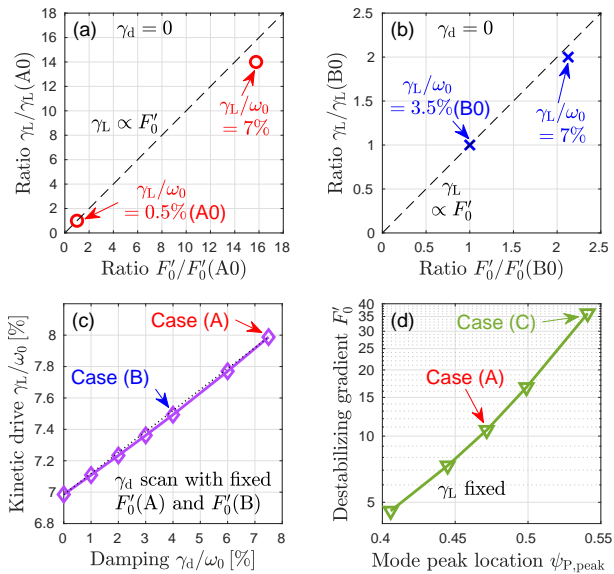}
  \caption{Dependence of the linear kinetic drive $\gamma_{\rm L}$ on (a,b) the radial density gradient $F_0' \equiv \partial_{P_\zeta}F_0$ and (c) the damping rate $\gamma_{\rm d}$. The results in panels (a) and (b) are obtained without damping ($\gamma_{\rm d} = 0$) and the axes are normalized by the values of $\gamma_{\rm L}$ and $F_0'$ in cases (A0) and (B0), respectively. The dashed lines indicate perfect linear scaling $\gamma_{\rm L} \propto F_0'$. Note that the $\gamma_{\rm d}$ scan in (c) is performed with fixed steep gradients $F_0'$, so the data points for $\gamma_{\rm d}=0$ do {\it not} correspond to cases (A0) and (B0), which have shallower gradients. Panel (d) shows how the destabilizing gradient $F_0'$ has to be increased exponentially in order to maintain the same linear drive $\gamma_{\rm L}/\omega_0 = 8\%$ when {\cred the mode's peak is shifted radially outward, away from the resonance} (cf.~Fig~\protect\ref{fig:model_fire-ABC_poink}(a)).}
  \label{fig:model_fire-AB_w-gr-damp}
\end{figure}

At first glance, Eqs.~(\ref{eq:evol}) and (\ref{eq:w}) suggest that the linear drive $\gamma_{\rm L} = {\rm d}\ln A/{\rm d}t + \gamma_{\rm d}$ should vary proportionally with the dominant destabilizing gradient, in our case $F_0' \equiv \partial_{P_\zeta}F_0$. Figure~\ref{fig:model_fire-AB_w-gr-damp}(a,b) confirms that this expectation is approximately satisfied in our simulations without damping. However, there is a small but noticeable deviation of about $10...13\%$. Furthermore, Fig.~\ref{fig:model_fire-AB_w-gr-damp}(c) shows that $\gamma_{\rm L}$ increases nearly linearly by about $15\%$ when the damping rate is raised to $\gamma_{\rm d}/\omega_0 = 7.5\%$, the value of case (A).

Another notable feature of our simulations is that the gradient $F_0'$ needed to obtain a certain value of $\gamma_{\rm L}$ depends on the location of the resonance relative to the modes peak. As one can see in Fig.~\ref{fig:model_fire-AB_w-gr-damp}(d), a 3.5 times steeper gradient is needed in case (C) in order to obtain the same value of $\gamma_{\rm L}/\omega_0 = 8\%$ as in case (A) after moving the peak of the mode from $\psi_{\rm P,peak} = 0.475$ to $0.54$ (cf.~Fig~\protect\ref{fig:model_fire-ABC_poink}(a)). Since its vertical axis has a logarithmic scale, Fig.~\ref{fig:model_fire-AB_w-gr-damp}(d) shows a nearly exponential dependence of $F_0'$ on $\psi_{\rm P,peak}$, which may be due to the Gaussian shape of the mode's radial profile.

Although we do not fully understand this behavior, we suspect that it is the result of a self-optimization process, where the particle-to-field energy transfer rate is maximized. In a self-consistent simulation as in Ref.~\cite{Bierwage17a}, both the spatial structure and the frequency of the MHD mode are optimized during the initial structure formation process. In the semi-perturbative model used here, the mode structure is held fixed, so only the frequency can self-optimize, depending on the local gradient $F_0'$, on the rate of damping $\gamma_{\rm d}$, and on the radial excursions of the GC drift orbits across the mode's profile.

\begin{figure}[tb]
  \centering
  \includegraphics[width=8cm,clip]{\figures/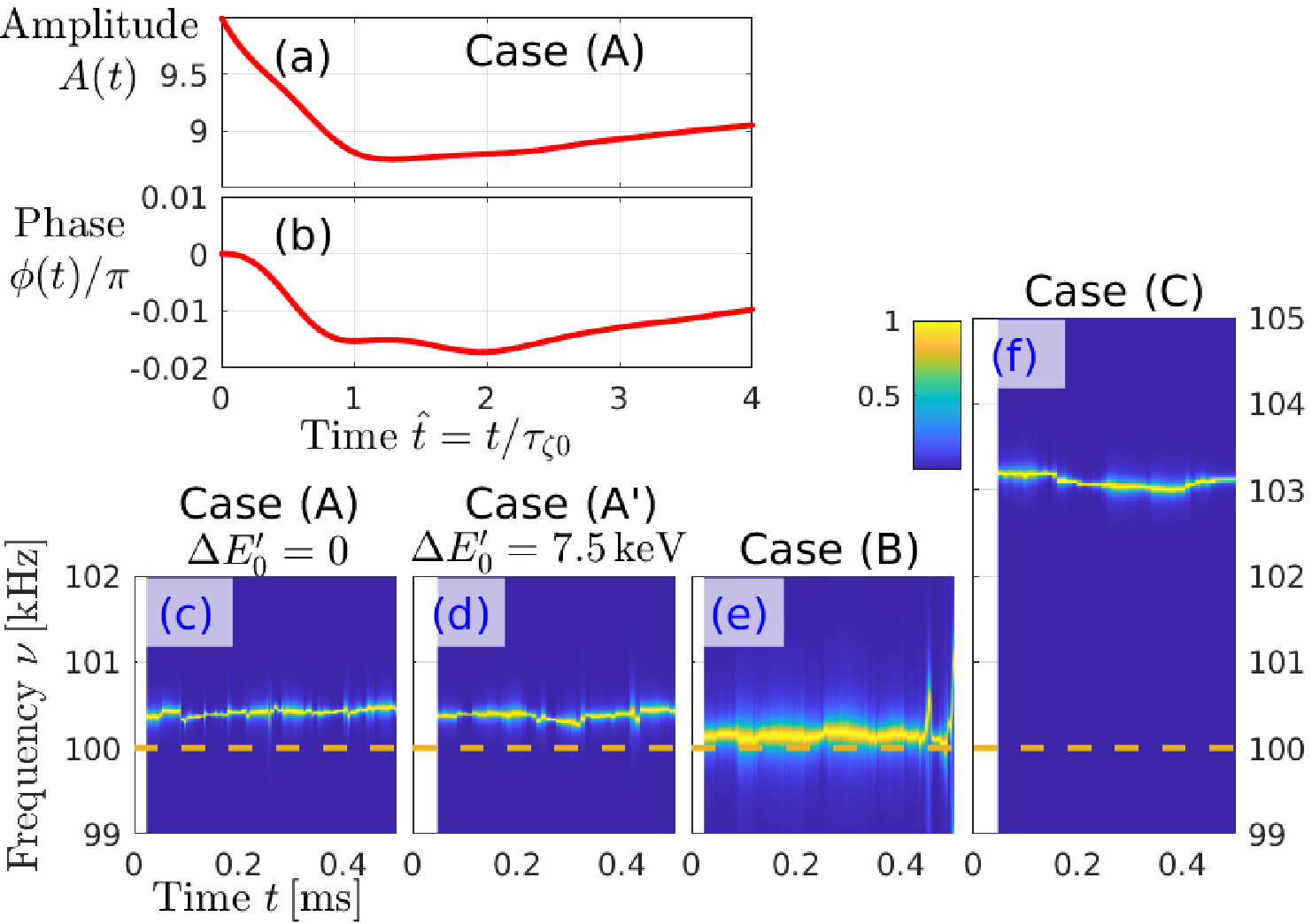}
  \includegraphics[width=8cm,clip]{\figures/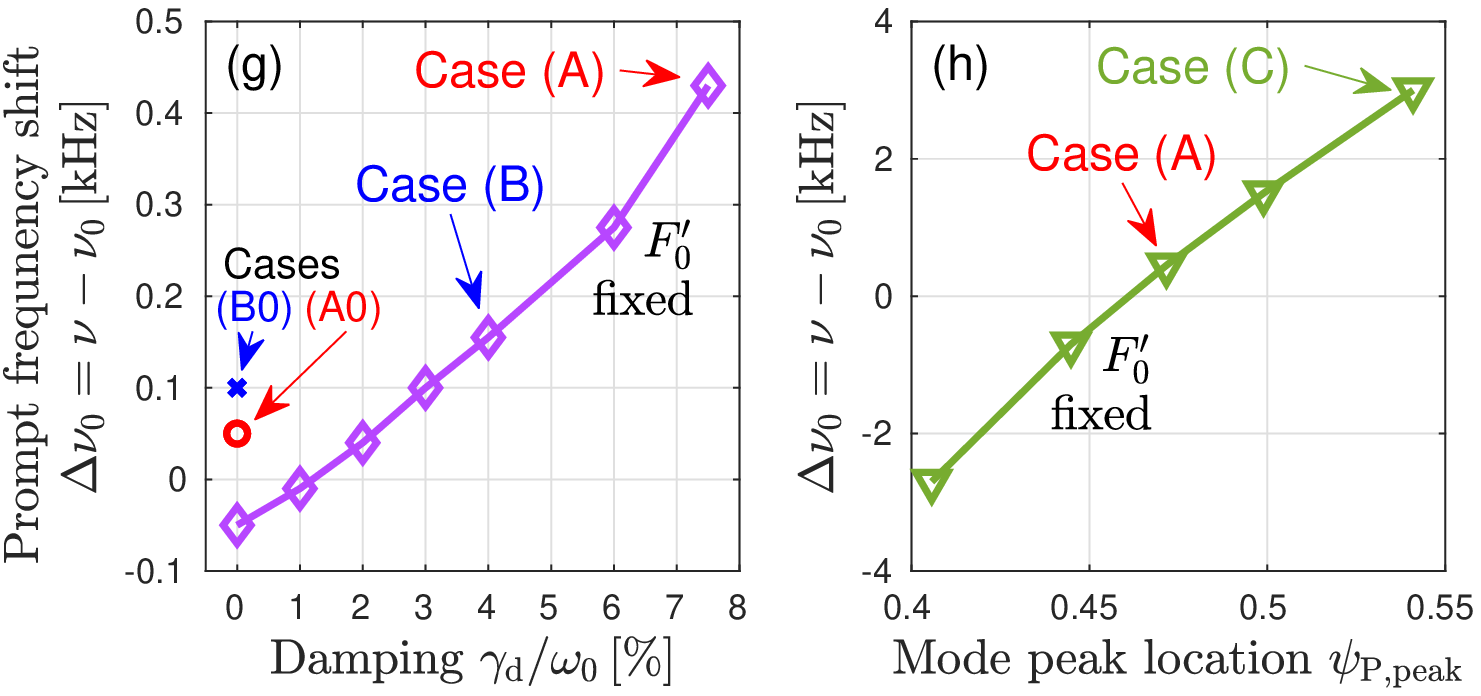}
  \caption{A prompt relaxation occurs during the first transit ($4.7\,\mu{\rm s}$), where the amplitude $A(t)$ and phase $\phi(t)$ drop transiently as shown in panels (a) and (b) for case (A). The relaxation is associated with a prompt frequency shift $\Delta\nu_0 = \nu-\nu_0$, which depends on the destabilizing gradient $F_0'$, the damping rate $\gamma_{\rm d}$ and the location of the resonance relative to the mode peak $\psi_{\rm P,peak}$ as shown in panels (g) and (h). The frequency shifts are measured using high-resolution DMUSIC spectrograms, some examples of which are shown in (c)--(f), where the initial frequency $\nu_0 = 100\,{\rm kHz}$ is indicated by a dashed horizontal line.}
  \label{fig:model_fire-ABC_prompt-freq-shift}
\end{figure}

This self-optimization hypothesis is supported by the observation of a prompt relaxation at the beginning of most simulations that we have performed. As an example, Fig.~\ref{fig:model_fire-ABC_prompt-freq-shift}(a,b) shows how the field amplitude $A(t)$ drops and the phase $\phi(t)$ adjusts transiently in the marginally unstable case (A). This relaxation lasts only about one transit time $\tau_{\zeta 0} \approx 4.7\,\mu{\rm s}$ and causes a prompt shift $\Delta\nu_0 = \nu - \nu_0$ in the frequency of the field oscillations, away from the prescribed value of $\nu_0 = 100\,{\rm kHz}$. The shifted frequencies can be seen in the high-resolution spectrograms plotted in panels (c)--(f) of Fig.~\ref{fig:model_fire-ABC_prompt-freq-shift}, which were obtained with the DMUSIC algorithm using a short time window $\Delta t_{\rm win} = 10\tau_{\zeta 0} = 47\,\mu{\rm s}$. Note that the frequency shift does no longer change after the instability has begun to grow exponentially. The small erratic fluctuations that can be seen in the spectrograms may be taken as a measure for the effect of statistical noise and an imperfect quiet start (see Appendix~\ref{apdx:model_qs}).

The measured frequency shifts in the marginally unstable case (A) and in the strongly unstable case (B) are
\begin{equation}
\Delta\nu_0 \approx \left\{\begin{array}{lcl}
+0.45\,{\rm kHz} &:& \text{Case (A)}, \\
+0.15\,{\rm kHz} &:& \text{Case (B)}.
\end{array}\right.
\end{equation}

\noindent As shown in Fig.~\ref{fig:model_fire-ABC_prompt-freq-shift}(g), the frequency shift is reduced to zero and becomes slightly negative when the damping rate $\gamma_{\rm d}$ is reduced to zero while keeping the values of the radial gradients $F_0'$ fixed. A reduction of the gradients to those in cases (A0) and (B0) increases the frequency shift to $\Delta\nu_0 \approx +0.05\,{\rm kHz}$ and $+0.1\,{\rm kHz}$, respectively, as indicated by the small symbols at $\gamma_{\rm d} = 0$ in Fig.~\ref{fig:model_fire-ABC_prompt-freq-shift}(g). Animations of the evolution of the phase space density perturbations $\delta f(\hat{P}_\zeta,\vartheta)$ in the undamped cases (A0) and (B0) exhibit a corresponding poloidal drift of the effective O- and X-points of the resonance in the frame rotating toroidally with the initial frequency $\omega_0/n$. The poloidal drift velocity is consistent with the values of $\Delta\nu_0$ obtained using DMUSIC, so we may rule out the possibility that this is merely a signal processing artifact. Although limited numerical accuracy does affect the value $\Delta\nu_0$, the phenomenon does not seem to be entirely numerical, at least for $|\Delta\nu_0| \gtrsim 0.1\,{\rm kHz}$.

As was noted at the beginning of this section, the location of the resonance relative to the mode peak matters. As one can see in Fig.~\ref{fig:model_fire-ABC_poink}(a) above, the resonance in our simulation setup is located in the region $\psi_{\rm P,res} \approx 0.13...0.33$, whereas the peak of the mode in cases (A) and (B) is located at $\psi_{\rm P,peak} = 0.475$. When we move the resonance further inward (away from the peak) or, equivalently, when we move the mode's peak outward (away from the resonance) while increasing $F_0'$ such that $\gamma_{\rm L}$ remains constant as in Fig.~\ref{fig:model_fire-AB_w-gr-damp}(d), we observe that the positive frequency shift increases significantly. This is demonstrated in Fig.~\ref{fig:model_fire-ABC_prompt-freq-shift}(h), where we scan $\psi_{\rm P,peak}$ up to the value of $0.54$ of case (C) and obtain a prompt frequency shift as large as $\Delta\nu \approx + 3\,{\rm kHz}$, much larger than in case (A), which has the same values of $\gamma_{\rm L}$ and $\gamma_{\rm d}$. 

The above-mentioned drift of the O- and X-points in the undamped cases suggests that the location of the resonance has shifted. In the cases studied here, we load particles only along a line in phase space, where $E'= E - \frac{\omega_0}{n}P_\zeta = {\rm const}$., and the numerical spread around this line is only about $\Delta E_{\cred \rm num} \lesssim 3\times 10^{-3}\,{\rm keV}$ (cf.~Fig.~\ref{fig:num_fire-A_dt-C}). This gives an upper bound on the possible resonance shift along $E'$. With Eq.~(\ref{eq:enr_rot}) written more conveniently as
\begin{equation}
E'{\rm [kHz]} \approx E{\rm [kHz]} - \frac{\nu{\rm [kHz]}}{n}\hat{P}_\zeta\times 0.85,
\end{equation}

\noindent and noting that the particle energy $E$ {\cred is effectively constant during the prompt relaxation phase due to the small field amplitude $A \lesssim 10^{-6}$,} we can estimate the radial shift of the resonance using the formula
\begin{equation}
\Delta\hat{P}_\zeta \approx \frac{n\times\Delta E_{\cred \rm num}{\rm [kHz]}}{0.85\times\nu_0{\rm [kHz]}} - \frac{\Delta\nu_0}{\nu_0}\hat{P}_{\zeta,{\rm res}}.
\label{eq:res-shift}
\end{equation}

\noindent For case (A) with $\Delta\nu_0 \approx 0.4\,{\rm kHz}$ and $\hat{P}_{\zeta,{\rm res}} = 0.719$, we obtain $\Delta\hat{P}_\zeta \approx 2\times 10^{-4} - 3\times 10^{-3} \approx -3\times 10^{-3}$. This is larger by a factor 3 than the observed shift $\Delta\hat{P}_\zeta \approx -10^{-3}$ to $\hat{P}_\zeta \approx 0.718$ of the resonant phase space structures in case (A). This means that the resonance does no longer lie in the region populated by marker particles and would imply that the field-particle interactions are occurring slightly off-resonant. More likely, however, is the possibility that our formula (\ref{eq:res-shift}) may not be sufficiently accurate. We suspect that the discrepancy may be due to the large magnetic drifts and the radial variation of the Gaussian mode structure in the domain occupied by the drift orbits (cf.~Fig.~\ref{fig:model_fire-ABC_poink}).

\begin{figure}[tb]
  \centering
  \includegraphics[width=8cm,clip]{\figures/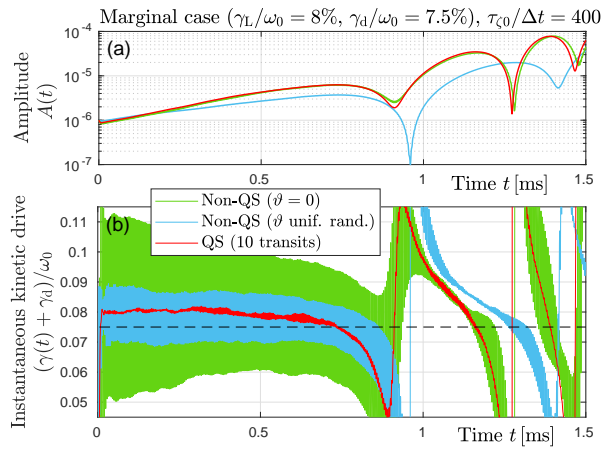}
  \caption{Noise reduction with quiet start (QS). Time traces of (a) the field amplitude $A(t)$ and (b) the {\cred instantaneous} kinetic drive $\gamma_{\rm k}(t) = \gamma(t) + \gamma_{\rm d}$ in the marginally unstable case (A) simulated without and with QS. Two non-QS simulations were performed: one with particles loaded only on the outer mid-plane ($\vartheta = 0$) and one with particles distributed uniformly {\cred randomized along $\vartheta$ (i.e., not consistent with the magnetic mirror force and drifts). The growth rates $\gamma(t)$ were measured using an exponential running average: $\gamma(t) = 0.01\times \frac{\Delta A(t)}{A(t)\Delta t} + 0.99\times\gamma(t-\Delta t)$}.}
  \label{fig:num_fire-A_qs-growth}
\end{figure}

Indeed, panels (c) and (d) in Fig.~\ref{fig:model_fire-ABC_prompt-freq-shift} show that the frequency shift seen in case (A) with $\Delta E'_0 = 0$ remains the same in case (A'), where we have loaded {\cred marker particles in} an energy band with a relatively large width $\Delta E'_0 = 7.5\,{\rm keV}$. In that case, the first term in Eq.~(\ref{eq:res-shift}) is approximately $0.4$, which would allow huge frequency shifts, but the system does not take advantage of his possibility. This insensitivity of $\Delta\nu_0$ with respect to $\Delta E'_0$ and the above-mentioned sensitivity with respect to the relative locations of the resonance and the mode's peak (Fig.~\ref{fig:model_fire-ABC_prompt-freq-shift}(f)) supports our hypothesis that the prompt relaxation is determined by the slope of the mode structure at the resonance (possibly in combination with drift effects \cite{Bierwage16a, Bierwage16b}). This is further corroborated by observations made in other case studies (not presented here) showing that (at least with sufficiently large $\gamma_{\rm d}$) the prompt frequency shift $\Delta\nu_0$ seems to be directed towards the peak of the mode when the resonance is located off-peak.

This {\cred subject} requires further investigation. In the meantime, we choose to speak of ``promptly shifted {\it effective} resonances'' in the present paper.

\begin{figure}[tb]
  \centering
  \includegraphics[width=8cm,clip]{\figures/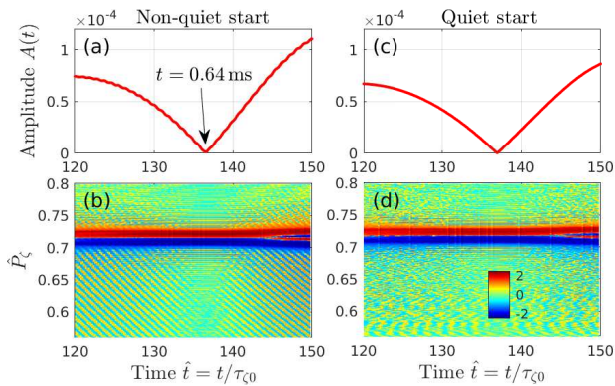}
  \caption{Noise reduction with quiet start (QS). Simulation of the case studied in Ref.~\protect\cite{White20}, without (left) and with quiet start (right). Panels (a) and (c) show time traces of the amplitude $A(t)$. The contour plots in (b) and (d) show $\left<\delta f - \left<\delta f\right>_{\rm transit}\right>_\vartheta(\hat{P}_\zeta,t)$ where $\left<\delta f\right>_{\rm transit}$ is $\delta f$ averaged over one transit time $\tau_{\zeta 0}$, and $\left<...\right>_\vartheta$ is an average over the poloidal angle. The quiet start smoothes the amplitude and eliminates the propagating micro-structures seen in (b) that were reported in Fig.~11 of Ref.~\protect\cite{White20}.}
  \label{fig:num_fire-A_micro_qs}
\end{figure}

\begin{figure}[tb]
  \centering
  \includegraphics[width=8cm,clip]{\figures/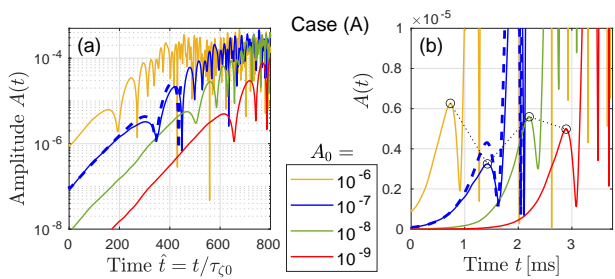}
  \caption{Sensitivity test with respect to {\cred the initial perturbation amplitude $A_0$. The diagrams show} the early evolution of the amplitude $A(t)$ in the marginally unstable case (A), starting from amplitudes in the range $A_0 = 10^{-6}...10^{-9}$. The log-scale panel (a) shows the nearly exponential growth. The circles in the linearly scaled panel (b) highlight the points of first saturation. The dashed blue curve shows the result for $A_0 = 10^{-7}$ with $N_{\rm p} = 10^7$ marker particles (otherwise $N_{\rm p} = 10^6$).}
  \label{fig:model_fire-A_a0-scan}
\end{figure}

\subsection{Quiet start and sensitivity}
\label{apdx:model_qs}

While loading particles only at the outer midplane ($\vartheta = 0$) suffices for kinetic Poincar\'{e} plots, this procedure introduces spurious oscillations in the mode evolution when applied in a simulation where Eq~(\ref{eq:evol}) is solved. Spreading the simulation particles uniformly along the poloidal angle $-\pi\leq\vartheta\leq\pi$ does not suffice either: due to the mirror force and the large magnetic drifts that can be seen in Fig.~\ref{fig:model_fire-ABC_poink}(b), the spurious oscillations remain noticeable throughout the simulation. Their effect is readily seen in the growth rate, which becomes extremely noisy as shown in Fig.~\ref{fig:num_fire-A_qs-growth}. Long-term effects can be observed in the EP distribution in the form of spurious radially propagating micro-structures as shown in Fig.~\ref{fig:num_fire-A_micro_qs}(b).

Although the evolution of the field remains more or less the same, such spurious oscillations are undesirable in detailed analyses of phase space dynamics as were performed in this work. Perhaps the most accurate way to deal with this problem is to load particles uniformly in time along precomputed unperturbed GC orbits \cite{Bierwage12a}. For the present study, the {\tt ORBIT} code has been extended with a routine that approximates this method in a simpler way as described in Section~\ref{sec:model_qs}. As one can see in Figs.~\ref{fig:num_fire-A_qs-growth} and \ref{fig:num_fire-A_micro_qs}(d), the growth rates become much smoother and the spurious micro-structures in EP phase space disappear.

Our quiet start procedure is not perfect and still generates {\cred a certain amount of ``statistical noise'' (including unphysical modulations)}. We believe that this is one reason for the results in Fig.~\ref{fig:model_fire-A_a0-scan}, where the amplitude at the time of first saturation in the marginally unstable case (A) changes in a somewhat irregular manner when we reduce the initial perturbation amplitude $A_0$ from $10^{-6}$ to $10^{-9}$. This case is very sensitive to statistical noise because its proximity to marginal stability causes the first saturation to occur at an extremely small amplitude, which corresponds to a narrow resonant structure in EP phase space (see Figs.~\ref{fig:intro_fire-A-freeze500_poink-df} and \ref{fig:stages_fire-A}).

\begin{figure}[tb]
  \centering
  \includegraphics[width=8cm,clip]{\figures/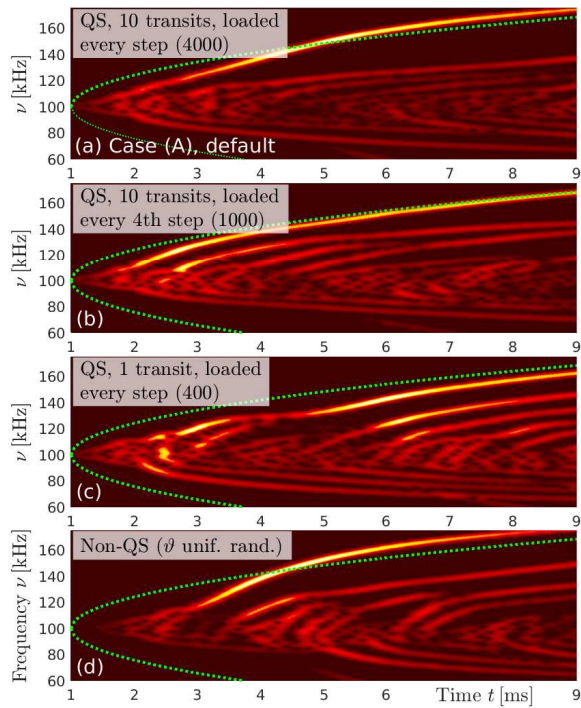}
  \caption{Sensitivity test with respect to quiet start (QS) parameters. Fourier spectrograms are shown for four simulations performed with the parameters of the marginally unstable case (A). Panels (a)--(c) show results of different realizations of the QS, and panel (d) shows the result with particles uniformly randomized along $\vartheta$ (non-QS). The size of the sliding FFT time window was $\Delta t_{\rm win} = 0.47\,{\rm ms}$. The green dotted parabola represents the BB model (\protect\ref{eq:bb}) and is shown here only as a reference {\cred for orientation and easier comparison of patterns} in different simulations.}
  \label{fig:num_fire-A_qs-spec}
\end{figure}

Statistical noise also affects the chirping patterns as can be seen in Fig.~\ref{fig:num_fire-A_qs-spec}, where we show results of case (A) simulated with different realizations of the quiet start (a)--(c) and with uniform particle loading (d). Our default procedure (a) was to load a fraction of the particles at every time step during the first 10 transits ($10\tau_{\zeta 0} = 47\,\mu{\rm s}$). With $\Delta t/\tau_{\zeta 0} = 1/400$, this makes 4000 loading steps. In (b), we have loaded particles only every 4th step. In (c), we loaded particles at each step, but only for 1 transit {\cred time} instead of 10. The onset of chirping is very similar in all cases, but after about $2\,{\rm ms}$, the patterns vary substantially, although the overall extent of the chirps remains similar.

\begin{figure}[tb]
  \centering
  \includegraphics[width=8cm,clip]{\figures/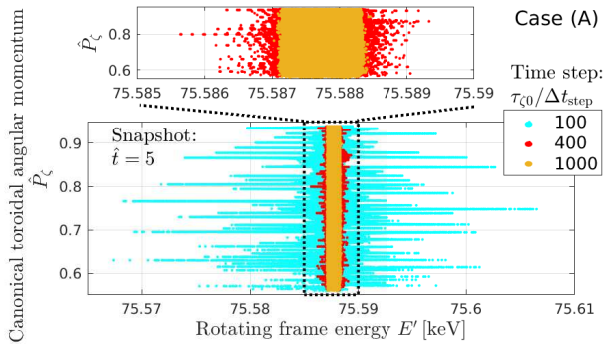}
  \caption{Numerical convergence with respect to time step $\Delta t_{\rm step}$. The particle distribution in the $(\hat{P}_\zeta,E')$-plane obtained after 5 transits is shown for simulations of case (A) with $\tau_{\zeta 0}/\Delta t_{\rm step} = 100$, $400$, $1000$ steps per transit. The upper diagram shows an enlarged view of the latter two cases.}
  \label{fig:num_fire-A_dt-C}
\end{figure}

\begin{figure}[tb]
  \centering
  \includegraphics[width=7.9cm,clip]{\figures/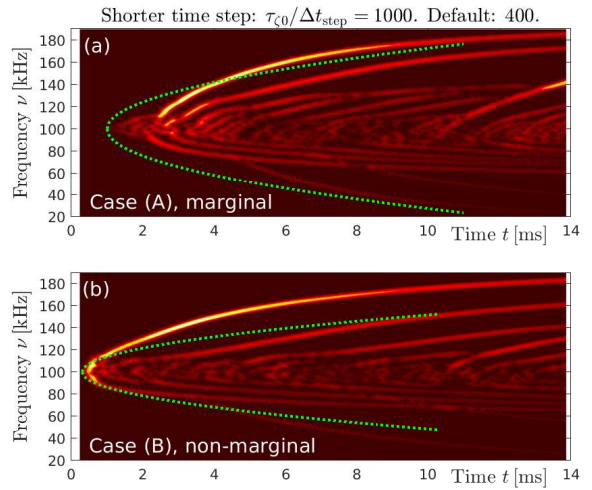}
  \caption{Numerical convergence with respect to time step $\Delta t_{\rm step}$. Fourier spectrograms are shown for cases (A) and (B) obtained in simulations with $\tau_{\zeta 0}/\Delta t_{\rm step} = 1000$ steps per transit. The results are similar to those in Fig.~\protect\ref{fig:result_fire-AB_tracers-spec} obtained with $400$ steps per transit. The green dotted parabola represents the BB model (\protect\ref{eq:bb}) and is shown here only as a reference for easier comparison between different cases.}
  \label{fig:num_fire-A_dt-spec}
\end{figure}

\subsection{Numerical convergence}
\label{apdx:model_convergence}

The {\tt ORBIT} code uses a 4th-order Runge-Kutta algorithm to push particles and evolve the fields, so there is a numerical error associated with the time step $\Delta t_{\rm step}$ that causes the energy conservation condition to be violated by a small amount. Figure~\ref{fig:num_fire-A_dt-C} shows snapshots of the particle distribution in the $(\hat{P}_\zeta,E')$-plane taken at a very early stage of a simulation of case (A) with quiet start, after only $\hat{t} = 5$ transit times $\tau_{\zeta 0}$. Results are shown for simulations performed with $\tau_{\zeta 0}/\Delta t_{\rm step} = 100$, $400$ and $1000$ steps per transit, where $400$ is the default value used in this paper. One can see that the spread in $E'$ is $\delta E'/E'_0 \approx 4\times 10^{-4}$ for $\tau_{\zeta 0}/\Delta t_{\rm step} = 100$, which decreases by an order of magnitude to $\delta E'/E'_0 \approx 4\times 10^{-5}$ for $\tau_{\zeta 0}/\Delta t_{\rm step} = 400$, and is nearly invisible on the scales shown for $\tau_{\zeta 0}/\Delta t_{\rm step} = 1000$.

Note that the spread occurs only along the $E'$ axis of Fig.~\ref{fig:num_fire-A_dt-C}. It does not affect $\hat{P}_\zeta$, so it is entirely limited to a (small) variation in the particle energy $E$. Moreover, the energy spread appears during the first few transit times of a simulation and remains more or less at the same level thereafter. This can be seen in the lower part $\hat{P} > 0.65$ of Fig.~\ref{fig:result_fire-AB_C-evol}(a). (The broadening of the $E'$ distribution in the upper part $\hat{P} \lesssim 0.65$ of Fig.~\ref{fig:result_fire-AB_C-evol}(a) has other reasons discussed in Section~\ref{sec:result_transport} and Appendix~\ref{apdx:model_pew}.) In other words, the energy conservation error is bounded and the bound is set by the size of the time step $\Delta t_{\rm step}$.

Figure~\ref{fig:num_fire-A_dt-spec} shows Fourier spectrograms of the long-time evolution in cases (A) and (B) simulated with $\tau_{\zeta 0}/\Delta t_{\rm step} = 1000$ steps per transit instead of the default $400$, spectrograms for which were shown in Fig.~\ref{fig:result_fire-AB_tracers-spec}. One can see that the results are very similar; especially, in the strongly driven case (B), which is less sensitive to statistical noise than the marginally unstable case (A). The rapid downward chirps in the lower part of both spectrograms in Fig.~\ref{fig:num_fire-A_dt-spec} are the spurious ``ghosts'' discussed in Section~\protect\ref{apdx:misc_ghost}.

\begin{figure}[tb]
  \centering
  \includegraphics[width=8cm,clip]{\figures/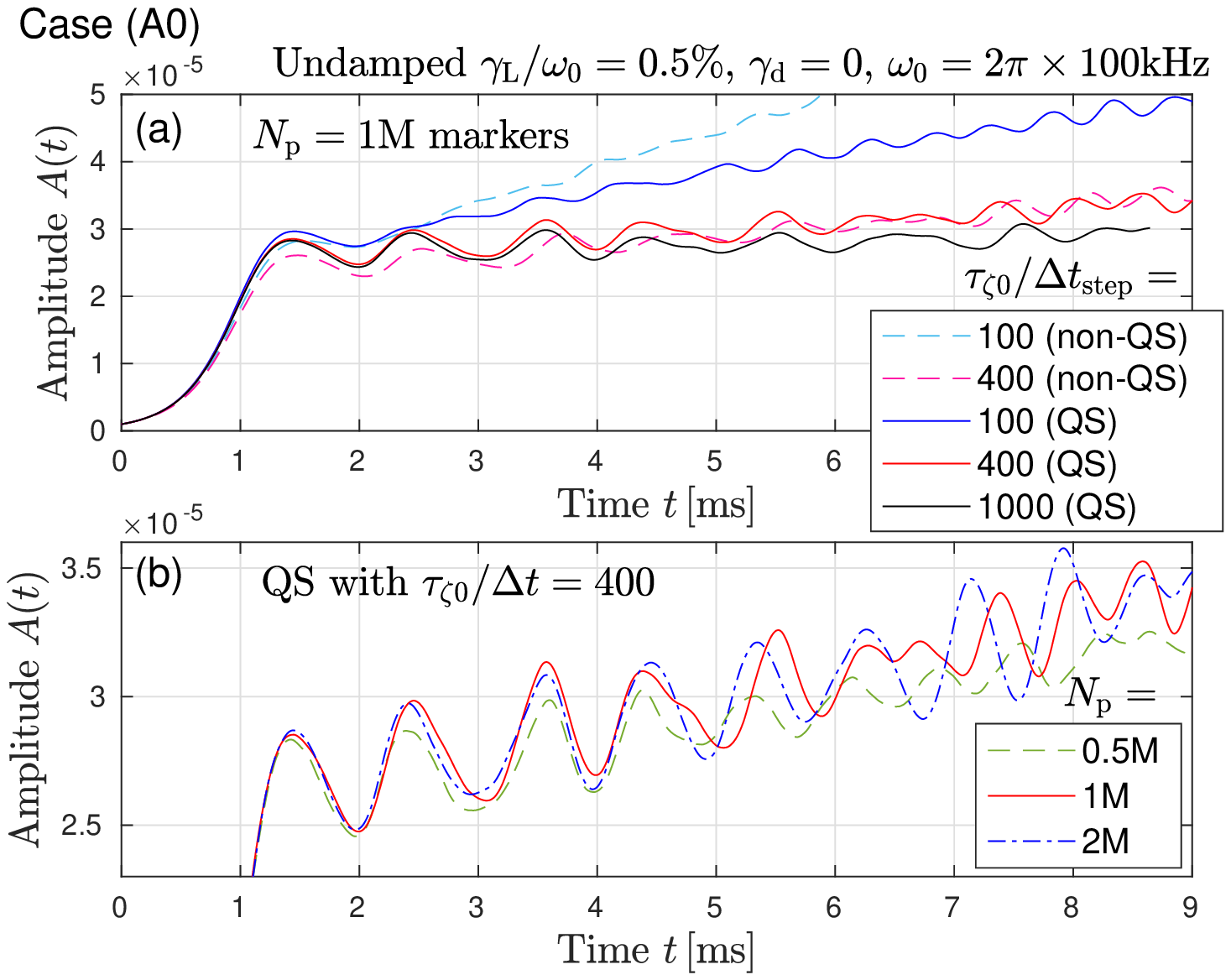}
  \includegraphics[width=8cm,clip]{\figures/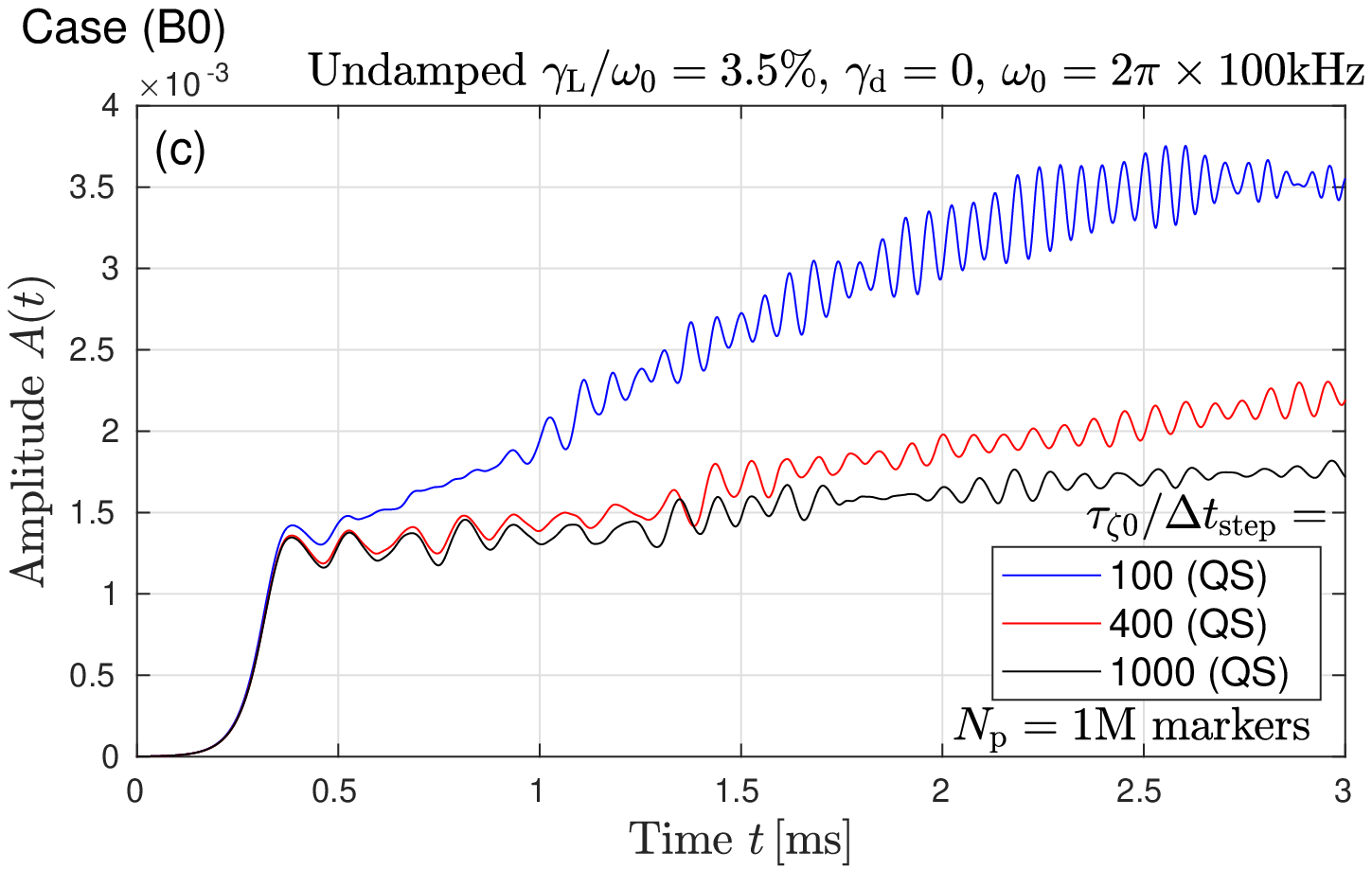}
  \caption{Numerical convergence test with respect to time step $\Delta t_{\rm step}$ and particle number $N_{\rm p}$. The diagrams show time traces of the field amplitude $A(t)$ in simulations without damping ($\gamma_{\rm d} = 0$). Panels (a) and (b) show results for the weakly driven case (A0) simulated with different time steps $\Delta t_{\rm step}$ and different numbers of marker particles $N_{\rm p}$, respectively. Panel (a) also shows results obtained with and without quiet start (QS). Panel (c) show results for the strongly driven case (B0) simulated with $N_{\rm p} = 10^6$ markers and different time steps $\Delta t_{\rm step}$.}
  \label{fig:num_fire-AB0_dt_qs}
\end{figure}

A higher (or more clearly visible) sensitivity with respect to the time step $\Delta t_{\rm step}$ is found in simulations of undamped instabilities ($\gamma_{\rm d} = 0$), where a quasi-steady state is expected to form. Figure~\ref{fig:num_fire-AB0_dt_qs}(a) for the weakly driven case (A0) and Fig.~\ref{fig:num_fire-AB0_dt_qs}(c) for the strongly driven case (B0) show that simulations with too large time steps fail to reproduce the saturation of the kinetic instabilities. In fact, this is a well-known problem and a standard method to test codes and choose suitable time steps. In addition, Fig.~\ref{fig:num_fire-AB0_dt_qs}(a) shows that unphysical fluctuations in simulations without quiet start can enhance the saturation problem.

In case (A0), taking $\tau_{\zeta 0}/\Delta t_{\rm step} = 1000$ steps per transit yields good saturation as one can see in Fig.~\ref{fig:num_fire-AB0_dt_qs}(a). Even $400$ steps per transit yield an acceptable result, with a very similar oscillation pattern. Figure~\ref{fig:num_fire-AB0_dt_qs}(b) shows that the number of simulation particles can influence the oscillation pattern but seems to have no significant influence on the saturation (or lack thereof).

In case (B0), simulations with $\tau_{\zeta 0}/\Delta t_{\rm step} = 1000$ steps per transit still suffer from spurious growth as one can see in Fig.~\ref{fig:num_fire-AB0_dt_qs}(c). However, the results for the first millisecond or so, which we have analyzed in Section~\ref{sec:saturation_damp} and Appendix~\ref{apdx:undamped}, seem to be acceptable.

\begin{figure}[tp]
  \centering
  \includegraphics[width=8cm,clip]{\figures/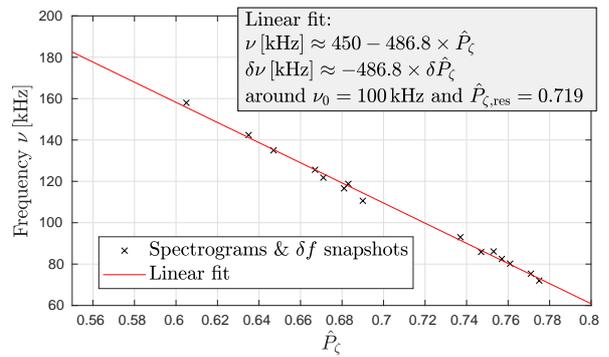}
  \caption{Calibration: Relation between the oscillation frequency $\nu$ and the radial location $\hat{P}_\zeta$ in EP phase space. Formulas based on a linear fit are given for the absolute value $\nu(\hat{P}_\zeta)$ and for the distance $\delta\nu(\delta\hat{P}_\zeta)$ from the seed resonance.}
  \label{fig:stages_fire_freq-pzeta}
\end{figure}\label{key}

\subsection{Calibration}
\label{apdx:model_calibration}

\paragraph{\it Ideal MHD displacement.} The magnitude of the ideal MHD displacement $\delta r = {\bm \xi}\cdot\nablab r \approx \xi^\Psi/\Psi' = \xi^\Psi/(q\Psi_{\rm P}')$ {\cred with $\Psi_{\rm P}' \equiv \partial_r\Psi_{\rm P}$} of the (nonresonant) bulk plasma can be estimated as follows. Using the relation $\psi_{\rm P} = \Psi_{\rm P}/\Psi_{\rm P,edge} \approx (r/a)^2$, making explicit the normalization of $\hat{\Psi}_{\rm P,edge} = \Psi_{\rm P,edge}/B_0$, and recalling that $\hat{\xi}^\Psi = |\xi^\Psi|/A$ is normalized to unity (so that $A$ has units of length squared), we obtain
\begin{align}
\frac{\delta r_{\rm mhd}}{R_0} &= \frac{A \hat{\xi}^\Psi}{R_0 q \Psi_{\rm P}'/B_0} \approx \frac{A \hat{\xi}^\Psi}{q \Psi_{\rm P,edge}/B_0 } \times \frac{a/R_0}{2 r_0/a} \nonumber
\\
&\lesssim \frac{10^{-3}}{1.15\times 0.276} \times \frac{1/3.7}{1.3} = 7\times 10^{-4}
\label{eq:dr_mhd}
\end{align}

\noindent for $A \lesssim 10^{-3}$. Meanwhile, the kinetic Poincar\'{e} plot in Fig.~\ref{fig:model_fire-ABC_poink}(b) shows an island with a half-width of about $\delta\psi_{\rm P} \sim \pm 0.03$ for $A = 10^{-3}$. This means that the resonant particles experience displacements 
\begin{align}
\frac{\delta r_{\rm res}}{R_0} \approx \frac{\delta\psi_{\rm P}}{2 r_0/a \times R_0/a} \approx \frac{0.03}{1.3\times 3.7} \approx 6\times 10^{-3} \hspace{-0.1cm}
\label{eq:dr_res}
\end{align}

\noindent that are one order of magnitude larger than $\delta r_{\rm mhd}/R_0$ for the nonresonant bulk.

\begin{figure*}[tb]
  \centering
  \includegraphics[width=16cm,clip]{\figures/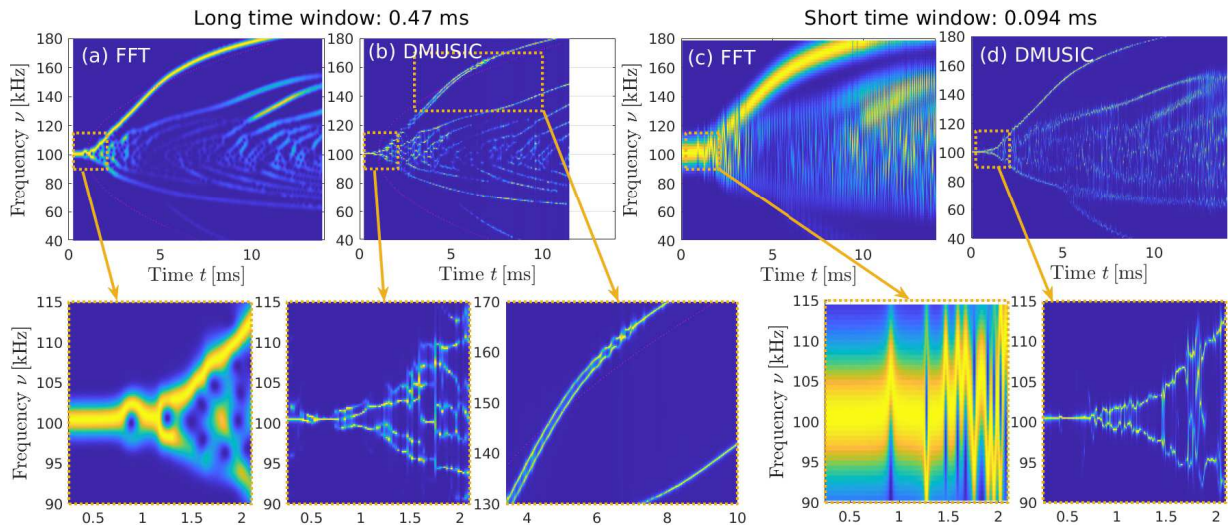}
  \caption{Comparison between FFT and DMUSIC spectrograms for case (A). Window sizes $\Delta t_{\rm win} = 0.47\,{\rm ms}$ (left) and $0.094\,{\rm ms}$ (right) were used. For better comparability, the FFT spectra were normalized to their maximal value at each time, so intensity information is lost (as in DMUSIC). The panels at the bottom show {\cred enlarged} portions of the full spectrograms in panels (a)--(d).}
  \label{fig:dmusic_fire-A_compare-fft}
\end{figure*}

\paragraph{\it Profile of EP reference state $F_0$.} The slope of the reference EP density profile $F_0(P_\zeta)$ is not known precisely, since our marker loading method gives only an approximately uniform density gradient and, thus, we do not know the exact form of $F_0/G_0 \approx {\rm const}$. The calibration of the EP profile was performed using Fig.~\ref{fig:intro_fire-A-freeze500_poink-df}(c), which shows the first snapshot of case (A') described in Appendix~\ref{apdx:model_freeze500}, where we have allowed the EP distribution to adapt to the field perturbation with fixed amplitude $A = A_0 = 10^{-6}$ and phase $\phi=0$. The O-point profile of such a relaxed distribution $\bar{F}(P_\zeta) = F_0(P_\zeta) + \delta\bar{f}(P_\zeta)$ is expected to be flat and we have calibrated the reference profile $F_0(P_\zeta)$ accordingly. The resulting profile $F_0(P_\zeta)$ is shown as a yellow shaded triangle in Fig.~\ref{fig:intro_fire-A-freeze500_poink-df}(c). The calibration was also confirmed independently using the undamped cases (A0) and (B0), where the perturbed profile $\bar{F}(P_\zeta)$ is also expected to become flat across the O-points (see Figs.~\ref{fig:result_fire-A-A0_saturation} and \ref{fig:result_fire-B-B0_saturation}).

\begin{figure}[H]
  \centering
  \includegraphics[width=8cm,clip]{\figures/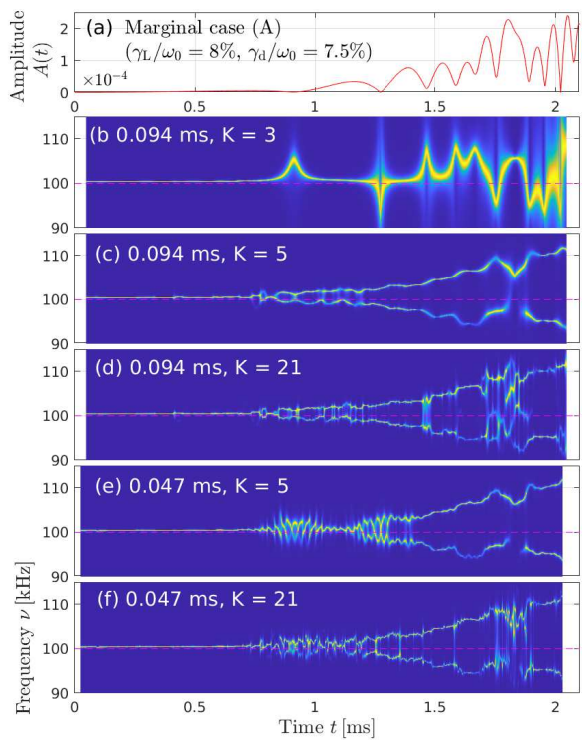}
  \caption{Characterization of DMUSIC spectrograms. Using the signal $s(t)$ from case (A) shown in panel (a), we show the form of the spectrogram during the onset of chirping for short time windows $\Delta t_{\rm win} = 0.096\,{\rm ms}$ or $0.047\,{\rm ms}$, and for different numbers $K = 3$, $5$ and $21$ of complex sinusoids used in the decomposition.}
  \label{fig:dmusic_fire-A_K-twin-scan}
\end{figure}

\paragraph{\it Pump frequency.} The relation between the radial location $\hat{P}_\zeta$ of a phase space structure and the corresponding oscillation frequency $\nu$ in a spectrogram of the field signal $s(t)$ is shown in Fig.~\ref{fig:stages_fire_freq-pzeta}. The data points in Fig.~\ref{fig:stages_fire_freq-pzeta} were roughly taken at the approximate centers of large hole and clump vortices in the advanced stages of the simulation, which is why there are no data points near the resonance $\hat{P}_{\zeta,{\rm res}} \approx 0.72$. A linear fit gives
\begin{equation}
\delta\hat{P}_\zeta \approx \delta\nu / 486.8\,{\rm kHz}.
\label{eq:pz_nu}
\end{equation}

\section{Characterization of diagnostics}
\label{apdx:dg}

\subsection{Fourier spectra}
\label{apdx:dg_fft}

The Fourier analysis was performed using a Hann-weighted sliding time window. The default window size was $\Delta t_{\rm win} = 100\,\tau_{\zeta 0} \approx 0.47\,{\rm ms}$ and the spectrum has been smoothed by extending the array length to $T_{\rm pad} = 16\times\Delta t_{\rm win}$ and padding with zeros.
\begin{equation}
Y(t,\omega) = \int_{-T_{\rm pad}/2}^{T_{\rm pad}/2}{\rm d}\tau\, y(t - \tau) H(\tau) e^{-i\omega\tau}.
\end{equation}

\noindent The Hann window function $H(\tau)$ eliminates artifacts from aperiodicity.

\begin{figure*}[tp]
  \centering
  \includegraphics[width=8cm,clip]{\figures/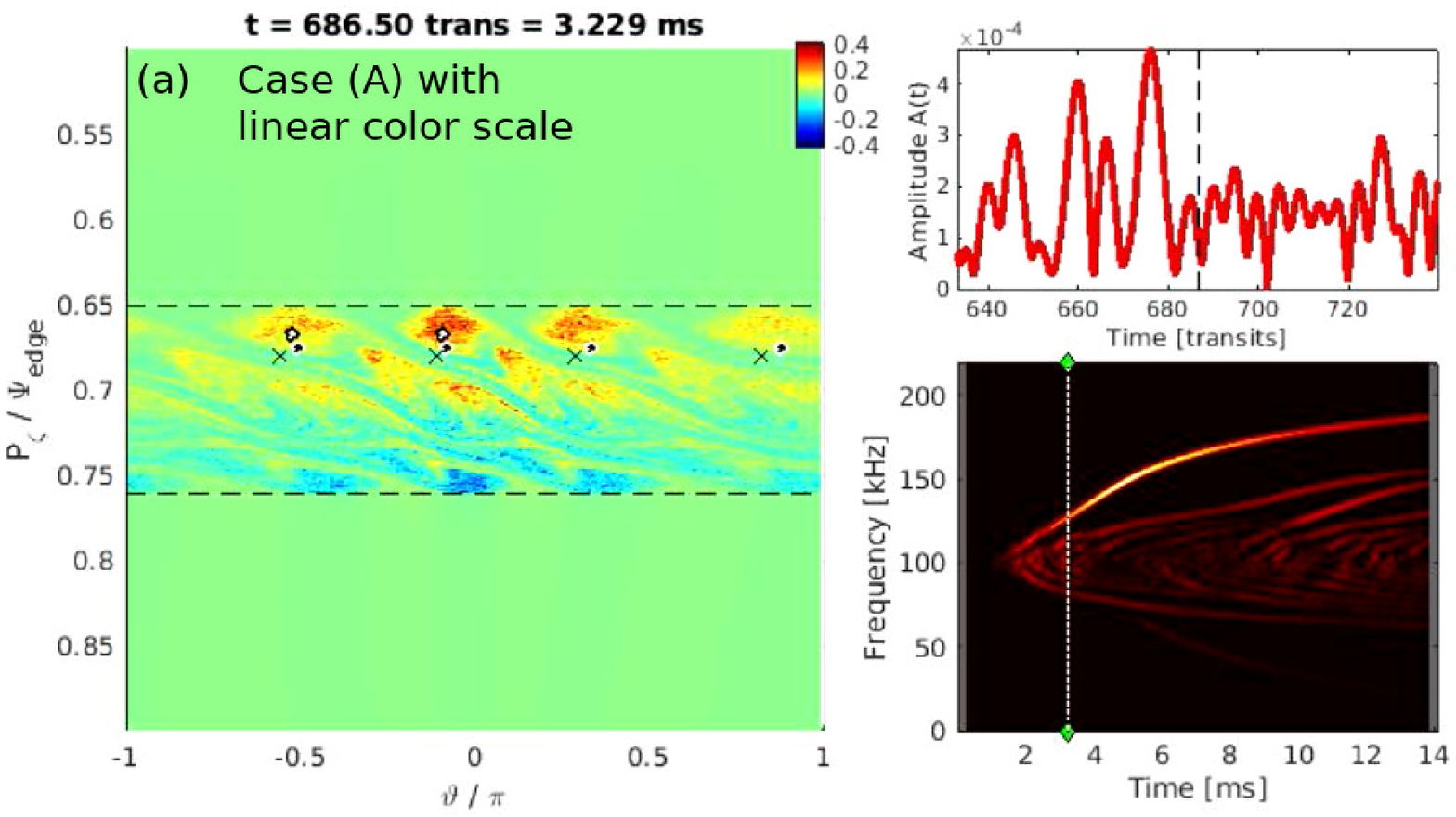}
  \includegraphics[width=8cm,clip]{\figures/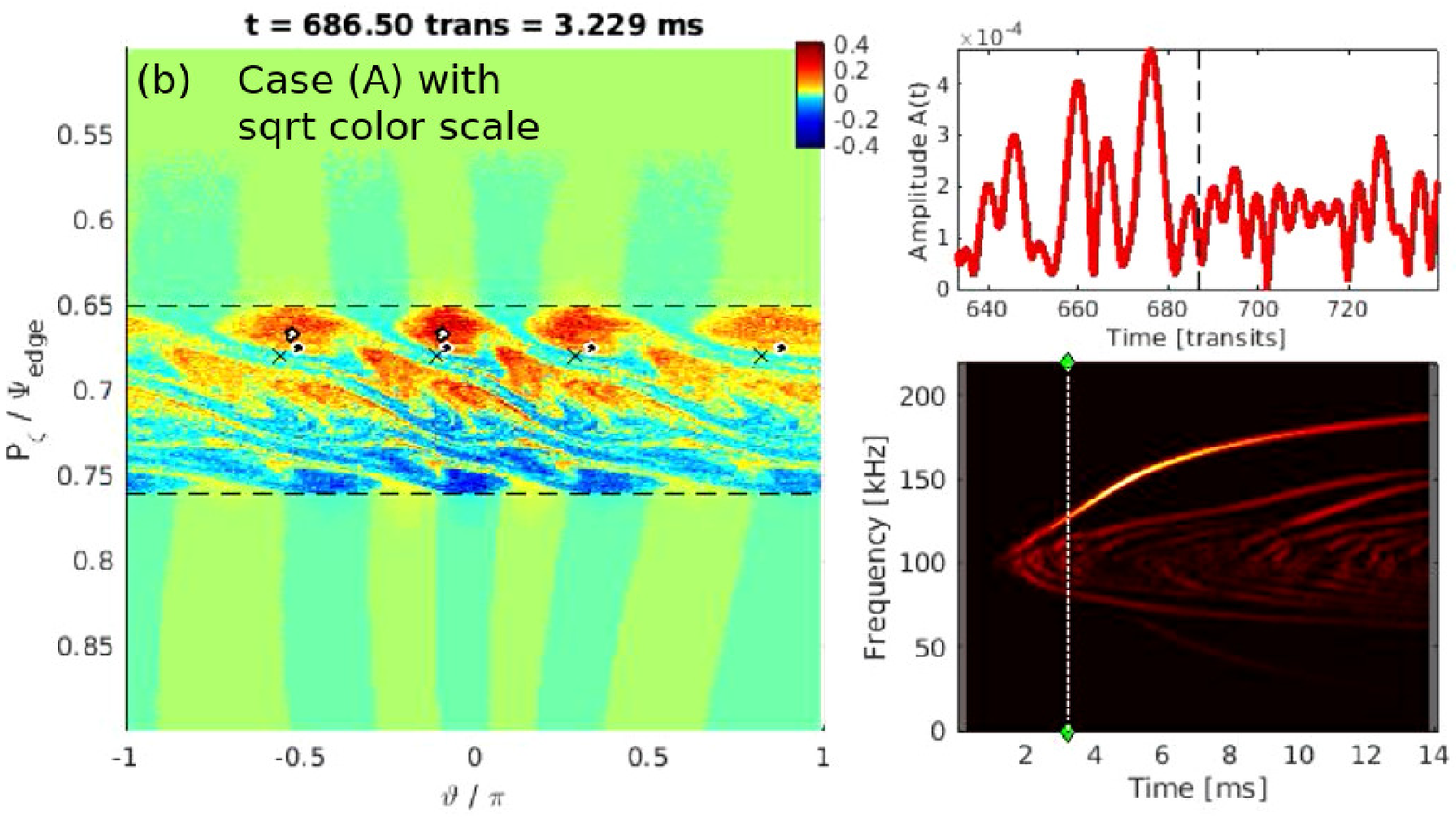} \\
  \includegraphics[width=8cm,clip]{\figures/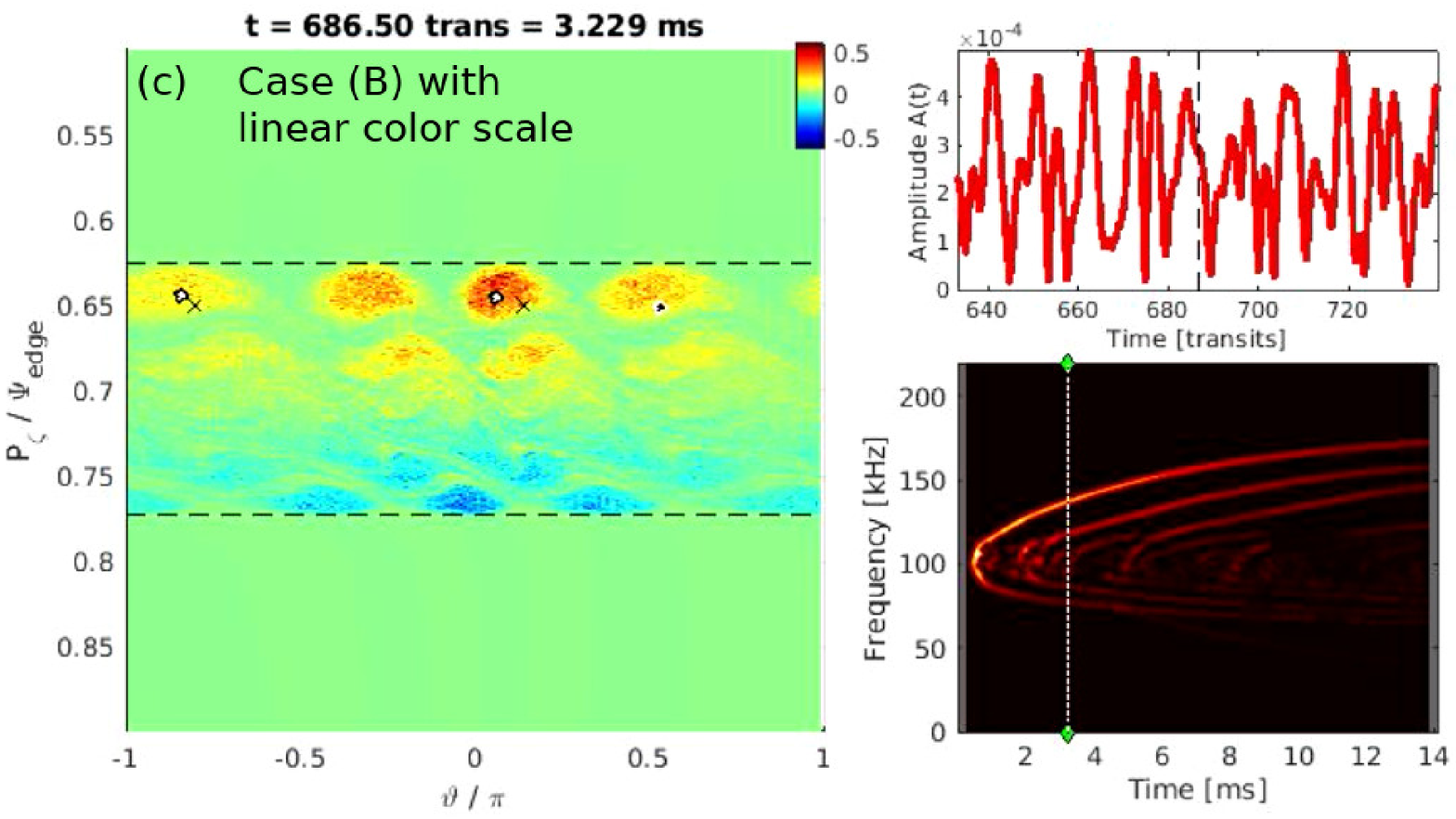}
  \includegraphics[width=8cm,clip]{\figures/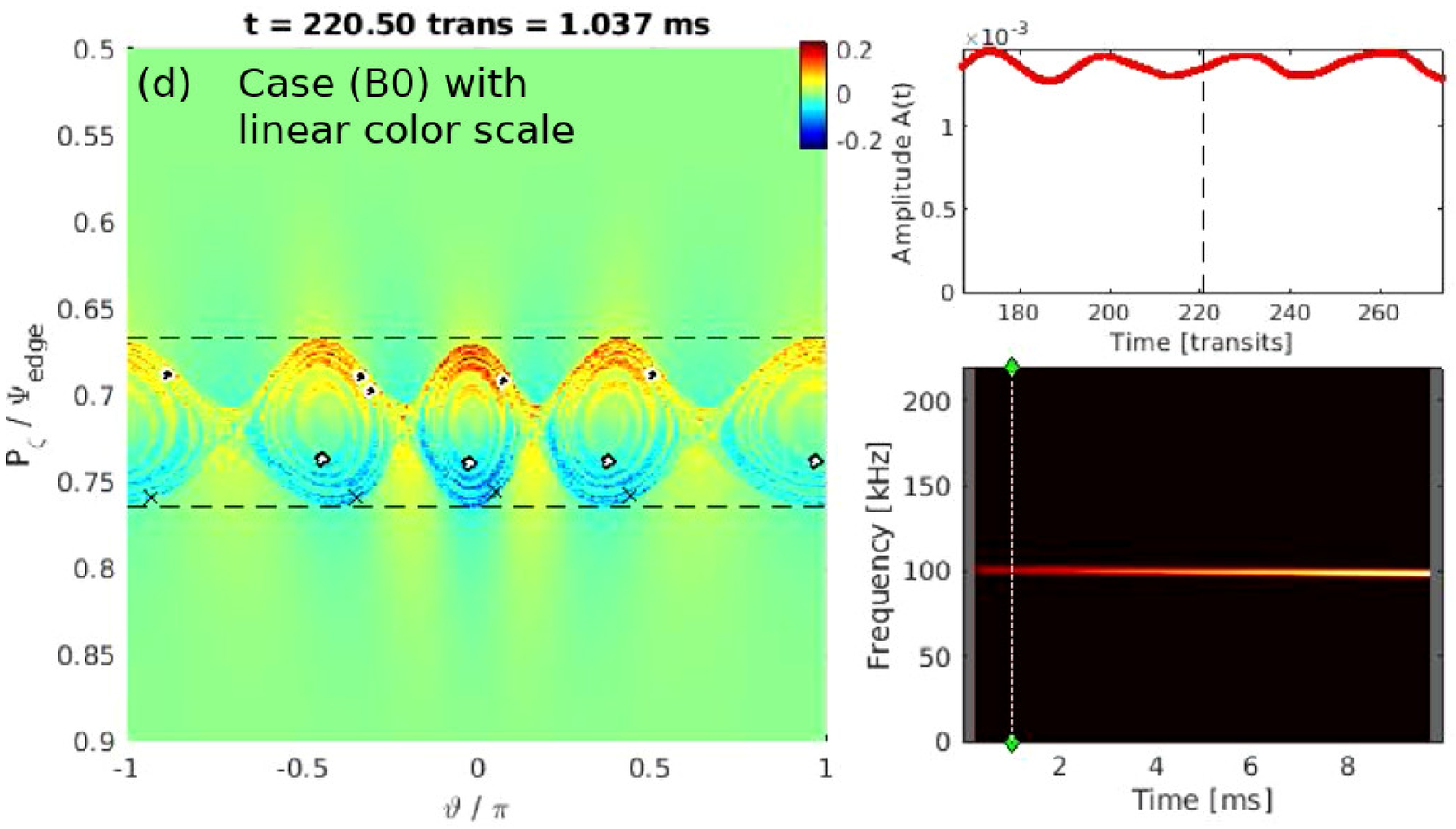}
  \caption{Click on panels (a)--(d) to view animated movies for cases (A), (B) and (B0). Playing the animations directly in the PDF file may require Adobe Acrobat Reader and QuickTime to be installed. On some systems, a separate media player may be launched.}
  \label{fig:movies}
\end{figure*}

\subsection{DMUSIC spectra}
\label{apdx:dg_dmusic}

DMUSIC is a spectral peak-finding algorithm originally developed for nuclear magnetoresonance (NMR) signal analysis by Li {\it et al}.\ \cite{Li98}. Our application of the method to fusion plasmas was inspired by Slaby {\it et al}.\ \cite{Slaby20}, who used it to visualize continuous spectra of shear Alfv\'{e}n waves in initial value simulations.

The strength of the algorithm lies in its applicability to noisy signals with phases of rapid growth or strong damping. Its weaknesses are that {\cred signal} intensity information is effectively lost and {\cred that} it can be difficult to attribute physical meaning to all the features appearing in a DMUSIC spectrogram. The spectral analyses {\cred in this study were performed with sliding time windows with sizes in the range $\Delta t_{\rm win} = (10...300)\tau_{\zeta 0} = (0.047...1.410)\,{\rm ms}$.} For each value of the real frequency $\nu$, we scanned the imaginary frequency axis $\gamma$ and selected the maximal value of the transform. At each time, the peak value of the spectrogram was normalized to unity.

Figure~\ref{fig:dmusic_fire-A_compare-fft} shows a comparison between FFT spectrograms and DMUSIC spectrograms obtained with long and short time windows. With the long time window $\Delta t_{\rm win} = 0.47\,{\rm ms}$, the spectra have a similar structure, although the DMUSIC spectra are much sharper and seem to reveal more detail. It is not clear how much of that detail bears physical meaning. For instance, the splitting of spectral lines seen in the upper part of Fig.~\ref{fig:dmusic_fire-A_compare-fft}(b) (enlarged in the third panel in the bottom row) remains to be {\cred understood} as discussed in Appendix~\ref{apdx:misc_split}. We use DMUSIC primarily for the purpose of spectral analysis with high temporal resolution (d), where FFT spectra become unintelligible (c).

Our default parameters for DMUSIC are as follows. We take a portion of $N = \Delta t_{\rm win}/\Delta t_{\rm sample}$ data points from a uniformly sampled signal, where $\Delta t_{\rm sample}$ is the sampling step. We choose $N$ to be even, and the parameter $J$ of Ref.~\cite{Li98} is chosen to be $J = N/2$ as recommended by the authors. Typically, we use $K=21$ complex sinusoids for the decomposition. Typically, a real frequency window covering the range $40\,{\rm kHz} \leq \nu \leq 180\,{\rm kHz}$ is sampled by at least $N_\nu = 201$ points. The window size for the imaginary component of the frequency is usually $-4\,{\rm kHz} \leq \gamma \leq 4\,{\rm kHz}$ and sampled by $N_\gamma = 31$ points (test runs with $N_\gamma = 121$ gave essentially identical results).

The role of the parameters $K$ and $\Delta t_{\rm win}$ can be seen in Fig.~\ref{fig:dmusic_fire-A_K-twin-scan}, which shows spectrograms of the onset of chirping during the first $2\,{\rm ms}$ of case (A). The spectrograms are computed with short time windows $\Delta t_{\rm win} = 0.094\,{\rm ms}$ and $0.047\,{\rm ms}$. When using a small number of only $K = 3$ complex sinusoids (b), the spectrogram effectively shows the instantaneous oscillation frequency. Phase jumps by $\pm\pi$ that occur around the amplitude minima and have the form of an arctan function (see Eq.~(\ref{eq:beat_pump_phase})) appear here in panel (b) as rapid up- and downward chirps. Frequency splitting becomes visible with $K = 5$ in panel (c), and becomes only a little richer in detail with $K = 21$ in (d). With a smaller time window as in panel (e) and (f), the phase jumps start to strongly affect the signals even for $K = 5$ and $21$. Further reduction of the window size would yield a result similar to that in panel (b) since the number of harmonics that can be captured decreases, so choosing a large value of $K$ has no effect for short time windows.

\subsection{Animations {\footnotesize\it [sorry, not included in arXiv.org version]}}
\label{apdx:dg_movie}

Figure~\ref{fig:movies} contains links to animations for cases (A), (B) and (B0). Each movie {\cred window} consists of three panels: one showing a $\delta f$-weighted Poincar\'{e} plot of the phase space structures (left), one for the field amplitude $A(t)$ (top right), and a Fourier spectrogrm (bottom right).

\section*{References}
\addcontentsline{toc}{section}{References}

\bibliographystyle{unsrt}
\bibliography{references}

\end{document}